\documentclass[a4paper,openright,twoside,cleardoublepage=plain, 12pt]{scrbook} 
\usepackage[utf8]{inputenc}
\usepackage{textcomp} 
\usepackage{amsmath}
\usepackage{float}
\usepackage{multicol}
\usepackage{amssymb}
\usepackage{float}
\usepackage{indentfirst}
\everymath{\displaystyle}
\usepackage{graphicx}
\usepackage{dcolumn}
\usepackage{makeidx}
\usepackage{geometry}
\usepackage{ae}
\usepackage{cite}
\usepackage{setspace}
\usepackage{subfigure}
\usepackage[dvipsnames]{xcolor}
\usepackage{appendix}
\usepackage{chngcntr}
\usepackage{etoolbox}
\usepackage[noabbrev]{cleveref}
\usepackage[normalem]{ulem} 

 \usepackage{booktabs}
 \usepackage{multirow}

\usepackage[T1]{fontenc}

\usepackage{msor_thesis} 

\newcommand*{\ORIGchapterheadendvskip}{}%
\let\ORIGchapterheadendvskip=\chapterheadendvskip
\renewcommand*{\chapterheadendvskip}{%
	{%
		\setlength{\parskip}{12pt}%
		\noindent\hspace{-.5\marginparwidth}\rule[.3\baselineskip]{6.15\marginparwidth}{1pt}\par
		\vspace{\baselineskip}%
	}%
	\ORIGchapterheadendvskip
}

\newcommand{\PR}[1]{\ensuremath{\left[#1\right]}}
\newcommand{\PC}[1]{\ensuremath{\left(#1\right)}}
\newcommand{\chav}[1]{\ensuremath{\left\{#1\right\}}}

\newcommand{\dbar}{\mathrm{d}\hspace*{-0.15em}\bar{}\hspace*{0.2em}}
\newcommand{\dif}{\mathrm{d}\hspace*{0.1em}} 
\def\d{{\mathrm{d}}}
\def\bigint{\hbox{\Large $\displaystyle\int$}}
\newcommand{\gsim}{\raisebox{-0.13cm}{~\shortstack{$>$ \\[-0.07cm]
      $\sim$}}~}
\def\D{{\mathcal{D}}}

\definecolor{violet}{rgb}{0.58, 0.0, 0.83}

\definecolor{azulzinho}{RGB}{242, 242, 242}

\colorlet{LightRubineRed}{RubineRed!70!}
\colorlet{Mycolor1}{green!10!orange!90!}
\definecolor{Mycolor2}{HTML}{00F9DE}



\usepackage[english]{babel}

\begin{document}

\frontmatter
\title{\Large{On the Connections between Thermodynamics and General Relativity}} 
\author{Jessica Santiago} 
\date{\today} 

\subject{Mathematics}
	\abstract{ 
	}

\ack{ 
} 

\phd
\maketitle

%

\tableofcontents 

\listoffigures 

\mainmatter


\newpage
\chapter{Introduction}

\vspace{-0.3cm}
We live in a curved space-time. You can call it gravity, you can call it curvature.
Whatever you choose to call it, you cannot escape it.
Gravity cannot be shielded. It is part of space-time itself, it is the metric.
Anything, any matter, any particle with mass or momentum will feel it. 
Gravity behaves very differently from the other forces because it is not a force. It is a property of the universe. 


On the other hand, we have matter. There are particles and quantum fields.
These particles and fields interact with each other, they agglomerate. They form structures and macroscopic systems.  
They exist and, as long as they exist in groups, it is possible to describe them thermodynamically.
You might have to be very careful doing so, looking for thermodynamic potentials varying in both space and time. You may find difficulties to do it consistently.
Nevertheless, distributions of particles will have statistical behaviours, from which at least some notion of a thermodynamic description can be derived.

\enlargethispage{15pt}
Given the universality of both situations, one looking at matter and its statistical character; the other facing the space-time wherein such particles move, it is absolutely natural to ask what happens when they meet (of course they have always met, but that does not mean we know how to describe that). This is the topic which we will explore in this thesis. 

As we are aware, both Thermodynamics and General Relativity were developed a long time ago, each having enough said about them that we could create specialized libraries for both.
However, since we are mainly interested in the connections between the two, let us start by describing a little fraction of their history from when they finally met. 

\paragraph{Relativistic Thermodynamics}

The first point we would like to address are the significant discussions and efforts made to formulate a \emph{special relativistic} version of thermodynamics, named simply relativistic thermodynamics. 
The main authors contributing to this quest were Max Planck --- who wrote several articles applying special relativity to basically all other theories known at the time, thermodynamics being one of them --- and Einstein himself. 
Some of their questions, however, are today understood to make very little sense.

Since special relativity brought about the knowledge that Galilean transformations were empirically useful only in the low speed approximation, the physics community became very excited with Lorentz transforming everything they could. Caught in this loop, a lot of effort ended up being put into finding the correct Lorentz transformations for temperatures and answering questions like: ``Does a moving body appears colder or warmer?''. 
There were numerous intense debates in the subject between Einstein, Planck, Ott, Landsberg and others \cite{Planck1, Planck2, Einstein1, Ott, Landsberg} when the answer was simply: this question doesn't quite make sense. Not when presented with these words anyway.
The first one to  clearly explain the reasons why this question is pointless was J.L. Anderson \cite{Anderson:1967} by clarifying how the concept of a ``moving temperature'' (the temperature of a moving frame) is ill defined. Unfortunately, even with Anderson's explanation, part of the scientific community still haven't understood and insist in believing that ``how to Lorentz transform a moving body'' is still an open question \cite{Boo, Pavani}.  In section \ref{S: How to measure temperatures} we will come back to this subject and discuss temperature measurements. We will define a way to ``measure'' temperature from a distance, concluding, however, that the outcome is observer dependent and gives us no meaningful information. 

\paragraph{Covariant Thermodynamics} Another very important stepping stone following the formulation of general relativity, was the ground breaking works published by Richard Chase Tolman. As we will see during this text, Tolman wrote several critically important articles \cite{tolman:1930,ehrenfest:1930,tolman:1928,tolman:1933a,tolman:1933b, weightofheat} and a book \cite{tolmanbook} dedicated to exploring how gravity changes thermodynamics and reformulating all thermodynamic laws in a covariant fashion. He was a true, dedicated pioneer to the cause and alongside the works of Israel, Stewart and others \cite{Israel:1980, Israel:1981, Israel:1986}, covariant thermodynamics is well established and used all over cosmology.
One of the major contributions of Tolman was to derive for the first time the concept of gravity-induced temperature gradients, which will be fully explored and extended in this thesis.



\paragraph{Black Hole Thermodynamics} Both relativistic and covariant thermodynamics are theories which aim to describe how the thermodynamic description of systems is affected by special and general relativity. There is, on the other hand, another way to connect both subjects, i.e., can thermodynamics tell us how gravitational systems have to behave? This side of the story had its beginning with the thoughts and proposal of Bekenstein. In 1972 he made the claim that, given the second law of thermodynamics, black holes must have entropy. His conclusion was purely based on the deep belief that the second law must hold in the whole universe and by noticing that black holes, as classically seen at the time, could work as a sink for entropy. 
For example, if you decide to throw a house or a roller coaster inside a black hole, all the information about what it was made of will vanish. 
The only information you can ``recover" from the outside is the mass, charge and angular momentum of what fell. Where did the entropy contained in those objects go? To fix this, Bekenstein proposed that black holes must have entropy themselves and his suggestion was given by 
\begin{equation*}
S_B \propto \frac{A}{L_P^2},
\end{equation*}
where $A$ is the area of the black hole and $L_P$ the Planck length. 

Then, in 1975, Stephen Hawking showed that black holes are indeed not quite as black. Adopting a semiclassical description, with quantum fields propagating through a classic background spacetime, he deduced that black holes do emit particles and, for Schwarzschild static black holes for example, this emission has exactly the spectrum of a black body with a temperature given by:
\begin{equation*}
T_H = \frac{\hbar\, c^3}{8\pi G M k_B}\,.
\end{equation*}
It then became possible to associate a temperature with black holes, to find the proportionality factor in Bekenstein's entropy proposal, 
\begin{equation*}
S_B = \frac{1}{4}\frac{A}{L_P^2},
\end{equation*}
and to formulate what we today call Black Hole Thermodynamics, which allow us to describe black holes as thermal machines and even extract work from them (theoretically, of course). 
 
Following such exciting events, W. G. Unruh proceeded to show what I believe to be the most unexpected and reality changing of these effects: that the concept of particle is observer dependent. Conducting calculations closely similar to those used by Hawking, Unruh adopted two observers, an inertial and a Rindler observer (accelerated from the infinite past until the infinite future) and showed that where the inertial observer sees vacuum, the Rindler observer sees a thermal bath of particles with a temperature proportional to its acceleration:
\begin{equation*}
T_U = \frac{\hbar \, a}{2\pi c\, k_B}.
\end{equation*} 
Particle creation by expanding universes were also deduced and a connection still not absolutely clear between space-time and thermodynamics could be seen through the fog. As a result, several researchers then decided to join the quest of revealing the link between both theories, each approaching the question from a different angle. We will now mention only a couple.

\paragraph{Statistical description of Gravitational Systems}
One of the researchers was Thanu Padmanabhan, who adopted a quite distinct and classical approach.
It seems important to mention the work of Padmanabhan in this area, especially given that when trying to develop a thermodynamical description of space-time and event horizons, we need to know how to do it for classical matter interacting gravitationally.
This can then be useful not only for its own purposes, but can also warn us of the possible changes that the long-range character of gravity might create.

In a couple of papers \cite{Pad1, Pad2}, Padmanabhan reproduces what he named ``statistical mechanics of gravitating systems''. 
In his approach no outstanding surrealistic inputs are made. 
The problem consists simply of statistically describing  matter 
(dust, classical particles,...) 
interacting gravitationally. Gravity is seen simply as a long range force (including, in some cases, the results for a cosmological background). 
It is indeed a statistical mechanical approach and the difficulties arise first -- due to the fact that we cannot bound a gravitational system and, second -- due to the non-extensivity of energy for gravitating systems.

Padmanabhan defines a Hamiltonian for the system and finds its behaviour using both the micro-canonical and canonical ensembles. He shows that in the limits of very low and high energies the results coincide for both ensembles. However, in the mean energy values, where the gravitational force is the main interaction between the particles, the micro-canonical ensemble gives a negative heat capacity while the canonical ensemble cannot deal with it, showing a phase-transition.

%

\paragraph{Thermodynamics of Spacetime}


In his famous paper \cite{Jacobson}, Ted Jacobson made a bold proposition about the thermodynamical behaviour of space-time itself. He states that if not only black holes, but any bifurcate Rindler horizon obeys the area law for entropy $dS \propto \delta\mathcal{A}$, then it is possible to derive 
Einstein equations (up to an undetermined cosmological constant) as an `equation of state' of the space-time thermodynamic system. 

The key elements involved in his work were the Rindler causal horizon, which defines the thermodynamic system; the Unruh temperature, which is imposed; Raychaudhuri's equation for geodesic deviations; and the demand that both the Clausius and Bekenstein definitions of entropy are valid.

The causal horizon plays the role of the barrier that separates the system from its surroundings. The entropy is assumed to exist due to the fact that part of the universe is inaccessible to the considered class of accelerated observers, being in this way defined as entanglement entropy in the paper, which also justifies the requirement that entropy must obey the area law.

\enlargethispage{10pt}
\vspace{0.8cm}

In summary, a lot has been said about the connections between thermodynamics and gravity and I believe a lot still remains undone. 
We hope to be able to answer, during this work, some of the questions and clarify some points of confusion.

\section{Outline}

This thesis will be organized as follows: We will initiate Chapter 2 by introducing and explaining some of the main topics in thermodynamics and general relativity. We will discuss the notion of thermal equilibrium and present each of the laws of thermodynamics. On the general relativity side we will discuss some topics which will be essential for the understanding of this work.

In Chapter 3, we will start exploring the connections between these two areas. We will introduce the work done by Tolman and Buchdahl on temperature distributions for equilibrium states, look at the physical interpretations of what it means to have a gravitationally induced thermal gradient and explain several different examples. We will finish by coming back to the laws of thermodynamics, now in a covariant formulation, pointing out the differences from the standard case, when they exist. 

Chapter four will be dedicated to reviewing two of the main theories of relativistic non-perfect fluids --- Classical Irreversible Thermodynamics and Extended Irreversible Thermodynamics. We will also study fluids following Born-rigid flows and show that Killing vectors are not a necessary condition for the existence of this kind of motion in general curved space-times. They are, however, a necessary condition for the existence of exact thermodynamic equilibrium states, as we will show from the equations of motion for  relativistic viscous fluids.  Finally, we will conclude with a discussion about the time-scales involved in the changes made on the system and compare it with the relaxation times for the system to settle into a new equilibrium state. What will be clear is that, besides perfect equilibrium not existing outside of Killing trajectories, sometimes the time scales involved are so distinct from each other that the approximate equilibrium could be considered equilibrium for all practical purposes. We will also show some other possible interesting cases which are allowed for non-Killing trajectories.
 
We then deviate somewhat from the work developed during the previous chapters and move into the semiclassical scenario of evaporating black holes. This is done in Chapter 5, where we propose a toy-model for resolving the trans-Planckian problem by looking at a spacetime formed by matching two Vaidya metrics together along a thin shell. The work presented will be almost entirely based on the kinematics of the model. This chapter is also related to Appendix A at the end of the thesis.

In chapter 6 we present the conclusions of this work.

\section{Notation and conventions}

We are adopting the $\PC{-,+,+,+}$ signature for the metric and set $G_N = c = \hslash = 1$ unless we say otherwise.

Also, when dealing with tensors, the completely symmetric part of a tensor $A_{\mu\nu}$ will be represented as $A_{(\mu \nu)}$, which is:
\begin{equation*}
A_{(\mu \nu)} = \frac{1}{2}\PC{A_{\mu\nu} + A_{\nu\mu}}.
\end{equation*}
The completely anti-symmetric part, on the other hand, will be represented as $A_{[\mu \nu]}$, given by:
\begin{equation*}
A_{[\mu \nu]} = \frac{1}{2}\PC{A_{\mu\nu} - A_{\nu\mu}}.
\end{equation*}

\newpage
\chapter{Thermodynamics and General Relativity}
\label{C: Thermodynamics and General Relativity}

This thesis, as explicitly suggested by its title, has its foundations in two distinct but still relatable theories -- thermodynamics and general relativity. These are the two pillars on which all the discussions of this thesis will be supported. So, in order to make this manuscript comprehensible, we need to make sure that the understanding of both these pillars is solid, trustable and guarantee that no subjectivities or ambiguities stand in our way. That is the mission of the next few sections.

\section{Thermodynamics}
\label{S: Thermodynamics}
We will now dedicate some time to review and discuss \emph{what is} classical thermodynamics. We will also present its pillars, the four laws, which later on, will be subtly rephrased in a covariant formulation. But first, as a good delayer, I would like to take the opportunity to point out how this future reformulation will simply be an incorporation of the new data that arises when general relativity is taken into account. The meaning, the message behind each law, behind what thermodynamics is, will not be even slightly changed. To understand \emph{why}, we need first to know what \emph{is not} thermodynamics.


The fact is, thermodynamics is not a mere subgroup of physical laws, limited to specific sets of systems and analysed over a range of specific conditions. Instead of a deterministic theory that dictates exactly what the final state of a system will be and how it will get there, thermodynamics works like constraints, imposed by nature itself, commonly called probability theory. 
As beautifully described by Lopez-Monsalvo \cite{Lopez-Monsalvothesis}, 

\emph{``The Laws of Thermodynamics are statements about nature which stem from the observation that certain phenomena - although allowed by the available theory of motion - simply do
	not occur. We need to impose these laws ‘on top’ of our more fundamental dynamical basis. Thus, the correct way to understand the role of thermodynamics is by regarding its
	laws as \emph{auxiliary hypotheses} which rule out entire classes of dynamical processes.''}

I would like to add to this description by explicitly coming back to the probabilistic nature of any system which contains a large enough number of constituents. As is well known from the history of science, all thermodynamic laws were developed considerably before the establishment and acceptance of the atomic theory. Maxwell and Boltzmann, the fathers of statistical mechanics -- who incorporated the concept of atoms and molecules into a statistical mechanics probability theory capable of describing the behaviour of gases --- were truly disbelieved by the majority of the physics community.
Indeed, it is possible to construct and use thermodynamics without the knowledge of probability or even without believing in the existence of atoms. To do so, however, is to reduce it to an ordinary theory, to kill its essence. 

Having said what thermodynamics is not, let us come back to what it is. 
It is simply the description of the equilibrium --- or local equilibrium --- emergent behaviour of any system composed by a large enough
number of particles. It arises from the coarse graining of the system's probabilistic behaviour and 
can easily be summarized by its four laws. Nevertheless, to fully understand this statement, we will need discuss what exactly is meant by ``equilibrium''.

\subsection{Thermodynamic Equilibrium}
\label{S: Thermodynamic Equilibrium}
Unless you are specifically studying non-equilibrium thermodynamics or non-equilibrium statistical mechanics, you will be always looking at systems in thermal equilibrium. Non-equilibrium thermodynamics is still a wide-open research area. For example, the question ``what is/does it make sense to talk about temperature for out of equilibrium systems?'' is still not satisfactorily answered. In this way, the concept of equilibrium defines the boundaries where the results from equilibrium thermodynamics can be applied. However, despite its importance, it is often taken for granted.

If you ask an instrumentalist to define thermodynamic equilibrium, they would probably be tempted to define it as the state with constant spatial distribution of temperature and whose macroscopic variables do not change spontaneously in time.
Such a scenario is, for most cases, valid. But for it to be a definition, it has to be valid in all possible cases.
When studying systems located in a curved spacetime, for example, we see that the spatial distribution of temperature does not obey such a constraint. 
But without it, the instrumentalist's definition would be reduced to staticity or quiescence. In this way, something else must be necessary.

To be honest, when trying to find a complete and general definition of thermodynamic equilibrium, one will encounter several authors reducing it to staticity, which is neither complete nor true. Callen \cite{Callen-book} is almost radical when emphasizing how quiescence doesn't define equilibrium and requiring \emph{absolutely no trace at all} of past history of forces that were previously applied in the system. He cites the following example: 

\emph{``[...] two pieces of chemically identical commercial steel may have very different properties imparted by cold-working, heat treatment, quenching, and annealing in the manufacturing process. Such systems are
clearly not in equilibrium.''}


A second possible definition is given by the microscopic approach, based on the validity of the Boltzmann--Gibbs probability distribution for equilibrium states. This can also not be accepted as a definition, however, firstly because it is not clear whether equilibrium systems in curved space-times will maintain Gibbs probability distributions and secondly given examples \cite{Godreche, Mariao2011} of spin models that are described by the Gibbs distribution but are not in thermodynamic equilibrium (in the sense that entropy is continuously being generated), as pointed out by Tome \& de Oliveira \cite{TaniaMario}.


Another interesting definition, also based on the information needed to fully characterize a system, can be found in the postulate of Callen:

\emph{Postulate: ``There exist particular states (called equilibrium states) of simple systems that, macroscopically, are characterized by the internal energy $U$, the volume $V$, and the mole numbers $N_1, N_2, ... N_r$ of the chemical components.''}

Such a definition leads us to start seeing the state of equilibrium as the state which can be fully characterized by its intrinsic parameters \emph{only}. 
However, as emphasized by Einstein and others, one of the most appealing features of thermodynamics is its universal character. In that way, given the increasing interest in applying its results to different branches of physics \emph{e.g.} information theory, black hole thermodynamics and so on, it seems useful to have as much flexibility as possible regarding the parameters used to describe different systems.
In this way, let us analyse a ``non-definition'' quote given by Callen when introducing the reader to what finally became his postulate as given above. His statement goes along these lines:

\emph{``In all systems there is a tendency to evolve toward states in which the properties are determined by intrinsic factors and not by previously applied external influences. Such simple terminal states are, by definition, time independent. They are called equilibrium states''}



I particularly like this quote from Callen since it does not specify the parameters that completely characterize the system, leaving it open simply as ``intrinsic factors''. For it to be truly universal, thermodynamics needs one to allow different systems to be characterized by different parameters, which might not include volumes or number of particles, \emph{e.g.} Schwarzschild black holes are fully characterized by their mass content only.


Another important feature of equilibrium states, however, is the lack of energy, mass and heat flows. When thinking about the evolution of stars, for example, from a low density cluster of dust until their bright shining state, no external forces were present. The system, impressively enough, was always evolving by itself, through self-gravitating forces and internal nuclear reactions. Nobody forced that fluid to become a star. Their history wasn't shaped by external forces. 
But no one believes a star is a system in \emph{absolute} thermal equilibrium\footnote{Since stars are constantly emitting energy, a steady state description can be seen as  a good approximation to study these objects. Some authors do use near-equilibrium approximations for stars, but the point here is that, besides the approximations being sufficiently valid for a short period, it is well known that they are not in thermal equilibrium, given its explicit time-dependent character.}. So, the question I finally want to ask here is what exactly is meant by ``intrinsic factors''? According to the Merriam Webster dictionary \cite{dictionary}, intrinsic means \emph{``belonging to the essential nature or constitution of a thing''}. Heat fluxes inside the star are definitely not essential, neither are the nuclear reactions, and so on. On the other hand, its mass and volume (or density) certainly are. In this way, we suggest an alternative postulate along the following lines:

\paragraph{Postulate:} When free of the influence of all external forces, all systems tend to evolve toward states which are \emph{fully} characterized by the lowest possible number of intrinsic parameters (\emph{i.e.} not dependent on past history nor on the microscopic constituents' characteristics). Such states are, by definition, time independent. They are called equilibrium states.

\vspace{0.7cm}

We do, nevertheless, recognize the lack of precision of what ``the lowest possible number of intrinsic parameters'' means. Again, different types of systems probably require distinct sets of parameters. So, for practical purposes, we will end up adopting a definition based on the entropy production of a state
\footnote{Even an entropy based definition might not be completely safe and accepted given the lack of knowledge about whether it makes sense to talk about and what entropy is for states far away from equilibrium. For Local Equilibrium Thermodynamics and Classical Irreversible Thermodynamics, both defined in chapter \ref{C: Can we still}, such a concept is well established and an entropy based definition is certainly well accepted. We will hope and believe that, although unable to quantify entropy for systems far from equilibrium, some version of the general concept of entropy and second law will remain valid. For a review on the subject, the reader is encouraged to look at reference \cite{Velasco}.}.

\paragraph{Thermodynamic Equilibrium:}  When free of the influence of all external forces, a system  is said to be in thermodynamic equilibrium when its probability distribution is time-independent and maximizes the entropy of that system. After equilibrium is reached, no more entropy will be generated.

\vspace{0.7cm}
Although the concept of entropy will only be introduced in the next section, particularly when discussing the second law, we assume the reader to have 
a sufficient background to understand the thermodynamic equilibrium definition just given. We invite the reader who is not used to the subject to re-read the definition above after reading \ref{SS: the laws of thermodynamics}.

\subsection{The Laws of Thermodynamics}
\label{SS: the laws of thermodynamics}
Let us now introduce the non-covariant version of the laws of thermodynamics. As previously mentioned, small but important differences will be made in the future, but let us, for now, focus on the classical standard version.
There are many different ways to formulate them and for the second law, for example, we will present more than one possibility of doing so.

\paragraph{The Zeroth Law:}
Formulated only after the completion of the other three laws, the role played by the zeroth law of thermodynamics is to establish the transitivity property of thermal equilibrium. Note that thermal and thermodynamic equilibrium are not the same. If we state that two systems A and B are in thermal equilibrium with each other, we are simply saying that they have the same temperature. Nothing is said about pressures, \emph{etc}. One can also affirm that the heat flow between A and B vanishes. In this way, the zeroth law can be stated as follows:

\emph{The zeroth law of thermodynamics states that if two thermodynamic systems A and B are separately in thermal equilibrium with a third system C, then they are in thermal equilibrium with each other. It defines thermal equilibrium as an equivalence relation between thermodynamic systems.}


\paragraph{The First Law:}
Conservation of energy is the message given by the first law of thermodynamics. It is as general as you can expect it to be, yet still extremely practical and useful. Given a certain system, we denote its internal energy content by $U$. Variations on this amount of energy, $\Delta U$, can originate from two different processes -- either due to some amount of heat $\Delta Q$ being injected or extracted from the system, or due to some amount of work $\Delta W$ being done on or by the system. 

However, besides changes $\Delta U$ being well defined, given that $U$ is a function of state, the same is not valid for $\Delta Q$ and $\Delta W$ separately. We can only know the value of the sum $Q+W$. How much each one contributed individually to the final sum is dependent on the path taken by the system to go from its initial to its final state.
In this way, the correct mathematical formulation of the first law is given by 
\begin{equation}
\label{E:first law}
\dif U =\dbar Q + \dbar W,
\end{equation}
where $\dbar$ is the inexact differential or imperfect differential, used to make the path dependence explicit.

\paragraph{Second Law:}
If one has accepted the role of thermodynamics as a set of constraints, or auxiliary hypotheses imposed by nature on the dynamical processes allowed, the second law of thermodynamics is certainly the least trivial of all such inviolable rules.

Unlike the other laws, which might have clearer ``reasons'' for us to understand and accept (although not obvious, \emph{e.g.} the conservation of energy took a long time and a lot of effort to be established), the concept of entropy is still misunderstood even by many modern-day physicists. The reason may lay in its subtlety which, in my point of view,  is due to its statistical origin.

When you think about it, the accomplishments of Carnot, Clausius and Kelvin of deriving the concept of entropy simply from macroscopic observations of thermodynamic systems seem quite remarkable.
The extension of the concept of entropy to more general systems, however, probably only took place after the work of Boltzmann and Caratheodory. To understand this, let us take a brief tour along the evolution of the second law\cite{Adkins}:

\emph{\underline{Carnot's principle }(pre-second law statement): 
	No engine operating between two given reservoirs can be more efficient than a Carnot engine operating between the same two reservoirs. 
}

Here, the thermal efficiency $\eta$ is given by:
\begin{equation}
\eta = \frac{\textsl{work out}}{\textsl{heat in}} = \frac{W}{Q_{in}}
\end{equation}
and, for reversible Carnot cycles, given that $W = Q_{in} - Q_{rej}$, with $Q_{rej}$ being the rejected heat, we have:
\begin{equation}
\eta =  1 - \frac{Q_{rej}}{Q_{in}}.
\end{equation}

\hspace{-0.43cm}Carnot concludes with the statement:

\emph{All reversible engines operating between the same reservoirs are equally efficient.}

\hspace{-0.43cm}His work was continued by Rudolf Clausis, who, amongst
 several other contributions, created the term ``entropy'' and used Carnot's statements to formulate the second law of thermodynamics :

\emph{\underline{Clausius' statement:} No process is possible whose sole result is the transfer of heat from a colder to a hotter body.}

This is probably one of the most intuitive statements of the second law and it emphasizes the existence of a preferred direction for heat and energy to flow. 
This direction being that which increases a quantity defined by Clausius as the entropy of a system. It is given by:
\begin{equation}
S= \frac{\Delta Q}{T}
\end{equation}
for reversible processes. Here $T$ is the temperature of the heat reservoir from which the heat amount $\Delta Q$ is put or taken out of the system. 

He also showed that such a quantity can only increase or stay the same in an isolated system, regardless of the processes occurring on it.
These were the first indications of the true importance of the second law.

The next in the line was Lord Kelvin, who proposed his own statement, and proved it to be the same as the one previously given by Clausius:

\emph{\underline{Kelvin's statement:} No process is possible whose sole result is the complete conversion of heat into work.}

The interpretation and understanding that we have today about the second law, however, would never be complete without the work of Boltzmann. All of the modern ``disorder'' interpretations of entropy simply wouldn't exist without it. The story, again, is not so straightforward. It can, though, be summarized with Boltzmann's proposal in 1872 of an equation that was thought to be able to describe the time development of a gas, valid even for out of equilibrium situations. 
Boltzmann then showed that his equation implied what he called the H-Theorem, which states that a quantity (equivalent to entropy in equilibrium) must always increase with time.

It was noticed, however, that his derivation could be run in reverse, due to the reversible time-symmetric character of molecular dynamics, implying with this the opposite result expected from the second law. Boltzmann dedicated himself to fix the situation and, in 1876, realized that, when dealing with systems composed by a large number of components, as a gas, the probabilities associated with the random, disorganized  distributions are tremendously higher than those for organized states.
This realization led Boltzmann to a remarkable equation, 
which relates the entropy $S$ of a specific state  with the number of macroscopically identical configurations $W$  accessible to the system when on that state. It is given by:
\begin{equation}
S = k_B \;\ln{W},
\end{equation}
where $k_B$ is the so called Boltzmann's constant. 
It becomes clearer now what was meant by our thermodynamic equilibrium definition given in the previous chapter. A system in thermodynamic equilibrium has the \emph{maximal number of indistinguishable states} allowed (at fixed energy) for that system~\footnote{This idea eventually led to the microcanonical ensemble derivation of statistical mechanics first derived by Gibbs \cite{Gibbs}.}. The particles inside it keep themselves in movement, occasionally colliding with each other, in a way that the system tends to visit all the possible configurations permitted, a property called ergodicity. This is the state of maximum entropy mentioned before.

\enlargethispage{10pt}
In this way, Boltzmann has not simply come up with a statement for the second law. He explained, based on probability theory, why the second law works; why energy flows in the directions that it does, and even more, what equilibrium and ergodicity mean, providing the tools which allowed Gibbs to create the ensemble statistical mechanics which is so well known and used today. 

For completeness, Boltzmann's explanation of entropy is normally understood on the basis of the phase-space description of a system, where the number of accessible states reduces to the hyper-volume of this same phase-space. 
We would just like to finish by pointing out that it is well understood today why Liouville's theorem for the conservation of phase-space volumes does not contradict the second law.
This essentially being due to the limit of precision in any measurement (including interactions between molecules), originating a coarse-graining which leads to an entropy increase \cite{AnaeMatt}. A longer discussion of this topic is, however, outside the scope of this thesis.


\paragraph{The Third Law:} The coldest natural place in the universe known up to now is the Boomerang Nebula, a protoplanetary nebula only $5,000$ light-years away from Earth, in the Centaurus constellation. Its temperature is measured at $1 K$ \cite{Sahai}, colder than the $2.72 K$ of the Cosmic Microwave Background (CMB). The third law of thermodynamics, however, imposes a limit not only on how cold the Boomerang Nebula can be, but on all structure and matter in the universe.

Its final form as known today was formulated as a ``new heat theorem'' by Walther Nernst and later used by Max Planck, who extended and rewrote it as the third law of thermodynamics. 
In the words of Wilks\cite{Wilks}, we can state the third law as:

\emph{It is impossible for any process, no matter how idealized, to reduce the entropy of a system to its absolute-zero value in a finite number of operations.}

As a remark, a zero entropy state would only hypothetically be possible in a perfect crystal, when all the atoms that form it are identical and positioned in perfectly symmetrical ways, with perfectly ordered magnetic moments and with no atomic motion at all, e.g. temperature at absolute zero. Any imperfections on the crystal would carry energy, resulting in a non-minimal entropy. So, from an entropic perspective, this can be considered to be part of the definition of what a ``perfect crystal'' is. 

But more than anything, this reveals the existence of a clear relation between absolute zero temperature and zero entropy states. Some believe that the third law could in principle be also described by the so called unattainability statement which, in the words of Zemansky \cite{Zemansky}, says:
	
	\emph{``By no finite series of processes is the absolute zero  attainable.''}
	
	\hspace{-0.45cm}Or, in the more careful words of Callen,
	
	\emph{``No reversible adiabatic process starting at
		nonzero temperature can possibly bring a system
		to zero temperature.''}
	
Nevertheless, although normally considered equivalent, there are disagreements about whether the third law and the unattainability statement are actually interchangeable \cite{Wheeler91} and, in principle, one could see systems that do not  have zero entropy at zero temperature as counter-examples of such equivalence. 
	
Just to mention a couple, we might look at systems which do not have a unique ground state, e.g. half-integer net spin systems, which have entropy at absolute zero of at least $k_B \ln{2}$. Crystalline systems with geometrical frustration, where the structure of the crystal lattice prevents the emergence of a unique ground state, are also an example.

This, however, does not disprove the unattainability statement,  nor necessarily separates it from the third law, which might very well impose limits both on the minimum temperature allowed as well as on the entropy content of matter.
A longer discussion of this topic, however, is far beyond the scope of this thesis.




\section{General Relativity}
\label{S: General Relativity}

The second pillar of the results presented in this thesis is the extremely successful gravitational theory, which interconnects matter and spacetime with its set of dynamical non-linear equations, that is, General Relativity.

We will, in the following section, discuss some selected issues in relativity which will be necessary for the understanding of the subsequent chapters. This, however, will be a focused introduction, consisting of refreshing reminders about specific topics rather than any attempt to actually explain all of relativity itself. For the reader who might need some extra concepts, we suggest the classic general relativity books \cite{Carroll, Hartle, Wald, MTW}.

\paragraph{The stress-energy tensor}\hfill

Given the aim of studying the thermodynamics of fluids in a curved space-time, we need a quantity capable of covariantly describing their matter and energy fluxes.
The most natural way to do so is throughout the \emph{stress-energy} or \emph{energy-momentum tensor}.
It consists in a tensorial description of all the energy, stresses and heat fluxes present in the system, which allows us to rewrite all the  hydrodynamics equations in a covariant way. When a coordinate system is defined, and once one chooses an orthonormal basis, the stress-energy can be expressed as a symmetric $4\times4$ matrix with contains 10 degrees of freedom. The physical interpretation of its components are as follows:

\vspace{0.5cm}
\hspace{1cm}
\colorbox{white}{
\begin{minipage}{11.5cm}
$T^{\hat{0}\hat{0}}\;$ represents the total energy density;

$T^{\hat{0}\hat{i}}\;$ represents the flux of energy density in the $\hat{i}$-th direction;

$T^{\hat{i}\hat{0}}\;$ represents the flux of $\hat{i}$-th momentum in the $\hat{0}$-th direction;

$T^{\hat{i}\hat{j}}\;$ represents the flux of $\hat{i}$-th momentum in the $\hat{j}$-th direction.
\end{minipage}
}
\vspace{0.5cm}

In the case of a perfect fluid, for example, where no anisotropies or energy fluxes exist, the energy-momentum is given by:
\begin{equation}
\label{E:perfect fluid}
T^{\mu\nu} = (\varrho + p)\;u^{\mu}u^{\nu} + p\, g^{\mu\nu},
\end{equation}
where $p$ is the isotropic pressure of the fluid, $u^{\mu}$ the fluid's 4-velocity and  $\varrho$ is the \emph{total energy density}, given by\footnote{Keeping the factor of $c$ this reads as:
\begin{equation}
\varrho = \rho\,(c^2 + \mathfrak{u}).
\end{equation}}:
\begin{equation}
\label{E: total energy density}
\varrho = \rho\,(1 + \mathfrak{u}).
\end{equation}
The quantity $\rho$ represents the \emph{rest mass density} of the fluid, defined in terms of the total mass $M$ and volume $V$ as:
\begin{equation}
M = \int \rho \,\d V,
\end{equation}
while $\mathfrak{u}$ is the \emph{specific internal energy}, given by:
\begin{equation}
U = \int \mathfrak{u}\,\rho\,\d V,
\end{equation}
with $U$ the internal energy of the fluid present in the first law \eqref{E:first law}. Whereas $\rho$ concerns mass, the quantity $\mathfrak{u}$ is actually related to internal movements of the fluid's particles, like vibrations and rotations. For monatomic fluids with atomic mass $m$, for example, $\mathfrak{u}$ is present in the famous relation:
\begin{equation}
\mathfrak{u} = \frac{3}{2}\frac{k_B\, T}{m}.
\end{equation}
The importance given to $\rho c^2$ in comparison to $\rho \mathfrak{u}$ then depends on the type of fluid being analyzed. While for dust the second term is practically negligible, the same is not true for ultra-relativistic fluids, where the first term can be discarded.  

As a side-note, in this thesis, every time a quantity is named \emph{specific}, for example specific internal energy, what is meant is ``internal energy per unit mass''. So, given a specific quantity $b$ related to an extensive quantity $B$, we have:
\begin{equation}
B = \int b\, \rho \,\d V.
\end{equation} 

Now, in order for $T^{\mu\nu}$ to represent the stress-energy contents of a fluid, it must also satisfy the hydrodynamic equations. These are the conservation of mass and conservation of energy and momentum equations. For a relativistic fluid the \emph{conservation of energy and momentum} can be shown \cite{Rezzolla} to be given by:
\begin{equation}
\label{E:conservation of energy}
\nabla_{\mu}T^{\mu\nu} =0.
\end{equation}

The conservation of mass, on the other hand, requires us to define the \emph{rest-mass density current}, which is given by:
\begin{equation}
J^{\mu} = \rho\,u^{\mu},
\end{equation}
where $u^{\mu}$ is the four-velocity of the observer ``measuring'' the fluid. In this way, conservation of mass or \emph{continuity equation} is given by:
\begin{equation}
\label{E:continuity equation}
\nabla_{\mu}J^{\mu} =0.
\end{equation}
In special relativity, for example, if in a certain coordinate system we have $J^{\mu} = (c\rho, j_x, j_y, j_z)$, where $j_i$ represents the mass-fluxes, then \eqref{E:continuity equation} reads:
\begin{equation}
c\;\frac{\partial \rho}{\partial t} + \mathbf{\nabla \cdot j} =0.
\end{equation}
Note that the rest-mass density current is not the same as the current defined as:
\begin{equation}
\tilde{J}^{\mu} = T^{\mu\nu}K_{\nu},
\end{equation}
which, for $K^{\mu}$ a Killing vector, is also conserved, as we will show in equation \eqref{E:newcurrent}.

Besides the focus in this section on perfect fluids, the physical interpretation of the energy-momentum components and equations \eqref{E:conservation of energy}--\eqref{E:continuity equation} will remain valid also for non-perfect fluids. These will be studied in  Chapter \ref{C: Can we still}.

\paragraph{Lie Derivatives}\hfill

An important concept which will be used in Chapter \ref{C: Can we still} is that of a Lie Derivative. Although less commonly used than covariant derivatives, Lie derivatives hold a very important role in general relativity and, backed up by concepts such as diffeomorphisms and isometries, lead naturally to the concept of symmetry.

To understand Lie derivatives, let us start by imagining we are sailing on the sea. You are inside a boat, which has a fixed mast holding the sail, and let us also imagine a bunch of loose boxes around, like a chiller bag with some refreshing drinks. The breeze is light and you smoothly drift through tropical waters. You look at some island a bit further away and notice that you are approaching its beautiful beaches. You look at the mast, which does not move in relation to you. The boxes don't move either, it is all too smooth. As you get closer to the beach, waves start shaking the boat. You look at the mast, not moving yet, but now the chiller bag is moving all around the boat.

Rather than making you feel relaxed and wanting to sail, the situation just described can help us to easily understand the concept of Lie derivatives. Rather than setting up a coordinate system with a connection, like we do with covariant derivatives, Lie derivatives do not require connections and not even a metric, only a vector field. In the situation described above, assume $\xi^{\mu}(t)$ to be the vector field tangent to the curve traced by the boat on its way to the island. 
Although there is absolutely no need to resort to the idea of a physically present observer when talking about Lie derivatives, I particularly like to keep this image of an observer with a certain four-velocity --- which generates the vector field --- in my mind, as it makes the understanding more intuitive.

Now, imagine a vector $M^{\mu}$ connecting you --- the observer inside the boat --- and the mast, plus another vector $C^{\mu}$ between you and the chiller bag. Before the approach of the waves, neither were moving in relation to you, although the boat was moving in relation to the island (which can be thought as setting up a fixed coordinate system which coincides with the boat's coordinate system at some fixed initial time $\tau_0$). So, we claim, the Lie derivative of the connecting vectors $M^{\mu}$ and $C^{\mu}$ in the direction of the boat's velocity $\xi^{\mu}$ is zero. Or, 
\begin{equation}
\mathcal{L}_{\xi} \;M^{\mu} \;= \;\mathcal{L}_{\xi} \;C^{\mu} \;= 0 \quad\;\; \text{(before the waves).}
\end{equation}
However, after the waves shake the boat, the chiller bag started to move around, whilst the mast kept still, giving us
\begin{equation}
\mathcal{L}_{\xi}\;M^{\mu} \;=0;\quad \;\mathcal{L}_{\xi} \;C^{\mu} \;\neq 0 \quad\;\; \text{(during the waves).}
\end{equation}
In this manner, if one wonders about the Lie derivative of a function, it is not hard to conclude that it consists simply of its directional derivative, i.e.,
\begin{equation}
\mathcal{L}_{\xi}\; f = \xi (f) = \xi^{\mu}\;\partial_{\mu} f.
\end{equation}

Keeping the tropical explanation in mind, let us add some more rigorous mathe- matics to these ideas. We will follow Anderson's Lie derivatives explanation \cite{Andersson2007} based on active and passive transformations. As it is implicitly put in the scenario above, the concept of Lie derivatives require a drag of the coordinate system along the vector field direction, such that after each infinitesimal displacement (of the boat) in the vector field direction, there is a displacement of the coordinate system following it. It is as if the coordinate system was set by the observer inside the boat. In mathematical terms this can be translated as the action of an \emph{active} coordinate transformation followed by a \emph{passive} transformation \cite{Yano}. To perform an active transformation, one initially has to fix, with respect to an external observer (island), the origin and orientations of a coordinate system. Given such a structure, one can move an object (boat) from point to point, without changing the reference system. In this way, imagining a curve $x^{\mu}(\lambda)$ connecting two points $x^{\mu}(\lambda=0)$ and $x^{\mu}(\lambda = \epsilon)$ which are infinitesimally away from each other, we have
\begin{equation}
x^{\mu}(\epsilon) \approx x^{\mu}(0) + \epsilon\; \xi^{\mu}, 
\end{equation}
where 
\begin{equation}
\xi^{\mu} = \frac{\dif x^{\mu}}{\dif \lambda}\bigg\vert_{\lambda = 0}
\end{equation}
represents the boat's 4-velocity, i.e., the vector tangent to the curve $x^{\mu}(\lambda)$.

For passive transformations, on the other hand, one fixes the object's position with relation to an external observer and then changes the coordinate system $x^{\mu} \to \bar{x}^{\alpha}(x^{\mu})$. Taking the particular case
\begin{equation}
\label{E:coord transf}
\bar{x}^{\mu} = x^{\mu} - \lambda\; \xi^{\mu}, 
\end{equation}
we have
\begin{equation}
\bar{x}^{\mu}(0) = x^{\mu}(0)
\end{equation}
and
\begin{equation}
\label{E:coord transf2}
\bar{x}^{\mu}(\epsilon) = x^{\mu}(\epsilon) - \epsilon\; \xi^{\mu} = \;  x^{\mu}(0)\; =\; \bar{x}^{\mu}(0),
\end{equation}
which represents the drag of the coordinate system mentioned above. Naturally, one now might want to evaluate changes in another vector, for example $C^{\mu}$, along the curve. In the active formulation, keeping the infinitesimal displacements assumption, we have
\begin{eqnarray}
\label{E: C and C0}
C^{\mu}(\epsilon) &\approx& C^{\mu}(0) +\epsilon \;\frac{\dif C^{\mu}}{\dif \lambda}\bigg\vert_{\lambda = 0} \nonumber \\
&=& C^{\mu}(0) +\epsilon \;\frac{\dif x^{\nu}}{\dif \lambda}
\frac{\dif C^{\mu}}{\dif  x^{\nu}}\bigg\vert_{\lambda = 0},
\end{eqnarray}
what gives us:
\begin{eqnarray}
C^{\mu}(\epsilon) \approx C^{\mu}(0) +\epsilon \;\xi^{\mu} \;\frac{\dif C^{\mu}}{\dif  x^{\nu}}\bigg\vert_{\lambda = 0}.
\end{eqnarray}
In the passive formulation, or, at the coordinates of the dragged reference system, we have
\begin{equation}
\bar{C}^{\mu}(\epsilon) = \frac{\dif \bar{x}^{\nu}}{\dif x^{\nu}} \; C^{\nu}\bigg\vert_{\lambda = \epsilon}.
\end{equation}
Using equation \eqref{E:coord transf2} and ignoring second order terms, we obtain 
\begin{eqnarray}
\label{E: C bar}
\bar{C}^{\mu}(\epsilon) \approx  C^{\mu}(\epsilon) - \epsilon\; C^{\mu}(0)\; \frac{\dif \xi^{\mu}}{\dif  x^{\nu}}\bigg\vert_{\lambda = 0}
\end{eqnarray}
Now we can define the Lie derivative as the difference between the initial vector state in the initial reference system and the final state in the dragged reference system. In this way we make sure that we are always comparing quantities at the ``observer's reference frame". Taking the limit $\epsilon\to 0$ we have:
\begin{equation}
\mathcal{L}_{\xi}\; C^{\mu} = \lim_{\epsilon\to 0}\; \frac{\bar{C}^{\mu}(\epsilon) -C^{\mu}(0)}{\epsilon}
\end{equation}
Now inserting equations (\ref{E: C and C0}) and (\ref{E: C bar}), we obtain:
\begin{equation}
\mathcal{L}_{\xi}\; C^{\mu} = \xi^{\nu}\; \frac{\partial C^{\mu}}{\partial x^{\nu}} - C^{\nu}\;\frac{\partial \xi^{\mu}}{\partial x^{\nu}},
\end{equation}
which can be rewritten in its most general well known form:
\begin{equation}
\mathcal{L}_{\xi}\; C^{\mu} = \xi^{\nu} \nabla_{\nu} C^{\mu} - C^{\nu} \nabla_{\nu} \xi^{\mu} = [\xi , C]^{\mu}\;.
\end{equation}
In contrast, for covariant vectors we have:
\vspace{-0.2cm}
\begin{equation}
\mathcal{L}_{\xi}\; C_{\mu} = \xi^{\nu} \nabla_{\nu} C_{\mu} + C_{\nu} \nabla_{\mu} \xi^{\nu}\;.
\end{equation}
It is also possible to then obtain Lie derivative's definition for tensors of arbitrary rank. A particularly important one is the Lie derivative of the metric tensor along a general vector $K^{\mu}$, given by:
\vspace{-0.2cm}
\begin{eqnarray}
\label{E:lie d metric}
\mathcal{L}_{K}\; g_{\mu\nu} &=& K^{\sigma}\nabla_{\sigma}\;g_{\mu\nu} + (\nabla_{\mu}K^{\sigma})\;g_{\sigma\nu} + (\nabla_{\nu}K^{\sigma})\;g_{\mu\sigma} \nonumber\\
&=& \nabla_{\mu}K_{\nu} + \nabla_{\nu}K_{\mu}\,.
\end{eqnarray}
\enlargethispage{20pt}
Notice that we can also choose a coordinate system $(y^1,...,y^n)$ such that $y^1$ is the parameter along the curve $x^{\mu}(\lambda)$, such that $\xi^{\mu} = \partial/\partial y^1$ and
\begin{equation}
\mathcal{L}_{\xi} \;C^{\mu} = \frac{\dif C^{\mu}}{\dif y^1}.
\end{equation}
In this way, having $\mathcal{L}_{\xi} C^{\mu} = 0$ implies a symmetry of $C^{\mu}$ along the $\xi^{\mu}$ direction, i.e., $C^{\mu}$ does not depend on the coordinate $y^1$. 

Now, extending this idea for the metric tensor, it is possible in some special cases to pick a coordinate system such that the metric does not depend on one or more of the coordinate directions, say $\zeta_i$. Let $K_i^{\mu} = \partial/\partial \zeta^i$. Then,
\vspace{-0.3cm}
\begin{equation}
\label{E:invariant metric}
\mathcal{L}_{K_i} \; g_{\mu\nu} = 0,
\end{equation}
which means that the metric is invariant under translations in the $K_i^{\mu}$ direction and we call $K_i^{\mu}$ a Killing field. From equation \eqref{E:lie d metric} we see that Killing vectors must satisfy
\vspace{-0.3cm}
\begin{equation}
\label{E: Killing Equation}
\nabla_{\mu}K_{\nu} + \nabla_{\nu}K_{\mu} = 0.
\end{equation}
Equation \eqref{E: Killing Equation} is the famous Killing's equation.


\paragraph{Symmetry and Killing Vectors}\hfill 

Symmetry is by far one of the most important concepts in physics. It is implicitly or explicitly required almost in any analytical calculation. Killing vectors, when they exist, are responsible for defining conserved quantities like energy, linear and angular momentum. For space-times without Killing vectors it becomes impossible or, in the best case, cumbersome to define such quantities, with the final result probably having its physical meaning reduced to local regions only. 

So, given equation (\ref{E: Killing Equation}), note that conserved currents can always be constructed whenever an energy-momentum tensor satisfying
\vspace{-0.3cm}
\begin{equation}
\label{E:energy-momentum divergence}
\nabla_{\mu} \; T^{\mu\nu} = 0
\end{equation}
\vspace{-0.3cm}
and a Killing vector $K^{\mu}$ exist. Such a current is given by:
\begin{equation}
\tilde{J}^{\mu}= K_{\nu}\;T^{\nu\mu}.
\end{equation}

The conservation property follows from (\ref{E:energy-momentum divergence}) and from the contraction of the anti-symmetric $\nabla_{\mu}K_{\nu}$ with the symmetric $T^{\mu\nu}$ tensor: 
\begin{equation}
\label{E:newcurrent}
\nabla_{\mu}\tilde{J}^{\mu} = (\nabla_{\mu}K_{\nu})T^{\nu\mu} + K^{\nu}(\nabla_{\mu}T^{\nu\mu})=0.
\end{equation}

\paragraph{Junction conditions}\hfill

\begin{figure}
	\begin{center}
		\includegraphics[scale=0.5]{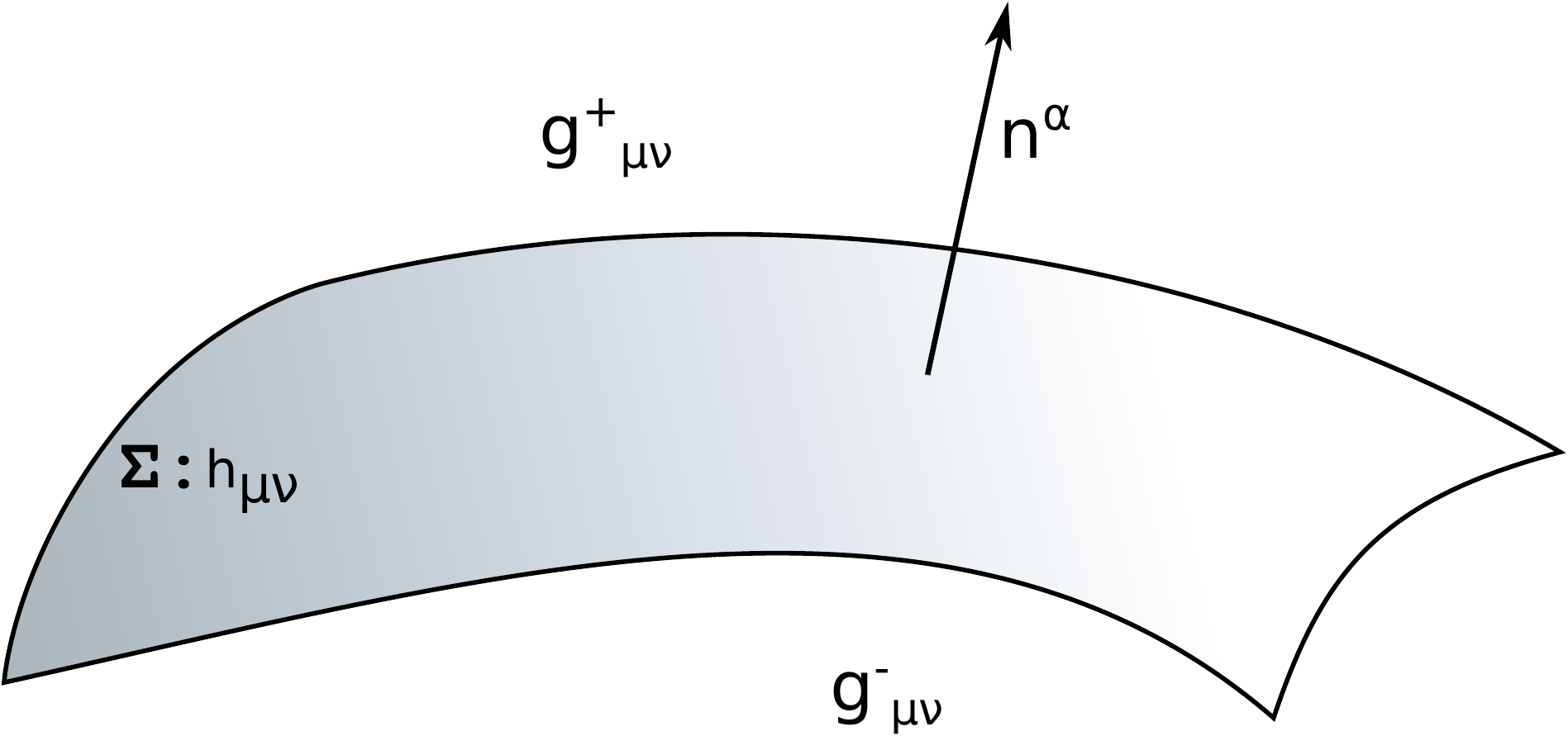}
		\caption[Junction conditions]{\label{F:matching} Two metrics $g^-_{\mu\nu}$ and $g^+_{\mu\nu}$ joined across the surface $\Sigma$.}
	\end{center}
\end{figure}

Another important concept, which we will use in Chapter \ref{C: Black Holes Thermodynamics} is that of junction conditions. Basically, depending on the type of question one might want to answer, it is sometimes useful to construct space-time metrics by patching two known metrics across a certain hypersurface $\Sigma$. In this situation, one would have something as shown in Figure \ref{F:matching}, where the metric on one side is $g^+_{\mu\nu}$ while on the other side of the surface it is $g^-_{\mu\nu}$. 
An example commonly found is to take a thin spherical shell, which represents the hypersurface $\Sigma$, separating the inside metric, described by a Minkowski space-time (or $g^-_{\mu\nu}$), from the outside region, which is described by a Schwarzschild metric (or $g^+_{\mu\nu}$). 

Although the idea is not complicated, we have to ensure that the overall metric will keep being a valid solution for Einstein's equations. For this to happen, we require that the two metrics $g^+_{\mu\nu}$ and $g^-_{\mu\nu}$ be joined smoothly across $\Sigma$. These are what the junction conditions will guarantee to happen. So, let us start by establishing how to describe the hypersurface $\Sigma$ and then move on to defining the first and second junction conditions. 
As we will be mainly interested on the thin shell case, we will focus on describing the junction conditions for this kind of situation.

Start with a thin hypersurface $\Sigma$ with a normal vector $n^{\alpha}$ defined at every point. Imagine, as well, a congruence of geodesics which crosses $\Sigma$ orthogonally. If we assume $l$ to be the proper distance along each geodesic, it is possible to parametrize the geodesic such that $l<0$ before the crossing (at the $g^-_{\mu\nu}$ region), $l=0$ at $\Sigma$ and $l>0$ after the crossing (at $g^+_{\mu\nu}$). By doing this, we can now write the overall metric as:
\begin{equation}
\label{E:heaviside metric}
g_{\mu\nu} = \Theta(l)\;g^+_{\mu\nu} + \Theta(-l)\;g^-_{\mu\nu},
\end{equation}
where $\Theta(l)$ is the Heaviside distribution given by
\begin{align*}
\Theta(l) =  \left\{ \begin{array}{ccc} + 1\quad \mbox{if}\;\; l>0\\  \;\; 0\quad \mbox{if}\;\; l\leq0 \end{array}\right.
\end{align*}
and it satisfies the following equations:
\begin{equation}
\Theta^2(l) = \Theta(l), \quad \Theta(l)\Theta(-l) = 0, \quad \frac{\d\Theta(\pm l)}{\d l} = \pm\delta(l),
\end{equation}
where $\delta(l)$ is the usual delta function distribution. 

Now, to have a valid metric, one must be able to compute the Riemann tensor and consequently the Christoffel symbols for such a space-time \cite{book}. These, however, contain derivatives of the metric. In this way, we must verify if the derivatives of \eqref{E:heaviside metric} are well behaved:
\begin{align*}
g_{\mu\nu,\gamma} &= \Theta(l)\;g^+_{\mu\nu,\gamma} + \Theta(l)\;g^-_{\mu\nu,\gamma} + \frac{\d l}{\d x^{\gamma}}\frac{\d \Theta(l)}{\d l}\; g_{\mu\nu}\\
&= \Theta(l)\;g^+_{\mu\nu,\gamma} + \Theta(l)\;g^-_{\mu\nu,\gamma} + \frac{\d l}{\d x^{\gamma}}\;\delta(l)\;(g^+_{\mu\nu} - g^-_{\mu\nu})|_{\Sigma}.
\end{align*} 
The $(g^+_{\mu\nu} - g^-_{\mu\nu})$ term comes from the derivative of $\Theta(\pm l)$ given above.
Now, given that the geodesics cross $\Sigma$ orthogonally, the gradient of their proper distance must be proportional to the normal vector $n^a$:
\begin{equation}
n_{\alpha} = \epsilon \;\partial_{\alpha} l , \quad \mbox{and}\;\; n^{\alpha}n_{\alpha} = \epsilon.
\end{equation}
Here $\epsilon$ takes the value $-1$ if the surface $\Sigma$ is space-like and $\epsilon = +1$ if the surface is time-like. This gives us:
\begin{align}
g_{\mu\nu,\gamma} &= \Theta(l)\;g^+_{\mu\nu,\gamma} + \Theta(l)\;g^-_{\mu\nu,\gamma} + \epsilon\; n_{\gamma} \;\delta(l)\; (g^+_{\mu\nu} - g^-_{\mu\nu})|_{\Sigma}\nonumber \\
&= \Theta(l)\;g^+_{\mu\nu,\gamma} + \Theta(l)\;g^-_{\mu\nu,\gamma} + \epsilon\; n_{\gamma} \;\delta(l)\; [g_{\mu\nu}],\label{E:delta}
\end{align} 
where we have adopted the notation of Poisson \cite{Poisson}:
\begin{equation*}
[A] \equiv A^+|_{\Sigma} - A^-|_{\Sigma}.
\end{equation*}

We then see that, in order for the Riemann tensor to be non-singular, the last term in equation \eqref{E:delta} containing the $\delta(l)$ must be zero, given the indefinite state of $\delta^2(l)$~\footnote{This term is problematic since it stops us from using linear distribution theory.}. For this to happen, we must impose that $[g_{\mu\nu}] =0$. However, as the subtraction is made at the crossing surface $\Sigma$, in a covariant way we can state that:
\begin{equation}
\label{E: first junction cond}
[h_{\mu\nu}] = 0,
\end{equation}
or, in words, this means that the metric across the surface is the same on both sides. This is known as the \emph{first junction condition}. Now, for the second junction condition we will only present the final results, since the derivation is quite long and will not be necessary in this work. We will, on the other hand, give the step by step procedure for the interested reader.

To find the second junction condition, one must write the Riemann tensor for the metric \eqref{E:heaviside metric}. It will be composed of three terms, one proportional to $\Theta(l)$, another to $\Theta(-l)$ and the remaining one proportional to $\delta(l)$, which represents a singularity for the curvature. From this, one can proceed and calculate the Ricci tensor and, using Einstein's equation, obtain the form of the energy-momentum tensor generating such a spacetime. This will be given by:
\begin{equation}
\label{E:shell}
T_{\mu\nu} = \Theta(l)\;T_{\mu\nu}^+ + \Theta(l)\;T^-_{\mu\nu} + \delta(l)\; S_{\mu\nu}.
\end{equation}
Here, $T_{\mu\nu}^+$ clearly is the energy-momentum tensor for the outside ($g^+_{\mu\nu}$ metric) region, while $T_{\mu\nu}^-$ is the equivalent for the inside region. The last term, however, is located at the thin shell. It then becomes clear that the $\delta(l)$ singular term present both in the Riemann as in the Ricci tensors are the ones which generate this surface mass distribution along the shell. 

 Furthermore, it can be shown that the tensor $S_{\mu\nu}$ is actually given by:
\begin{equation}
\label{E:surface tensor}
S_{\mu\nu} = -\frac{\epsilon}{8\pi}\PC{[K_{\mu\nu}] - [K]\;h_{\mu\nu}}.
\end{equation}
We can then state the \emph{second junction condition} as follows: 

In the absence of mass or energy present in the shell, i.e., for $S_{\mu\nu}=0$, we must have:
\begin{equation}
\label{E:second junction condition}
[K_{\mu\nu}] = 0,
\end{equation}
implying that the extrinsic curvature must be the same at both sides of $\Sigma$. In this case, both $T_{\mu\nu}$ and the Riemann tensor can be proved to be non-singular. If \eqref{E:second junction condition} is not satisfied, then this means that the thin shell \emph{must} contain some mass or energy distribution given by:
\begin{equation}
T^{\Sigma}_{\mu\nu} = \delta(l)\; S_{\mu\nu},
\end{equation}
with $S_{\mu\nu} \neq 0$ and given by \eqref{E:surface tensor}. So, all the singularities present in \eqref{E:shell} are justified by the presence of a thin shell of matter/energy at $\Sigma$.

\newpage

\chapter{Gravity-induced temperature gradients}
\label{C: Tolman chapter}

How does the action of gravity affect classical non-relativistic thermodynamics? This chapter will be dedicated to understanding how general relativity, the equivalence principle and curved spacetimes not only interact with thermodynamic systems, but can also shape their equilibrium states. 

We will start by reviewing the first results in this context, obtained by Tolman, in 1930, in a beautifully written paper called ``On the weight of heat and thermal equilibrium in General Relativity"\cite{weightofheat}. There, Tolman concluded that systems in thermal equilibrium under the action of a gravitational field do not have a constant temperature.  The local temperature distribution is position dependent, a result today very well known by relativists and cosmologists and used in several applications in both areas.

This chapter is organized in the following way: Section \ref{S: The weight of heat} starts by introducing some of Tolman's original thoughts and giving a historical and physical background. 
In section \ref{S: The physics behind Tolman temperature gradients}, we will discuss the physics behind gravity induced temperature gradients. 
Section \ref{S: The general case extension} will be dedicated to  extending Tolman's results to observers with general 4-velocities in any stationary spacetime. This generalization leads to several interesting examples, which we expound in section \ref{S: The rotating universe example}, where we  analyse the rotating universe case, and  \ref{S: Some other examples} where we present some results for observers outside of a black hole.

\section{The weight of heat}
\label{S: The weight of heat}

\vspace{-0.5cm}
Let us start by reviewing some of the main points of Tolman's results. 
As is argued in \cite{weightofheat}, 
heat is just another source of energy and, given Einstein's theory of relativity, it must be affected by the action of gravitational fields. Heat must have weight. Inspired by this idea, Tolman decides to analyse the equilibrium state of a perfect fluid in a spherically symmetric spacetime. Assuming the metric to be in the form
\begin{equation}
\label{E: sherically symmetric metric}
ds^2 = e^{\nu(r)}dt^2 - e^{\mu(r)}\PC{dr^2 + r^2 d\Omega^2},
\end{equation}
where $\nu (r)$ and $\mu(r)$ are functions of the radial coordinate, and taking a perfect fluid with energy-momentum tensor given by \eqref{E:perfect fluid}:
\begin{equation}
\tag{\ref{E:perfect fluid}}
T^{ab} = (\varrho + p)\;u^{a}u^{b} + p\; g^{ab},
\end{equation}
we can impose conservation of the fluid stress-energy tensor, i.e., $\nabla_a T^{ab} = 0$,
from which we obtain the following:
\begin{equation}
\nabla_a \left[(\varrho + p)\;u^a u^b + p g^{ab}\right] = (\varrho + p)\nabla_a \PC{u^a u^b}
+(u^a u^b)[\nabla_a p + \nabla_a \varrho] + g^{ab} \nabla_a p  = 0.
\end{equation}
Projecting this result in the direction orthogonal to $u^a$, i.e, multiplying the above equation by $h_{bc}=(u_bu_c + g_{bc})$, we obtain:
\begin{equation}
\label{E:eq.motion1}
(\varrho + p) \;h_{bc}\nabla_a \PC{u^a u^b}
+h_{b}{}^ {a}\; \nabla_a p = 0,
\end{equation}
implying
\begin{equation}
\label{E: relativistic euler}
(\varrho + p)\;a^b \;+\;h_{b}{}^ {a}\; \nabla_a p = 0,
\end{equation}
where $a^b$ is the four-acceleration of the fluid. Making use of the projected covariant derivative $\D$, which will be more precisely defined in Chapter \ref{C: Can we still}:
\begin{align}
\mathcal{D}_{a}\phi := h^{b}_{a}\;\nabla_{b} \phi,
\end{align}
we have:
\begin{equation}
\label{E: fluid a and p}
(\varrho + p)\;a^b \;+\;\D_a \;p = 0.
\end{equation}

In Chapter \ref{C: Can we still} we will explore non-perfect fluids with anisotropies and show that, when those fluids achieve equilibrium, their equations of motion assume exactly the same form as \eqref{E: fluid a and p}.
Now, assuming the fluid to have the 4-velocity $u^a = (1,0,0,0)$, and using the Christoffel symbols associated with the metric \eqref{E: sherically symmetric metric} to unwrap the covariant derivative, we finally obtain the following result:
\begin{equation}
\label{E: continuity eq}
\frac{\partial p}{\partial r} = -\frac{\varrho +p}{2}\;  \frac{\partial \nu}{\partial r}.
\end{equation}
This is simply the general relativistic version of the  Euler equation for this specific situation.

Focusing on the black body radiation case, for example, it is easy to see that applying Stephan-Boltzmann's law, $\varrho = a T^4$, together with the equation of state $p = (1/3) \;\varrho$, in equation \eqref{E: continuity eq} we arrive at:
\begin{equation}
\frac{d \ln T}{dr} = -\frac{1}{2}\;\frac{d\nu}{dr}.
\end{equation}
This leads us to the temperature dependence on the metric 
\begin{equation}
T(r) = T_0\; e^{-\nu(r) /2}.
\end{equation}
Here $T_0$ is an integration constant that physically corresponds
to the temperature seen by an observer at $r = \infty$, assuming asymptotic flatness $\nu(\infty) = 0$ in the metric given by \eqref{E: sherically symmetric metric}.

For massive fluids, the Euler equation \eqref{E: continuity eq} is still valid, but the equations of state are missing. 
In order to fill in this gap, one can resort to the second law of thermodynamics in its covariant formulation. 
The first to rewrite all the laws of thermodynamics in a covariant notation was Tolman \cite{tolman:1928}, who also introduces the entropy four-vector, which is still used today and will be further analysed in the next chapter.
Furthermore, given the complication in developing the massive case in this way and, given the possibility of obtaining the same result via more direct routes, we will simply give a very brief guideline on how Tolman proceeds. 
For the full details, the reader is encouraged to go to reference \cite{tolman:1930}. 

One can start by defining an entropy vector 
\begin{equation}
S^{\mu} = s\rho\,u^{\mu},
\end{equation}
where $u^{\mu}$ refers to the  matter (or energy) velocity at the point in question and $s$ is the \emph{specific entropy density} as measured by an observer moving with that matter. From this, one might require  the vanishing of the entropy variation in equilibrium states
\begin{equation}
\delta S = 
4\pi \int (\nabla_{\mu}S^{\mu})\sqrt{-g}\; d^4 x =0.
\end{equation}

For the specific case of a perfect fluid in the spherically symmetric spacetime (\ref{E: sherically symmetric metric}), this can be rewritten as:
\begin{equation}
\label{E: entropy symm case}
\delta S = \frac{\delta U + p\delta V}{T} =
 \int_{r_1}^{r_2} \PR{\frac{\delta(\varrho e^{3\mu /2})}{T} +\frac{p }{T}\delta(e^{3\mu /2})} 4\pi r^2 dr =0.
\end{equation}
Applying equation (\ref{E: entropy symm case}) to (\ref{E: continuity eq}) and performing several algebraic steps, together with assumptions about the behaviour of the temperature and pressure at the center of the sphere, Tolman is able to obtain the following result:
%
\begin{equation*}
\frac{d \ln T}{dr} = -\frac{1}{2}\;\frac{d\nu}{dr} \;\;\rightarrow\;\; T = T_0\; e^{-\nu(r) /2},
\end{equation*}
which is the same that followed from the radiation gas analysis.

This result can be extended for other static space-times by noticing that, without any loss of generality, one can always write static metrics in a block diagonal form: 
\begin{equation}
\label{E: block diagonal static metric}
ds^2 = g_{00}\; dt^2 + g_{ij}\;dx^i dx^j.
\end{equation}

Additionally, when dealing with static space-times, the notion of a preferred 4-velocity always exists. In this way, taking the preferred block diagonal form \eqref{E: block diagonal static metric} given above, we have a \emph{unique} naturally defined 4-velocity,
\begin{equation}
\label{E: killing velocity}
V^a = \hat K^a =  {K^a\over||K||},
\end{equation} 
where $K^a$ is the Killing vector given by 
\begin{equation}
\label{E: KillingK}
K^a = (\partial_t)^a = (1,0,0,0)^a, \qquad ||K|| = \sqrt{|g_{00}|}.
\end{equation}
For such observers, Tolman's temperature gradient reads:
\begin{equation}
\label{E: tolman temp}
T(x) = T_0 \; \sqrt{|g^{00}|} = \frac{T_0}{\sqrt{|g_{00}|}}.
\end{equation}
Combining the above result \eqref{E: KillingK} with \eqref{E: tolman temp}, we obtain:
\begin{equation}
T(x) = {T_0\over||K||}.
\end{equation}
This holds for fluids moving along the worldlines generated by \eqref{E: killing velocity} in a spacetime metric given by \eqref{E: block diagonal static metric}.
This is a slightly different way of expressing \eqref{E: tolman temp}  and it is probably the most well known present day formulation of Tolman's temperature gradient. 
The generalization of these relations to the stationary metric case will be provided in section \ref{S: The general case extension}.

\section{The physics behind Tolman temperature gradients}
\label{S: The physics behind Tolman temperature gradients}

Now, before providing more general results,
we would like to discuss the physical aspects of gravity-induced temperature gradients. There is a lot to be discussed, from the magnitude of such an effect, to whether temperature is an observer dependent quantity or not. This is our aim in this section.

\subsection{The static weak field approximation}

Let us start by analysing the static weak field approximation. Factors of $c$ will be kept along this part for greater clarity.  Let us now specifically look to the weak field spherically symmetric spacetime metric, given by:
\begin{equation}
\label{E: flat earth}
ds^2 = -\PC{1 + \frac{2\Phi}{c^2}}\;(c^2\;dt^2) + \PC{1 - \frac{2\Phi}{c^2}}\;\PR{dr^2 + r^2(d\theta^2 + \sin^2\theta d\phi^2)}.
\end{equation}
As we can see, combining equations (\ref{E: flat earth}) and (\ref{E: tolman temp}), we obtain:
\begin{equation}
T(z) \simeq\; T_0 \; \PC{1 - \frac{\Phi}{c^2}},
\end{equation}
which is the weak-field formula for the temperature gradient. 
In the flat-Earth approximation we have $ \nabla \Phi  \longrightarrow g$, giving us:
\begin{equation}
\label{E:weak field tolman}
T(z) \simeq T_0 \; \PC{1 - \frac{gz}{c^2}}.
\end{equation}
Near the surface of the Earth we have $\nabla T(z)/T(z) \approx 10^{-16} \mathrm{m}^{-1}$, which is negligible in almost all experimental settings.  
Another way of phrasing this is that the ``scale height''\footnote{A scale height is a distance over which a quantity decreases by a factor of $e$.} over which the Tolman effect becomes appreciable is $\ell = c^2/g$, which for 1 ``gee'' of acceleration is approximately $\ell_* \approx 9 \times 10^{15}$ metres, a little under one light-year.

\vspace{0.6cm}
\subsection{Planck's blackbody spectrum}
\label{S:planck}

\vspace{1cm}
One of the assumptions made by Tolman when deriving \eqref{E: tolman temp} was the validity of the Stefan--Boltzmann law, i.e. $\varrho = a T^4$, regardless of the presence or absence of a gravitational field.
We shall now explore such an assumption and check whether it is indeed correct.
This analysis will follow the paper \cite{Santiago:2018-ejp} written by the author in collaboration with Matt Visser.

Let us start by simply applying the known gravitational redshift formula of photons to Planck's spectral law. According to Planck, the energy density of a photon gas is given by the integral
\begin{equation}
\label{E:photon gas}
u = \int \frac{b\; \nu^3}{e^{h\nu/k_B T} -1} \; d\nu = a\, T^4, \quad \hbox{where} \quad a = \frac{8\,\pi^5\, k_B^4}{15 \,(hc)^3}\quad \hbox{and}
\quad b= \frac{8\,\pi \,h}{c^3}.
\end{equation}

If this gas is situated in a gravitational field, each individual photon will be subjected to gravitational redshift in a way that, if $\nu_0$ is the frequency of the photon 
at some reference height $z=0$, the frequency seen by one observer at a random height $z$ will be given by:
\begin{equation}
\label{E:redshift}
\nu(z) \simeq \nu_0 \PC{1 - gz/c^2}.
\end{equation}
Consequently, by substituting \eqref{E:redshift} into \eqref{E:photon gas}, we have:
\vspace{0.3cm}
\begin{equation}
\label{E:P1}
u(z) = \bigint \;\; \frac{{b}\; \PC{\nu_0 \PC{1 - gz/c^2}}^3}{e^{h\nu_0 \PC{1 - gz/c^2}/k_B T} -1} \; d\PC{\nu_0 \PC{1 - gz/c^2}}.
\end{equation}
It is possible to directly perform the integration on equation \eqref{E:P1}, and so immediately obtain the Stefan--Boltzmann law. However, we will instead use equation (\ref{E:weak field tolman}) to rewrite the temperature in terms of $T_0$:
\vspace{0.3cm}
\begin{equation}
u(z) = \bigint \;\; \frac{{b}\; \PC{\nu_0 \PC{1 - gz/c^2}}^3}{e^{h\nu_0/k_B T_0} -1} \; d\PC{\nu_0 \PC{1 - gz/c^2}}.
\end{equation}
Dividing the system into horizontal slices, we can focus on specific fixed heights $z$, in a way that $z$ can be treated as a constant.  Doing so, we obtain: 
\vspace{0.3cm}
\begin{eqnarray}
u(z) &=& \bigint \frac{{b}\;\nu_0^3}{e^{h\nu_0/k_B T_0} -1}  \PC{1 - gz/c^2}^4  \; d\nu_0 \nonumber\\
&=& \PC{1 - gz/c^2}^4 \int  \frac{{b}\; \nu_0^3}{e^{h\nu_0/k_B T_0} -1} \;   d\nu_0 \nonumber\\
&=& \PC{1 - gz/c^2}^4 \; a \,T_0^4 = a\, T(z)^4.
\end{eqnarray}
This might naively be misinterpreted as a circular argument, but there is an important physics point here --- self-consistently demonstrating that the validity of the Stefan--Boltzmann law is not affected by the presence of a temperature gradient due to gravity. 
Indeed the argument also shows that the Tolman effect can in principle be fully explained by the gravitational redshift --- which is a purely kinematic effect in any metric theory of gravity.
We will further explore this link between temperature gradients and redshifts in the upcoming section.
But, in short, Tolman's result is completely consistent  with the Stefan--Boltzmann law.

\subsection{How to measure temperatures}
\label{S: How to measure temperatures}

Given this extended technical discussion about thermodynamics and general relativity, one might ask what precise definition of temperature is being used. We will now discuss not only what we mean by temperature but also how to measure it when gravitational gradients are present.
During this section we will be following the discussion in reference \cite{Santiago:2018-ejp} written by the author.

Let us start by introducing the definition of temperature being used in this thesis. Let $S$ be the entropy  and $U$ the internal energy of a small fluid element located at position $x$. The spatially dependent temperature from (\ref{E: tolman temp}) is defined as~\cite{frolov}:
\begin{equation}
\label{temperature}
T(x)^{-1} = \frac{dS}{dU}.
\end{equation} 
Or, in terms of the specific units, we have:
\begin{equation}
\label{E: specific temp}
\frac{1}{T} = \PC{\frac{\d s}{\d \mathfrak{u}}}_{\rho}.
\end{equation}
Here $\mathfrak{u}$ is the specific internal energy defined in section \ref{S: General Relativity}. 

An important question that might arise is this: Temperature, entropy and energy measured by whom? Given that $T(x)$ is normally referred to as ``the locally measured temperature'', the answer must be: Those are the thermodynamic quantities measured by a local observer. But what if another observer, not quite local, decides to do the same measurements? What will she see?

Before answering that question, is it important to know how to calibrate thermometers. Given Tolman's result, $T(x) = T_0 /\sqrt{g_ {00}(x)}$, it is clear that the measurements of each thermometer 
will explicitly depend on their positions. We  might then, in a manner similar to clock synchronization in general relativity, attempt to ``synchronize thermometers''. But to do so, it is necessary to either set the zero of the temperature scale by placing all the thermometers at the same position (or on the same equipotential surface) or to use controlled physical processes at each height to establish the temperature there. Otherwise the temperature gradient (or lack thereof) might merely be an artefact of thermometer calibration. 

\begin{figure}
	\begin{center}
		\includegraphics[scale=0.7]{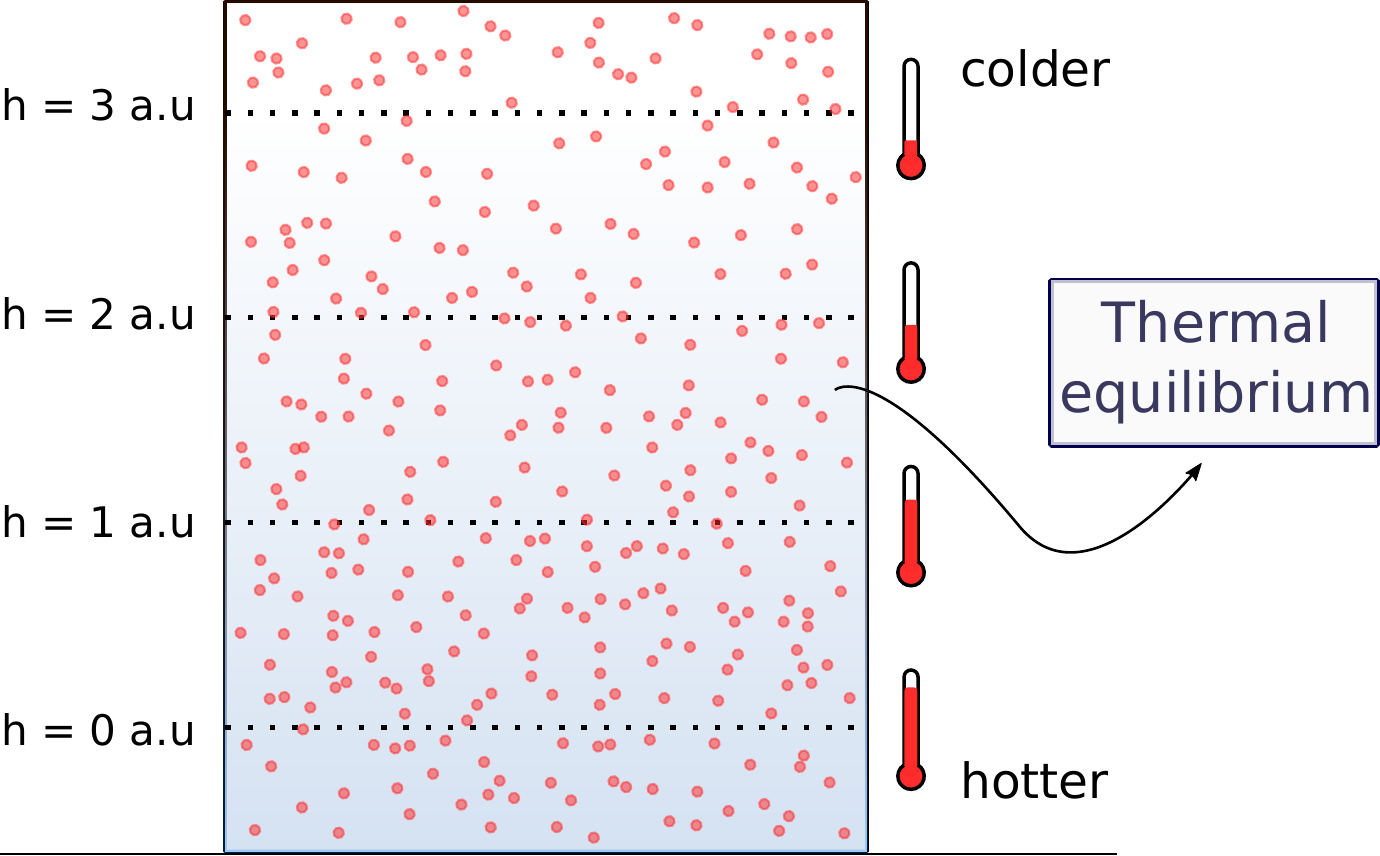}
		\caption[Temperature gradient caused by a gravitational field ]{\label{F:thermometers} Representative picture in arbitrary units (a.u) of the temperature gradient caused by a gravitational field.}
	\end{center}
\end{figure}

Now, let us assume we place the carefully calibrated thermometers at different heights in a gas column, as shown in figure \ref{F:thermometers}. They will keep track of what the local observers are measuring, the position-dependent $T(x)$. However, assume also that there is an observer outside the box that wishes to know what the internal temperature distribution of the gas is, without making any local measurement.

She might do that, for example, by placing some device which opens a small cavity at the desired position, in a way that a sample of the black body radiation of the gas at that height will be sent to her. 
However, in the process of travelling towards the observer, the light frequency will be modified due to gravitational redshift [eq. \eqref{E:redshift}], which will exactly cancel the metric dependence factor in the temperature $T(x)$. 

\begin{figure}
	\center
	\includegraphics[scale=0.7]{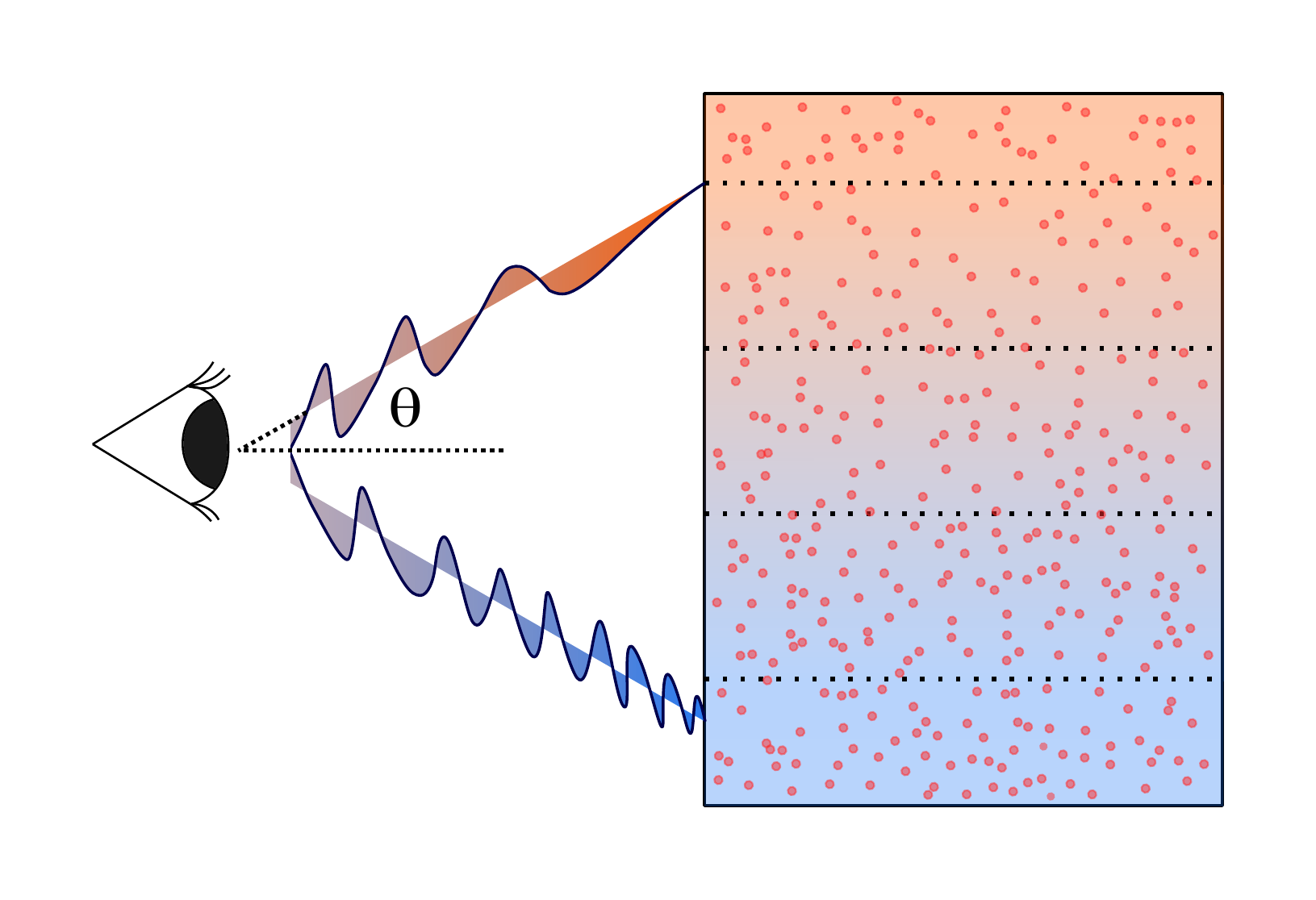}
	\caption[External observer looking at blackbody radiation from a box]{\label{F:thermometers2}External observer looking at photons leaking from the box containing the photon gas,	with the photons arriving at some angle $\theta$ to the vertical.}
\end{figure}

\enlargethispage{0.5cm}

To understand this better, consider the observer to be located at $z=0$ for convenience, looking in a direction which has an angle $\theta$ with respect to the horizontal plane (see figure~\ref{F:thermometers2}). Photons coming from a distance $r$ away from her are coming from a height $z=r \sin\theta$. Suppose, for argument's sake, the Tolman effect was \emph{not present}, (that is, if locally measured temperatures were constant), then a Planck spectrum emitted from $z=r \sin\theta$ would be redshifted/blue\-shifted by a factor $(1+gz/c^2)$ by the time the photons arrive at the observer at $z=0$. However, light rays coming from distinct places will redshift/blueshift differently, in a way that the observer at $z=0$ would see not a simple Planck spectrum, but rather a \emph{superposition} of Planck spectra of different temperatures. But then the radiation gas is not at equilibrium at $z=0$, and we have a \emph{reductio ad absurdum}. 
It is worthwhile to point out that this argument is not valid only for an outside observer, but also for the photons and particles inside the box. Particles composing the fluid are constantly moving both sideways as well as vertically. When moving throughout the fluid this temperature redshifts/blueshifts will also inevitably take place. 
Hence, the only way to avoid inconsistency is if the radiation gas has a position dependent temperature $T(z)=T_0/(1+gz/c^2)$, since then the gravitational redshift guarantees that all these Planck spectra, when seen by the observer at $z=0$ will have the \emph{same} temperature $T_0$. Again, Tolman's result is completely consistent with the Stefan--Boltzmann law and the Planck spectrum for a photon gas in internal equilibrium.

As expected from the universality of free fall, the black body radiation, as seen by an external observer, will never directly ``reveal''  local accelerations in space-time. 
In this way, it also becomes clear that temperature always has to be measured locally (or at worst quasi-locally).

Another interesting point is that the constant temperature seen by the outside observer will depend on the observer's position as well, 
since they will only ``see'' the temperature that is relevant to the equipotential slice on which they are located.
That is included in the physical meaning of $T_0$ in equation \eqref{E: tolman temp}. In the constant gravity case, for example, the higher the observer's position, the smaller the measured $T_0$. In such manner, $T_0$ is indeed a constant for each fixed external observer, but it may vary from one external observer to another. Concluding this discussion, we see that temperature, just as time, has to be measured locally or quasi-locally, even when a system is in thermal equilibrium.

\section{Electrically induced temperature gradients?}

Gravity can change the locally measured temperature distribution of systems in thermal equilibrium. This must now be clear. One question, on the other hand, might remain: can other forces, like electromagnetism, also induce temperature gradients in equilibrium states or not? 

We will, during this section, start with an argument, given by Maxwell in 1868, and use its logic to construct a \emph{gedankenexperiment} (thought experiment) that will answer this question not only for electric forces, but for any non-universal force. 

\subsection{Maxwell's argument}\label{S:maxwell}

Let us start with the argument given by Maxwell~\cite{maxwell:1868} some 150 years ago, regarding the equilibrium temperature of a vertical column of gas. It is based on the second law of thermodynamics and, as we will discus, it is subtly misleading when applied to gravity, although it is fully valid for other forces \cite{Santiago:2018-ijmpd}. Using the more recent 1902 presentation~\cite{maxwell:1902}, the first part of Maxwell's argument goes along these lines, and is certainly valid in all generality: 

\begin{quote}
	\emph{``[...]
		if two vertical columns of different substances stand on the same perfectly conducting horizontal
		plate, the temperature of the bottom of each column will be
		the same; and if each column is in thermal equilibrium of
		itself, the temperatures at all equal heights must be the same. 
		In fact, if the temperatures of the tops of the two columns
		were different, we might drive an engine with this difference of temperature, 
		and the refuse heat would pass down the colder column, through the conducting plate, and up the warmer
		column; and this would go on till all the heat was converted
		into work, contrary to the second law of thermodynamics.''
	}
\end{quote}

This first part of Maxwell's argument establishes that temperature gradients in equilibrium states, if present at all, must be \emph{universal}, otherwise the Clausius version of the second law is violated. (Temperature differences at the \emph{same height} will certainly drive heat fluxes, and would allow one to construct a \emph{perpetuum mobile}.) 

Now this is not exactly what Maxwell originally concluded, because he was primarily interested in \emph{non-relativistic} atomic and molecular gases. The second part of his original argument went as follows:


\begin{quote}
	\emph{``But we know that if one of the columns is gaseous, 
		its temperature is uniform \emph{[from the kinetic theory of gases]}.    
		Hence that of the other must be
		uniform, whatever its material.''
	}
\end{quote}
This second part of Maxwell's argument is now known to be incomplete once one includes relativistic effects. 

As we have seen, to obtain his \emph{reductio ad absurdum} result Maxwell made two quite specific assumptions: 1) that the (non-relativistic) kinetic theory result regarding the temperature of vertical gas column is true, so gases have zero temperature gradient when in thermal equilibrium regardless of the presence or absence of gravity, and, 2) that the temperature gradient, if it exists, is different for distinct substances. These two strong assumptions, when put together, indeed do not leave enough space for evading a \emph{perpetuum mobile}.

Another possible version of this argument, which does not use the kinetic theory result \emph{a priori}, but keeps the substance dependence assumption, can be formulated as follows: Assume you have a vertical column of gas in a gravitational field and suppose that, after equilibrium is reached, a vertical temperature gradient is present. If this is true, we can use a wire or some other heat permeable material to connect the upper and lower parts of the gas container and create, just like in Maxwell's scheme, a \emph{perpetuum mobile} of the second kind. 

The reason this second argument is again misleading is based on the universality of general relativity, which translates to the statement that any form of mass or energy is equally subjected to gravity. With the development of general relativity we became aware that gravity does not concern forces between bodies. It is about space-time, curvatures and geodesics. So, it doesn't matter whether we are looking at a gas, a piece of lead or photons. They will all experience the same metric and the effects that arise from it. 

In this way, we see that if we use a wire to connect the top and the bottom of the gas container, all the atoms comprising the wire (and the phonons within the wire) will also be suffering gravity's influence, in exactly the same way as the atoms in the gas. So the wire itself will exhibit a vertical temperature gradient, which is exactly the same as that in the gas, making the idea of a thermal machine impossible, since all its components would be in thermal equilibrium at every individual horizontal slice. The same argument is valid for Maxwell's two-column system.

Given all the discussion presented in this chapter up to now, we can even rewrite a relativistic version of Maxwell's final conclusion as:

\begin{quote}
\emph{But we know that if one of the columns is a photon gas, its temperature
must be position dependent, as given by Tolman’s relation. Hence that of
the other must be position dependent as well, whatever its material. }
\end{quote}


To conclude, it is important to point out that Maxwell's argument is only \emph{evaded} due to gravity's universality. In that fashion, one might still possibly apply Maxwell's argument  to other forces, as we will do in the following section.

\subsection{The impossibility of electrically induced temperature gradients}\label{S:electric}

Now that we have Maxwell's argument available, we can proceed with the question of whether temperature gradients in equilibrium states could also be generated by other forces or not. Is there a similar effect for some external potential that, for example, break isotropy and homogeneity of space? Or is it specific to general relativity (possibly special relativity) and its many peculiar features?
To clarify this point, we will consider an electric analogue of the gas column in a gravitational field, and analyse some consequences that an electrically induced thermal gradient would create. From them, we will be able to infer something about the plausibility of an electric temperature gradient.  (Spoiler alert: No, it is not plausible.)

Consider an electron gas inside a box. An external electric field will be assumed to act on the whole system for long enough so that the particles already have had sufficient time to rearrange themselves into an equilibrium situation. Assume also that the gas density is very low, so that the force exerted by the external field is much stronger than the interactions between individual electrons
(although they \emph{do} interact in order for thermal equilibrium to be achieved).
If any temperature gradient occurs, it will be aligned with the  direction of the external electric field. For simplicity, assume \emph{no gravitational field is present}.

Let us now (for the sake of the argument) assume that a temperature gradient in the equilibrium configuration does exist and ask what the possible thermodynamic consequences might be? A possible way to answer that question is to take the same path that Maxwell's argument followed. Two columns of different materials are placed on top of a conducting plate. One of the columns is the box with the electron gas inside, while the other will be 
filled with electrically neutral particles, \emph{i.e.}, photons, neutrons, \emph{etc.}
Due to its neutrality, this second column will not interact with the electric field, thus having no reason at all to present a temperature gradient. Continuing the argument on the same lines as before, we might allow heat to flow from the top of one column to the other.
If electrically induced temperature gradients exist, the top of the electron column will have a different temperature from the top of the electrically neutral column. This would then create a heat flow,  enabling the possibility of constructing a perpetual motion machine of the second kind. 
In this way, the existence of electrically induced temperature gradients would violate the second law of thermodynamics. The fact that Maxwell's argument works in this case relies on the fact that, unlike gravity, electric fields are \emph{not universal}, given that the effect it will cause on a particle depends on the particle's electric charge.

\begin{figure}
		\includegraphics[width=1.0\textwidth]{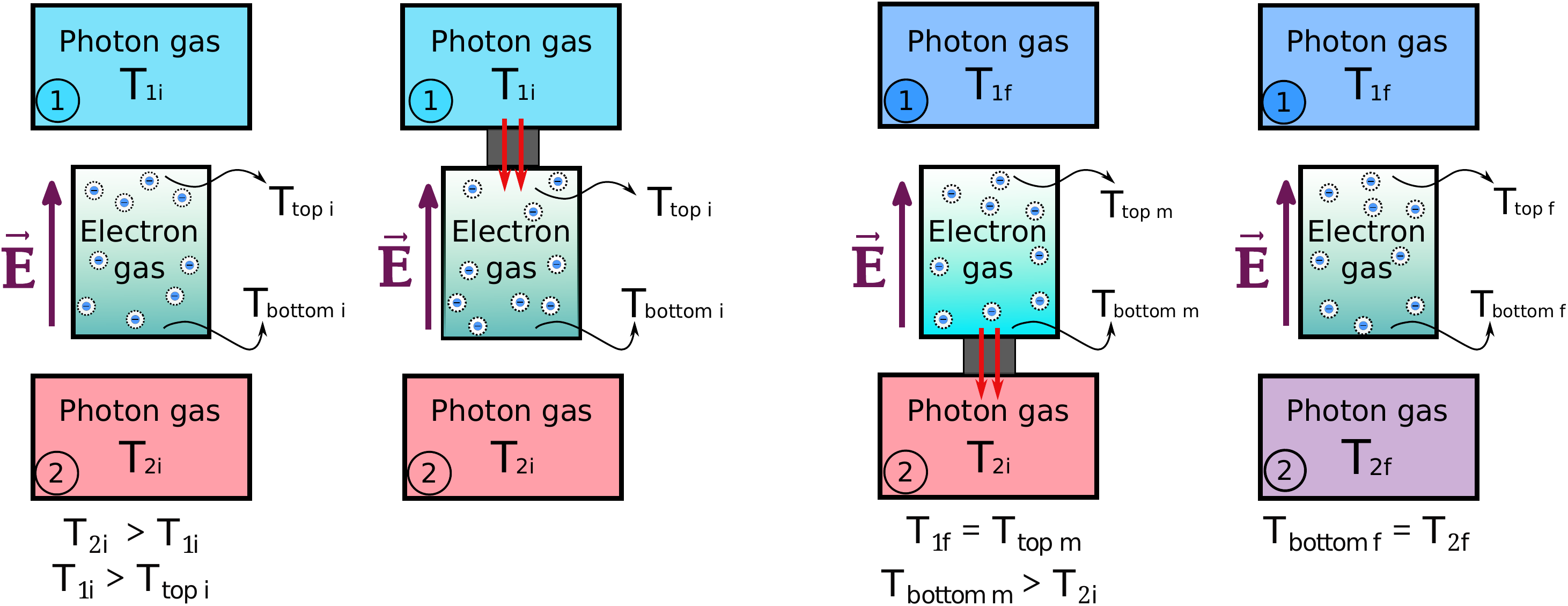}
		\caption[\emph{Gedankenexperiment:} Electric heat engine]{\emph{Gedankenexperiment:} Heat engine showing how heat is being transferred from the cold to the hot photon gas. Since heat flows depend only on the quasi-local distributions of temperature, it is possible to transfer heat from Box 1 to the electron gas, followed by a heat transfer from the electron gas to Box 2. In the final stage we have $T_{1f} < T_{1i}$ and $T_{2f} > T_{2i}$, which violates the second law of thermodynamics.}
		\label{F:heat}
\end{figure}

For the sake of clarity, we will now explicitly show that, if electric fields are able to produce temperature gradients in a gas in thermal equilibrium, then heat engines that violate the second law can be easily created. We will use a 
\emph{gedankenexperiment} to do so.
In the system presented in figure~\ref{F:heat} we have three boxes aligned in the direction of an external constant electric field (vertically). The boxes labelled 1 and 2 contain radiation gas (or any other electrically neutral gas) while the middle container is filled with an electron gas. As the external electric field is applied everywhere, if it can indeed create temperature gradients, the temperatures at the top and at the bottom of the electron gas will be such that $T_{top\; i}  < T_{bottom\; i}$. The temperatures of the photon gases are constant (remember that no gravitational field is present). 

Now we will choose the temperatures of the boxes wisely. Box 1 will be colder than 2, but it will be hotter than the top temperature of the electron box: $T_{top\; i} < T_{1\,i} < T_{2\,i}$. In this way, if we connect  Box 1 with the electron gas, the laws of thermodynamics tell us that heat will flow to the latter  until the top temperature of the electron gas equalizes with Box 1's temperature. The only assumption we are making here is that heat transfer depends on the local temperatures where the boxes touch. So, although the electron gas has (by assumption) a position-dependent temperature, it is the temperature where the contact is made with the photon gas that will tell us whether a heat flow will occur or not. 

After this step, Box 1 will be colder and the electron gas warmer than its initial state, with $T_{1 f} = T_{top\; m} > T_{top\; i}$, and $T_{bottom\; m} > T_{bottom\; i}$ after equilibrium is reached. Additionally, we demand that the temperature of Box 2 be such that after the first heat transfer, $T_{bottom\; m} > T_{2 i}$. In this way, if we now connect Box 2 with the electron gas, given the temperature differences, heat will flow to Box 2 until its temperature equalizes with the bottom temperature of the electron gas. In the final picture we have temperatures satisfying $T_{1f} < T_{1i}$ and $T_{2f} > T_{2i}$. The final average temperature of the electron gas will depend on its own heat capacity as well as on the heat capacity of both photon boxes.

But this means that heat was transferred from a colder to a warmer body, without any work being done on or by the system, which is a clear violation of the second law of thermodynamics. As the construction of the argument is extremely simple and depends only on the non-universal character of the electric force, it is easy to extend it to any force that is not universal.

We might state the conclusion of this argument as: 

\emph{Given that temperature gradients created by any force that is not universal (e.g. dependent on charge, mass, spin,...) allows the creation of heat machines that violate the second law of thermodynamics, these temperature gradients must not exist. }

Going even further, up to date no force other than gravity seems to act on all sources of matter or energy in the same way, regardless of composition or charges. So, if desired, we might even state this as:

\emph{Gravity, and via the equivalence principle, uniform acceleration, are the only effects capable of creating temperature gradients in thermal equilibrium states without violating the laws of thermodynamics.}

\section{The general case extension}
\label{S: The general case extension}

Up to this point, all the discussions remained restricted to the mathematical result obtained by Tolman in 1930. Now, we wish to continue extending the validity of those results to fluids with generic 4-velocities in generic stationary spacetimes. 

The first to attempt an extension of Tolman's results to stationary spacetimes was Buchdahl~\cite{buchdahl:1949} in 1949. His generalization, although valid for any stationary spacetime, 
 kept the requirement that the fluid should be following an orbit of the \emph{specific} timelike Killing vector
\begin{equation}
\label{E: killing vector}
K^a = (\partial_t)^a = (1,0,0,0)^a.
\end{equation}
The result, which we will derive during this section with a modern  calculation, states that if one chooses the fluid to follow the integral curves of the Killing vector (\ref{E: killing vector}), i.e.,
\begin{equation}
\label{E: killing-flow}
V^a = \hat K^a =  {K^a\over||K||},
\end{equation}
in a stationary spacetime, then the equilibrium temperature gradient present in such a fluid is given by:
\begin{equation}
\label{E: temp killing}
T(x) = {T_0\over||K||}.
\end{equation}
Remember that the equivalence between \eqref{E: temp killing} and the result originally obtained by Tolman \eqref{E: tolman temp} was already shown for the static metric case on section \ref{S: The weight of heat}.
Let us now extend this formulation for a broader class of 4-velocities.

\subsubsection{Photon gas}
\label{s:photon}

We will start our analysis by focusing on the simple case of a photon gas. Later, we will extend the validity of the results here obtained to other fluids. 

For a photon gas in internal equilibrium, the following  equations of state are satisfied:
\begin{equation}
\varrho = 3 p = \tilde{a} \,T^4.
\label{E: eos}
\end{equation}
Here the $\varrho=3p$ condition comes from the fact that photons have zero rest mass, 
while $\tilde{a}$ is the radiation constant coming from the Stefan--Boltzmann law. Now consider the relativistic Euler equation for a perfect fluid, given by equation \eqref{E: relativistic euler}:
\begin{equation}
\tag{\ref{E: relativistic euler}}
(\varrho+p) \, a_b =  - h_b{}^c \; \nabla_c p.
\label{E: euler}
\end{equation}
Restricting now our attention to a photon gas, equation (\ref{E: euler}) simplifies to
\begin{equation}
a_b = - h_b{}^c\,\nabla_c \ln T = - (\delta_b{}^c + V_b V^c)\, \nabla_c \ln T.
\label{E: euler2}
\end{equation}
This equation, besides being here obtained for the specific case of a photon gas, will be shown, in section \ref{S:CIT}, to be one of the necessary conditions for any relativistic viscous fluid to be in thermal equilibrium. This will be shown for the theories of Classical Irreversible Thermodynamics (see equation \eqref{E:eq.forces2}) and Extended Irreversible Thermodynamics (equation \eqref{E:EIT equilibrium}). Equation \eqref{E: euler2} is then valid well beyond the perfect photon gas case.

We will now proceed by making the further assumption that, in perfect equilibrium states, the temperature distribution must not vary along the proper time of an observer comoving with the fluid, that is:
\begin{equation}
\label{E: eq condition}
V^b \, \nabla_b T =0.
\end{equation}
Note, however, that \eqref{E: eq condition} is a necessary but not sufficient condition. To properly define equilibrium other state functions will have to be taken into account, as largely discussed in section \ref{S: Thermodynamic Equilibrium}, and as will be fully mathematically analyzed in Chapter \ref{C: Can we still}. For now, though, let us apply \eqref{E: eq condition} into \eqref{E: euler2}:
\begin{equation}
a_b =  - \nabla_b \ln T. 
\label{e:euler3}
\end{equation}

This relation now intimately connects thermal gradients with the 4-acceleration of the photon fluid. One key point is this: Temperature is certainly a scalar, but defining a heat bath also requires you to specify the 4-velocity (and therefore the 4-acceleration) of the heat bath.

Specifically, for any photon gas in free-fall we have $a=0$, and so $T(x)$ is actually a position-independent constant, as expected. Tolman temperature gradients are zero for any fluid following a geodesic path.

In counterpoint, if the heat bath is accelerating, (that is, the 4-acceleration is non-zero), then expanding around some fiducial point $x_0^a$, to lowest order we have
\begin{equation}
T(x) = T(x_0) \; \left\{ 1 + a_b (x^b-x_0^b) + O([\Delta x]^2) \right\}.
\end{equation}
Therefore, for any accelerating thermal bath, we \emph{do} expect temperature gradients in thermal equilibrium. 

\subsubsection{Extension for general fluids}

\enlargethispage{10pt}
As mentioned, equation \eqref{E: euler2} is one of the necessary conditions for relativistic viscous fluids to be in thermal equilibrium (see section \ref{S:Non-Homogeneous Fluids} for the full discussion). Furthermore, as shown by Tolman and Ehrenfest \cite{ehrenfest:1930}, and discussed in section \ref{S:maxwell}, Maxwell's two-column argument shows that, for systems in thermodynamic equili- brium,  the temperature gradient must not depend on the substance, nor on the state of matter. Therefore this result, equation (\ref{e:euler3}), is automatically extended to arbitrary systems in internal thermal equilibrium.

Making this statement clearer: Equation \eqref{e:euler3} tells us the relation between the 4-acceleration and its temperature gradient, regardless of the fluid's composition or whether the space-time is Minkowski, or Schwarzschild,  or Kerr--Newman. The space-time can be flat, curved, stationary, static, whatever --- if the 4-acceleration of the fluid (assumed to obey the relativistic Euler equation and to be in internal equilibrium) is given, the temperature gradient can be obtained.

As it will be discussed in the next chapter, the trickiest part one may find in being able to use equation \eqref{e:euler3} will concern defining and making sure that the notion of thermal equilibrium is still valid for general non-Killing trajectories. We will show that true perfect and eternal thermal equilibrium states can, in fact, only be defined for fluids following Killing flows. On the other hand, situations of near equilibrium or of extremely slow evolution (when compared to the relaxation times of the system) are plentiful. So, keeping those in mind, let us, for now, assume that some notion of equilibrium (or near-equilibrium) exists and make full use of equation \eqref{e:euler3}. This topic, concerning the validity of equilibrium outside of Killing trajectories, will be fully investigated in Chapter \ref{C: Can we still}. For the time being, let us look at some special cases.

\subsubsection{Tolman 1930: Killing flow}

For completeness, let us now see how a simplified derivation of Tolman's result can be obtained. Here, simplified is meant in the sense that this derivation makes it clear that the Einstein equations are not necessary for obtaining relativistic temperature gradients.

Consider a static spacetime with the metric presented in the block-diagonal form of equation
\begin{equation}
\d s^2 = g_{00}\;\d t^2 + g_{ij}\d x^1\d x^j.
\end{equation}
It is a standard well-known result that world-lines ``at rest'', i.e. observer following the Killing trajectories of 
\begin{equation}
V^a = {K^a\over||K||} = {(1,0,0,0)^a\over ||K||}
\end{equation} 
are subject to a non-zero 4-acceleration given by
\begin{equation}
\label{E:aaaa}
a_b = \nabla_b \ln \sqrt{|g_{00}|}.
\end{equation}
A formal proof of this result can be found in the more general Buchdahl result discussed below.
Now, combining \eqref{E:aaaa} with equation~\eqref{e:euler3}  immediately leads to the condition $T(x) \sqrt{|g_{00}|} = \hbox{(constant)}$, which is Tolman's key result \eqref{E: tolman temp}.

\subsubsection{Buchdahl 1949: Killing flow}

From a modern perspective Buchdahl's 1949 result can be extended as follows:
Suppose we have some arbitrary timelike Killing vector (not necessarily the time-translation Killing vector; neither does it need to be hypersurface orthogonal) in a spacetime which is either static or stationary.
Now assume a fluid following some world-line in this metric. We want to know whether this system will exhibit Tolman-like temperature gradients or not.
If we choose the fluid to follow integral curves of the Killing vector, as in 
\begin{equation}
\label{e:killing-flow}
V^a = \hat K^a =  {K^a\over||K||},
\end{equation}
then the fluid 4-acceleration can be easily computed. We start by noting that
\begin{eqnarray}
K^a \nabla_a(g_{bc} K^b K^c) &=& 2 g_{bc}(K^a \nabla_a K^b) K^c 
\nonumber\\
&=& 2 K^a \nabla_{(a} K_{c)} K^c = 0.
\end{eqnarray}
We now compute:
\begin{eqnarray}
a_b &=& V^c \nabla_c V_b =  V^c \nabla_c \left(K_b\over||K||\right) 
= {V^c \nabla_c K_b\over||K||}.\qquad
\end{eqnarray}
Here we have used the fact that $g_{ab}K^aK^b = - ||K||^2$, so  $K^b \nabla_b ||K|| =0$. 
Applying Killing's equation, 
\begin{eqnarray}
a_b &=&  - {V^c \nabla_b K_c\over||K||} = {1\over2} {\nabla_b( ||K||^2)\over ||K||^2}.
\end{eqnarray}
Then
\begin{equation}
\label{e:A-killing}
a_b =  \nabla_b \ln ||K||.
\end{equation}
This purely kinematic result, valid for any Killing flow, is the key part of the calculation. Combining it with equation~\eqref{e:euler3}, this immediately leads to
\begin{equation}
\label{e:tolman eq 1}
T(x) = {T_0\over||K||}.
\end{equation}
Here $K$ is now \emph{any} timelike Killing vector, \emph{as long as the fluid follows integral curves of that same Killing vector}.

It is then clear how temperature gradients depend on the system's 4-velocity. For a distorted rotating space-time (without axial symmetry) there will only be one time-like Killing vector.
For a stationary axisymmetric space-time (for example the Kerr or Kerr--Newman space-times), on the other hand, there are \emph{two ``fundamental''} Killing vectors --- the time-translation and rotational Killing vectors.
Any (constant) linear combination of these Killing vectors  is again a Killing vector --- so there are \emph{infinitely many} time-like Killing vectors to choose from, each one with a different norm, resulting in distinct internal temperature gradients.

The physics message here is this: When applying the Tolman temperature gradient argument in stationary spacetimes, even if you restrict attention to Killing flows,  you have to specify the 4-velocity of the particular thermal bath you are interested in.

\subsubsection{Equilibrium Normal flow}

Given a general stationary spacetime, it can always be locally decomposed into its ADM-like form:
\begin{equation}
\label{e:stationary}
ds^2 = - N^2 dt^2 + h_{ij} \,(dx^i-v^i \,dt)\, (dx^j-v^j\, dt), 
\end{equation}
with inverse
\begin{equation}
g^{ab} = \left[\begin{array}{c|c} 
- 1/N^2 & -v^j/N^2\\ \hline -v^i/N^2 & h^{ij} - v^i v^j/N^2
\end{array}\right],
\end{equation}
For such a spacetime, there is no unique naturally defined 4-velocity. One possible option, as we know, is to keep using the Killing flow, though even the Killing flow will not be unique.

Another appealing option, on the other hand, is to consider the ``normal flow'', which is orthogonal to the constant time slices, such that  $V^a \propto -g^{ab}\, \nabla_b t$:
\begin{equation}
\label{e:normal}
\hat N_a = - {\nabla_a t \over ||\nabla t||} = N\; (-1,0,0,0)_a.
\end{equation}
In static spacetimes the normal flow and Killing flow can be made to coincide, but not otherwise. 
Explicitly, the 4-velocity is given by:
\begin{equation}
\label{e: normal flow}
V^a = \hat N^a = {(1; v^i)\over N},
\end{equation}
or even 
\begin{equation}
\label{e: normal flow2}
V^a =  -{\nabla^a t \over||\nabla t||};
\qquad
||\nabla t|| = \sqrt{-g^{tt}}= {1\over N}.
\end{equation}
Here the minus sign is introduced to keep $V^a$ future-directed. 
To obtain the temperature gradient for a fluid with 4-velocity given by \eqref{e: normal flow}, let us first notice that, since we want the fluid travelling along the normal flow to be in internal equilibrium, the fluid should see a ``time-independent'' environment. We must, in this way, demand the two (somewhat non-trivial) compatibility conditions, 
\begin{equation}
\label{e:compatibility1}
V^a\nabla_a N =0;  
\end{equation}
and
\vspace{-0.3cm}
\begin{equation}
\label{e:compatibility2}
V^a\nabla_a p = 0.
\end{equation}
The second compatibility condition is actually the natural extension of the previously imposed thermal equilibrium condition $V^a\nabla_a T = 0$, originally applied to a photon gas to obtain \eqref{e:euler3}, but now extended to general fluids. But the motivation for all such compatibility conditions is basically the same: If a fluid is in thermal equilibrium, it should not have its state variables changing along its proper time. Again, true perfect equilibrium states can only be defined for fluids following Killing trajectories. But, as previously mentioned, we will assume that quasi-equilibrium states exist for now, and further discuss this subject in the next chapter.

\enlargethispage{15pt}
Also, for such an equilibrium-compatible normal flow, calculating the 4-acceleration is easy but  slightly different from the calculation for a Killing flow:
\begin{eqnarray}
a_b &=& V^c \nabla_c V_b =  -V^c \nabla_c \left(\nabla_b t\over||\nabla t||\right)
=
-{V^c \nabla_c \nabla_b t\over||\nabla t||}.\qquad
\end{eqnarray}
We cannot apply Killing's equation anymore. Instead, we can use $\nabla_b \nabla_a t = \nabla_a \nabla_b t$, so that
\vspace{-0.3cm}
\begin{eqnarray}
a_b &=& - {V^c \nabla_b \nabla_c t\over||\nabla t||} 
= -{1\over2} {\nabla_b( ||\nabla t||^2)\over ||\nabla t||^2}.
\end{eqnarray}
In this way, for a normal flow satisfying the compatibility condition \eqref{e:compatibility1}, we have the following purely kinematic result:
\vspace{-0.3cm}
\begin{eqnarray}
\label{e:A-normal}
a_b &=&   - \nabla_b \ln ||\nabla t||.
\end{eqnarray}
Given equation \eqref{e: normal flow2}, this is equivalent to
\vspace{-0.3cm}
\begin{eqnarray}
\label{e:A-normal2}
a_b &=&  \nabla_b \ln N.
\end{eqnarray}
Note $V^b a_b =0$.  This is formally somewhat similar to Buchdahl's result for Killing flows, see equation \eqref{e:A-killing},
with $||K||\to N$. 

\newpage
In \emph{static} spacetimes (in block diagonal form) we have $g_{00} \, g^{00} = 1$, implying that for the time translation Killing vector 
$||\nabla t|| \, ||K|| =1$. Therefore, for static spacetimes, both Tolman's original computation for 4-acceleration as the normal flow calculation just shown can be made to coincide. For \emph{stationary} spacetimes, on the other hand, they can and typically will be physically different.

Combining equation \eqref{e:A-normal2} with equation~\eqref{e:euler3} immediately leads to
\begin{equation}
\label{e:new}
T(x) = T_0 \; ||\nabla t|| = T_0 \; \sqrt{-g^{tt}} = {T_0\over N}.
\end{equation}
This is the analogue of Buchdahl's 1949 result, but now applied to (equilibrium compatible) normal flows.
Note this is a very different physical setup from the Buchdahl 1949 result~\cite{buchdahl:1949}, even if the final result superficially looks very similar.

\section{The rotating universe example}
\label{S: The rotating universe example}

\begin{center}
	\begin{figure*}[!htb]
		\center
		\includegraphics[width=0.55\textwidth]{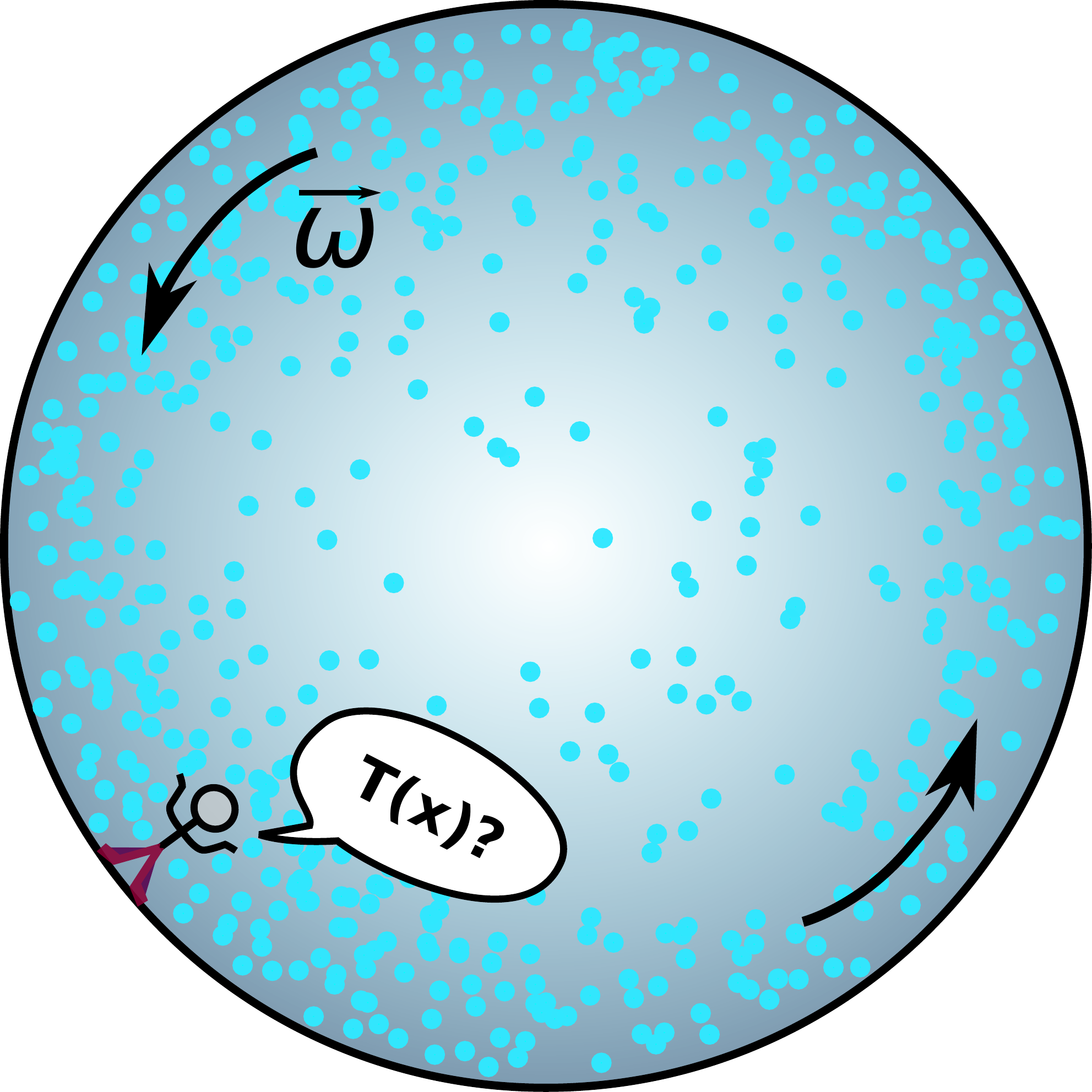}
		\caption[Rotating cylinder]{\emph{Gedankenexperiment:} What temperature gradient is seen inside a rotating cylinder that has come to internal thermal equilibrium? }
		\label{F:disk-1}
	\end{figure*}
\end{center}

We will now explore an example where we evaluate and understand thermodynamic equilibrium states for thermal baths seen by observers in a rotating cylinder.
We will do the calculations from the point of view of an outside observer and interpret the final results from both the external as well as from the internal observer's point of view.
In this case we have two important coordinate systems, Cartesian $(t, x, y, z)$ coordinates for the external observer and co-moving $(t, r, \theta, z)$ coordinates for the observer moving with the rotating cylinder.
We will consider a thermodynamic system which will be placed inside the rotating cylinder for long enough in order for thermodynamic equilibrium to be achieved. The questions we will answer here are  i) what is the temperature distribution inside such system and ii) what observers inside the cylinder will actually observe (see figure \ref{F:disk-1}).

\subsubsection{Temperature distribution}

As we have seen, 
if the fluid is following some Killing trajectory, given the norm of the relevant Killing vector, it is possible to obtain the Tolman-like temperature gradient present in thermal equilibrium states.

Let us start with the metric seen by the co-moving observers, in rotating cylindrical polar coordinates:
\begin{equation}
\label{E: rotating metric}
ds^2 = -dt^2 + dr^2 +r^2 (d\phi-\omega dt)^2 + dz^2,
\end{equation}
which is obtained simply by performing a coordinate transformation $\phi_{new} = \phi_{old} + \omega t$ on the static cylindrical polar coordinates. It is nice to keep in mind that this is just flat Minkowski space, written in co-rotating cylindrical polar coordinates. The Riemann tensor is still zero and, from a modern perspective, this is just special relativity in disguise.  Rearranging the terms we get:
\begin{equation}
ds^2 = -dt^2(1-\omega^2 r^2) -2r^2\omega d\phi dt  + dr^2 +r^2 d\phi^2 + dz^2.
\end{equation}

In this coordinate system the gas follows trajectories of the Killing field $K^a = (1,0,0,0)$, with $V^a = (1,0,0,0)/||(1,0,0,0)||$. Specifically,
\begin{equation}
||K|| = \sqrt{1-\omega^2 r}.
\end{equation}
Applying this result to \eqref{e:tolman eq 1},
we can easily obtain the temperature distribution across the rotating cylinder:
\begin{equation}
\label{E: temperature grad}
T(x) = {T_*\over||K||} = {T_* \over \sqrt{1-\omega^2 r^2}}.
\end{equation}

This equation tell us that any system which is in thermodynamic equilibrium in a rotating cylinder will have an internal temperature gradient which depends both on its angular velocity as well as on the radial distance from the axis  of rotation. $T_*$ is the temperature at the center of the cylinder and it drops for larger radius positions.


\vspace{0.5cm}
\subsubsection{Redshift} 

\vspace{0.5cm}
We may as well ask what an observer inside the disk will see. In order to answer this question, let us assume the thermal system to be emitting photons with a blackbody radiation spectrum. By receiving these photons, the internal observer is able to know the temperature distribution throughout the system. For simplicity, we will consider only the case where the thermal bath and the observer are at rest in respect with each other, so they are co-rotating with the cylinder. 
Given that the metric seen by the co-moving observer is given by equation (\ref{E: rotating metric}), from their point of view photons will suffer redshifts/blueshifts when moving around. In this way, to know the thermal spectrum measured by them, we need to take these details into account. 

\begin{figure*}
	\center
	\includegraphics[width=0.5\textwidth]{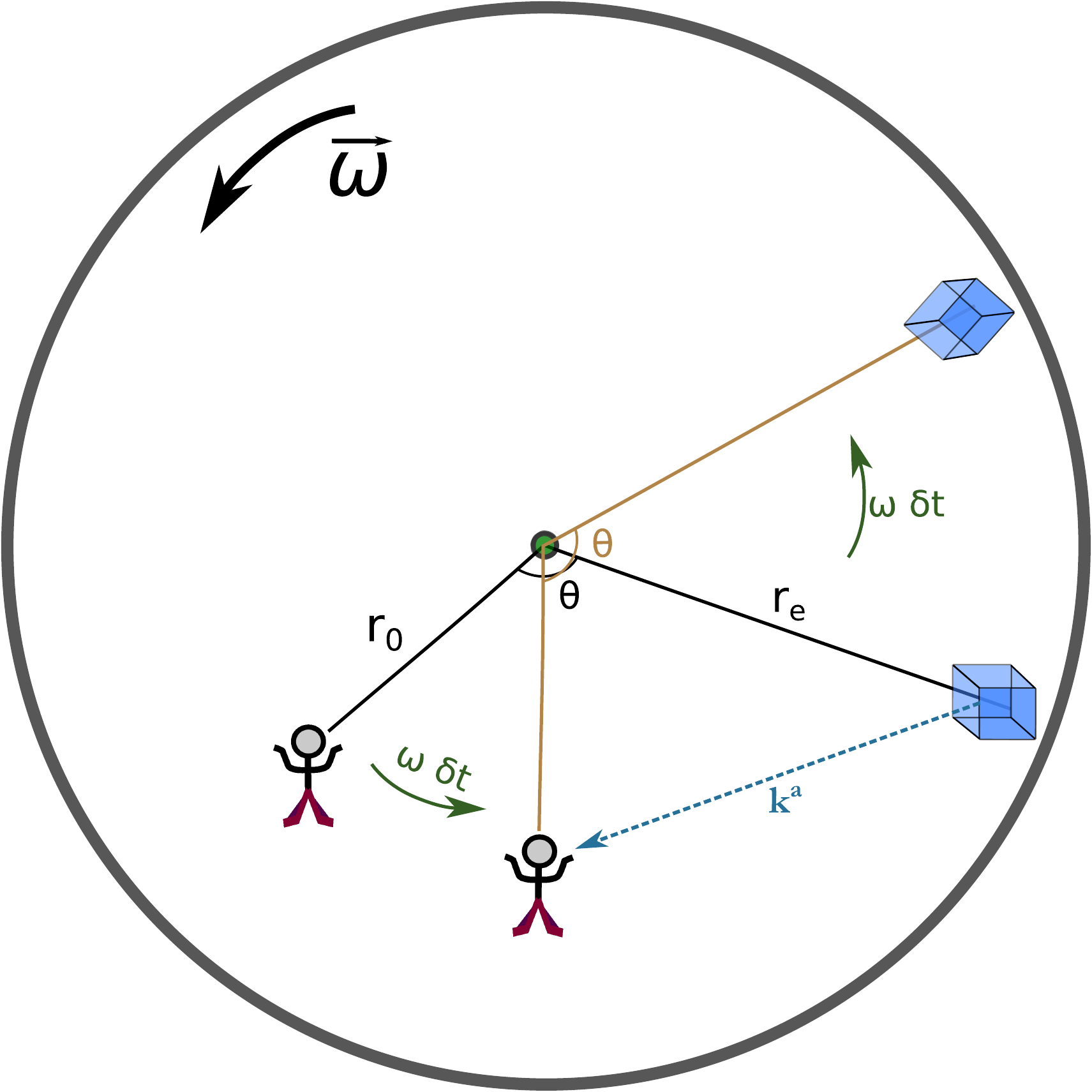}
	\caption[Photon emission/absorption in a rotating cylinder]{The view from an outside observer on the emission and absorption of a light ray $k^a$ from a certain fluid element at radial position $r_e$ to an observer with $r=r_o$. The two configurations (yellow and black lines) are separated by time $\delta t$.}
	\label{F:light}
\end{figure*}

Fortunately, this is one of those happy moments where a change in the reference frame can make calculations simpler. In this way, we will adopt the external (static) observer point of view to calculate the redshift factors. The reason being that, since the external metric is flat Minkowski spacetime in Cartesian coordinates, the path followed by the emitted photons will simply be straight lines as seen by an external observer. In this case, the redshifts/blueshifts will be interpreted as being due to Doppler effects, given that from the external observer's point of view, the thermal bath and the internal observer are moving away/towards the emitted photons (see Figure \ref{F:light}). 

So, defining  $V_e^a$ to be the 4-velocity of the emitter (thermal bath), $V_o^a$ the 4-velocity of the internal (co-moving) observer, and $k^a$ the null vector connecting source and observer, the redshift is given by the standard formula
\begin{equation}
1+z = {(g_{ab} V_e^a k^b)_e\over (g_{ab} V_o^a k^b)_o} = \frac{\nu_e}{\nu_o},
\end{equation}
where $\nu_e$ and $\nu_o$ are the emitted and observed frequencies respectively.
Using $(t,x,y,z)$ coordinates, let the emission event take place at
\begin{equation}
X_e^a = (0,r_e,0,0);  \qquad   V_e = \gamma_e (1,0,\omega r_e,0)
\end{equation}
and let the observation event take place at
\begin{equation}
X_o^a = (\delta t,r_o\cos\theta,r_o\sin\theta,0);  \qquad   
V_o = \gamma_o (1,-\omega r_o\sin\theta,\omega r_o\cos\theta,0),
\end{equation}
as shown in Figure \ref{F:light}. Then we have for the light displacement 
\begin{equation}
\delta X^a = (\delta t, r_o\cos\theta-r_e,r_o\sin\theta,0).
\end{equation}
In this way, we can calculate the 4-vector tangent to the light ray leaving the emitter and arriving at the observer: 
\begin{equation}
k^a = {\delta X^a\over \delta t} = \left(1; {r_o\cos\theta-r_e\over \delta t},{r_o\sin\theta\over \delta t},0\right).
\end{equation}
For completeness we note that as $\,k^a k_a =0$, we have
\begin{equation}
\delta t^2 = (r_o\cos\theta-r_e)^2+ (r_o\sin\theta)^2 = r_o^2+r_e^2-2r_or_e\cos\theta.
\end{equation}
That is
\begin{equation}
\delta t = \sqrt{r_o^2+r_e^2-2r_or_e\cos\theta}.
\end{equation}
\enlargethispage{10pt}
So the photon's time-of-flight is particularly simple and exactly what one would expect (from the law of cosines).
Now, remembering that the metric for the external observer is simply $\eta = \mathrm{diag}(-1,1,1,1)$,  the emitted frequency $\nu_e$ will be proportional to
\begin{eqnarray}
(g_{ab} V^a k^b)_e &=& \gamma_e (1,0,\omega r_e,0)\; \eta  \left(1; {r_o\cos\theta-r_e\over \delta t},{r_o\sin\theta\over \delta t},0\right)
\nonumber\\
&=&
\gamma_e \left(-1 +{\omega r_er_o\sin\theta\over \delta t}\right).
\end{eqnarray}
Similarly,  the observed frequency $\nu_o$ will be proportional to
\begin{eqnarray}
(g_{ab} V^a k^b)_o &=& \gamma_o (1,-\omega r_o\sin\theta,\omega r_e\cos\theta,0)\; \eta  \left(1; {r_o\cos\theta-r_e\over \delta t},{r_o\sin\theta\over \delta t},0\right)
\nonumber\\
&=&
\gamma_o \left(-1 +{\omega r_er_o\sin\theta\over \delta t}\right).
\end{eqnarray}
Explicitly:
\begin{equation}
(g_{ab} V^a k^b)_e
=
\gamma_e \left(-1 +{\omega r_er_o\sin\theta\over \delta t}\right);
\quad
(g_{ab} V^a k^b)_o
=
\gamma_o \left(-1 +{\omega r_er_o\sin\theta\over \delta t}\right).
\end{equation}
This then gives us the redshift formula for any two points inside the cylinder:
\begin{equation}
\label{E: redshift}
1+z =  {(g_{ab} V_e^a k^b)_e\over (g_{ab} V_o^a k^b)_o}= {\gamma_e\over\gamma_o} = \sqrt{1-\omega^2 r_o^2\over1-\omega^2 r_e^2},
\end{equation}
which now explicitly shows  how (as expected)  the redshift factor depends on both positions, that of the emitter and the observer, as well as on the angular velocity.

\subsubsection{What is the temperature seen by the co-moving observer?}

Combining the temperature distribution results from equation (\ref{E: temperature grad}) together with the redshift factor just obtained in equation (\ref{E: redshift}), we can now calculate the light spectrum seen by the co-moving observer. 

As the blackbody spectrum is emitted by the rotating gas, which is in internal thermal equilibrium, we have for its temperature distribution
\begin{equation}
\label{E: temp emitter}
T(x_e) = {T_* \over \sqrt{1-\omega^2 r_e^2}}.
\end{equation}

Given Wien's displacement law, if $\nu_*$ is the maximum emission frequency at $r_e=0$, the frequency at a random emission point will be: 
\begin{equation}
\nu_e = \frac{\nu_*}{\sqrt{1-\omega^2 r_e^2}}.
\end{equation}
Now, given that $\nu_e / \nu_o = 1 + z$, we have
\begin{equation}
\nu_o = \frac{\nu_e}{1 + z} = \frac{\nu_*}{\sqrt{1-\omega^2 r_e^2}}\; \sqrt{\frac{1-\omega^2 r_e^2}{1-\omega^2 r_o^2}} = \frac{\nu_*}{\sqrt{1-\omega^2 r_0^2}}.
\end{equation}
Again using Wien's law, we know that the temperature seen by the co-moving observer coming from any point inside the rotating cylinder will be given by:
\begin{equation}
T(x_{o}) = {T_* \over \sqrt{1-\omega^2 r_{o}^2}},
\end{equation}
which is exactly the equilibrium temperature at the observer's location. 
This is the result we were aiming for. This shows that the temperature seen by the rotating observer is constant, regardless of the presence of the internal temperature gradient present in the thermal bath. This is the same as was the case for a static observer in a constant gravitational field as explained in section \ref{S: How to measure temperatures}, particularly in Figure \ref{F:thermometers2}. 
Furthermore, generalizations of this idea can be used to capture the position-dependence of the locally measured Hawking temperature for rotating Kerr black holes~\cite{Santiago:2018-prd}. We will explore this in the next section.

\subsubsection{Consistency checks}
\enlargethispage{10pt}

Let us now look at some simple cases to evaluate whether our results are in internal agreement or not.
First, assume the case where the observer is on the axis of rotation of the cylinder ($r_o=0$). In this case, the redshift will be given by:
\begin{equation}
1+z = {\gamma_e} = {1\over\sqrt{1-\omega^2 r_e^2}}.
\end{equation}
This result exactly agrees with what you would expect based on the transverse Doppler shift, so all good up to now.

Second, it is easy to see that a  gas with a position dependent angular velocity cannot be in internal thermal equilibrium. Consider what happens if we try to replace $\omega \to \omega(r)$.  The velocities become:
\begin{equation}
V_e = \gamma_e \left(1,0,\omega_e r_e,0\right);
\qquad
V_o = \gamma_o \left(1,-\omega_o r_o\sin\theta,\omega_o r_o\cos\theta,0\right).
\end{equation}
Then we see
\begin{equation}
(g_{ab} V^a k^b)_e
=
\gamma_e \left(-1 +{\omega_e r_er_o\sin\theta\over \delta t}\right);
\end{equation}
and
\begin{equation}
(g_{ab} V^a k^b)_o
=
\gamma_o \left(-1 +{\omega_o r_er_o\sin\theta\over \delta t}\right).
\end{equation}
Whenever $\omega_e \neq \omega_0$ there is no longer a nice factorization, instead we have
\begin{equation}
1+z = {\gamma_e \left(-1 +{\omega_e r_er_o\sin\theta\over \delta t}\right)\over\gamma_o \left(-1 +{\omega_o r_er_o\sin\theta\over \delta t}\right)} .
\end{equation}
This implies
\begin{eqnarray}
1+z =& \sqrt{1-\omega_o^2 r_o^2\over1-\omega_e^2 r_e^2}
\left( {\sqrt{r_o^2+r_e^2-2r_or_e\cos\theta} -{\omega_e r_er_o\sin\theta}\over\sqrt{r_o^2+r_e^2-2r_or_e\cos\theta} -{\omega_o r_er_o\sin\theta}} \right)\\
=& \sqrt{1-\omega_o^2 r_o^2\over1-\omega_e^2 r_e^2}
\chav{1 -\frac{\PC{\omega_o -\omega_e}}{\omega_o} \PC{1-\frac{\delta t}{\omega_o r_er_o\sin\theta}\, }^{-1}},
\end{eqnarray}
where we have used the time-of-flight of the photon $\delta t$. This is nowhere near as nice as the case $\omega_e=\omega_0$. Once  $\omega_e\neq \omega_0$, you can no longer nicely separate the effects of emitter and observer.
Also, once $\omega\to\omega(r)$ you cannot do the simple coordinate transformation $\phi_{new} = \phi_{old} + \omega t$ which allowed you to use the modified Buchdahl result. But the worst problem is this: Suppose one somehow finds a formula for the emission temperature that depends only on the properties of the emission point, ($r_e$ and $\omega_e$). Then one must have a relation of the form
\begin{equation}
T_e = T_* \; f(r_e\omega_e).
\end{equation}
But this implies that the observer will see not one temperature, but a \emph{superposition} of blackbody spectra of different temperatures
\begin{eqnarray}
T_o(r_o,\omega_o;r_e,\omega_e,\theta) &=& {T_e\over 1+z} = {T_* \; f(r_e\omega_e)\over1+z}  
\\
&=&
T_*  f(r_e\omega_e) \sqrt{1-\omega_e^2 r_e^2\over1-\omega_o^2 r_o^2} \chav{1 -\frac{\PC{\omega_o -\omega_e}}{\omega_o}\PC{1-\frac{\delta t}{\omega_o r_er_o\sin\theta} }^{-1}}.
\nonumber
\end{eqnarray}
This inextricable entangling of emitter and observer implies that for $\omega\to\omega(r)$ the spectrum seen at the observer cannot be Planckian, so the gas \emph{cannot} be in internal thermal equilibrium. Of course this dis-equilibrium could also be derived from the fact that differential rotation implies shear, which, for viscous fluids, implies friction. The redshift argument is however purely kinematic and does not need to appeal to any dynamics.
In particular, this is an elementary way of seeing that a differentially rotating (classical Newtonian gravity) star cannot be in internal thermal equilibrium. Probably, with a bit more work this type of argument can be extended to fully general relativistic stars.

\enlargethispage{10pt}
\section{Black Hole examples}
\label{S: Some other examples}

 \vspace{0.5cm}
Let us now take some time to evaluate some other applications of equation \eqref{e:euler3}. We will start with the free-fall cases and then look at Killing flows in Kerr spacetimes, finishing with an analysis of the normal flow also for the Kerr metric.

 \vspace{0.5cm}

\paragraph{Schwarzschild/Reissner--Nordstrom: Free-fall normal flow} \hfill
 
 For either Schwarzschild or Reissner--Nordstrom spacetimes let us choose to use Painleve--Gullstrand coordinates~\cite{painleve,gullstrand,unruh:1981,unexpected,acoustic,LRR}, given by
\begin{equation}
ds^2 = -dt^2 + \left(dr - \sqrt{2m(r)\over r} \; dt\right)^2 + r^2(d\theta^2+\sin^2\theta\,d\phi^2).
\end{equation}
Such a spacetime is static, but not \emph{manifestly} static, since we have chosen to write the metric in non-diagonal form. 

Consider the normal flow $V^a \propto - g^{ab}\,\nabla_b t$. Given equation \eqref{e:stationary}, we see that $N=1$, and $||\nabla t|| = 1/N =1$, from which equation \eqref{e:A-normal2}, i.e.,
\begin{equation*}
a_b = \nabla_b \ln N
\end{equation*} 
implies a zero 4-acceleration. That is, our ``reference fluid'' is in free-fall. Using then equation \eqref{e:euler3}, $a_b = \nabla_b \ln T$, we obtain that $T(x)=\hbox{(constant)}$.

So we explicitly see that a fluid in a freely falling box (in Schwarzschild or Reissner--Nordstrom spacetime) will \emph{not} exhibit a Tolman temperature gradient, as expected from the equivalence principle.
Furthermore, this particular normal flow automatically satisfies the compatibility conditions~\eqref{e:compatibility1} and ~\eqref{e:compatibility2} \emph{a priori}.

For completeness, the explicit expression for the 4-velocity of the relevant thermal bath is given by:
\begin{equation}
V^a =  \left( 1; \sqrt{2m(r)/r},0,0\right).
\end{equation}

\paragraph{Static spherically symmetric spacetimes}\hfill
 
Any  static spherically symmetric spacetime can (at least locally) be put in the form
\begin{equation}
ds^2 = -dt^2 + h(r) \left(dr - v(r) \; dt\right)^2 + r^2(d\theta^2+\sin^2\theta\,d\phi^2).
\end{equation}
This spacetime is static, but not manifestly static, since we have chosen to write the metric in non-diagonal form.
The normal flow is given by:
\begin{equation}
V^a =  \left( 1; v(r),0,0\right).
\end{equation}
This case is again about a geodesic flow.  A freely falling fluid following this trajectory will not see any Tolman temperature gradient.

\paragraph{Kerr/Kerr--Newman: Free-fall normal flow}\hfill

For the Kerr or Kerr--Newman spacetime, let us choose to work in the Doran coordinate system~\cite{doran,river}:
\begin{eqnarray} 
ds^2 = &-&dt^2 + (r^2 +a^2\cos^2\theta)^2 d\theta^2 + (r^2 +a^2) \sin^2\!\theta \, d\phi^2\\
&+& \left[\frac{ r^2 +a^2\cos^2\theta}{r^2+a^2}\right] \left( dr + 
\frac{\sqrt{2mr(r^2 +a^2)}}{r^2 +a^2\cos^2\theta}( dt - a \sin^2\!\theta \, d\phi) \right)^2.\nonumber
 \end{eqnarray}
The normal flow, in these Doran coordinates, is
\begin{eqnarray}
\hat N_a &=& - {\nabla_a t} = {(-1;0,0,0)_a}.
\end{eqnarray}
We have $||\nabla t|| = N^{-1} =1$.  
From equation \eqref{e:A-normal2} this implies $a=0$. That is, our ``reference fluid'' is now in free-fall, obeying the compatibility conditions \eqref{e:compatibility1} and \eqref{e:compatibility2}, and we again deduce $T(x)=\hbox{(constant)}$.

Thus, again we see that a gas confined in a freely falling box (in Kerr or Kerr--Newman spacetime) will \emph{not} exhibit a Tolman temperature gradient which, as in the Schwarzschild case, is exactly what you should expect based on the equivalence principle.

\paragraph{Kerr/Kerr-Newman: Some Killing flows}\hfill

In the Boyer-Lindquist coordinate system
\begin{eqnarray}
\label{E:boyer-linquist}
		ds^2 = &-& \left[ 1- {2mr\over r^2+a^2\cos^2\theta}\right] dt^2 
		- {4mra\sin^2\theta\over r^2+a^2\cos^2\theta}\; dt\;d\phi 
		+ \left[{r^2+a^2\cos^2\theta\over  r^2-2mr+a^2}\right] dr^2 
		\nonumber
		\\
	&+& (r^2+a^2\cos^2\theta) \;d\theta^2
		+ \left[r^2+a^2+ {2mr a^2 \sin^2\theta\over r^2+a^2\cos^2\theta}\right] \sin^2\theta\;d\phi^2.
\end{eqnarray}
we have the ``natural'' timelike Killing vector $(1,0,0,0)$ plus the rotational Killing vector $(0,0,0,1)$. In this way, any vector of the form $(1,0,0,\Omega)$ will also be timelike Killing vectors. 
Looking at some interesting cases.
\begin{itemize}
	\item 
	The $\Omega=0$ Killing vector $(1,0,0,0)$ is well behaved at spatial infinity, giving us:
	\begin{equation}
	T(x) = {T_0\over\sqrt{-g_{tt}}} = {T_0\over\sqrt{N^2-h_{ij} v^i v^j}},
	\end{equation}
	where $v^i$ and $h_{ij}$ is defined in \eqref{e:stationary}.
	However, for both Kerr or Kerr--Newman, its norm $||(1,0,0,0)||$ is zero \emph{at the ergosurface} --- not \emph{at the horizon}. This, clearly, is not surprising, since the physical property which defines the ergosphere is the impossibility of entering that region of spacetime without spinning in the direction of the black hole. For an observer to keep a 4-velocity  $(1,0,0,0)$, they would need to have infinite acceleration, explaining the ``infinite temperature'' that they would see in case this was possible.
	
	\item
	For Kerr or Kerr--Newman, setting  $\Omega\to\Omega_H$ the angular velocity of the horizon, the Killing vector $(1,0,0,\Omega_H)$ has a norm $||(1,0,0,\Omega_H)||$, which is zero \emph{at the horizon} --- not \emph{at the ergosurface}.  
	But this Killing vector has the annoying feature that its norm also vanishes in the exterior asymptotic region, near $r\sin\theta \approx1/\Omega_H$. (This is merely an ``annoyance'', not a ``problem'', the same thing happens for a rotating coordinate system in flat Minkowski space.)
	In this situation
	\begin{equation}
	T(x) = {T_0\over\sqrt{N^2-h_{\phi\phi} (v_\phi-\Omega_H)^2}}.
	\end{equation}
	This clearly is a different generalization of Tolman's result.
\end{itemize}
So Killing vectors in Kerr/Kerr-Newman spacetimes  are either well behaved at spatial infinity, but problematic at the ergosurface; or are well-behaved at the horizon but problematic sufficiently far from the axis of rotation. Worse, if we take a generic constant $\Omega$ such that $0\neq \Omega\neq \Omega_H$ then the resulting Killing vector $K^a=(1,0,0,\Omega)$ has null surfaces (and so formally infinite local Tolman temperatures) that correspond neither to the horizons nor to the ergosurfaces. This now leads us to analyze what happens when the flows are not generated by Killing vectors.

\paragraph{Kerr/Kerr-Newman: ZAMO normal flow}\hfill

In the specific case of axial symmetry, the normal flow $V^a \propto - g^{ab} \nabla_b  t$ is often referred to as a ZAMO flow;  the ``Zero Angular Momentum Observer'' flow. Now let us further specialize to Boyer--Lindquist coordinates (\ref{E:boyer-linquist}), where (under mild technical conditions) we can, using $(t,r,\theta,\phi)$ coordinates, block diagonalize the metric into the form~\cite{Wald,MTW}:
\begin{equation}
g_{ab} = \left[\begin{array}{c|cc|c} 
g_{tt}  & 0 & 0& g_{t\phi}\\
\hline
0 & g_{rr} &0 &0\\
0&0&g_{\theta\theta}&0\\
\hline
g_{t\phi}&0&0&g_{\phi\phi}
\end{array}\right].
\end{equation}
The inverse metric is easily computed
\begin{equation}
g^{ab} = 
\left[\begin{array}{c|cc|c} 
g_{\phi\phi}/ g_2  & 0 & 0& - g_{t\phi} /g_2 \\
\hline
0 & 1/g_{rr} &0 &0\\
0&0&1/g_{\theta\theta}&0\\
\hline
-g_{t\phi} /g_2 &0&0&g_{tt} /g_2
\end{array}\right].
\end{equation}
Here $g_2 = g_{tt} \,g_{\phi\phi}-g_{t\phi}^2$, and $\det(g_{ab})= g_2\,g_{rr}\,g_{\theta\theta}$.

Note that $g_{tt}=0$ defines the ergosurfaces, where the time translation Killing vector $(1;0,0,0)^a$ becomes null. In contrast, horizons are defined by the condition $g^{tt}=\infty$, equivalent to $(g^{tt})^{-1}=0$. If $g_{t\phi}\to 0$, then horizons and ergosurfaces coalesce, but for $g_{t\phi}\neq 0$ they are distinct.

The normal flow, in these Boyer--Lindquist coordinates, is then
\begin{eqnarray}
\hat N_a &=& - {\nabla_a t \over ||\nabla t||} = {(-1;0,0,0)\over \sqrt{-g^{tt}}} = 
\sqrt{-g_2\over g_{\phi\phi}} \; (-1;0,0,0)
\nonumber\\
&=& \sqrt{-g_{tt} +{g_{t\phi}^2\over g_{\phi\phi} }} \;(-1;0,0,0).
\end{eqnarray}
The corresponding flow vector (contravariant vector) is 
\begin{equation}
V^a = \hat N^a =  \sqrt{g_{\phi\phi}\over -g_2 } \; \left(1;0,0,-{g_{t\phi}\over g_{\phi\phi}}\right).
\end{equation}
In terms of the time translation and axial Killing vectors,
(and now defining $\varpi = - g_{t\phi}/g_{tt}$), we have
\begin{equation}
V^a = {[K_T]^a + \varpi [K_\phi]^a\over ||K_T + \varpi K_\phi||}.
\end{equation}
This is not a (normalized) Killing vector, because $\varpi$ is not a constant, it still has $(r,\theta)$ dependence. Indeed we have
\begin{eqnarray}
||K_T + \varpi K_\phi||^2 &=&  - (g_{tt} + 2 \varpi g_{t\phi} +\varpi^2 g_{\phi\phi})
\nonumber\\
&=& - \left(g_{tt} - {g_{t\phi}^2\over g_{\phi\phi}}\right)  
\nonumber\\
&=& - {g_2\over g_{\phi\phi}} =  -{1\over g^{tt}} = N^2.
\end{eqnarray}
This particular normal flow automatically satisfies the compatibility conditions~\eqref{e:compatibility1} and \eqref{e:compatibility2}. (Because both $N$ and $p$ are functions of $(r,\theta)$ only, whereas the vector $V^a$ lies in the $(t,\phi)$ plane.)
Since this is a special case of a normal flow we still find
\begin{equation}
T(x) = T_0 \, ||\nabla t|| =  {T_0\over N} = T_0 \sqrt{-g^{tt}}.
\end{equation}
In terms of these Boyer-Linquist coordinates and the free parameters $m$ and $a$,
\begin{eqnarray}
	T(x) = T_0\;	\sqrt {1 + \; {\frac { 2mr \; (r^2 +a^2)		
			}{ \left( {a}^{2}-2\,mr+{r}^{2} \right)  \;\left( {r
				}^{2}+{a}^{2} \cos^2 \theta \right)
	}}}.
\end{eqnarray}
Noticing that $\;(a^2 -2\,mr+r^2) = 0\;$ defines the event horizon, we have:
\begin{eqnarray}
	T(x) = T_0\;	\sqrt {1 + \; {\frac { 2mr \; (r^2 +a^2)		
			}{ (r - r_+) (r - r_-)  \left( {r
				}^{2}+{a}^{2} \cos^2 \theta \right) 
	}}},
\end{eqnarray}
where $r_{\pm}$ represent the outer and inner horizons for a Kerr black hole.
So for this particular ZAMO gradient flow, which is definitely not a Killing flow, the redshifted temperature is well behaved from just above the horizon all the way out to spatial infinity 
with
\begin{equation}
T(x) \rightarrow T_0 \;\;\; \text{for} \;\;\; r \to \infty
\end{equation}
and diverging only at the event horizon.
This observation is useful for thinking about how to redshift the Hawking temperature for Kerr and Kerr--Newman black holes from the horizon (where the locally measured Hawking temperature diverges) out to spatial infinity (where the locally measured Hawking temperature is finite).

Note that the choice of coordinates (eg, Boyer--Lindquist versus Doran) does not change the physics; rather the choice of coordinates guides one as to choosing some physically appropriate 4-velocity for the heat-bath; and it is this physical choice of 4-velocity for the heat-bath that is responsible for physical differences in the Tolman temperature gradient.

\section{Covariant Thermodynamics}
\label{S: Covariant Thermodynamics}

In section \ref{S: Thermodynamics} we started a long discussion about thermodynamics, thermodynamic equilibrium and the four laws. There, we mentioned how part of what was presented would be rephrased in a covariant formulation. This is exactly what we will present now, incorporating the gravitational thermal gradients into each one of the four laws of thermodynamics. The thermodynamic equilibrium topic, due to its complexity, will be explored in the next chapter.

\paragraph{The Zeroth Law}
During section \ref{S: Thermodynamics} we presented the zeroth law with the following statement:

\emph{The zeroth law of thermodynamics states that if two thermodynamic systems A and B are separately in thermal equilibrium with a third system C, then they are in thermal equilibrium with each other. It defines thermal equilibrium as an equivalence relation between thermodynamic systems.}

\begin{figure}
	\center
	\includegraphics[width=0.6\textwidth]{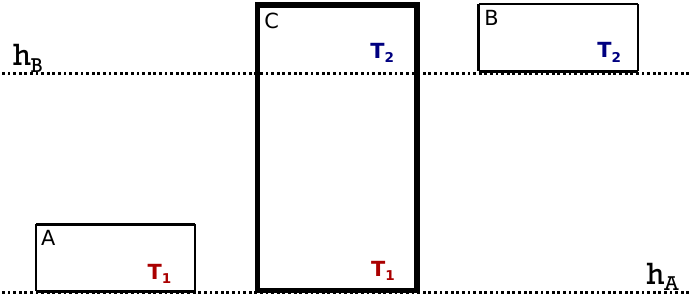}
	\caption[Covariant zeroth law]{Representation of systems A, B and C. Systems A and B are in thermal equilibrium with system C. The surfaces $h_{A/B}$ represent the slices of constant gravitational potential.}
		\label{F: zeroth law}
\end{figure}

This statement, as previously mentioned, establishes the transitivity property of thermal equilibrium. Regardless of the action of gravity, what thermal equilibrium means (no energy flows whatsoever) remains unchanged, though the conditions for thermal equilibrium can be altered. In this way we could, in principle, keep this formulation of the zeroth law. On the other hand, given a) the great opportunity to further explore the physics of thermal systems in the presence of gravity and b) the unfortunate possibility that, despite the issues being relatively clear, confusion  may still arise from these concepts, let us consider the example described in Figure \ref{F: zeroth law}. 

Start by assuming an initial situation where two systems A and B are separately in thermal equilibrium with a third system C. Now, let us assume that the three systems are under the action of a gravitational field. Without loss of generality, we might assume systems A and B to be thin enough that 
they can be considered to be at a constant temperature. In this way we might assume system A to lie on a slice of constant gravitational potential  $h_A$ and system B to lie on slice $h_B$. 
System C, however, is assumed to be extensive enough to cross several slices of constant potential, presenting significant differences between its top and bottom temperatures. 
The point of this discussion is to emphasize how the thermal equilibrium configurations of such systems depend on their relative positions. In this way, although systems A and B are in thermal equilibrium with system C (at the particular space slices $h_A$ and $h_B$ respectively), they have different temperatures from each other. This might not a problem, since we now understand that thermal equilibrium does not mean equal temperatures. But how can we check this?

\begin{figure}
	\center
	\includegraphics[width=0.8\textwidth]{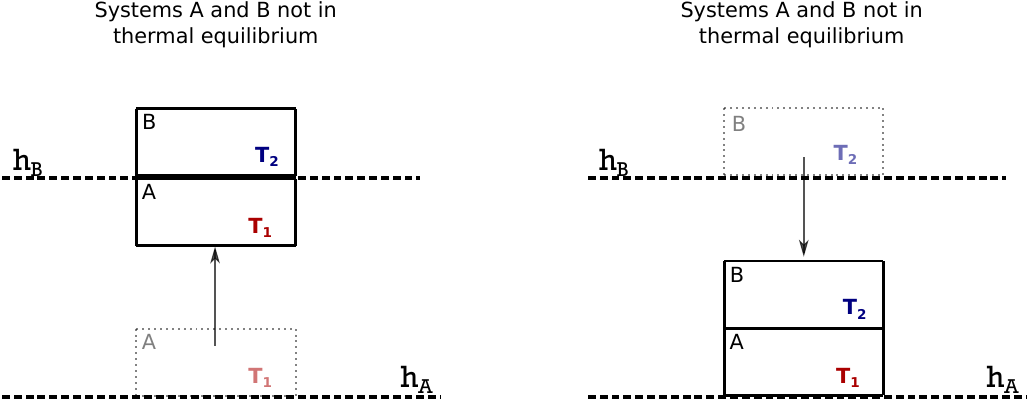}
	\caption[Covariant zeroth law - What not to do]{Representation of systems A and B in direct thermal contact, where either system B was lowered or system A was raised. In this configuration A and B are clearly not in thermal equilibrium.}
		\label{F: zeroth law2}
\end{figure}

\begin{figure}[h]
	\center
	\includegraphics[width=0.35\textwidth]{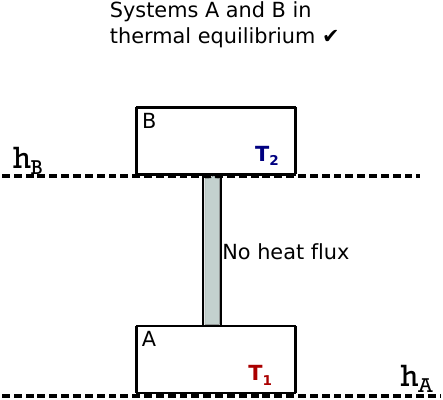}
	\caption[Covariant zeroth law - Correct configuration]{Representation of systems A and B in thermal contact via some conductive material. Their positions are kept the same in relation with each other and with the metric. In this configuration both systems are in thermal equilibrium with each other.}
		\label{F: zeroth law3}
\end{figure}

One way to do so is to allow them to exchange heat and measure whether there are heat fluxes or not. And then we have two possibilities, from which only one gives you the correct result. You can, as pictured in Figure \ref{F: zeroth law2}, lower or raise one of the systems, bringing them in thermal contact with each other. Once in contact, the experimenter would certainly measure heat fluxes and declare that systems A and B were not in thermal equilibrium. Another experimenter, on the other hand, could proceed as pictured in Figure \ref{F: zeroth law3}, introducing a wire or rod or anything that could conduct heat from A to B and vice versa without changing their positions. This observer would see no heat fluxes between the two systems, confirming the validity of the zeroth law in the presence of gravity.

The error of the first experimenter, of course, was in failing to consider the systems' positions in relation to the gravitational field and each other as an important characteristic for describing the thermodynamic system. In this way we can, in order to avoid confusion, reformulate the zeroth law statement to include the exceptional circumstances that arises from the gravitational action:

\emph{Assume three thermodynamic systems A, B and C to be placed in space-time, holding a certain configuration in relation to the metric and to each other. The zeroth law of thermodynamics states that if A and B are separately in thermal equilibrium with system C, then, keeping the same spatial configuration, they are in thermal equilibrium with each other. It defines thermal equilibrium as an equivalence relation between thermodynamic systems.}

\hspace{0.5cm}

\vspace{-0.3cm}

\paragraph{The First Law}
As is well known, the first law is concerned with the conservation of energy.
The problem now is: How do we express this conservation in a covariant way? This, gladly is not a new question for physicists at all. The solution requires you to firstly, describe your system (which can be a fluid, solid, etc) in a covariant way. As mentioned in section \ref{S: General Relativity} from the previous chapter, this can done by the energy momentum tensor $T^{\mu\nu}$ describing that system. Given this tensor, the zero divergence of such an object already gives us the covariant version of the classical energy-momentum principle:
\begin{equation}
\label{E: conservation}
\nabla_{\mu}T^{\mu\nu} =0.
\end{equation}

This also guarantees that, when the system is following the trajectory generated by a Killing vector $K^{\mu}$, it is possible to create conserved currents defined as 
\begin{equation}
\tilde{J}^{\mu} = - T^{\mu\nu}K_{\nu},
\end{equation}
which satisfy
\begin{equation}
\nabla_{\mu}\tilde{J}^{\mu} = 0,
\end{equation}
as already shown in section \ref{S: General Relativity}. 

Another possible way to state the first law can be obtained by defining the energy-momentum tensor density, given by:
\begin{equation}
\mathfrak{T}^{\nu}{}_{\mu} = T^{\nu}{}_{\mu}\;\sqrt{-g},
\end{equation}
with which we can rewrite \eqref{E: conservation} as
\begin{equation}
\label{E: old fashion first law}
\mathfrak{T}^{\nu}{}_{\mu\nu} = \nabla_{\nu}T^{\nu}{}_{\mu}\;\sqrt{-g} = \frac{\partial \mathfrak{T}^{\nu}{}_{\mu}}{\partial x^{\nu}} -\frac{1}{2} \mathfrak{T}^{\alpha\beta} \frac{\partial g_{\alpha\beta}}{\partial x^{\mu}} = 0.
\end{equation}
Which one of these formulas one decides to use is a matter of taste. We believe, however, that the notation used in \eqref{E: old fashion first law} is a bit outdated and we have simply included it for completeness. To be fair, a possible benefit from using \eqref{E: old fashion first law} is the  fact that it does not include any Christoffel symbol.

\paragraph{Second Law}
To solve the problem of covariantly stating the second law, Tolman \cite{tolman:1928, tolmanbook} defined, as mentioned in section \ref{S: The weight of heat}, an entropy vector given by
\begin{equation}
S^{\mu} = s\rho\; u^{\mu},
\end{equation}
where, again, $u^{\mu}$ refers to the macroscopic motion of the matter (or energy) at the point in question and $s$ is the specific entropy density as measured by a comoving observer. 

Note that both in relativistic as in classic thermodynamics, all definitions are made in a macroscopic level. The proper density seen by the comoving observer can be obtained, in their reference system, as one would normally do in any laboratory and the velocity at the point in question is the macroscopic velocity of the fluid. This is a very important point to always keep in mind. 

The covariant formulation of the second law can then be postulated as:
\begin{equation}
\label{E: second law density}
\delta S = \int \nabla_{\mu} S^{\mu} \,\sqrt{-g} \, d^4x  \geq 0,
\end{equation}
which can be interpreted as the vanishing of the variation of entropy $\delta S$ in equilibrium states (for reversible processes) and increase for irreversible processes.
It states, as in classical thermodynamics, that the total entropy of a closed system must not decrease. 
Another way of seeing this is to note that:
\begin{equation}
\nabla_{\mu} S^{\mu} = \nabla_{\mu}\PC{s\rho\; u^{\mu}} = \rho\, u^{\mu}\nabla_{\mu}s + s \,\nabla_{\mu}\PC{\rho u^{\mu}} = \rho \,\dot{s},
\end{equation}
where the term $\nabla_{\mu}\PC{\rho u^{\mu}}$ is zero due to the continuity equation for fluids. Hence, in terms of the specific entropy, equation \eqref{E: second law density} reads:
\begin{equation}
\dot{s} \geq 0.
\end{equation}
We see, in this way, that the message behind each law is being kept unaltered, as one would expect.

\paragraph{The Third Law} 
The third law is probably the easiest one to reformulate covariantly of all the laws. Firstly because, when requiring the entropy of a system to never reach zero (or the temperature to never reach absolute zero), the reference system is automatically defined. It is the one comoving with the system in question. 

But, also, given the validity of the third law in classical thermodynamics, and given the covariant formulation of the second law, it is clear that in any reference system this must be true. 

Even with the presence of temperature gradients, the third law is protected by the fact that objects at absolute zero would not emit any radiation at all, having no light to be redshifted. If one observer ``sees'' an object at absolute zero, all observers will measure the same.  The third law, in this way, can be trivially  interpreted in a covariant fashion.

\newpage
\chapter{Can we still define thermal equilibrium for non-Killing flows?}
\label{C: Can we still}

In the previous chapter we have discussed in depth the structure of thermodyna- mics in the presence of gravitational fields. We have introduced the Tolman temperature gradient and have also extended its definition for fluids following a wider class of four-velocities, which do not necessarily have to be proportional to a Killing vector.

At this stage, we would like to remind the reader of an assumption made to obtain such a generalization and raise several points about it. First, the assumption: To derive equation \eqref{e:euler3} given in the last chapter, i.e.:
\begin{equation}
\tag{\ref{e:euler3}}
a_b =  - \nabla_b \ln T,
\end{equation}
we had to explicitly assume the fluid to be in thermal equilibrium (or at least in local thermal equilibrium). Using this result, we then proceeded by presenting some examples (in section \ref{S: Some other examples}) of fluids with different four-velocities in distinct black hole space-times. This was a very important exercise, especially in order to highlight how the internal state of a fluid can be influenced not only by the metric of its surroundings, but also by its own four-velocity and four-acceleration.  All the examples given in that section, however, were for observers following non-Killing trajectories.

Now, given that observers following non-Killing trajectories experience a space-time which is varying along their proper time, how can we expect a fluid to be maintained in equilibrium  (or ``close to equilibrium") when it keeps being disturbed by a changing space-time?
One might then see this as a contradiction of the assumption that the fluid must be in thermal equilibrium in the first place. Keeping this in mind, we would like to discuss such a point of view by looking at it from two different angles.

The first one follows a pragmatic line of thought, in the sense that we do know that situations of eternal and exact equilibrium in the real world are extremely unlikely to naturally occur. All we really have are good approximations to equilibrium. This, however, hasn't stopped us from assuming ``thermal equilibrium'' in a number of situations. More than this, even in flat Minkowski space-time we barely know how to do thermodynamics for systems completely out of equilibrium. 
Surely, a lot of effort has been put into the area, much of which is actually focused on perturbation schemes, for which some underlying equilibrium state exists, but we still have not got ourselves comfortable outside of near-equilibrium situations. With this in mind, we cannot deny how important it is to be able to define and talk about thermal equilibrium for a more general class of observers. Furthermore, coming back to the non-Killing examples from the previous chapter, even if the fluids following such trajectories are not in equilibrium themselves, it is important to know how their temperature distribution would be in case they were, so that we can use all the machinery from near-equilibrium pertubation theory in our favor.

On the other hand, we also cannot deny the importance of the question, which is in fact the main point of this chapter: Can we still talk about thermal equilibrium for fluids following non-Killing flows? 
What are the limitations of this concept? What are the time scales of variations in space-time against the relaxation time of the systems?
In this chapter, our mission will be to tackle these non-obvious questions. 

We will introduce and study fluids following what is called Born-rigid trajectories, and show explicit examples of Born-rigid congruences which are not generated by Killing vectors. Such special exact solutions exist -- and are not perturbed by the metric along its evolution. Below, we will turn our attention to non-perfect fluids, analyzing them from the point of view of two of the main theories for relativistic thermodynamics --- Classical Irreversible Thermodynamics and Extended Irreversible Thermodynamics. We will then show that Born-rigidity is actually one of the main conditions for a fluid to be in thermal equilibrium.

We will conclude this chapter by asking the question: 
Is it possible for fluids following non-Killing Born-rigid trajectories to keep a state of thermal equilibrium while moving through space-time? Or, more generally: Can we still define thermal equilibrium for specific non-Killing flows? We will see that, although the straight answer to this question is actually \emph{No}, interesting approximate scenarios do exist for fluids belonging to non-Killing congruences. They will be analyzed case by case.



\section{Relativistic Fluids}

Having talked about fluids a few times in the previous chapters, it was sufficient, at that stage, to simply assume that a certain fluid existed. The analysis that will be presented in this chapter, however, will require a bit more care and delicacy when describing the fluid and four-velocities involved. Due to this, it seems a good idea to start by asking: what exactly is a fluid?

All matter is formed by subatomic particles, like protons, electrons, neutrons and so on. In this way, what are the requirements that must be fulfilled such that we can simply put aside all the ``granular'' components of matter and focus on the large scale emergent behaviour?  The first two quantities that will give us this answer are $l$, the \emph{typical inter-particle separation} and $\lambda_B$, the \emph{de Broglie wavelength} associated with the particles involved. The reason must be quite clear: if $\lambda_B \gsim l$, individual particles wave packets will overlap, and the system will have to be described quantum mechanically by an $N$-particle Schroedinger equation. On the other hand, if $\lambda_B \ll l$, then each particle will be described by an isolated Schroedinger equation and, as shown by Ehrenfest theorem, will on average move like classical particles.

The other important quantity is the size of the system. In order to describe $N$ classical particles as a fluid, the number $N$ must be very large so that, statistically speaking, what each individual particle does has no importance compared to the bulk of all particles. In this situation, the best approach is to adopt the the statistical description by the means of a \emph{distribution function} $f(t, x, u)$, where $x$ and $u$ represent, respectively, the space coordinates and the velocities of the particles.
Boltzmann was the first person to suggest such a description, besides coming up with the correct equations which actually describe the dynamics of these complicated systems; hence the name \emph{Boltzmann equation}.

Apart from $l$, $\lambda_B$ and $N$, let us, to conclude, quickly introduce the so called \emph{Knudsen number}, given by $K_n = l/L$, where $L$ is length-scale of the system. 
It is another important quantity, which helps us to know when it is acceptable or not to adopt the fluid description for a set of particles. For example, for Knudsen numbers above 0.1, the typical inter-particle distance will be $10\%$ of the length scale of the system. For such cases it does not make sense to talk about a fluid continuum and most of the gas flow must be characterized using statistical methods \cite{Knudsen}. On the other hand, for $K_n \ll 1$, (while $\lambda_B \gsim l$ and $N$ very large) the dynamics of individual particles cannot be described even statistically, and this is what is called a fluid continuum.  When a fluid description is possible, one can then depict the system in terms of the so called \emph{fluid elements}, which are big enough in order to contain a large  number $n$ of particles, but small enough to be considered homogeneous. This is the assumption we will make in the next section, when talking about congruences moving through space-time with a certain 4-velocity.

\section{Kinematics of fluids and spacetime optics}
\label{S: Kinematics bitch}
Given a certain fluid, let us now focus on its movement throughout space-time. As argued in the last section, when a fluid description is allowed, it is possible to describe the movement of the whole system by focusing only on its fluid elements (sometimes called cells). Let us start by assuming our fluid to occupy a total volume $\Sigma(t)$ of a 3 dimensional space-like surface (representing a frozen instant of time $t$), and divide this volume into $\mathcal{N}$ fluid elements with 3-d volumes $\epsilon (t, x_i) = \epsilon_i(t)$, where $x_i$ is the position of the center of mass of each element. In this way, we have
\begin{eqnarray}
\sum_{i=1}^{\mathcal{N}} \epsilon_i(t) = \Sigma(t). 
\end{eqnarray}

\begin{figure}
	\center
	\includegraphics[scale=0.35]{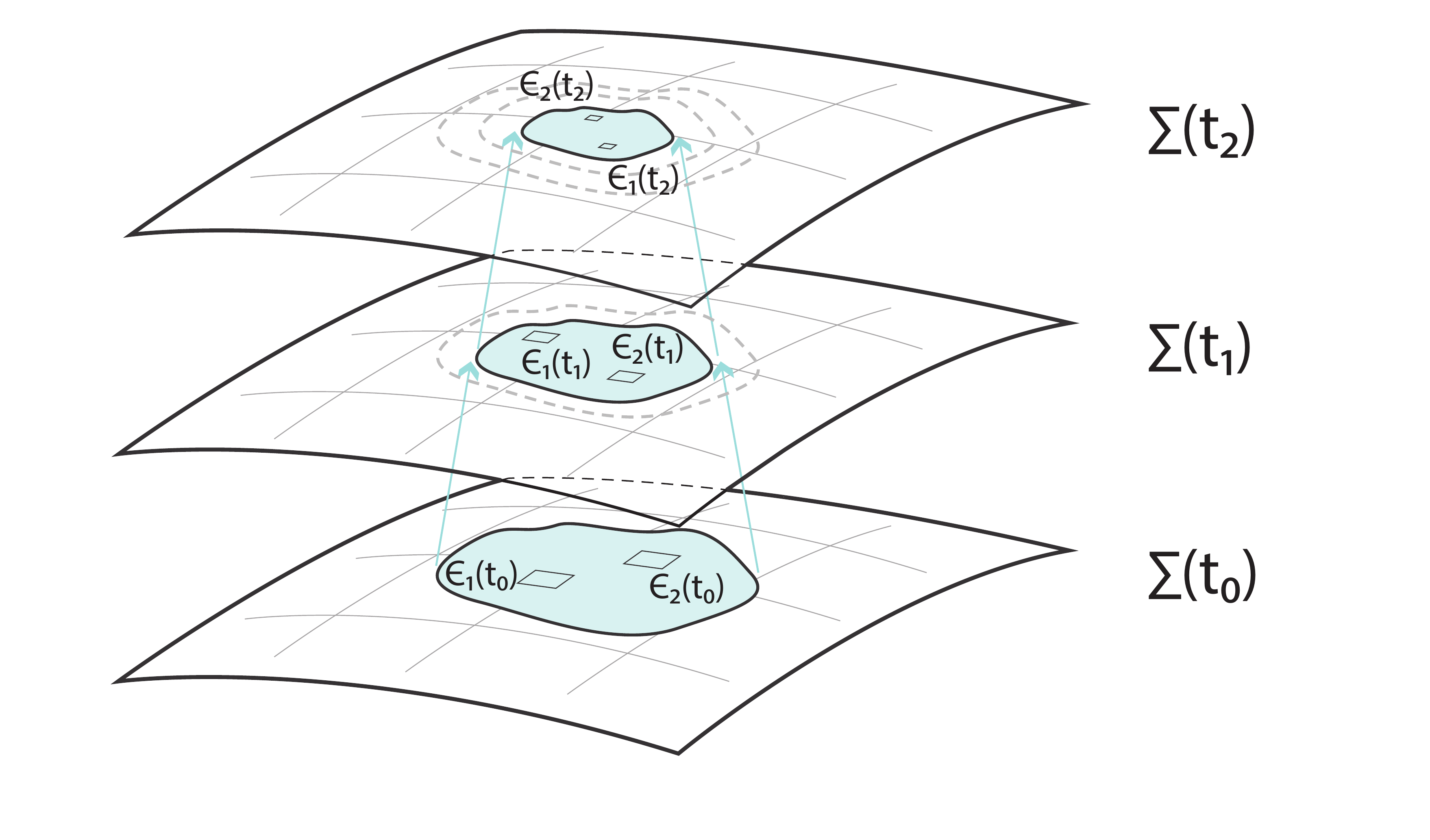}
	\caption[Representation of a contracting fluid]{Representation of a contracting fluid at constant time slices $\Sigma(t_0)$, $\Sigma(t_1)$ and $\Sigma(t_2)$. It also shows how the element's volume also evolves in time.}
	\label{F:evolvingblobs}
\end{figure}

As the fluid evolves, both its total volume as well as the elements' volume might change (see Figure \ref{F:evolvingblobs} for a clearer understanding). The total number of particles and the total number of elements, though, are fixed. In this way, the fluid's and the elements' particle density will be a function of time. The advantage of taking the element description is that while the fluid's density will depend both on the spatial coordinates as well as on time, the individual cell's density will be assumed spatially homogeneous, but time dependent.

\begin{figure}
	\center
	\includegraphics[scale=0.45]{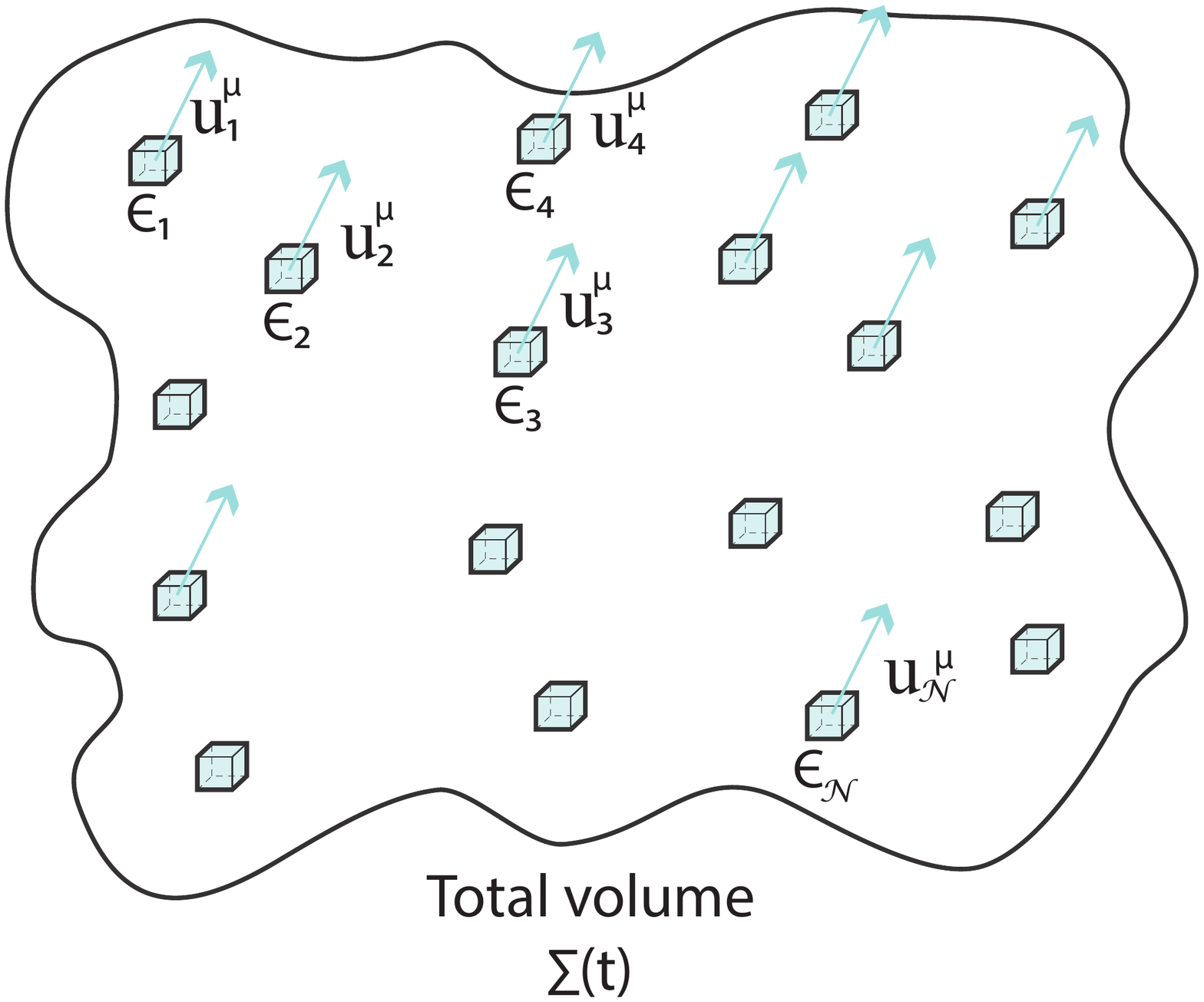}
	\caption[Fluid composed by $\mathcal{N}$ fluid elements]{\label{F:elemnts} Fluid interpretation in terms of fluid elements, where $u_i^{\mu}$ and $\epsilon_i$ are the elements' four-velocity and volume respectively.}
\end{figure}

Now, considering that the total mass of the fluid is big enough to compose a macroscopic system but small enough such that self-gravitating effects are negligible, let us associate a four-velocity $u^{\mu}(x_i)=u^{\mu}_i$ with each fluid element, as shown in Figure \ref{F:elemnts}.
Hence, knowing the initial position of the elements' center of mass and their four-velocities, it is possible to follow each element's world line as it evolves. 

We will also assume that exchanges of mass and energy are allowed between neighbouring cells. In this way, particles belonging to a certain element volume $\epsilon_i$ at time $t_0$ may belong to another element $\epsilon_j$ at a different time due to diffusion. In this way, the system is maintained connected, what will allow us to keep talking about its thermodynamic properties, including thermal equilibrium. This fluid description, on which we are following the fluid as it moves, is called the Lagrangian description.
Furthermore, the description adopted here exactly coincides with the scenario described by the \emph{Local Equilibrium Hypothesis}, which is one of the cornerstones of the Classical Irreversible Thermodynamics theory. We will return to this subject in the future but, summarizing, local equilibrium assumes that, at a given instant of time, equilibrium is achieved in each individual element. The state of equilibrium, however, being possibly different from one cell to the other, i.e, local Gibbs equations are assumed valid for each individual element:
\begin{equation}
\d\,s(x_i, t) = \frac{1}{T}\,\d\mathfrak{u} + \frac{p}{T}\,\d v - \frac{1}{T}\,\mu\, dn,
\end{equation}
where $\mu$ is the chemical potential and $n$ the particle number.
In this way, in each cell the equilibrium state is not frozen, but changes in the course of time \cite{Lebon}.
Furthermore, since we are not worrying about how individual particles are behaving,  it is possible to assume that the union of the four-velocities of all the elements form  what is called a time-like congruence.
In general relativity, a congruence is described as follows~\cite{Wald}: 

\enlargethispage{20pt}
\emph{``Let $\mathbb{M}$ be a manifold and let $\mathcal{O} \in  \mathbb{M}$ be open. A congruence in $\mathcal{O}$ is a family of curves such that through each $p \in \mathcal{O}$ there passes precisely one curve in this family. Thus, the tangents to a congruence yield a vector field in $\mathcal{O}$, and, conversely [...] every continuous vector field generates a congruence of curves.''}

Keeping this definition in mind, we see that, as long as our elements don't collide, they specify a continuous vector field eligible to define a time-like congruence. Let us now see how to connect the dynamic properties of such a congruence with the fluid's behaviour.

\vspace{-0.3cm}	
\subsection{Shear, expansion and vorticity}

Given a time-like congruence generated by a four-velocity $u^{\mu}$, let us now define some quantities, namely its \emph{shear}, \emph{expansion} and \emph{twist} tensors, which will help us to visualize the physics of a system connected to such a congruence.
Let us start by introducing what is called the  \emph{induced metric} on the plane orthogonal to $u^{\mu}$:
\begin{equation}
\label{E:induced metric}
h_{\mu\nu} = g_{\mu\nu} + u_{\mu}u_{\nu}.
\end{equation}
The orthogonality relations being given by:
\begin{equation}
h_{\mu\nu}u^{\mu} = h_{\nu\mu}u^{\mu} =0.
\end{equation}
We can also define the \emph{expansion scalar} of the congruence by:
\vspace{-0.2cm}
\begin{equation}
\label{E:expansion}
\theta = \nabla_{\mu}u^{\mu}.
\end{equation}
This scalar gives us information about the separation between the curves generated by $u^{\mu}$. If they are spreading we have $\theta >0$ and, of course, $\theta<0$ implies that the curves are focusing. 
The other two tensors to be defined will also give us information about how the curves generated by $u^{\mu}$ will behave. Before defining them, let us start by rewriting $\nabla_{\mu}u_{\nu}$ in a useful way:
\begin{eqnarray}
\nabla_{\mu}u_{\nu} &=& \PC{- u_{\mu}u^{\xi}  + g_{\mu}{}^{\xi} + u_{\mu}u^{\xi}}\nabla_{\xi}u_{\nu} \nonumber \\
&=& -u_{\mu}\;u^{\xi} \nabla_{\xi}\;u_{\nu}  + h_{\mu}{}^{\xi}\nabla_{\xi}\;u_{\nu}. 
\end{eqnarray}
In terms of the 4-acceleration $a^{\mu} = u^{\nu}\nabla_{\nu} \;u^{\mu}$ this gives us:
\begin{equation}
\label{E: B tensor}
\nabla_{\mu}u_{\nu} = -u_{\mu}\;a_{\nu} + h_{\mu}{}^{\xi}\nabla_{\xi}\;u_{\nu},
\end{equation}
We also know that any tensor can be written as the sum of its symmetrical and anti-symmetrical parts, so we have:
\vspace{-0.3cm}
\begin{eqnarray}
\nabla_{\mu}u_{\nu}  = \nabla_{[\mu}u_{\nu]} + \nabla_{(\mu}u_{\nu)}.
\end{eqnarray}
The symmetric part of \eqref{E: B tensor} gives:
\begin{eqnarray}
	\nabla_{(\mu}u_{\nu)} &=& -u_{(\mu}\;a_{\nu)} + \frac{1}{2}\PC{h_{\mu}{}^{\xi}\nabla_{\xi}u_{\nu} + h_{\nu}{}^{\xi}\nabla_{\xi}u_{\mu}},
\end{eqnarray}
which allows us to write:
\begin{equation}
\label{E:symmetric tensor}
\frac{1}{2}\PC{h_{\mu}{}^{\xi}\nabla_{\xi}u_{\nu} + h_{\nu}{}^{\xi}\nabla_{\xi}u_{\mu}} = \nabla_{(\mu}u_{\nu)} + u_{(\mu}\;a_{\nu)}.
\end{equation}
Note that the right hand side is a completely symmetric tensor. Given matrix decomposition rules, we know that any completely symmetric tensor can be decomposed in a traceless symmetric tensor plus its trace. But the trace of \eqref{E:symmetric tensor} is:
\begin{equation}
\nabla_{\mu}u^{\mu} + u_{\mu}\;a^{\mu} = \theta .
\end{equation}
This gives us:
\begin{equation}
\frac{1}{2}\PC{h_{\mu}{}^{\xi}\nabla_{\xi}u_{\nu} + h_{\nu}{}^{\xi}\nabla_{\xi}u_{\mu}} =
\PR{\frac{1}{2}\PC{h_{\mu}{}^{\xi}\nabla_{\xi}u_{\nu} + h_{\nu}{}^{\xi}\nabla_{\xi}u_{\mu}} - \frac{1}{3}\theta h_{\mu\nu}} + \frac{1}{3}\theta h_{\mu\nu},
\end{equation}
We can, in this way, define the shear as:
\begin{equation}
\sigma_{\mu\nu} = \frac{1}{2}\PC{h_{\mu}{}^{\rho}\nabla_{\rho}u_{\nu} + h_{\nu}{}^{\rho}\nabla_{\rho}u_{\mu}} - \frac{1}{3}\theta h_{\mu\nu}.
\end{equation}
This, finally, can be rewritten using \eqref{E:symmetric tensor} in its most well known form:
\begin{equation}
\label{E:shear}
\sigma_{\mu\nu} =
\nabla_{(\mu}u_{\nu)} + u_{(\mu}\;a_{\nu)} -\frac{1}{3}\theta h_{\mu\nu}.
\end{equation}
In this way, it is clear that the shear is a traceless symmetric tensor. It tells us about \emph{distortions} which do not change the ``volume'' of the congruence of lines. You may visualize the action of the shear as what happens to ``blobs'' in phase space while evolving in time: they change shape, but given Liouville's theorem, the volume must be kept constant. 

Now, let us finally look to the anti-symmetric part:	
\begin{eqnarray}
	\nabla_{[\mu}u_{\nu]} = -u_{[\mu}\;a_{\nu]} + \frac{1}{2}\PC{h_{\mu}^{\xi}\nabla_{\xi}u_{\nu} - h_{\nu}^{\xi}\nabla_{\xi}u_{\mu}}.
\end{eqnarray}
This can be rearranged as:
\begin{eqnarray}
\frac{1}{2}\PC{h_{\mu}^{\xi}\nabla_{\xi}u_{\nu} - h_{\nu}^{\xi}\nabla_{\xi}u_{\mu}} = \nabla_{[\mu}u_{\nu]} + u_{[\mu}\;a_{\nu]}.
\end{eqnarray}
The vorticity tensor is then simply defined as:
\begin{equation}
\omega_{\mu\nu} = \omega_{[\mu\nu]} = \frac{1}{2}\PC{h_{\mu}^{\rho}\nabla_{\rho}u_{\nu} - h_{\nu}^{\rho}\nabla_{\rho}u_{\mu}},
\end{equation}
another possible definition being:
\begin{equation}
\label{E:vorticity}
\omega_{\mu\nu} = \nabla_{[\mu}u_{\nu]} + u_{[\mu}\;a_{\nu]}.
\end{equation}
The vorticity is the tensor which contains all the information about \emph{rigid rotations} of the fluid, without distorting its internal structure. Naturally, however, given Raychaudhuri's equation, the presence of rotation can in turn drive other types of distortion on the fluid, like expansion and even shear. In the next section we will talk about rigid bodies and the Ehrenfest paradox, which concerns the attempt to impose rotation while keeping null shear and expansion, and see that it is possible to obtain quite non-obvious conclusions.

Note now that we can obtain a time-scale from the variations imposed by expansion, shear and vorticity by:
\begin{equation}
\label{E:timescales}
\tau_{\theta} = \frac{1}{|\theta|}\, , \qquad \tau_{\sigma} = \frac{1}{\sqrt{\sigma_{\mu\nu}\,\sigma^{\mu\nu}}}\, , \qquad \tau_{\omega} = \frac{1}{\sqrt{\omega_{\mu \nu}\,\omega^{\mu\nu}}}\, .
\end{equation}
Such time-scales allow us to compare how quickly a system is changing due to its movement through space-time with the time-scale of internal processes, for example their relaxation time. Another interesting possibility is to use such time-scales to define macroscopic lengths associated with the fluid following such a congruence. Returning to the Knudsen number introduced in the last section, for example, we might use for $L$, the length scale of the system, the minimum amongst 
\begin{equation}
L_{\theta} = c\,\tau_{\theta} = \frac{c}{|\theta|}\, , \quad L_{\sigma} = c\,\tau_{\sigma} = \frac{c}{\sqrt{\sigma_{\mu\nu}\,\sigma^{\mu\nu}}}\, ,
\end{equation}
or $L$ given by the system's characteristic spatial length. In this way, if the system has zero expansion and shear, both $L_{\theta}$ and $L_{\sigma}$ are infinite and the minimum characteristic length is the system's regular spatial length. For systems evolving fast, however, this might not be the case anymore.

\subsection{The rate of deformation tensor}

Let us now define what we will call the \emph{rate of deformation tensor}\footnote{This name was initially given by V. P. Frolov and I. D. Novikov \cite{frolovnovikov}.}. It can be defined in terms of the shear and expansion and gives us the total amount of deformation which is being imposed on the system. Note that the vorticity tensor $\omega_{\mu\nu}$ is not part of this quantity. This is the case since rotation, by itself, might be rigid and not change the internal state of the system (such rotations are the ones which can be eliminated by a coordinate change). The rate of deformation tensor is defined as:
\begin{equation}
\label{E:rate}
D_{\mu\nu} = \sigma_{\mu\nu} + \frac{1}{3}\theta h_{\mu\nu}.
\end{equation}
It is a completely symmetric tensor. Using equation \eqref{E:shear} for the shear, we see that it can also be rewritten as:
\begin{equation}
\label{E:D(u,a)}
D_{\mu\nu} = \nabla_{(\mu} u_{\nu)} + u_{(\mu}a_{\nu)}.
\end{equation}

Another very interesting way of defining the rate of deformation tensor, which actually gives a deeper physical reason for its name, is in terms of the Lie derivative of the induced metric $h_{\mu\nu}$. To do so, note that:
\begin{eqnarray}
\mathfrak{L}_u h_{\mu\nu} &=& \mathfrak{L}_u (g_{\mu\nu} +u_{\mu} u_{\nu}) =  \nabla_{\nu} u_{\mu} +  \nabla_{\mu} u_{\nu} + u_{\mu}\;a_{\nu} + u_{\nu}\;a_{\mu} \nonumber\\ 
&=& 2\PC{\nabla_{(\mu} u_{\nu)} + u_{(\mu}a_{\nu)}}. 
\end{eqnarray} 
This can be obtained straightforwardly from the definition of Lie derivatives (see Section \ref{S: General Relativity}). In this way, we have:
\begin{equation}
\label{E:D}
D_{\mu\nu} =  \frac{1}{2}\mathfrak{L}_u h_{\mu\nu}.
\end{equation} 
\vspace{-0.2cm}
We can see that such a quantity is actually measuring how much an observer with four-velocity $u^{\mu}$ will see the 3-space around them change as they evolve in time. If such an observer is inside a fluid, $D_{\mu\nu}$ will contain information about the movement of the fluid as seen by the co-moving observer.
Note that we can also write:
\begin{align}
D_{\mu\nu} = \;\; h_{\mu}{}^{\alpha}h_{\nu}{}^{\beta}\;\nabla_{(a}u_{\beta)}
= \;\;\frac{1}{2} h_{\mu}{}^{\alpha}h_{\nu}{}^{\beta} \PC{\mathfrak{L}_u g_{\mu\nu}}.
\end{align}
 
So, if $u^{\mu}$ is a Killing vector, this implies that $D_{\mu\nu}=0$, but the reverse is not true.
Let us now derive a final way of rewriting the deformation tensor. We can start with the definition of $D_{\mu\nu}$ given by equation \eqref{E:D}:
\begin{eqnarray}
	D_{\mu\nu} &=& \frac{1}{2}\mathfrak{L}_u h_{\mu\nu} = \frac{1}{2}\PR{\mathfrak{L}_u g_{\mu\nu} + \mathfrak{L}_u (u_{\mu}u_{\nu})}\\
	&=& \nabla_{(\mu}u_{\nu)} + \frac{1}{2}\PR{ u^{\alpha}\nabla_{\alpha}(u_{\mu}u_{\nu}) + u^{\alpha}u_{\mu}\nabla_{\nu}u_{\alpha}
		+ u^{\alpha}u_{\nu}\nabla_{\mu}u_{\alpha} }\nonumber\\
	&=& \nabla_{(\mu}u_{\nu)} + \frac{1}{2}\PR{ u_{\mu}u^{\alpha}\nabla_{\alpha}(u_{\nu}) + u_{\nu}u^{\alpha}\nabla_{\alpha}(u_{\mu}) + u^{\alpha}u_{\mu}\nabla_{\nu}u_{\alpha}
		+ u^{\alpha}u_{\nu}\nabla_{\mu}u_{\alpha} }.\nonumber
\end{eqnarray}
But now, combining the first and the third terms  inside the brackets plus combining the second and last terms again inside the brackets, we have:
\begin{equation}
\label{E:D in terms of u}
D_{\mu\nu} = \nabla_{(\mu}u_{\nu)} + u^{\alpha}u_{\mu}\nabla_{(\nu}u_{\alpha)}
+ u^{\alpha}u_{\nu}\nabla_{(\mu}u_{\alpha)}.
\end{equation}
\enlargethispage{10pt}
This result is interesting since all its terms contain the tensor $\nabla_{(\mu}u_{\nu)}$. In the next section we shall explore the physical interpretations of having a congruence with a vanishing rate of deformation tensor and when can that be achieved.



%

\section{Rigid body motion}

An interesting scenario for fluid dynamics in general relativity comes from the notion of a \emph{Born-rigid body motion}. Born-rigid bodies, as one may easily guess by their name, are \emph{bodies which do not suffer deformations when disturbed or moved}. The definition was firstly given by Max Born in 1909. 
A big surprise, however, came when studying such a concept within the framework of special relativity, by trying to impose rigidity while maintaining a causal theory. It was shown that Born-rigidity, as normally phrased, is too restrictive and must be abandoned in some situations --- the most famous example probably coming from what is called the \emph{Ehrenfest paradox}\cite{paradox1, paradox2}. It considers the example of an ideal Born-rigid cylinder at rest which starts to rotate. The rigidity assumption imposes that the cylinder must neither expand nor contract during this process, keeping all its dimensions constant. The problem comes from the realization that, if one imagines little measuring rods along the cylinder's circumference, those will suffer a Lorentz contraction which will depend on the tangent velocity at each point. This, however, contradicts the first assumption that the cylinder would keep a constant radius all along, showing that  Born-rigidity is not generally compatible with special relativity. It is important to point out, however, that once the cylinder has reached a constant angular velocity the rigid motion presents no problem whatsoever. The Born-rigidity condition can be conflicting only in situations of accelerated motion, as one should expect.

Given the discussion above, an interesting problem then comes from analyzing which motions do not disturb a body or fluid in the sense of Born-rigidity. As we will shown in this section, any fluid whose motion can be described by a time-like congruence which has a four-velocity field proportional to a Killing vector, naturally satisfies the Born-rigidity conditions. We will see, however, some other explicit examples that are not generated by Killing vectors. 

Let us start by mathematically defining rigid motions. Using the rate of deformation tensor from last chapter, we can say that a fluid is moving rigidly if 
\begin{equation}
\label{E: rigid motion}
D_{\mu\nu} =0.
\end{equation}
Based on the definitions from the last section, this means that such a fluid will suffer no expansion nor shear while evolving along its trajectory. Moreover, given the relation between $D_{\mu\nu}$ and the induced metric, \eqref{E: rigid motion} implies:
\vspace{-0.2cm}
\begin{equation}
\mathfrak{L}_u h_{\mu\nu} = 0,
\end{equation}
meaning that distances in the local rest-frame must be preserved along each world-line. We hope the physical intuition behind this concept to be sufficiently clear. For further readings on the subject we recommend the paper by Williams \& Pirani \cite{Pirani}.

Let us now turn to the question on which we will focus during this section: What are the possible sets of four-velocity plus metric which allows \eqref{E: rigid motion} to be satisfied? We will begin to answers this question by looking at the flat space-time case.

\subsection{Herglotz-Noether theorem}

\enlargethispage{10pt}
Let $\xi^{\mu}$ be a Killing vector. Then, by definition we have:
\vspace{-0.2cm}
\begin{equation}
\mathfrak{L}_{\xi}\, g_{\mu\nu} = 0, 
\end{equation}
\vspace{-0.2cm}
which implies 
\begin{equation}
\mathfrak{L}_{\xi}\, h_{\mu\nu} = \mathfrak{L}_{\xi}\, (g_{\mu\nu} + \xi_{\mu}\xi_{\nu}) =0.
\end{equation}
This can also be directly obtained by assuming $u^{\mu}$ to be a Killing vector in equation \eqref{E:D in terms of u}. So, if the four-velocity is given by a Killing vector, then we have a solution for rigid body motion already. 
Besides being a bit more complicated, due to the normalization factor necessary to keep $u^{\mu}u_{\mu}=-1$, we will show in \ref{S: special cases} that the result remains valid for 4-velocities given by:
\begin{equation}
\label{E:killvel}
u^{\mu} =  \frac{\xi^{\mu}}{\|\xi\|}, 
\end{equation}
$\xi^{\mu}$ being a Killing vector.  The interesting question, though, is: are there any other solutions? As simultaneously proved in 1910 by Fritz Noether \cite{Noether}~\footnote{Emmy Noether's brother.}  and Herglotz \cite{Herglotz}, the answer for flat space-times is no. Any rotational rigid motion in flat space must be a Killing motion. Note that their proof came before the theory of general relativity was developed. 
The next significant development on the subject came only in 1967, when Wahlquist extended the Herglotz-Noether theorem for conformally flat space-times \cite{conformaltheorem}. 

Before that period, on the other hand, people discussed the subject and some of them truly believed that an extension of the Herglotz-Noether theorem should exist for general curved space-times as well \cite{Boyer, SalzmanTaub, Wahlquist}. However, as suggested in \cite{Pirani},  while Killing motion implies rigidity, the converse is not true. The condition for $u^{\mu}$ satisfying $D_{\mu\nu}=0$ to be generated by a Killing flow can be shown to be:
\begin{equation}
\label{E:non-Killing condition}
\nabla_{[\mu}a_{\nu]} = 0,
\end{equation}
where $a^{\mu}$ is the four-acceleration of the congruence.
This can be shown by noticing that, if the 4-velocity of the flow is given by \eqref{E:killvel}, then
\begin{eqnarray}
a_{\mu} &=& u^{\nu} \nabla_{\nu} u_{\mu} = ||\xi||^{-1} \xi^{\nu}\nabla_{\nu} \PC{\frac{\xi_{\mu}}{||\xi||}} \nonumber\\
&=& ||\xi||^{-2} (\xi^{\nu}\nabla_{\nu} \xi_{\mu}) - ||\xi||^{-3} (\xi^{\nu}\nabla_{\nu} ||\xi||) \xi_{\mu} 
\end{eqnarray}
But since $\xi_{\mu}$ is Killing, we have $\xi^{\nu}\nabla_{\nu} ||\xi||=0$. 
In this way we have:
\begin{eqnarray}
a_{\mu} &=& ||\xi||^{-2} (\xi^{\nu}\nabla_{\nu} \xi_{\mu}) = -||\xi||^{-2} (\xi^{\nu}\nabla_{\mu} \xi_{\nu}) = {1\over2} ||\xi||^{-2} \nabla_{\mu} ||\xi||^2 \nonumber \\
&=& ||\xi||^{-1} \nabla_{\mu} ||\xi|| = \nabla_{\mu} \ln||\xi||.
\end{eqnarray}
So $a_{\mu} = \nabla_{\mu} \phi$, where $\phi$ is a scalar function. But this implies \eqref{E:non-Killing condition}.

Moreover, we will show in this section explicit examples of rigid motions along non-Killing vectors, showing that indeed the Herglotz-Noether theorem is not valid for the general case. 
Let us first start with a proper derivation of $D_{\mu\nu}$ for some special cases. We will then move to the non-Killing examples.


\subsection{Killing, conformal Killing and geodesic congruences}
\label{S: special cases}

\paragraph{Killing congruence}\hfill

Supposing our space-time has the required symmetries, it is possible to pick a Killing congruence, in which the 4-velocity is given by
\begin{equation}
u^{\mu} = \frac{\xi^{\mu}}{\|\xi\|} \qquad \text{and} \qquad \nabla_{(\mu}\xi_{\nu)} =0.
\end{equation}
For this situation, we have
\begin{eqnarray}
	2\;\nabla_{(\mu}u_{\nu)} &=& \nabla_{\mu}\PC{\frac{\xi_{\nu}}{\|\xi\|}} + \nabla_{\nu}\PC{\frac{\xi_{\mu}}{\|\xi\|}}\\
	&=&  \frac{1}{\|\xi\|}\nabla_{\mu}\PC{\xi_{\nu}} + \frac{1}{\|\xi\|}\nabla_{\nu}\PC{\xi_{\mu}}  + 
	\xi_{\mu}\nabla_{\nu}\PC{\frac{1}{\|\xi\|}} + \xi_{\nu}\nabla_{\mu}\PC{\frac{1}{\|\xi\|}}\nonumber
\end{eqnarray}
The two first terms cancel since we are dealing with a Killing field. We then have:
\begin{eqnarray}
	2\;\nabla_{(\mu}u_{\nu)} &=& -\frac{\xi_{\mu}}{\|\xi\|^2}\nabla_{\nu}\|\xi\| - \frac{\xi_{\nu}}{\|\xi\|^2}\nabla_{\mu}\|\xi\|\\
	&=& -\frac{\xi_{\mu}}{2\|\xi\|^3}\nabla_{\nu}\|\xi\|^2 - \frac{\xi_{\nu}}{2\|\xi\|^
		3}\nabla_{\mu}\|\xi\|^2\\
	&=& -\frac{\xi_{\mu}}{2\|\xi\|^3}\nabla_{\nu}\PC{-\xi_{\alpha}\xi^{\alpha}} - \frac{\xi_{\nu}}{2\|\xi\|^3}\nabla_{\mu}\PC{-\xi_{\alpha}\xi^{\alpha}}\\
	&=& \frac{\xi_{\mu}}{\|\xi\|^3}\;\xi^{\alpha}\nabla_{\nu}\PC{\xi_{\alpha}} + \frac{\xi_{\nu}}{\|\xi\|^3}\;\xi^{\alpha}\nabla_{\mu}\PC{\xi_{\alpha}}
\end{eqnarray}
Then, using the Killing equation:
\begin{eqnarray}
	2\;\nabla_{(\mu}u_{\nu)}
	&=& -\frac{\xi_{\mu}}{\|\xi\|^3}\;\xi^{\alpha}\nabla_{\alpha}\PC{\xi_{\nu}} - \frac{\xi_{\nu}}{\|\xi\|^3}\;\xi^{\alpha}\nabla_{\alpha}\PC{\xi_{\mu}}\\
	&=& -\frac{\xi_{\mu}\xi^{\alpha}}{\|\xi\|^2}\PC{\frac{1}{{\|\xi\|}}\nabla_{\alpha}\PC{\xi_{\nu}}} - \frac{\xi_{\nu}\xi^{\alpha}}{\|\xi\|^2}\PC{\frac{1}{{\|\xi\|}}\nabla_{\alpha}\PC{\xi_{\mu}}}
\end{eqnarray}
This then gives us:
\begin{eqnarray}
	2\;\nabla_{(\mu}u_{\nu)} &=& -\frac{\xi_{\mu}\xi^{\alpha}}{\|\xi\|^2}\PR{\nabla_{\alpha}\PC{\frac{\xi_{\nu}}{\|\xi\|}} -  \xi_{\nu}\nabla_{\alpha}\PC{\frac{1}{\|\xi\|}}} - \frac{\xi_{\nu}\xi^{\alpha}}{\|\xi\|^2}\PC{\frac{1}{{\|\xi\|}}\nabla_{\alpha}\PC{\xi_{\mu}}}\nonumber\\
	&=& -\frac{\xi_{\mu}}{\|\xi\|}\; a_{\nu}  -\frac{\xi_{\nu}}{\|\xi\|}\; a_{\mu} = -u_{\mu}\; a_{\nu}  - u_{\nu}\; a_{\mu} = -2\; u_{(\mu}\; a_{\nu)},
\end{eqnarray}
where we have used the fact that $\xi^{\alpha}\nabla_{\alpha}\|\xi\| = 0$ if $\xi$ is a Killing vector.
So, given equation \eqref{E:D(u,a)}, we have:
\begin{equation}
D_{\mu\nu}  = 2\PR{\nabla_{(\mu} u_{\nu)} + u_{(\mu}a_{\nu)}} = 0 .
\end{equation}

\paragraph{Conformal Killing congruence}\hfill

This is actually \emph{not} a rigid body motion case. As we will see, this class of observers have null shear, but non-zero expansion. It seems, however, worthy of being included in this list of examples. It has interesting applications for congruences in FLRW cosmological backgrounds and, given that it has no shear, it is simple enough in order to give us a unique time-scale for the system's evolution, obtained via the expansion scalar $\theta$, as given in equation \eqref{E:timescales}. 
Let us then calculate what is $D_{\mu\nu}$ for conformally Killing congruences.
 
When following a conformal Killing congruence, the fluid's 4-velocity is given by:
\begin{equation}
\label{E:conformal killing velocity}
u^{\mu} = \frac{\xi^{\mu}}{\|\xi\|}, \qquad \text{where} \qquad \nabla_{(\mu}\xi_{\nu)} =\frac{\epsilon}{2}\; g_{\mu\nu}.
\end{equation}
Now, notice that, by taking the trace of the second equality in \eqref{E:conformal killing velocity}, we obtain
\begin{equation}
\epsilon = \frac{1}{2}(\nabla_{\mu} \xi^{\mu}).
\end{equation}
\enlargethispage{10pt}
Now our aim will be to use the definition of $D_{\mu\nu}$ in terms of the divergence of the four-velocity, given in equation \eqref{E:D in terms of u}, since it makes the calculation much more straightforward. For this case we have:
\begin{eqnarray}
\nabla_{\mu}u_{\nu} = \nabla_{\mu}\PC{\frac{\xi_{\nu}}{\|\xi\|}} = \; \; \frac{1}{\|\xi\|}\nabla_{\mu}\PC{\xi_{\nu}} - 
\frac{\xi_{\nu}}{\|\xi\|^2}\nabla_{\mu}\|\xi\|\,.
\end{eqnarray}
This gives us:
\begin{eqnarray}
2\; \nabla_{(\mu}u_{\nu)} &=&  2 \frac{1}{\|\xi\|}\nabla_{(\mu}\xi_{\nu)} -  2\;\frac{\xi_{(\nu}\nabla_{\mu)}\|\xi\|}{\|\xi\|^2}. 
\end{eqnarray}
Then, using \eqref{E:conformal killing velocity}, we obtain:
\begin{eqnarray}
\label{E:conf symm div}
2\; \nabla_{(\mu}u_{\nu)} &=&  \frac{\epsilon}{\|\xi\|} \;g_{\mu\nu} -
2\;\frac{\xi_{(\nu}\nabla_{\mu)}\|\xi\|}{\|\xi\|^2}.
\end{eqnarray}

Applying to \eqref{E:D in terms of u} we have:
\vspace{-0.5cm}
\begin{eqnarray}
	D_{\mu\nu} &=& \nabla_{(\mu}u_{\nu)} + u^{\alpha}u_{\mu}\nabla_{(\nu}u_{\alpha)}
	+ u^{\alpha}u_{\nu}\nabla_{(\mu}u_{\alpha)}\nonumber\\
	\;\nonumber\\
	&=&  \frac{\epsilon}{2\|\xi\|} \;g_{\mu\nu} -
	\frac{1}{\|\xi\|^2}\xi_{(\nu}\nabla_{\mu)}\|\xi\|
	+  \frac{\epsilon}{2} \frac{u_{\mu}u^{\alpha}}{\|\xi\|} \;g_{\alpha\nu} -
	\frac{u_{\mu}u^{\alpha}}{\|\xi\|^2}\xi_{(\alpha}\nabla_{\nu)}\|\xi\|\nonumber\\
	&& + \; \frac{\epsilon}{2} \frac{u_{\nu}u^{\alpha}}{\|\xi\|} \;g_{\alpha\mu} \; -    \frac{u_{\nu}u^{\alpha}}{\|\xi\|^2}\xi_{(\alpha}\nabla_{\mu)}\|\xi\|,
\end{eqnarray}
which, expanding the terms gives us:
\begin{eqnarray}
	D_{\mu\nu} &=& \quad \frac{\epsilon}{2\|\xi\|} (g_{\mu\nu} + 2\;u_{\mu}u_{\nu})
	-\frac{1}{\|\xi\|} u_{(\nu}\nabla_{\mu)}\|\xi\|\\
	&& \; - \frac{1}{2\|\xi\|}\PC{ u_{\mu}u^{\alpha}u_{\alpha}\nabla_{\nu}\|\xi\|
		+ u_{\mu}u^{\alpha}u_{\nu}\nabla_{\alpha}\|\xi\|
		+  u_{\nu}u^{\alpha}u_{\mu}\nabla_{\alpha}\|\xi\|
		+  u_{\nu}u^{\alpha}u_{\alpha}\nabla_{\mu}\|\xi\|\nonumber
	}.
\end{eqnarray}
This can be arranged as:
\begin{eqnarray}
	D_{\mu\nu} =\; \frac{\epsilon}{2\|\xi\|} (g_{\mu\nu} + 2\;u_{\mu}u_{\nu}) 
	- \frac{1}{2\|\xi\|}\PC{ 2\; u_{\mu}u_{\nu}\;u^{\alpha}\nabla_{\alpha}\|\xi\|}
\end{eqnarray}
Now, using the fact that 
\begin{equation}
\label{E:xi derivative}
u^{\alpha}\nabla_{\alpha}\|\xi\| = \frac{\epsilon}{2},
\end{equation}
we have:
\begin{eqnarray}
D_{\mu\nu} &=& \frac{\epsilon}{2\|\xi\|} (g_{\mu\nu} + 2\;u_{\mu}u_{\nu} - u_{\mu}u_{\nu}) \\
&=& \frac{\epsilon}{2\|\xi\|} h_{\mu\nu}\quad = \quad 
\frac{(\nabla_{\delta} \xi^{\delta})}{\|\xi\|} \frac{h_{\mu\nu}}{4}.
\end{eqnarray}
And, given that 
\begin{equation}
\sigma_{\mu\nu} = D_{\mu\nu} -\frac{1}{3}\theta h_{\mu\nu},
\end{equation}
the shear will be given by
\begin{eqnarray}
	\sigma_{\mu\nu} &=& \frac{(\nabla_{\mu} \xi^{\mu})}{\|\xi\|} \frac{h_{\mu\nu}}{4} - \frac{1}{3}\nabla_{\alpha}\PC{\frac{\xi^{\alpha}}{\|\xi\|}}h_{\mu\nu} \\
	\;\nonumber\\
	&=& \PC{\frac{(\nabla_{\mu} \xi^{\mu})}{4\|\xi\|} -\frac{1}{3\; \|\xi\|} \nabla_{\alpha}\xi^{\alpha}
		+ \frac{1}{3\; \|\xi\|^2}\xi^{\alpha}\nabla_{\alpha}\|\xi\|  }h_{\mu\nu},
\end{eqnarray}
which, using \eqref{E:xi derivative} again, we have:
\begin{eqnarray}
	\sigma_{\mu\nu} &=& \PC{\frac{(\nabla_{\mu} \xi^{\mu})}{4\|\xi\|} -\frac{1}{3\; \|\xi\|} \nabla_{\alpha}\xi^{\alpha}
		+ \frac{1}{3\; \|\xi\|} \frac{\nabla_{\alpha}\xi^{\alpha}}{4}  }h_{\mu\nu} = 0.
\end{eqnarray}

In this way, we have obtained, as expected, that the shear of a congruence with 4-velocity defined by a conformal Killing vector is zero. Such congruences have only expansion, which is given by
\begin{equation}
\theta = \nabla_{\alpha}\PC{\frac{\xi^{\alpha}}{\|\xi\|}} =
\frac{3}{4}\;\frac{(\nabla_{\mu} \xi^{\mu})}{\|\xi\|}
\end{equation}
or, in terms of $\epsilon$,
\begin{equation}
\theta = \frac{3}{2}\;\frac{\epsilon}{\|\xi\|}.
\end{equation}
As we see, $\theta$ depends both on the norm of the conformal Killing vector $\xi$ as well as in $\epsilon$, the conformal factor. For example, for a FLRW metric given by
\begin{equation}
\d s^2 = -\d t^2 + a^2(t)\PR{\frac{\d r^2}{1 - kr^2} + r^2\PC{\d \theta^2 + \sin^2\d\phi^2}},
\end{equation}
an observer with 4-velocity $u^{\mu} =(1,0,0,0)$ would have
\begin{eqnarray}
\theta = 3\;\frac{\dot{a}}{a}\,,
\end{eqnarray}
with $\sigma_{\mu\nu} = \omega_{\mu\nu} = 0$.
This result will be used in section \ref{S: EIT} to compare the rate of expansion of FLWR universes with the relaxation time of some systems.

\paragraph{Geodesic Congruences}\hfill 

For the specific case when dealing with geodesic congruences, let us see what $\theta = \sigma_{\mu\nu} = 0$ implies. From \eqref{E:D(u,a)} and $a^{\mu}=0$ we obtain:
\begin{equation}
D_{\mu\nu} = \nabla_{(\mu}u_{\nu)} = 0,
\end{equation}
giving us:
\begin{equation}
\nabla_{\mu}u_{\nu} + \nabla_{\nu}u_{\mu} = 0,
\end{equation}
which tell us that $u^{\mu}$ has to be  a Killing vector. This is not unexpected since, when following a geodesic trajectory, our congruence will simply be moving along the ``natural curves'' of space-time. So, in order to have no expansion or shear happening in the congruence, it is necessary that space itself will not change along such a direction. This naturally implies a symmetry and the existence of a Killing vector. We are in this way not surprised by such a result. The question, however, is: when not following geodesics, can we find four-velocities for which $D_{\mu\nu} =0$ without the need for a Killing vector?

\subsection{Non-geodesic non-Killing congruences:}
Let us now present the results found when trying to obtain rigid motion solutions for congruences not generated by Killing vectors. This is exciting as it shows that indeed the Herglotz-Noether theorem is not valid for general curved space-times. We will present specific examples that could be found through a case-by-case analysis. We adopted such a method because -- as known up until now --  the non-Killing solutions for
\begin{equation}
\mathfrak{L}_u h_{\mu\nu} =0
\end{equation} 
cannot be generally found. There is no general strategy to do so.  This is the case since each solution will depend both on the metric as well as on the four-velocity chosen, in a way that we end up with an under-determined system of PDEs. We could, however, also find a ``general'' time-dependent solution --- ``general'' in the sense that we could find the most general time-dependent metric for which the specific four-velocity $u^{\mu} = (1,0,0,0)$ is a rigid motion. We will also present some of the cases where no `non-Killing generated solution' could be found. 

\paragraph{Universe expanding in the x-coordinate}\hfill

Let us start with the simplest example. It consists of a Bianchi type I universe with expansion in only one of the spatial coordinates. The metric is given by~\footnote{Of course the result does not depend on which spatial coordinate is chosen, as long as the four-velocity is adapted.}:
\begin{equation}
g_{\mu\nu} = \begin{pmatrix}
-1 & 0 & 0 & 0\\
0 & a(t) & 0 & 0\\
0 & 0 & b_1 & 0\\
0 & 0 & 0 & b_2
\end{pmatrix}\; ,
\end{equation} 
where $b_1$ and $b_2$ are constants. The translation Killing vectors for this metric are given by:
\begin{equation}
\xi^{\mu} = \PR{\;(0,1,0,0)\;, \;(0,0,1,0)\;, \;(0,0,0,1)\;}.
\end{equation}
Now, if we pick a four-velocity of the form
\begin{equation}
U_{\pm}^{\mu} = \PC{\pm\sqrt{\frac{d}{a(t)}} \;,\; \frac{\sqrt{d - a(t)}}{d}\;, \;0\;,\; 0},
\end{equation}
where $d$ is a constant, one can show that\footnote{The calculations were developed on Maple software.}
\begin{equation}
D_{\mu\nu} =\nabla_{(\mu} U_{\nu)} + U_{(\mu}A_{\nu)} = 0,
\end{equation} 
where $A^{\mu}$ is the 4-acceleration of the congruence given by:
\begin{equation}
A_{\pm}^{\mu} = \PC{- \frac{\dot{a}(t)}{2a(t)} \;,\; \mp\; \frac{\dot{a}(t)}{2a(t)}\sqrt{\frac{d}{a(t)(d-a(t))}}\;,\;0\;,\;0\;}
\end{equation}
is the congruence's four-acceleration, which satisfies:
\vspace{0.3cm}
\begin{equation}
\nabla_{[\mu}A_{\nu]} = \sqrt{\frac{d}{a^3}}\PR{\frac{2 a \ddot{a}(d -a) - \dot{a}^2(d-2a)}{4\;(d - a)^{3/2}}}
\begin{pmatrix}
0 & -1 & 0 & 0\\
1 & 0 & 0 & 0\\
0 & 0 & 0 & 0\\
0 & 0 & 0 & 0
\end{pmatrix}.
\end{equation}

The fact that $\nabla_{[\mu}A_{\nu]} \neq 0$ proves that such a trajectory is not generated by a Killing vector, as discussed around equation \eqref{E:non-Killing condition}.
For completeness, note that by comparing this four-velocity with the metric's Killing vectors, there is no linear combination of $\xi$'s that can form $U^{\mu}$. Something to be pointed out, though, is the need for $d$ to be greater than $a(t)$ for all $t$, otherwise such a four-velocity will not be in the real domain. In this way, the only condition for $g_{\mu\nu}$ to have a non-Killing rigid motion congruence is $a(t)$ to be a bounded function of time. It is also worthy to point out that $U_{\pm}^{\mu}$ are the only two four-velocities which generate non-Killing rigid motion for this metric.

\paragraph{The ``oblate'' universe in the x-y plane}\hfill 

\vspace{-0.3cm}
\enlargethispage{10pt}
Our second example is also a Bianchi type I, with the $x$ and $y$ expansion factors correlated in a specific way. The metric is given by:
\begin{equation}
g_{\mu\nu} = \begin{pmatrix}
-1 & 0 & 0 & 0\\
0 & a(t) & 0 & 0\\
0 & 0 & \frac{b_1{}^2}{a(t) - d} & 0\\
0 & 0 & 0 & b_2{}^2
\end{pmatrix}
\end{equation} 
where $b_1$ and $b_2$ and $d$ are constants. Again we see, now from the metric, that the function $a(t)$ must be bounded in time. The physical interpretation of this universe is interesting because, if $a(t)$ is increasing with time, the $x$ coordinate will expand, while the $y$ direction will contract, like a two-dimensional cigar. The $t$ and $z$ directions, on the other hand, keep constant all through.
The Killing vectors for this spacetime are:
\vspace{-0.4cm}
\begin{equation}
\label{E:kill}
\xi^{\mu} = \PR{\;(0,1,0,0)\;, \;(0,0,1,0)\;, \;(0,0,0,1)\;}.
\end{equation}
\enlargethispage{20pt}
The chosen four-velocity this time will be given by:
\begin{equation}
U^{\mu} = \PC{-\sqrt{\frac{d - a(t)}{a(t)}} \;,\; \frac{\sqrt{d - a(t)}}{a(t)}\;, \; - \frac{\sqrt{d - a(t)}}{b_1} \;,\; 0},
\end{equation}
which leads to the four-acceleration:
\begin{equation}
A^{\mu} = \PC{ \frac{\dot{a}(t)\;d}{2 a(t)\,(a(t)-d)}\;,\; \frac{\dot{a}(t)}{2 \,\sqrt{a(t)}} \;,\; \frac{\dot{a}(t)}{2\,b_1 \,a(t)^{1/2}}\; \;,0\;}.
\end{equation}
This four-acceleration satisfies $\nabla_{[\mu}A_{\nu]} \neq 0$. Explicitly, taking $a(t)\to a$, we have:
\begin{equation*}
\small
\nabla_{[\mu}A_{\nu]} = 
\begin{pmatrix}
0 & \frac{2 \ddot{a} a -\dot{a}^2}{4\; a^{3/2}} & 
-\frac{b_1\PR{2\, a \ddot{a} (c-a) -\dot{a}^2(c-3a)}}{4\, \|c-a\| \, a^{3/2}} & 0\\
- \frac{2 \ddot{a} a -\dot{a}^2}{4\; a^{3/2}}  & 0 & 0 & 0\\
\frac{b_1\PR{2\, a \ddot{a} (c-a) -\dot{a}^2(c-3a)}}{4\, \|c-a\| \, a^{3/2}} & 0 & 0 & 0\\
0 & 0 & 0 & 0
\end{pmatrix}.
\end{equation*}
This four-velocity can again be shown to satisfy
\begin{equation}
D_{\mu\nu} = 0,
\end{equation} 
while being linearly independent of the set of Killing vectors \eqref{E:kill}, guaranteeing that this is indeed a non-Killing rigid motion congruence.

\paragraph{A similar solution for Bianchi type I}\hfill

Given the similarity with the last case just presented, we will quickly display the other solution found for Bianchi type I universes. It is given by the metric:
\begin{equation}
g_{\mu\nu} = \begin{pmatrix}
-1 & 0 & 0 & 0\\
0 & e^{-t} & 0 & 0\\
0 & 0 & \frac{b_1{}^2\; (e^t -1)}{2\,d\, (d\, e^t -1)} & 0\\
0 & 0 & 0 & b_2{}^2
\end{pmatrix}
\end{equation} 
and has four-velocity
\begin{equation}
U^{\mu} = \PC{ e^{t/2} \sqrt{\frac{d\;e^t -1}{e^t -1}}\; ,\; e^{t/2}\sqrt{d\;e^t -1} \;,\; 
e^{t/2}\; \frac{\sqrt{2\,d (d-1) (d\;e^t -1)}}{b_1\, (e^t -1)}, 0} .
\end{equation}
The Killing vectors for this metric are also given by 
\begin{equation}
\xi^{\mu} = \PR{\;(0,1,0,0)\;, \;(0,0,1,0)\;, \;(0,0,0,1)\;},
\end{equation}
implying that this is a non-Killing rigid motion congruence. A brief calculation then shows that $D_{\mu\nu}=0$ for this case. Furthermore, one can show that the four-acceleration for this congruence, which is given by:
\begin{equation}
A^{\mu} = \PC{ \frac{d\, e^{2t} -2\,d\, e^t +1}{2[d\, e^{2t} - e^t (d +1) +1 ]} \;,\; \frac{e^t}{2\sqrt{e^t -1}} \;,\; - \frac{e^t\;\sqrt{2\,d(d-1)(e^t -1)}}{2\,b_1\,(e^{2t} -2\,e^t +1)}\;,0\;}
\end{equation}
also satisfies  $\nabla_{[\mu}A_{\nu]}\neq 0$.


\paragraph{Rigid motion in a general time-dependent case}\hfill

Now, instead of looking at a predetermined space-time, we will keep the structure of the metric free and fix only the four-velocity vector. The reason for inputting $u^{\mu}$ being that the metric, on its own, already gives us ten unknown functions of the coordinates $X = (t,x,y,z)$, which would require 10 equations to be completely solved\footnote{It is actually 6 degrees of freedom due to the coordinate freedom. However, this freedom is lost when we fix the four-velocity, as we will. So we will keep the 10 degrees of freedom in the end.}. 
Furthermore, the fluid's four-velocity (already imposing the normalization condition) adds extra 3 degrees of freedom to the problem. On the other hand, the number of equations obtained by imposing
\begin{equation}
\mathfrak{L}_u h_{\mu\nu} = 0 
\end{equation}
is only 6. It is then clear that some assumptions must be made. We will, in this way, assume the four-velocity to have a specific form, namely:
\begin{equation}
\label{E:U(t)}
u^{\mu} =\PC{k(t),0,0,0}.
\end{equation}
Let us now check how $D_{\mu\nu}=0$ looks like for this four-velocity and see what conditions the components of $g_{\mu\nu}$ will have to satisfy.
We will do this analytically. Start by noticing that:
\begin{eqnarray}
D_{\alpha\beta} &=& h^{\mu}_{\alpha} h^{\nu}_{\beta} \;\nabla_{(\mu}u_{\nu)}\nonumber\\
&=& h^{\mu}_{\alpha} h^{\nu}_{\beta}\PR{ \partial_{(\mu}u_{\nu)} - \Gamma^{\gamma}_{\mu\nu}u_{\gamma} }
\end{eqnarray}
Using \eqref{E:U(t)}, we have:
\begin{eqnarray}
\label{E:Dint}
D_{\alpha\beta} &=& h^{\mu}_{\alpha} h^{\nu}_{\beta}\PC{ \partial_{(\mu}g_{\nu)\gamma}u^{\gamma} - \Gamma^{\gamma}_{\mu\nu}g_{\gamma\kappa}u^{\kappa} }.
\end{eqnarray}
Letting $\stackrel{*}{=}$ represent an equality valid only in the specific coordinate system here adopted, we have: 
\begin{equation}
\label{E:uunder}
u_{\mu} = g_{\mu\nu}\; u^{\nu} \stackrel{*}{=} g_{\mu\nu}\;k(t) \;\delta^{\nu}_0 \stackrel{*}{=} g_{\mu 0}\; k(t).
\end{equation}

Applying \eqref{E:uunder} to \eqref{E:Dint}, we obtain:
\begin{eqnarray}
D_{\alpha\beta} &\stackrel{*}{=}& h^{\mu}_{\alpha} h^{\nu}_{\beta}\PC{ \partial_{(\mu}g_{\nu) 0}\;k - \Gamma^{\gamma}_{\mu\nu}g_{\gamma 0}\;k }\\
&\stackrel{*}{=}&  \PC{g^{\mu}_{\alpha} + u^{\mu}u_{\alpha}} \PC{g^{\nu}_{\beta} + u^{\nu}u_{\beta}}\PC{ \partial_{(\mu}g_{\nu) 0}k - \Gamma^{\gamma}_{\mu\nu}g_{\gamma 0}\;k}.
\end{eqnarray}
Distributing and arranging the terms:
\begin{align}
D_{\alpha\beta} \stackrel{*}{=} & \quad
g_{0 (\alpha} \partial_{\beta)} k + k\; \partial_{(\alpha} g_{0 \beta)} -k\;\Gamma^{\gamma}_{\alpha\beta} g_{\gamma 0}  \\ &
+ k\; u_{(\alpha}\partial_{\beta)}(g_{00}\; k)  -2 k^2\; u_{(\alpha}\Gamma_{\beta) 0}^{\gamma} g_{\gamma 0} + k\; g_{0 (\alpha}u_{\beta)} \partial_0 k \nonumber \\ &
+ k^2\; u_{(\alpha}g_{\beta) 0,0} + k^2 \;u_{\alpha}u_{\beta} (k g_{00})_{,0} - k^3\; u_{\alpha}u_{\beta} \Gamma^{\gamma}_{00}\; g_{\gamma 0}.\nonumber
\end{align}
Rewriting the Christoffel symbols in terms of the metric,
\begin{eqnarray}
\Gamma^{\alpha}_{\mu\nu} = \frac{g^{\alpha\beta}}{2}\PC{\partial_{\mu}g_{\beta\nu} + \partial_{\nu}g_{\mu\beta} - \partial_{\beta}g_{\mu\nu}}
\end{eqnarray}
 and again rearranging the terms, we have:
\begin{align}
\label{E:D*}
D_{\alpha\beta} \stackrel{*}{=} & \PR{1 + k^2\; g_{00}} g_{0( \alpha}\partial_{\beta)} k + 
\PR{1 + k^2\; g_{00}} k^2 g_{0 (\alpha} g_{\beta) 0} \;\partial_0 k \\
& + \frac{k}{2} \; g_{\alpha\beta ,0} + k^3\; g_{0( \alpha}\;g_{\beta)0, 0} + \frac{k^5}{2} g_{0( \alpha}\;g_{\beta)0} \; g_{00,0}.\nonumber
\end{align}
Note now that
\begin{equation}
-1 = g_{\mu\nu} u^{\mu}u^{\nu} \stackrel{*}{=} g_{00} \;k^2,
\end{equation}
which implies that the first two terms in equation \eqref{E:D*} are zero. The remaining non-zero terms can be rewritten as:
\begin{equation}
D_{\alpha\beta} \stackrel{*}{=} \frac{k}{2}\; \partial_0 \PR{g_{\alpha\beta} - \frac{g_{0\alpha}\;g_{0\beta}}{g_{00}}}.
\end{equation} 
So, the conditions that the metric components have to satisfy for rigid motion can be written as:
\begin{equation}
\label{E:conditions metric}
\partial_0 \PR{g_{\alpha\beta} - \frac{g_{0\alpha}\;g_{0\beta}}{g_{00}}} \stackrel{*}{=} 0.
\end{equation}
 Furthermore, using \eqref{E:uunder}, we see that this can also be expressed as:
\begin{equation}
\label{E:aiaiai}
D_{\alpha\beta}  \stackrel{*}{=}  \partial_0 \PC{g_{\alpha\beta} + u_{\alpha}\;u_{\beta}} \stackrel{*}{=}  \partial_0 \PC{h_{\alpha\beta}} \stackrel{*}{=} 0.
\end{equation} 
If, for reasons of simplicity, we impose $g_{00} = -1$ and $k(t) = 1$, the most general metric which satisfies \eqref{E:conditions metric} is given by:
\begin{equation}
\label{E:metric 1}
g_{\mu\nu} = \begin{pmatrix}
-1   & a(X) & b(X) & c(X) \\
a(X) & c_1 -a(X)^2 &  c_4 - b(X) a(X) & c_5 - c(X) a(X)\\
b(X) & c_4 - b(X) a(X) & c_2 - b(X)^2 & c_6 - b(X) c(X)\\
c(X) & c_5 - c(X) a(X) & c_6 - b(X) c(X) & c_3 - c(X)^2
\end{pmatrix},
\end{equation} 
where $X = (t,x,y,z)$. The induced metric $h_{\mu\nu}$ for this case is given by: 
\begin{equation}
\label{E:R 1}
h_{\mu\nu} = \begin{pmatrix}
0   & 0 & 0 & 0 \\
0 & c_1 &  c_4  & c_5 \\
0 &  c_4  & c_2  & c_6 \\
0 & c_5  & c_6  & c_3 
\end{pmatrix},
\end{equation}  
clearly satisfying \eqref{E:aiaiai}.  One can then check that with the metric given by \eqref{E:metric 1}, the four-velocity given by:
\begin{equation}
U^{\mu} = (\pm 1, 0, 0, 0)
\end{equation}
generates a rigid motion.
On the other hand, if one decides to impose only that $g_{00}(X) = g_{00}(t)$ and
\begin{equation}
\label{E:vel 2}
U^{\mu} = (\pm k(t), 0, 0, 0),
\end{equation} 
then the most general metric is given by:
\vspace{0.2cm}
\begin{equation}
\label{E:metric 2}
g_{\mu\nu} = \begin{pmatrix}
-\frac{1}{k(t)^2}   & a(X) & b(X) & c(X) \\
a(X) & c_1 -a(X)^2 &  c_4 - b(X) a(X) & c_5 - c(X) a(X)\\
b(X) & c_4 - b(X) a(X) & c_2 - b(X)^2 & c_6 - b(X) c(X)\\
c(X) & c_5 - c(X) a(X) & c_6 - b(X) c(X) & c_3 - c(X)^2
\end{pmatrix},
\end{equation} 
where the functions $a(X)$, $b(X)$ and $c(X)$ are now not absolutely free as before, but constrained to assume the form:
\vspace{0.3cm}
\begin{align}
a(X) = \frac{f_1(x,y,z)}{\sqrt{k(t)^2 -1}}\;;\quad 
b(X) = \frac{f_2(x,y,z)}{\sqrt{k(t)^2 -1}}\;;\quad
c(X) = \frac{f_3(x,y,z)}{\sqrt{k(t)^2 -1}}\;.
\end{align}
We still have, in this way, a lot of freedom. This is more or less expected, since, as discussed in the beginning of this section, our initial problem contained 10 unknown functions and only 6 constraint equations, leaving us with the remaining free functions $k(t)$, $f_1(x_i)$, $f_2(x_i)$ and $f_3(x_i)$, where $x_i = (x,y,z)$. For completeness, the tensor $h_{\mu\nu}$ for this case is given by:
\begin{equation}
\label{E:R 2}
h_{\mu\nu} = \begin{pmatrix}
0   & 0 & 0 & 0 \\
0 & c_1 +(f_1(x_i))^2 &  c_4 + f_1(x_i)f_2(x_i) & c_5 + f_1(x_i)f_3(x_i)\\
0 &  c_4 + f_1(x_i)f_2(x_i) & c_2 +(f_2(x_i))^2 & c_6 + f_3(x_i)f_2(x_i)\\
0 & c_5 + f_1(x_i)f_3(x_i) & c_6 + f_3(x_i)f_2(x_i) & c_3 + (f_3(x_i))^2
\end{pmatrix},
\end{equation}  
which again satisfies \eqref{E:aiaiai}. In this way,  \eqref{E:vel 2} and \eqref{E:metric 2} also generate rigid motions.


\paragraph{The `no-solution' cases}\hfill

To conclude this section, we would like to quickly display the most interesting `no-solution' situation we have encountered along the process of finding non-Killing solutions for $D_{\mu\nu}=0$. 

One example of a metric that we could not find any solution for was the general three dimensional Bianchi type I universe, i.e.,
 \begin{equation}
 \d s^2 = -\d t^2 + a(t)\; \d x^2 + b(t) \;\d y^2 + c(t)\; \d z^2,
 \end{equation}
with $a(t)$, $b(t)$ and $c(t)$ non-constants. This was actually very surprising, given the existence of solutions for both  $a(t)$ and $b(t)$ non-constants (oblate and exponential cases) and for the $a(t)$ non-constant case (one dimensional expansion case), which were presented above. On the other hand, for $a(t)$ and $b(t)$ non-constant, all the solutions found presented a metric which is expanding in one dimension while contracting in the other dimension. The possibility of meeting this requirement (in case it is indeed a necessary condition on the metric) with a (3+1) diagonal metric does not seem obvious.
To investigate this, one further case that we wish to analyze in the future is a truly oblate universe, axisymmetric, expanding in the $z$ axis direction, for example, and contracting on the orthogonal $x$--$y$ plane. 

Furthermore, we have tested all the FLRW cases given by
\begin{equation}
\d s^2 = -\d t^2 + a(t)^2 \PC{\frac{\d r^2}{1 - k r^2} + r^2 \d\Omega^2},
\end{equation}
and all the four-velocities satisfying $D_{\mu\nu}=0$ were proportional to a Killing vector. This is exactly the expected outcome since FLRW universes are conformally flat for all possible curvatures $k = (-1, 0, 1)$ and, accordingly to the extension of the Herglotz-Noether theorem this must be the case.

Given the limitations of the software used during this process, these are all the cases that we could fully analyze. Possibly in the future, as mentioned, we will de- dicate some effort into investigating more complex space-times and four-velocities.

\section{Non-Perfect Fluids}
\label{S:Non-Homogeneous Fluids}

Let us now leave the kinematics and move on to the dynamics of fluids. In section \ref{S: General Relativity} we introduced the energy-momentum tensor and explicitly showed its form for a perfect fluid. Now we would like to do the same for non-perfect fluids. The motivation must be clear: perfect fluids are oversimplified quantities which can only describe homogeneous systems in thermal equilibrium. It is then natural to desire to describe situations more complex than this.

When generalizing the energy-momentum to non-perfect fluids, a reasonable approach adopted is to consider that the contributions from the perfect and non-perfect parts of the fluid do not couple, i.e., that they can be separated in a way that:
\begin{align}
T^{\mu\nu} = T^{\mu\nu}_{PF} + T^{\mu\nu}_{NPF};\\
J^{\mu} = J^{\mu}_{PF} + J^{\mu}_{NPF},
\end{align}
where ``PF'' and ``NPF'' refer to perfect fluid and non-perfect fluid. One of the first attempts given to the non-perfect fluid part was originally given by:
\begin{equation}
\label{E:begNPF}
T^{\mu\nu}_{NPF} = \PR{-2\eta \sigma^{\mu\nu} - \PC{\zeta -\frac{2}{3}\eta}\theta g^{\mu\nu}} +\PR{q^{\mu}u^{\nu} + q^{\nu}u^{\mu}}.
\end{equation}
Here $\eta$ and $\zeta$ refer to shear and bulk viscosity respectively, and $q^{\mu}$ is the heat flux vector. It is such that, with $u^{\mu}$ being the fluid's four-velocity, we have:
\begin{equation}
q^{\mu}u_{\mu} =0.
\end{equation} 

In this way, the first square brackets of \eqref{E:begNPF} is responsible for describing stresses and shear inside the fluid while the second square brackets describes the heat flows. 
In reference  \cite{Rezzolla} it was shown, however, that such form of the energy-momentum tensor implies superluminal propagation speeds. This then lead physicists to look for a new way of formulating $T^{\mu}_{NPF}$. 
Nowadays, the definition most typically used is the one given by:
\vspace{0.2cm}
\begin{equation}
T^{\mu\nu}_{NPF} = \PR{\pi^{\mu\nu} + \Pi\; h^{\mu\nu}} + \PR{q^{\mu}u^{\nu} + q^{\nu}u^{\mu}}.
\end{equation}

As one can notice, the heat flux terms did not change from the previous energy-momentum tensor presented. The shear and stress terms, on the other hand, are now described by two other less clear quantities, namely $\pi^{\mu\nu}$ the \emph{anisotropic stress tensor} and $\Pi$, the \emph{viscous bulk pressure}. We would also like to point out that, just as the heat flux $q^{\mu}$, the anisotropic stress tensor is also orthogonal to the four-velocity of the fluid, i.e.,
\begin{equation}
u_{\nu}\; \pi^{\mu\nu} = 0 \; .
\end{equation}

Hence, using the previously introduced energy momentum tensor for the perfect fluid, given by equation \eqref{E:perfect fluid}, for $T^{\mu\nu}_{PF}$, the complete final form of the energy-momentum tensor for a relativistic viscous fluid is given by:
\vspace{0.2cm}
\begin{equation}
\label{E:energymom}
T^{\mu\nu} = \varrho\, u^{\mu}u^{\nu} + \PC{p + \Pi}\;h^{\mu\nu} + \pi^{\mu\nu} + q^{\mu}u^{\nu} + q^{\nu}u^{\mu}.
\end{equation}
The hydrodynamic equations are then given by:
\begin{align} 
\nabla_{\nu}T^{\mu\nu} = &\; u^{\mu}u^{\nu}\nabla_{\nu}\PC{\varrho + p +\Pi} + \PC{\varrho + p +\Pi}\PC{u^{\mu}\theta +a_{\mu}} + \nabla_{\nu}\pi^{\mu\nu} \nonumber \\& + g^{\mu\nu}\nabla_{\nu}\PC{p +\Pi} 
+ q^{\mu}\theta + u^{\nu}\nabla_{\nu}q^{\mu} +
q^{\nu}\nabla_{\nu}u^{\mu} + u^{\mu}\nabla_{\nu}q^{\nu} =0
\end{align}
Now, making use of the projected covariant derivative $\D_{\mu}$, given by:
\begin{align}
\mathcal{D}_{\mu}\phi := h^{\nu}_{\mu}\;\nabla_{\nu} \phi \; ;\\
\D_{\mu}A_{\nu} := h^{\alpha}_{\mu}h^{\beta}_{\nu}\;\nabla_{\alpha}A_{\beta}\; ;\\
\D_{\rho}A_{\mu\nu} := h^{\alpha}_{\rho}h^{\beta}_{\mu}h^{\gamma}_{\nu} \; \nabla_{\alpha} A_{\beta\gamma} \; 
\end{align}
we can write the projections of \eqref{E:energymom} along 
the four-velocity direction as \cite{Rezzolla}:
\begin{equation}
\label{E:divTu}
u^{\mu}\nabla_{\mu} \varrho + \PC{\varrho + p + \Pi}\theta + 2 q_{\mu}a^{\mu} + \D_{\mu}q^{\mu} +\pi^{\mu\nu}\sigma_{\mu\nu} =0;
\end{equation}
and its projection along the plane orthogonal to $u^{\mu}$ as:
\begin{align}
\PC{\varrho + p + \Pi}\;a_{\mu} +\D_{\mu}\PC{p + \Pi} +\D_{\nu}\pi^{\nu}_{\mu} + a_{\nu}\pi_{\mu\nu} & \nonumber\\
+\;h^{\nu}_{\mu}\;u^{\lambda}\nabla_{\lambda} q_{\nu} +\PC{\omega_{\mu \nu} +\sigma_{\mu\nu} +\frac{4}{3} \theta h_{\mu\nu}}q^{\nu} & =0.
\label{E:divTh}
\end{align}

\vspace{-0.2cm}
Furthermore, as we will discuss in this section, the level of complexity in order to causally describe non-perfect fluids escalates significantly when compared to the perfect-fluid case. Since now energy fluxes and stresses are present in the fluid, we have to calculate how such quantities will increase the entropy of the system. Such information is given by the \emph{entropy current} $S^{\mu}$, which satisfies:
\begin{equation}
\nabla_{\mu} \;S^{\mu} \geq 0. 
\end{equation}
Sadly, the entropy current is not straightforwardly given by $T^{\mu\nu}$, but it has to be separately imposed. Using $\rho$, the rest mass density, the entropy current is then given by:
\begin{equation}
\label{E:entropy current}
S^{\mu} = s \rho u^{\mu} + \frac{R^{\mu}}{T},
\end{equation}
where $R^{\mu}$ is a four-vector with a non-zero divergence.
Nowadays the two main theories that stand out
assume very distinct forms for $R^{\mu}$.
\emph{Classical Irreversible Thermodynamics} is the simplest of the two and, for most purposes, it describes the dynamics sufficiently well. Unfortunately, it is not complete since it allows superluminal signals, breaking the causality requirement. In view of this fact, the more complex \emph{Extended Irreversible Thermodynamics} was formulated to correct this situation and it is considered, at the moment, to be the correct description for out of equilibrium fluids. It contains, however, quite a few extra free parameters, many of which have to be imposed by a kinetic theory. In this way, both theories have their benefits and drawbacks. Keeping hold of the equations presented in this part, let us now properly introduce the theories Classical and Extended Irreversible Thermodynamics.

\subsection{Classical Irreversible Thermodynamics}
\label{S:CIT}

The Classical Theory of Irreversible Processes, or Classical Irreversible Thermodynamics, is a thermodynamic theory for non-isotropic viscous fluids originally developed by Eckart in 1940 \cite{Eckart}. Before further exploring this theory, however, we would like to first talk about one of its main assumptions, which we mentioned in the beginning of this chapter, namely, the Local Equilibrium Hypothesis (LEH). According to it, ``the local and instantaneous relations between thermodynamic quantities in a system out of equilibrium are the same as for a uniform system in equilibrium.''\cite{Lebon}. In this way, the mental visualization of a fluid described by elements discussed in section \ref{S: Kinematics bitch} is valid and necessary for CIT. An important  consequence of LEH is that entropy remains a valid state function even for systems somewhat out of equilibrium. The temperature also remains well defined. The difference is that now both are allowed to vary in space and time. A condition for LEH to be valid is given by the Deborah number. This is defined as the ratio between the relaxation time for the elements to achieve thermal equilibrium and a macroscopic time, related to the time of an experiment, $De := t_r/t_E$. In this way, for $De \ll 1$,  LEH is perfectly valid, since variations of time scales $t_r$ are not perceived by the experiment. This is not true for high frequency systems, shock waves and ultrasound propagation, where a new theory which does not assume local equilibrium has to substitute CIT.

Eckart's work is part of the set of thermodynamic theories called \emph{first-order theories}, the reason being that it assumes the entropy current
\begin{equation}
S^{\mu} = s \rho u^{\mu} + \frac{R^{\mu}}{T}
\end{equation}
to have a linear dependence on the thermodynamic fluxes and ignores any higher order contributions. In order for this to happen, $R^{\mu}$ must clearly be a  linear function of $\Pi$, $q^{\mu}$ and $\pi^{\mu\nu}$. However, if we want $R^{\mu}$ to be in agreement with what is already known for thermodynamic systems while keeping the linearity requirement, the options actually reduce to one. It can be shown\cite{Rezzolla} that the most general first-order form $R^{\mu}$ can assume is given by: 
\begin{equation}
R^{\mu} = q^{\mu} \,,
\end{equation}
being $q^{\mu}$ the heat flux. In this way, we have:
\begin{equation}
S^{\mu}_{CIT} = s \rho u^{\mu} + \frac{q^{\mu}}{T}.
\end{equation}

Now, once we have the entropy current for CIT, it is possible to proceed and  calculate the entropy production rate. This will, consequently, supply us with all the necessary conditions for a fluid to be in thermal equilibrium according to such a theory. The rate of entropy production is given by:
\begin{align}
T\nabla_{\mu} S^{\mu} &= T\; \nabla_{\mu}\PC{s \rho u^{\mu} + \frac{q^{\mu}}{T}}\nonumber\\
&= T \rho \;u^{\mu}\nabla_{\mu} s + T s \nabla_{\mu} \PC{ \rho u^{\mu}} + \nabla_{\mu}q^{\mu} -q^{\mu} \nabla_{\mu} \ln T.
\label{E:divS}
\end{align}
Note, however, that the second term on the right-hand side of \eqref{E:divS} is actually the same as $\nabla_{\mu}J^{\mu}$ for an observer co-moving with the fluid and, given the continuity equation, we have:
\begin{equation}
\label{E:continuity eq.2}
\nabla_{\mu}J^{\mu} =\nabla_{\mu} \PC{\rho\, u^{\mu}} = 0.
\end{equation}
Plus, given the relativistic specific enthalpy defined as
\begin{equation}
h := \frac{\varrho + p}{ \rho},
\end{equation}
where, again, $\varrho = \rho \,(1 + \mathfrak{u})$, we can write the first law of thermodynamics \eqref{E:first law} in terms of the specific quantities as:
\begin{equation}
\d \varrho = h\; \d\rho + T  \rho \;  \d s
\end{equation}
This then allows us to rewrite the first term of \eqref{E:divS} as:
\begin{equation}
T \rho \;u^{\mu}\nabla_{\mu} s\; =\; u^{\mu}\nabla_{\mu} \varrho - h\; u^{\mu}\nabla_{\mu} \rho,
\end{equation}
which gives us:
\begin{align}
T\nabla_{\mu} S^{\mu} =u^{\mu}\nabla_{\mu} \varrho - h\; u^{\mu}\nabla_{\mu} \rho  +\; \nabla_{\mu}q^{\mu} \;-\; q^{\mu} \nabla_{\mu} \ln T.
\end{align}
Now, using \eqref{E:divTu} and the fact that
\begin{align}
\D_{\mu}q^{\mu} &= h^{\mu\nu}\nabla_{\mu}q_{\nu} = 
\PC{g^{\mu\nu} + u^{\mu}u^{\nu}}\nabla_{\mu}q_{\nu}\\ &=
\nabla_{\mu}q^{\mu} - q_{\nu}u^{\mu}\nabla_{\mu}u^{\nu}\\
&= \nabla_{\mu}q^{\mu} - q_{\nu}a^{\nu}, 
\end{align}
we have:
\begin{align}
T\nabla_{\mu} S^{\mu} =&\;  - \PC{\varrho + p + \Pi}\theta - q_{\mu}a^{\mu} -\pi^{\mu\nu}\sigma_{\mu\nu} 
- h\; u^{\mu}\nabla_{\mu}  \rho  \;-\; q^{\mu} \nabla_{\mu} \ln T .
\end{align}
Since $q^{\mu}u_{\mu}=0$, we can rearrange the terms as:
\begin{align}
T\nabla_{\mu} S^{\mu} =\; - \PC{\varrho + p + \Pi}\theta - h\; u^{\mu}\nabla_{\mu}  \rho  -\pi^{\mu\nu}\sigma_{\mu\nu}  -\PC{\D_{\mu}\ln T + a_{\mu}}q^{\mu} .\label{E:divS2}
\end{align}
Note as well that the two first terms of  \eqref{E:divS2} can be rewritten as:
\begin{align}
- \PC{\varrho + p + \Pi}\theta - h\; u^{\mu}\nabla_{\mu}  \rho =& 
- \PC{\varrho + p + \Pi}(\nabla_{\mu}u^{\mu}) - \frac{\varrho + p}{ \rho}\; u^{\mu}\nabla_{\mu}  \rho \\
=&  -\PC{ \frac{\varrho + p}{ \rho}} \PC{ \rho\,\nabla_{\mu}u^{\mu}+ \; u^{\mu}\nabla_{\mu}  \rho} - \Pi\; \theta\\
=&  -\PC{ \frac{\varrho + p}{ \rho}} \nabla_{\mu}\PC{ \rho\, u^{\mu}} - \Pi\; \theta \\
=& \; -\Pi\; \theta,
\end{align}
where we have used the continuity equation on the last step. This then gives us:
\begin{align}
\label{E:divSfinal}
T\nabla_{\mu} S^{\mu} = -\Pi\; \theta -\pi^{\mu\nu}\sigma_{\mu\nu}  -\PC{\D_{\mu}\ln T + a_{\mu}}q^{\mu} .
\end{align}
Accordingly to the second law of thermodynamics, the entropy production must be such that:
\begin{equation}
\nabla_{\mu}S^{\mu} \geq 0.
\end{equation}

Given \eqref{E:divSfinal} and keeping in mind that we are dealing with a first-order theory, we see that the simplest way for this to happen is to assume a linear relation between the flux terms and the pressure/forces:
\begin{align}
\label{E:flux&forces1}
\Pi =& -\zeta\; \theta;\\
\label{E:flux&forces2}
q_{\mu} =& -\kappa T\PC{\D_{\mu}\ln T + a_{\mu}};\\
\label{E:flux&forces3}
\pi_{\mu\nu} =& -2\eta\; \sigma_{\mu\nu},
\end{align}
where $\eta$ and $\zeta$ are the \emph{shear} and \emph{bulk viscosity},   and $\kappa$ is the \emph{thermal condictivity} of the fluid.  Equations \eqref{E:flux&forces1}--\eqref{E:flux&forces3} are known  as the \emph{transport} or  \emph{constitutive equations of CIT}. 
They are relativistic generalizations of the corresponding Newtonian laws:
\begin{align}
\Pi =& -\zeta\; \vec{\nabla}\cdot\vec{v} &&\text{(Stokes law),}\\
\vec{q} =& -\kappa \vec{\nabla}T &&\text{(Fourier),}\\
\pi_{\mu\nu} =& -2\eta\; \sigma_{\mu\nu} && \text{(Newton)} .
\end{align}

Furthermore, it is then clear that for a viscous system in thermal equilibrium we must have no fluxes present, which implies:
\begin{align}
\label{E:eq.forces1}
\theta =& \; 0\\
\label{E:eq.forces2}
\D_{\mu}\ln T + a_{\mu} =& \;0\\
\label{E:eq.forces3}
\sigma_{\mu\nu} =&\;0 .
\end{align}

We will discuss the implications of equations \eqref{E:eq.forces1}--\eqref{E:eq.forces3} in  section \ref{S:equlibrium non Killing} and connect them with everything developed in this and previous chapters of this thesis. But first, let us see what are the problems present in CIT.

\paragraph{Limitations of Classical Irreversible Thermodynamics}\hfill

Despite its numerous successes and many applications, there are a number of problems presented by Classical Irreversible Thermodynamics. Some are more severe than others. We will briefly present them here:

\begin{itemize}
	\item \emph{It is a first-order theory:} As we know, this theory is limited to look at entropy production rates which are only first-order dependent on the fluxes (and forces). Of course, this is the purpose of the theory, but it can not be denied that it does limit the variety of applications severely. 
	
	\item \emph{Local Equilibrium Hypothesis:} As previously mentioned, LEH puts a limit for CIT, since it is valid only for systems which vary under the condition of $De\ll 1$. For quickly changing systems, local equilibrium cannot be assumed and CIT is not able to perform good predictions. For this type of system, a new theory has to be used. This theory, which we will introduce in the next section, is called Extended Irreversible Dynamics and does not make use of LEH to make predictions. 
	
	\item \emph{Superluminal propagation speed:} When we look at the constitutive equations of CIT, given by eq. \eqref{E:flux&forces1}--\eqref{E:flux&forces3}, we can see that the relation between fluxes and forces is instantaneous. As soon as a force appears/disappears, a flux is generated or ceases, instantly.  This is due to the linear condition of the theory, not allowing more complex terms which incorporate relaxation times. Furthermore, equations \eqref{E:flux&forces1}--\eqref{E:flux&forces3}  can be shown to form a set of \emph{parabolic} equations (not hyperbolic), which present all the superluminal velocity pathologies as well. Since this is an approach to a relativistic hydrodynamics theory, the implications of such a flaw must be clear. We would like to emphasize, however, that the practical applications of this theory are still vast and as long as the characteristic time of the system is much longer than the propagation time of the signals, no major problem should occur\cite{Lebon}.
	\item \emph{Instability:} Another problem present in the first-order theories is their unstable character. The fluid can be shown to exhibit exponentially growing instabilities when slightly disturbed under reasonable conditions \cite{InstabilityCIT}.
\end{itemize}

So, given all these pitfalls, it was natural the wish to ``extend'' such theory for higher-order flux terms, and eventually correct its flaws.  Let us now introduce the theory designed to do this job.

\subsection{Extended Irreversible Thermodynamics}
\label{S: EIT}

\vspace{0.4cm}
The Extended Theory of Irreversible Processes, or Extended Irreversible Thermodynamics (EIT), was originally developed by Israel (1976) and Stewart (1977) and it is a second-order theory in the sense that allows the entropy production rate to depend on second-order terms. Its general form is assumed to be given by:
\vspace{0.3cm}
\begin{align}
S^{\mu} = s  \rho u^{\mu} + \frac{q^{\mu}}{T} - \PC{\beta_0 \Pi^2 +\beta_1 q_{\nu}q^{\nu} +\beta_2 \pi_{\alpha\beta}\pi^{\alpha\beta}}\frac{u^{\mu}}{2T} 
+\alpha_0 \Pi \frac{q^{\mu}}{T} +\alpha_1\frac{q_{\nu}\pi^{\mu\nu}}{T}
\end{align}
here $\Pi$, $q^{\mu}$ and $\pi^{\mu\nu}$ are the standard flux terms defined on section \ref{S:Non-Homogeneous Fluids}.
This entropy is composed by the first order terms present in CIT, followed by the squared terms in the round brackets -- which are multiplied by the thermodynamics coefficients $\beta_0$, $\beta_1$ and $\beta_2$ -- and the last two terms -- multiplied by the coefficients $\alpha_0$ and $\alpha_1$. This forms the complete set of all possible algebraic combinations of  $\Pi$, $q^{\mu}$ and $\pi^{\mu\nu}$  up to second order.

As we can notice, even going only as high as second order terms, the entropy production rate already shows extremely more complicated behaviour than before. One interesting point, however, is that while in the CIT theory all the terms responsible for producing entropy had a very clear physical meaning, this is not exactly true for EIT, with the physical origin of some of the terms being a bit obscure. On the other hand, this must be expected when leaving the clean predictable world of equilibrium situations and moving towards more real and complex systems.

Now, we can proceed just like we did for CIT and calculate the entropy production rate in order to find the constitutive equations for EIT. In this way, taking the divergence of $S^{\mu}$ and using the equations of motion we have \cite{Hiscock}:
\begin{align}
\small
T\;\nabla_{\mu}S^{\mu} = 
-\Pi &\PR{\theta + \beta_0\dot{\Pi} -\alpha_0\nabla_{\mu}q^{\mu} -\gamma_0 T q^{\mu}\nabla_{\mu}\PC{\frac{\alpha_0}{T}} +\frac{1}{2} T \nabla_{\mu}\PC{\frac{\beta_0}{T}u^{\mu}}\Pi
} \nonumber \\
-q^{\mu}&\left[\nabla_{\mu}\ln T +a_{\mu} +\beta_1 \dot{q}_{\mu} - \alpha_0\nabla_{\mu}\Pi  -(1-\gamma_0)T\;\nabla_{\mu}\PC{\frac{\alpha_0}{T}}\Pi \right. \label{E:divSEIT}\\
&\left. -\alpha_1\nabla_{\nu} \pi^{\nu}_{\mu} -(1-\gamma_1)\;T\; \pi^{\nu}_{\mu}\nabla_{\nu}\PC{\frac{\alpha_1}{T}} +\frac{1}{2}\; T q_{\mu}\nabla_{\nu}\PC{\frac{\beta_1}{T}u^{\nu}}\right] \nonumber\\
-\pi^{\mu\nu}&\PR{ \nabla_{\mu}u_{\nu} +\beta_2\dot{\pi}_{\mu\nu} -\alpha_1\nabla_{\mu}q_{\nu} -\gamma_1 T q_{\mu}\nabla_{\nu}\PC{\frac{\alpha_1}{T}} +\frac{1}{2}T \pi_{\mu\nu}\nabla_{\gamma}\PC{\frac{\beta_2}{T} u^{\gamma}}} \nonumber
\end{align}
Note that equation \eqref{E:divSEIT} contains two new thermodynamic coefficients $\gamma_0$ and $\gamma_1$. According to \cite{Hiscock}, these were introduced due to the ambiguity involved in factoring the terms with the products $(\Pi\; q^{\mu})$ and $(\pi^{\mu\nu} q_{\nu})$ on the right hand side of \eqref{E:divSEIT}. And, since the magnitudes of the $\gamma$'s are not known a priori, they could in principle be large. Now, applying the second law of thermodynamics, the simplest way to ensure that  $\nabla_{\mu} S^{\mu} \geq 0$ is satisfied is to assume the following  \emph{constitutive equations for EIT}:
\begin{align}
\small
\Pi &= -\zeta \PR{\theta + \beta_0\dot{\Pi} -\alpha_0\nabla_{\mu}q^{\mu} -\gamma_0 T q^{\mu}\nabla_{\mu}\PC{\frac{\alpha_0}{T}} +\frac{1}{2} T \nabla_{\mu}\PC{\frac{\beta_0}{T}u^{\mu}}\Pi
},\\
q^{\nu}& = -\kappa T h^{\mu\nu}\left[\nabla_{\mu}\ln T +a_{\mu} +\beta_1 \dot{q}_{\mu} - \alpha_0\nabla_{\mu}\Pi  -(1-\gamma_0)\Pi \;T\;\nabla_{\mu}\PC{\frac{\alpha_0}{T}} \right. \nonumber\\
& \quad\qquad\qquad\left. -\alpha_1\nabla_{\nu} \pi^{\nu}_{\mu} -(1-\gamma_1)\;T\; \pi^{\nu}_{\mu}\nabla_{\nu}\PC{\frac{\alpha_1}{T}} +\frac{1}{2}\; T q_{\mu}\nabla_{\nu}\PC{\frac{\beta_1}{T}u^{\nu}}\right],
\\
\pi_{\mu\nu} &= -2\eta \PR{ \nabla_{\mu}u_{\nu} +\beta_2\dot{\pi}_{\mu\nu} -\alpha_1\nabla_{\mu}q_{\nu} -\gamma_1 T q_{\mu}\nabla_{\nu}\PC{\frac{\alpha_1}{T}} +\frac{1}{2}T \pi_{\mu\nu}\nabla_{\gamma}\PC{\frac{\beta_2}{T} u^{\gamma}}} .
\end{align}
As we can see, this set of equations is not nearly as clear as the ones obtained by CIT. It is indeed a very complicated group of interconnected differential equations which can not easily be solved. We can, however, proceed with some simplifications which might make things a bit clearer. A simplification made both by \cite{Rezzolla} and \cite{Maartens} was to assume 
\begin{equation}
\label{E:simplify}
\alpha_0 =0\,,\quad \alpha_1 =0\,,\quad \gamma_0 = 0\,,\quad \text{and}\quad \gamma_1 =0.
\end{equation}
The justification to do so, given by Maartens \cite{Maartens}, is based on the fact that this assumption is consistent with linearisation in a perturbed FRW universe, since the coupling terms lead to non–linear deviations from the FRW background. They say, however, that such an assumption may not be reasonable for non–uniform stellar models and other situations where the background solution is inhomogeneous. Rezzolla \emph{et. al} \cite{Rezzolla}, on the other hand, simply apply the simplification without further explanations. As we will proceed with applying even further simplifications to the system, we will follow the strategy adopted by these authors. We do believe, however, that the subject deserves further investigation. So, using equation \eqref{E:simplify}, the constitutive equations become:
\begin{align}
\Pi &= -\zeta \PR{\theta + \beta_0\dot{\Pi}  +\frac{1}{2} T \nabla_{\mu}\PC{\frac{\beta_0}{T}u^{\mu}}\Pi
},\\
q^{\nu}& = -\kappa T h^{\mu\nu}\left[\nabla_{\mu}\ln T +a_{\mu} +\beta_1 \dot{q}_{\mu}   +\frac{1}{2}\; T q_{\mu}\nabla_{\nu}\PC{\frac{\beta_1}{T}u^{\nu}}\right],
\\
\pi_{\mu\nu} &= -2\eta \PR{ \nabla_{\mu}u_{\nu} +\beta_2\;\dot{\pi}_{\mu\nu} +\frac{1}{2}T \pi_{\mu\nu}\nabla_{\gamma}\PC{\frac{\beta_2}{T} u^{\gamma}}} .
\end{align}
Furthermore, these terms can be rearranged as \cite{Maartens}:
\begin{align}
\label{E:Pi}
\tau_0\dot{\Pi} + \Pi = -\zeta\theta - \PR{\frac{1}{2}\zeta T \nabla_{\mu} \PC{\frac{\tau_0}{\zeta T}u^{\mu}}\Pi}\;,\\
\label{E:q}
\tau_1 h_{\mu}{}^{\nu} \dot{q}_{\mu}  +q_{\mu}= -\kappa T\PC{\D_{\mu}\ln T +a_{\mu}} -\PR{\frac{1}{2}\kappa T^2 \nabla_{\nu}\PC{\frac{\tau_1}{\kappa T^2}u^{\nu}}q_{\mu}}\;,\\
\label{E:pi}
\tau_2 h_{\mu}{}^{\alpha}h_{\nu}{}^{\beta}\;\dot{\pi}_{\alpha\beta} + \pi_{\mu\nu} = -2\eta\; \sigma_{\mu\nu} -\PR{\frac{1}{2}\eta T \nabla_{\gamma}\PC{\frac{\tau_2}{\eta T}u^{\gamma}}\pi_{\mu\nu}},
\end{align}
where
\begin{align}
\label{E:tau}
\tau_0 := \zeta \beta_0,\quad \tau_1 := \kappa T \beta_1, \quad\text{and} \quad\tau_2 := 2 \eta \beta_2
\end{align}
will be shown to be the relaxation times of the different stresses present in the system. If one desires to further simplify the constraint equations, 
in many situations the terms organized in the square brackets are actually significantly smaller when compared to the other terms in the equations. Besides noticing that both \cite{Rezzolla} and \cite{Maartens} adopted such a simplification, neither of them explained why this is indeed a reasonable thing to do. Let us now see why this is so. If we take a closer look to these square brackets terms, we see that for \eqref{E:Pi} we have:
\begin{equation}
\label{E:Pii}
\PR{\frac{1}{2}\zeta T\; \nabla_{\mu}\PC{\frac{\tau_0\; u^{\mu}}{T \zeta}}\Pi} = \frac{1}{2}
\zeta T\;\Pi \; u^{\mu}\nabla_{\mu}\PC{\frac{\tau_0}{ T \zeta}} + \frac{\tau_0}{2} \theta\;\Pi.
\end{equation}
The first term on the right-hand side of \eqref{E:Pii} is related to changes along the fluid's proper time of the system's relaxation time, temperature and transport coefficient $\zeta$. It is then a very reasonable assumption to assume it to be negligible when compared even with the second term in this equation. By assuming this term to be zero and substituting \eqref{E:Pii} into \eqref{E:Pi} we then obtain:
\begin{align}
\tau_0\dot{\Pi} + \PC{1 + \frac{\tau_0}{2}\;\theta}\Pi = -\zeta \theta.
\end{align}
So, as long as $\tau_0 \ll \tau_{\theta}  = \theta^{-1}$, it is a good approximation to remove the square brackets terms from \eqref{E:Pi}. This does seem a reasonable assumption, given that the relaxation time for most fluids is a naturally small quantity. On the other hand, this approximation would not be valid for quickly expanding fluids. If we proceed with exactly the same analysis for \eqref{E:pi} and \eqref{E:q}, we obtain:
\begin{align}
\tau_1 h_{\mu}{}^{\nu} \dot{q}_{\mu}  + \PC{1 + \frac{\tau_1}{2}\theta}q_{\mu}=& -\kappa T\PC{\D_{\mu}\ln T +a_{\mu}} \\
\tau_2 h_{\mu}{}^{\alpha}h_{\nu}{}^{\beta}\;\dot{\pi}_{\alpha\beta} +\PC{1+ \frac{\tau_2}{2}\theta} \pi_{\mu\nu} =& -2\eta\; \sigma_{\mu\nu}\,,
\end{align}
where, again, we have assumed that the time derivative of $\kappa$, $\eta$, $T$ and $\tau_i$ are negligible. We then see whenever $\tau_i \ll \theta^{-1}$, these simplifications are rather reasonable.

As an example, for an observer with 4-velocity $u^{\mu}=(1,0,0,0)$ in a FLRW universe, we have:
\vspace{-0.4cm}
\begin{eqnarray}
\theta = 3\;\frac{\dot{a}}{a}\;= 3 \;H(t),
\end{eqnarray}
where $H(t)$ is the Hubble parameter given by $74.03\pm1.42$ (km/(s$\cdot$Mpc)).
This condition, applied to this situation would mean
\begin{align}
\frac{1}{\tau_i} \ll 74.01  \PC{\frac{10^3}{s}}\frac{1}{3,08\cdot10^{22}} \approx 24.03 \cdot 10^{-19} \,s^{-1} = 7.6 \cdot 10^{-11} year^{-1} 
\end{align}
which gives us:
\begin{align}
\tau_i \ll 6\;\text{(age of the universe)}
\end{align}
which probably makes it clear how good the approximation is in that particular case.

One might also be wondering about the curious fact that, to simplify all three equations \eqref{E:Pi}--\eqref{E:pi}, the only force term producing a constraint was $\theta$. We are not sure, at this point, whether this is a natural consequence from assuming all the $\alpha$'s and $\gamma$'s to be zero or not and, as mentioned, we believe that these approximations deserve further investigation in the future. In any case, now we are comfortable and aware of when the square brackets terms of \eqref{E:Pi}--\eqref{E:pi} can be neglected. We may then proceed and finally obtain a new set of equations which are in the so called \emph{Maxwell-Cattaneo} form, and are given by:
\begin{align}
\label{E:EITflux&forces1}
\tau_0\;\dot{\Pi} + \Pi =& -\zeta\; \theta\;,\\
\label{E:EITflux&forces2}
\tau_1\;h_{\mu}{}^{\nu}\;\dot{q}_{\nu} + q_{\mu} =& -\kappa T\PC{\D_{\mu}\ln T + a_{\mu}}\;,\\
\label{E:EITflux&forces3}
\tau_2\;h_{\mu}{}^{\alpha}h_{\nu}{}^{\beta}\;\dot{\pi}_{\alpha\beta} + \pi_{\mu\nu} =& -2\eta\; \sigma_{\mu\nu}\;.
\end{align}
With the constraint equations displayed in that form, the physical interpretation of $\tau_0$, $\tau_1$ and $\tau_2$ as relaxation times for the system becomes much clearer. Note also that the equilibrium state in EIT is achieved when no fluxes or stresses are present, just like we had in CIT, i.e.:
\begin{equation}
\theta =0, \;\; a_{\mu}=-\D_{\mu}T,\;\; \text{and}\;\; \sigma_{\mu\nu}=0\;\;\text{when the system is in equilibrium}.
\end{equation}
However, putting this back into equation \eqref{E:EITflux&forces1}, for example, we obtain:
\begin{align}
\label{E:taupi}
\tau_0\;\dot{\Pi} =-\Pi \implies
\Pi(t) = \Pi(0)\;e^{-\frac{t}{\tau_0}},
\end{align}
showing that the system exponentially settles into the equilibrium state. Equations \eqref{E:EITflux&forces2} and \eqref{E:EITflux&forces3}, besides looking more complicated, work exactly in the same way as \eqref{E:taupi}. The projection operators are simply guaranteeing that all the terms in the equations belong to the same plane (orthogonal to $u^{\mu}$). 
The presence of these relaxation times is a big difference between CIT and EIT. The theory, to be causal, needs to take into account the time taken by the signals to propagate to different parts and also the time taken by the system to settle into a new equilibrium state. This is what makes EIT causal while CIT is not. Of course, the downside is the presence of new parameters $\tau_i$ which, as well as the transport coefficients $\zeta$, $\kappa$ and $\eta$, must be given by the kinetic theory  and vary for different materials. It adds a new complexity to the system, making it harder to conduct simple calculations.

To complete this discussion, we would like to point out that other theories for relativistic hydrodynamics exist, some of them based on EIT. The main motivation for their formulation being the possibility of causality violation for systems strongly far away from equilibrium \cite{Hiscock, Hiscock1988} described by EIT theory. As an example, an alternative theory which does not contain such undesirable features was proposed by Liu \cite{Liu} and it is called the divergence-type formulation of Extended Irreversible Thermodynamics, also known as \emph{Rational Extended Thermodynamics}. The exploration of this topic is, however, well beyond the aim of this thesis.

\section{The possibility of equilibrium along non Killing flows}
\label{S:equlibrium non Killing}

We would like now to discuss whether Killing flows are absolutely necessary to define equilibrium states or not. We will start this analysis by focusing on the association between Born-rigid motion and equilibrium. As we have seen, a condition for a system to be in equilibrium according to both CIT and EIT is that no heat fluxes or stresses exist:
\begin{equation}
\Pi =0 \;, \quad q_{\mu}=0\;, \quad \pi_{\mu\nu}=0.
\end{equation}
Now, given the constitutive equations \eqref{E:flux&forces1}--\eqref{E:flux&forces3} for CIT and \eqref{E:EITflux&forces1}--\eqref{E:EITflux&forces3} for EIT, we see that for this to be satisfied we need to have 
\begin{equation}
\label{E:EIT equilibrium}
\theta =0 \quad\text{and}\quad\sigma_{\mu\nu}=0 \implies D_{\mu\nu} =0 \quad \text{plus}\quad T\;a_{\mu} = -\D_{\mu}T.
\end{equation}
Hence, systems in equilibrium must be moving in a Born-rigid way. Now, the converse is not true by itself, since $D_{\mu\nu}=0$ only implies no stresses, but heat flows are still allowed to occur. The extra condition is given by Tolman temperature gradients, equation \eqref{E: euler2}, derived in the previous chapter.

At a first glance, it seems that the condition of a Killing flow is not strictly necessary to define equilibrium states. Up to this point in the argument, no obvious reason to demand the fluid to be moving along a Killing orbit has been pointed out. 
This condition appears, however, once one starts to  substitute the just mentioned conditions back into the equations of motion of the fluids. By substituting $\theta =0=\Pi$, $q_{\mu}=0$ and $\sigma_{\mu\nu}=0=\pi_{\mu\nu}$ into \eqref{E:divTu}, \eqref{E:divTh} and into the continuity equation \eqref{E:continuity eq.2}, we obtain:
\begin{align}
u^{\mu}\nabla_{\mu} \varrho =0\;,\label{E:const rho}\\
\PC{\varrho + p }a_{\mu} +\D_{\mu}p =0\;,\\
u^{\mu}\nabla_{\mu} \rho =0. \label{E:const n}
\end{align}
These are the equations of motion for a relativistic viscous fluid in equilibrium moving rigidly (in the Born sense). Furthermore, the definition of temperature being adopted here is the one given in section \ref{S: How to measure temperatures}, namely:
\begin{equation}
\tag{\ref{temperature}}
\frac{1}{T(x)} = \frac{\d S}{\d U},
\end{equation}
with $S$ being the entropy and $E$ the energy of a small fluid element located at position $x$. Again, in specific units we have:
\begin{equation}
\tag{\ref{E: specific temp}}
\frac{1}{T} =  \PC{\frac{\d s}{\d \mathfrak{u}}}_{\rho}.
\end{equation}
One point that we did not explicitly mention yet, regards the free variables adopted for both CIT and EIT. Both these theories make the assumption that the entropy is a state function which depends only on $\rho$ and $\mathfrak{u}$, i.e., 
\begin{equation}
s = s(\rho,\mathfrak{u}),
\end{equation}
implying that the same is also valid for the temperature:
\vspace{-0.2cm}
\begin{equation}
T = T(\rho,\mathfrak{u}).
\end{equation}

\enlargethispage{20pt}
So, assuming this to be the case, we see that the conditions \eqref{E:const rho} and \eqref{E:const n} naturally imply that
\begin{equation}
\label{E:const T}
u^{\mu}\nabla_{\mu} T =0,
\end{equation}
since $\varrho = \rho\,(1+\mathfrak{u})$. This was indeed one of the conditions assumed in Chapter 3 when deriving Tolman temperature gradient for equilibrium compatible flows given by equation \eqref{e:euler3}, which, in the notation of this chapter is:
\begin{equation}
\tag{\ref{e:euler3}}
a_{\mu} =  - \nabla_{\mu} \ln T.
\end{equation}
The interesting part comes when we decide to join condition \eqref{E:const T} with $D_{\mu\nu}=0$. Starting with the definition for the rate of strain tensor, we have:
\begin{equation}
2\frac{D_{\mu\nu}}{T} = \frac{\nabla_{\mu}u_{\nu}}{T} + \frac{\nabla_{\nu}u_{\mu}}{T} + \frac{a_{\mu}u_{\nu}}{T} + \frac{a_{\nu}u_{\mu}}{T}.
\end{equation}
Using Tolman's temperature gradient, $a_{\mu}=-\D_{\mu}\ln T$, we get:
\begin{align}
2\frac{D_{\mu\nu}}{T} = &\nabla_{\mu}\PC{\frac{u_{\nu}}{T}} + \nabla_{\nu}\PC{\frac{u_{\mu}}{T}} 
+ \frac{u_{\nu}}{T^2}\nabla_{\mu}T + \frac{u_{\mu}}{T^2}\nabla_{\nu}T
- \frac{u_{\nu}}{T^2}\D_{\mu}T - \frac{u_{\mu}}{T^2}\D_{\nu}T \nonumber\\
=& \nabla_{\mu}\PC{\frac{u_{\nu}}{T}} + \nabla_{\nu}\PC{\frac{u_{\mu}}{T}} 
+ \frac{u_{\nu}}{T^2}\PC{\nabla_{\mu}T -\D_{\mu}T} +\frac{u_{\mu}}{T^2}\PC{\nabla_{\nu}T - \D_{\nu}T}.
\label{E:Dwith T}
\end{align}
On the other hand, we have:
\begin{equation}
\D_{\mu}T = h_{\mu}{}^{\nu}\nabla_{\mu}T = \nabla_{\mu}T + u_{\mu}u^{\nu}\nabla_{\nu}T, 
\end{equation}
which implies
\vspace{-0.2cm}
\begin{equation}
\label{E:hi}
\D_{\mu}T - \nabla_{\mu}T = u_{\mu}\;u^{\nu}\nabla_{\nu}T.
\end{equation}
So, by imposing condition \eqref{E:const T}, the right hand side of equation \eqref{E:hi} reduces to zero. Substituting this back into equation \eqref{E:Dwith T}, we obtain:
\begin{align*}
2\frac{D_{\mu\nu}}{T} = \nabla_{\mu}\PC{\frac{u_{\nu}}{T}} + \nabla_{\nu}\PC{\frac{u_{\mu}}{T}}.
\end{align*}

\begin{figure}
	\begin{center}
		\includegraphics[scale=0.22]{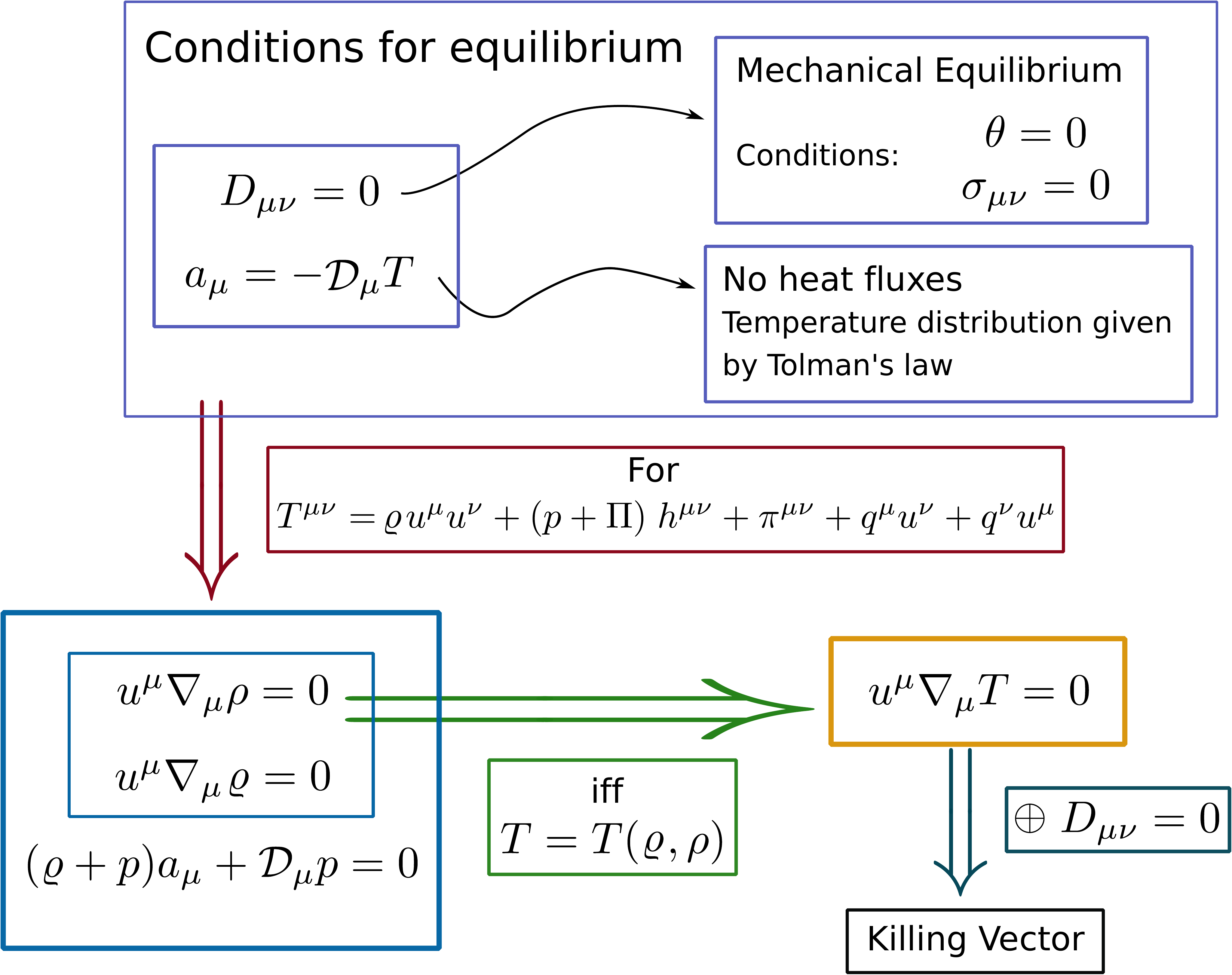}
		\caption{Diagram showing the relationship between the conditions for thermal equilibrium.}
		\label{F:diagram}
	\end{center}
\end{figure}

Now, if we are moving along a rigid body congruence, $D_{\mu\nu}=0$ and
\begin{align}
\label{E:Killing vector}
\nabla_{\mu}\PC{\frac{u_{\nu}}{T}} + \nabla_{\nu}\PC{\frac{u_{\mu}}{T}} =0,
\end{align}
which is the Killing equation for the vector
\begin{equation}
\label{E:vkill}
\xi_{\mu} = \frac{u_{\mu}}{T}.
\end{equation}
This result can also be obtained by looking at the relativistic Boltzmann equation and deriving from them the equilibrium conditions for the system \cite{Kremer}. The outcome is again that \eqref{E:vkill} must be a Killing vector.

Let us now take a moment to step back and appreciate the big picture. To help with this, we have exposed all the equilibrium requirements in a clear diagram in Figure \ref{F:diagram}.

As we can see, for a system to achieve and maintain itself in a exact thermodynamic equilibrium state, Killing vectors are unavoidable. When adding all of the conditions obtained for the energy momentum given by \eqref{E:energymom} -- currently accepted as the correct one -- and assuming $T=T(\rho, \mathfrak{u})$, one necessarily obtains the Killing vector condition \eqref{E:Killing vector}. On the other hand, there are interesting possibilities outside of the Killing vector scenario for perfect equilibrium which still may be worthy to analyze, especially for situations of adiabatic evolution happening outside of Killing orbits (for example the FLRW case presented on section \ref{S:CIT}). We summarize them in the following table:


\vspace{0.5cm}
\begin{tabular}{|c|c|}
	\hline
	\textbf{\underline{Killing vectors}}   & \underline{\textbf{No Killing vectors required}} \\
	& \\
	$\mathbf{D_{\mu\nu}=0}$ & \textbf{Case 1. } $\mathbf{D_{\mu\nu}=0 \;\;\oplus\;\; a_{\mu} = -\D_{\mu} T}$\\
	$\mathbf{\oplus}$  &   $\quad$ but $\;u^{\mu}\nabla_{\mu}T\neq 0$  \\
	$\mathbf{a_{\mu} = -\D_{\mu}T}$ & \\
	$\mathbf{\oplus}$ & \textbf{Case 2. } $\mathbf{a_{\mu} = -\D_{\mu} T \;\;\oplus\;\; u^{\mu}\nabla_{\mu}T=0}$\\
	$\mathbf{u^{\mu}\nabla_{\mu}T=0}$   & but $\;D_{\mu\nu} \neq 0$   \\
	$\mathbf{\Downarrow}$ & \\
	\emph{Exact thermodynamic} &  \textbf{Case 3. } $\mathbf{D_{\mu\nu}=0 \;\;\oplus\;\;u^{\mu}\nabla_{\mu}T=0}$ \\
	\emph{equilibrium set-up} &  $\quad$ but $\; a_{\mu} \neq -\D_{\mu}T$   \\
	\hline
\end{tabular}
\vspace{1.5cm}

Let us now physically interpret all three non-Killing vector cases:
\begin{itemize}
	\item \textbf{Case 1. } This case focuses on the situation of a rigidly moving fluid with a time dependent temperature. 
	It satisfies, in this way, all the \emph{initial} conditions for equilibrium presented in Figure \ref{F:diagram}. 
	Its temperature, however, does not remain constant as time passes. 
	For this case, according to \eqref{E:Dwith T}, we have:
	\begin{equation}
	\label{E:case1 D}
	D_{\mu\nu} = \nabla_{\mu}\PC{\frac{u_{\nu}}{T}} + \nabla_{\nu}\PC{\frac{u_{\mu}}{T}} 
	+  u_{\nu}u_{\mu}\;\PC{\frac{2}{T^2} \;u^{\alpha}\nabla_{\alpha}T} =0.
	\end{equation}
	The question to be made, though, is whether this is a physically possible scenario or not. Equation \eqref{E:case1 D} seems like it could be satisfied for certain four-velocities $u^{\mu}$. But the truth is that this is a tricky question since a non-constant temperature $u^{\mu}\nabla_{\mu} T\neq 0$ implies either $(u^{\mu}\nabla_{\mu} \rho \neq 0$ or $u^{\mu}\nabla_{\mu} \varrho \neq 0)$ \underline{or} $T \neq T(\rho, \mathfrak{u})$.
	As the constancy of $\rho$ and $\varrho$ are imposed by the fluid's equations of motion, we are left with $T \neq T(\rho, \mathfrak{u})$.
	So, technically, by allowing the temperature to depend on more variables than only $\rho$ and $\varrho$, this could be possible. But, if this is the case, then the entropy would also have to be generalized, implying that, in principle,  the equilibrium definition could change for something other than $D_{\mu\nu}=0$ and $a_{\mu}=-\D_{\mu}T$. Furthermore, having  $T$ depending on more variables would imply in adding extra terms to the first law of thermodynamics, for example charge dependent terms. In summary, a lot would have to be changed and it is improbable that the new set-up would be physically correct or represent an equilibrium state. 
	
	Another possibility, which is much more likely, is to recognize Case 1 as an \emph{out of equilibrium situation} and interpret it as an isochoric (constant volume) thermodynamic process, i.e., a constant volume transformation for which all thermodynamic state functions maintain a reasonable physical interpretation at all stages. This could then represent a system which is not in thermodynamic equilibrium, but with an evolution that (just like in ordinary thermodynamics) assumes infinite intermediate equilibrium states between the initial and final configurations.

	\item \textbf{Case 2. } This is a situation of thermal but \emph{not thermodynamic equilibrium}, since mechanical equilibrium does not exist for such congruences. It can, though, be interpreted as a generalized description of an \emph{isothermal  process} for inviscid (zero viscosity) fluids in curved space-times. The conditions for this case can be seen to be exactly the ones adopted along chapter 3 when calculating Tolman temperature gradients for non-Killing flows.
	
	Something to be pointed out, however, is that clearly not all $D_{\mu\nu}\neq 0$ will generate physically solvable  isothermal processes. The additional requirements which would have to be satisfied by such a congruence's four-velocity is left unanswered for the moment. It is important, however, to keep in mind that out of equilibrium states with constant temperatures do exist and that the temperature distribution for such processes must be given by the generalized Tolman temperature gradient developed in the previous chapter:
	\begin{equation}
	a_{\mu} = -\nabla_{\mu} T.
	\end{equation}

	\item \textbf{Case 3. } This is probably the most enigmatic case of the three. It represents a rigid body with constant temperature but non-zero heat fluxes. Is that a physically possible scenario? Well, yes, but for open systems only. These are the conditions for a system under the action of a stationary current passing through it. It describes a stationary (or steady) state. This can be achieved, for example, by coupling the physical system with two heat baths at different temperatures.  Again, this is \emph{not an equilibrium situation}, but a physically interesting one that could be used to describe  steady states for fluids following both Killing and non-Killing flows.
\end{itemize}

\vspace{0.5cm}
As we can see from the three cases presented above, no true equilibrium scenario can occur for non-Killing flows. Case 1, as mentioned, does satisfy the initial conditions for equilibrium, but these come from maximizing a specific entropy current which would probably change if $s\neq s(\rho,\mathfrak{u})$ anymore. Another possible way, though, is to keep the entropy current but change the energy momentum tensor, what would modify the fluid equations, possibly avoiding the conclusion that $u^{\mu}\nabla_{\mu}T=0$ for equilibrium. Again, though not obviously impossible, it seems a bit of a stretch. More likely this scenario does represent slow isochoric transformations. Case 2 and 3 are certainly not in equilibrium but can represent physically interesting situations, namely, isothermal transformations and steady states. 

\vspace{1cm}

\paragraph{Time-scales}\hfill
\vspace{0.2cm}

Another interesting point to add before concluding this chapter is the fact that the evolution of sufficiently slowly changing systems can always be described as a transition between a sequence of equilibrium states. Given the time scale ($10^{10}$ years) implied by the magnitude of the Hubble parameter, this approach can certainly be adopted for observers in FLRW space-times, for example. In this way, even when outside of a Killing orbit, perturbative schemes around the equilibrium state can readily be implemented. This possibility is also backed up by the stability of EIT equilibrium states, analysed by Hiscock and Lindblom \cite{Hiscock}, who have shown that, under reasonable assumptions, these fluids are both stable and exhibit subluminal propagation speeds.  

\newpage
So, in order to obtain a better feeling of the orders of magnitude of the time-scales $\tau_{\theta}$ and $\tau_{\sigma}$ given in equation \eqref{E:timescales}, let us analyse the simple case of a radially falling geodesic in Schwarzschild space-time. This will then help us to comprehend more clearly when one might be able to conduct approximations for treating certain fluids in a out of equilibrium context.

Schwarzschild space-time in Painleve-Gullstrand coordinates assumes the form:
\begin{equation}
\d s^2 = -\d t^2 +\PC{\d r + \sqrt{\frac{2M}{r} \d t}}^2 + r^2 \PC{\d \theta^2 + \sin^2\theta\, \d\phi^2}.
\end{equation} 
The four-velocity of a free falling system dropped from infinity with zero initial velocity is given by:
\begin{equation}
u^{\mu} = \PC{\,1\, ,\, -\sqrt{\frac{2M}{r}}\, , \, 0 \, , \, 0\,}.
\end{equation}
Calculating the four-acceleration we obtain $a^{\mu} = (0,0,0,0)$, as it must be for a geodesic motion. Now we can proceed and calculate the rate of deformation tensor for a congruence with four-velocity given by $u^{\mu}$. It is given by:
\begin{equation}
\label{E:DforSchw}
D_{\mu\nu} = \begin{pmatrix}
\frac{M}{r^2}\,\sqrt{\frac{2M}{r}} & \frac{M}{r^2} & 0 & 0 \\
\frac{M}{r^2} & \frac{1}{2r}\,\sqrt{\frac{2M}{r}} &  0 & 0\\
0 & 0 & -r\,\sqrt{\frac{2M}{r}} & 0\\
0 & 0 & 0 & -r\sin^2\theta\,\sqrt{\frac{2M}{r}}
\end{pmatrix}.
\end{equation} 
Furthermore, the expansion coefficient for this congruence is:
\begin{equation}
\label{E:theta}
\theta = \nabla_{\mu}u^{\mu} = - \frac{3}{2r}\,\sqrt{\frac{2M}{r}}.
\end{equation}
Now, we can calculate the shear tensor via equation \eqref{E:rate}:
\begin{equation}
\sigma_{\mu\nu} = D_{\mu\nu} - \frac{1}{3}\theta \,h_{\mu\nu},
\end{equation}
by  using the fact that the induced metric on the surface orthogonal to $u^{\mu}$  is, in this case, given by the following matrix:
\begin{equation}
h_{\mu\nu} = \begin{pmatrix}
\frac{2M}{r}   & \sqrt{\frac{2M}{r}} & 0 & 0 \\
\sqrt{\frac{2M}{r}} & 1 &  0 & 0\\
0 &0 & r^2& 0\\
0 & 0 & 0 & r^2\sin^2\theta
\end{pmatrix}.
\end{equation} 

Substituting the values for $\theta$ and $h_{\mu\nu}$, we obtain:
\begin{equation}
\sigma_{\mu\nu} = \begin{pmatrix}
\frac{2M}{r^2}\,\sqrt{\frac{2M}{r}} & \frac{2M}{r^2} & 0 & 0 \\
\frac{2M}{r^2} & \frac{1}{r}\,\sqrt{\frac{2M}{r}} &  0 & 0\\
0 & 0 & -\frac{r}{2}\,\sqrt{\frac{2M}{r}} & 0\\
0 & 0 & 0 & -\frac{r\sin^2\theta}{2}\,\sqrt{\frac{2M}{r}}
\end{pmatrix}.
\end{equation} 
Note that $g^{\mu\nu}\sigma_{\mu\nu} =0$, as expected. The time-scales for this congruence are then given by:
\begin{equation}
\tau_{\theta} = \frac{1}{|\theta|} =  \frac{2}{3}\sqrt{\frac{r^3}{2M}}  \quad\text{and}\quad \tau_{\sigma} = \frac{1}{\sqrt{\sigma^{\mu\nu}\sigma_{\mu\nu}}} = \sqrt{\frac{r^3}{3M}}  \,.
\end{equation}
Reinserting the factors of $G$ and $c$, we have:
\begin{equation}
\tau_{\theta} =  \frac{2}{3}\sqrt{\frac{r^3}{2GM}}  \quad\text{and}\quad \tau_{\sigma} =  \sqrt{\frac{r^3}{3GM}}  \,.
\end{equation} 

Note also that this can be rewritten as:
\begin{equation}
\tau_{\theta} =  \frac{2r}{3}\sqrt{\frac{r}{2GM}} = \frac{2}{3}\frac{r}{ v_{esc}} \quad\text{and}\quad \tau_{\sigma} =   \sqrt{\frac{2}{3}}\,\frac{r}{v_{esc}}  \,,
\end{equation}
where $v_{esc}$ is the escape velocity. Let us now substitute values for $r$ and $M$ to obtain the order of magnitude for some cases of interest. For a system free-falling near the surface of the Earth, i.e., taking $M=M_{\oplus}$ and $r=R_{\oplus}$, we have:
\begin{equation}
\tau_{\theta\; \oplus} \approxeq 3.8 \cdot 10^2 \,s    \qquad\text{and}\qquad \tau_{\sigma\; \oplus} \approxeq 4.6 \cdot 10^2\, s   .
\end{equation}
Near the surface of the Sun, we obtain:
\begin{equation}
\tau_{\theta\; \odot} \approxeq 7.4 \cdot 10^2\, s   \qquad\text{and}\qquad \tau_{\sigma\; \odot} \approxeq 9.1 \cdot 10^2\, s   ,
\end{equation}
where $\oplus$ represents values associated with the Earth and $\odot$ with the Sun. For the masses and radius we have used the following values:
\begin{align}
M_{\oplus} &= 5.9 \cdot 10^{24} \,\text{kg}\,; \qquad R_{\oplus} = 6.4 \cdot 10^6 \,\text{m}\,; \\
M_{\odot} &= 2.0 \cdot 10^{30} \,\text{kg}\,; \qquad R_{\odot} = 7.0\cdot 10^8\,\text{m}\,.
\end{align}
Note that the values for the time-scales obtained for a body falling into Earth and into the Sun have the same order of magnitude.
For an observer free-falling into a black hole, with $r=2GM/c^2$, we obtain:
\begin{equation}
\tau_{\theta_{BH}} =  \frac{4}{3}\frac{G M}{c^3}  \quad\text{and}\quad \tau_{\sigma_{BH}} =  \sqrt{\frac{8}{3}}\frac{G M}{c^3}   \,,
\end{equation} 
which gives us:
\begin{equation}
\tau_{\theta} \approxeq 6.6 \cdot 10^{-6}\, s     \qquad\text{and}\qquad \tau_{\sigma} \approxeq 8.2 \cdot 10^{-6}\, s  
\end{equation}
for a solar mass black hole. For a supermassive black hole with mass $M$ of 10 million $M_{\odot}$, on the other hand, we obtain:
\begin{equation}
\tau_{\theta}  \approxeq 66 \, s     \qquad\text{and}\qquad \tau_{\sigma} \approxeq 82 \, s  \,.
\end{equation}

Some of the time scales here obtained were quite small, while others not so much. For other non-Killing orbits, for example a stable elliptical orbit around a massive body, these time scales are expected to be larger, allowing even more flexibility for the relaxation time of the system being studied. It must be clear, however, that different non-Killing orbits will clearly have different time-scales associated with them. In this way, it is important to not simply throw out of the window anything regarding quasi or near equilibrium states for systems following non-Killing flows. Depending on the relaxation time and on the orbit being analysed, it might well be that for all practical purposes some notion of equilibrium can be adopted.

Now, another question to be asked is: What are the relaxation times for typical everyday fluids? The answer to this question is not as simple as one might expect. It will depend on the magnitude of the fluctuations (like temperature and pressure differences), as well as on the average temperature of the fluid (at high temperatures molecular interactions are sufficiently small that the relaxation is exponentially fast)~\cite{relax, relax3, relax4}.
The truth is that relaxation times are extremely variable and can be quite difficult to estimate.  According to reference \cite{relax2}, the time for a system to relax into a new equilibrium state might be as short as $10^{-6}\,s$ for some systems, while it might be a century or longer for others. The range given in \cite{relax} goes from $10^{-10}\, s$ up to $10^4\,s$. 
In this way, a true notion of whether or not approximations may be applied will have to be done case by case.

To  conclude, we would like to point out that the literature concerning CIT and EIT for fluids  in curved space-times is still significantly smaller than that for flat space (Newtonian and special relativistic fluids). Due to its increased complexity, a lot of simplifications normally have to be implemented, either for the fluid or for the background space-time. There are still a lot of questions to be answered that in the future could help astrophysicists to improve star formation models and galaxy evolution scenarios. This is still a very live area of research, with a lot to be investigated, and we hope that the results from this chapter, as well as from Chapter 3, have added some new information to the subject.

%
%
%
%
%
%
%
%
%
%

\newpage
\chapter{The trans-Planckian Problem}
\label{C: Black Holes Thermodynamics}

Let us now turn our attention to black hole thermodynamics.
As is well known by now, black holes are extremely special astronomical objects in the sense that, due to the coarse graining created by the presence of a horizon, they are classically characterized by a very small number of degrees of freedom. Focusing on $(3+1)$ dimensional black holes, any static classic black hole will necessarily be described by a Schwarzschild or Reissner–Nordström metric, as proven by the no-hair theorems \cite{Carter, HawkingEllis, Israelvacuum, Israelelvacuum, Robinson}. In this way, the value of their mass and electric charge are all the information necessary to describe such systems. For stationary black holes, on the other hand, the variables are mass, angular momentum and charge. Kerr and Kerr-Newman are the possible final metrics for such black holes.  

On the other hand, after Hawking's renowned paper \cite{Hawking1975}, it became clear that in the presence of a quantum field background (semiclassical scenario), black holes do not behave as immutable eternal objects as suggested by the classical theory. Hawking's calculation has shown us how, from birth, a black hole interacts with such fields. The evolution of a star, culminating in the formation of a horizon, changes the vacuum state of the background field in such a way that not only a big initial burst of particles is created, but there is also a steady flux to observers infinitely far away from this energetic event.

This has changed the physical status of black holes. Where before you would have a no-return three dimensional barrier from which nothing would ever escape, now you have astronomical objects that are genuinely seen as thermodynamical systems, from which you can even extract work. Black holes are not changeless inflexible structures, they can evolve and the mechanism through which they evolve is the Hawking radiation. 

However, despite its importance, some questions still remain unanswered about details of the calculation. Probably the main one being the trans-Planckian problem, which we will introduce and explore in this chapter. 
Although some of the results obtained so far in this thesis will be applied, the approach adopted in this chapter will be rather different from what we have presented so far. In this current chapter, our aim will be focused on finding a purely kinematical toy model (although very much simplified) that captures enough of the key behaviour of Hawking radiation, while still remaining reasonably tractable, that would make it obvious how to evade the so-called ``trans-Planckian'' problem during early and intermediate stages of the Hawking evaporation process. This chapter will be based on reference \cite{Jess-Vaidya} written by the author together with Ivan Booth, Bradley Creelman and Matt Visser.


\section{Introduction}
\begin{figure}[!htbp]
	\begin{center}
		\includegraphics[height=7.9cm]{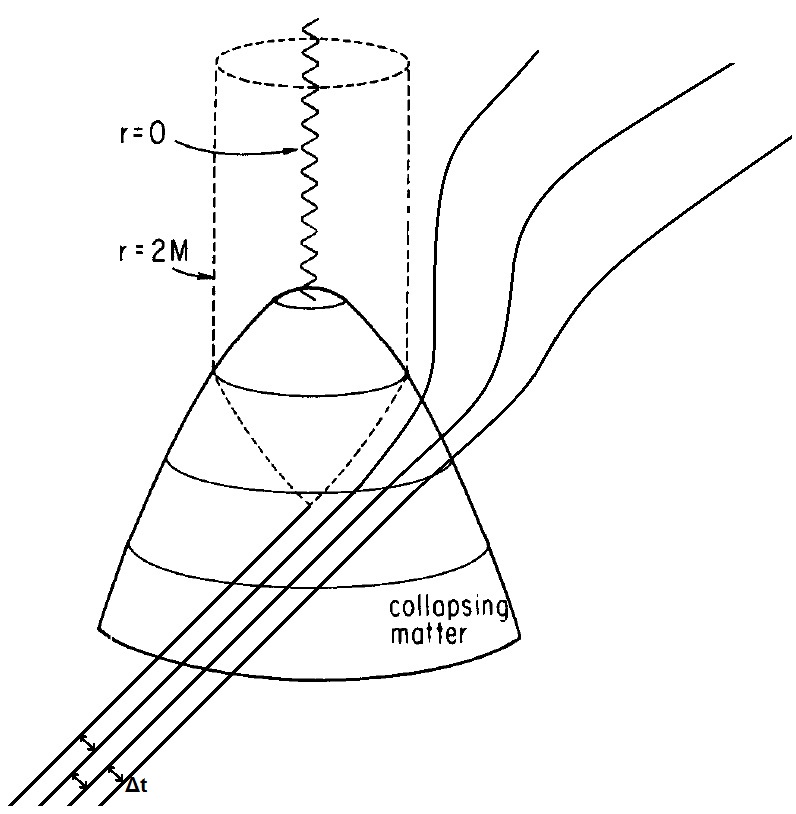}
		\caption[Ray optics Hawking radiation]{The formation of a black hole event horizon acting on wave packets emitted at a constant rate $\Delta t$. The closer they pass to the horizon formation, the bigger the time interval $\Delta t_0$ measured by an observer at infinity, until the limit where $\Delta t_0 \to\infty$ (they never arrive).} 
		\label{F:lightrays}
	\end{center}
\end{figure}

Imagine a star collapsing and forming a black hole. Also imagine an observer emitting light rays at a constant rate $\Delta t$ before the black hole forms. Suppose such light rays can pass through the star and eventually reach infinity   (see Figure \ref{F:lightrays}).
In Hawking's original derivation of the radiation emitted by a black hole \cite{Hawking1975}, the \emph{light rays} that were passing through the star just \emph{after} the formation of the horizon obviously fall into the black hole and never arrive at infinity. The light rays that passed
at the \emph{exact moment} when the horizon forms suffer the fate of being eternally trapped along the horizon worldline. 
On the other hand, the light rays that have managed to pass just \emph{before} the formation of the horizon can actually escape, but with an extremely large redshift factor, the closer it passed to the horizon formation time the bigger the redshift factor. This is basically where the problem lies, since a Hawking photon near future null infinity,  if back-tracked to the immediate vicinity of the horizon, is hugely blue-shifted and found to have once had trans-Planckian energy. And, if back-tracked all the way to the horizon, the photon is formally infinitely blue-shifted, and formally acquires infinite energy. This is the so called \emph{trans-Planckian problem}.

\vspace{-0.2cm}
 Unruh has repeatedly emphasized that  Hawking's original 1973 calculation is a \emph{ray optics} calculation~\cite{explosions}, not a \emph{wave optics} calculation and that it can give us results which are not physically meaningful. For example, if you take a Hawking photon arriving at future null infinity and (in the ray optics approximation) back-track its null geodesic to a region close to the horizon, once the back-tracked null geodesic gets closer and closer to the horizon, the (locally measured) energy of the photon is gravitationally blue-shifted to extremely large energies. These large energies exceed not only the Planck energy, but in fact easily exceed the total mass-energy of the known universe.  Clearly, something is missing.
We should, in this way, look carefully at what escapes to future null infinity, \emph{and what falls into the black hole}.

Indeed, the well-known textbook by Birrell \& Davies~\cite{B&D} presents a discussion on exactly this point: they indicate how to calculate the renormalized stress energy tensor (static approximation, scalar field, no back reaction), and argue that at future null infinity there is an outgoing positive energy flux, whereas near the horizon there is a \emph{ingoing negative energy flux}. This negative energy flux is, of course, how we are able to get around the classical area increase theorem for black holes, since the classical energy conditions are violated sufficiently close to the horizon~\cite{gvp4, gvp3, Bardeen:2017}.

\begin{figure}
	\begin{center}
		\includegraphics[height=9cm]{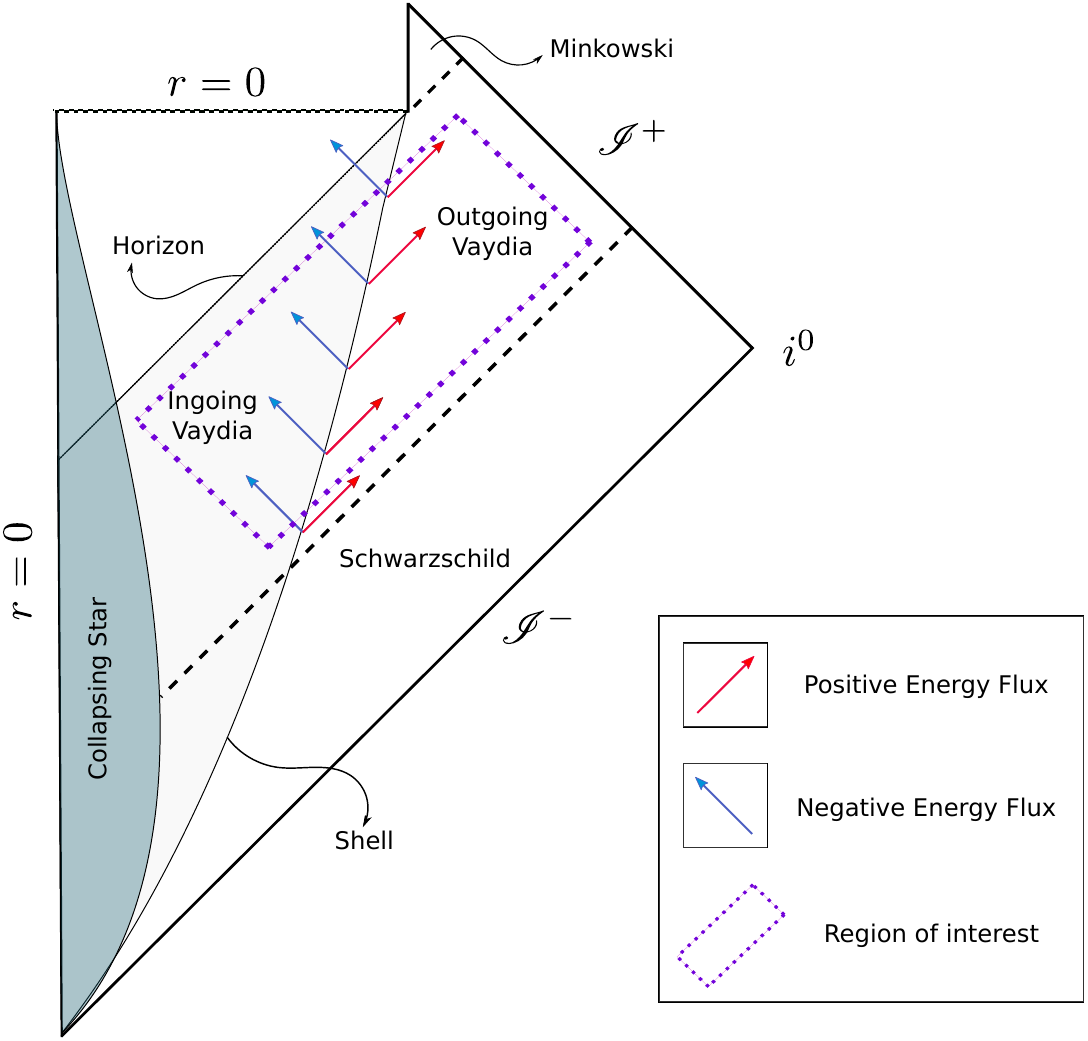}
		\caption[Carter--Penrose diagram for matched outgoing and ingoing Vaidya space-times]{Possible Carter--Penrose diagram for matched outgoing and ingoing Vaidya space-times as a model for Hawking evaporation.} 
		\label{F:Vaidya}
	\end{center}
\end{figure}

As suggested by some authors, a possible solution might be that Hawking photons are actually emitted from some region exterior to the horizon. But, where from exactly?
We seek to make this idea more precise and somewhat explicit by building a purely kinematical model for Hawking evaporation. Our model will be based on two Vaidya space-times (outer and inner) joined across  a  time-like boundary layer (see Figure~\ref{F:Vaidya} for one  of many possible Carter--Penrose diagrams). 
The kinematics of this model will be shown to be rich enough, so that we shall defer consideration of its dynamics for subsequent work. 

Taking into account Unruh and Birrell \& Davies ideas, we will consider, at large distances, a (positive energy flux) outgoing Vaidya ``shining star'' solution~\cite{Vaidya, Vaidya2} and, near the horizon, consider a (negative energy flux) ingoing Vaidya solution. We will then match these two space-times in some intermediate region. We have, then, to choose between two possible options:
\begin{itemize}
	\itemsep-3pt
	\item Matching these two Vaidya regions across a thick shell;
	
	\item or matching across a thin shell using the Israel--Lanczos--Sen junction condition formalism~\cite{Israel, Lanczos, Lanczos0, book}.
 
\end{itemize}
	
	Since the choice of a thick shell would very much depend on its internal dynamics and unlikely lead to interesting physical insights, we will adopt the thin shell possibility. 
	A benefit of such a choice is that, given its simplicity, we will be able to focus and explore the \emph{kinematics} of such a model, leaving the \emph{dynamics} for future work, as previously mentioned. 
	So, for the time being, we will only impose the first junction condition, which establishes the continuity of the metric, and avoid discussing the second junction condition involving extrinsic curvatures (the second fundamental forms).
	
	Another advantage of the thin-shell model is its simplicity, which still allows enough complexity to capture the key physics. However, we clearly still have free parameters to determine:
	\begin{itemize}
		\item 
		We need to decide \emph{where} the transition layer is to be located;
		\item
		we need to make some choices regarding the \emph{internal dynamics} of  the transition layer;
		\item 
		we need to make choices regarding how the \emph{coordinates} are set up. 
	\end{itemize}

Another important remark is that, for understanding the trans-Planckian problem, there is neither a real need for, nor advantage in, using generalized Vaidya space-times~\cite{Wang:1998qx}.
These all involve extra matter fields, which for our purposes would only result in more complications, without any extra physical insight.

Let us start by first considering the static approximation case, temporarily ignoring back-reactions and with the Hawking flux treated in the test-field limit. 
Subsequently we shall add back reaction, kinematics, and even some dynamics.

\section{Static approximation case}
\label{S: Static approximation --- no back reaction}
Let us first consider the static approximation, in which one ignores back-reaction from the Hawking flux and treats the space-time geometry as purely Schwarzschild. 
This is  exactly the situation described in Hawking's 1973 calculation, with the Hawking flux seen as a steady flux and with its effects on the black hole space-time being ignored~\cite{explosions}. To construct our toy model, we will introduce, outside of the horizon, a thin layer located at some \emph{fixed}~\footnote{In the non-static cases the position of the shell will be allowed to move as the black hole evolves.} radial coordinate $r_s=2G_Nm+\epsilon$, from which we shall assume the Hawking radiation is emitted. Conserving energy for the test-flux implies that an equal but opposite ingoing negative energy flux is emitted from the inside of this thin layer, falling into the black hole. We set $c=1$ and $G_N=L_P/m_P$. 

We can, in this way, calculate the total gravitational blueshift factor from spatial infinity down to the static thin shell at $r_s$:
\begin{equation}
\label{E: redshift vaidya}
Z = 1+z ={1\over\sqrt{1-{2G_Nm\over r_s}}}  = \sqrt{r_s\over\epsilon} \approx \sqrt{2G_Nm\over\epsilon}.
\end{equation}
Now, given that a typical Hawking photon has energy $m_P^2/(8\pi m)$ at spatial infinity, when blue-shifted down to the thin shell this becomes a locally measured  energy of order $[m_P^2/(8\pi m)] \sqrt{2G_Nm/\epsilon}$.  For this blue-shifted energy to not exceed the Planck scale (and so avoid the trans-Planckian problem), we require
\begin{equation}
{m_P^2\over8\pi m} \sqrt{2G_Nm\over\epsilon} \lesssim m_P.
\end{equation}
That is
\begin{equation}
\epsilon \gtrsim  {1\over32\pi^2} {G_Nm_P^2\over m} =  {1\over32\pi^2} \;{m_P\over m}\; L_P.
\end{equation}
Now notice that $\epsilon = r_s-2G_Nm\;$ is a coordinate distance, not a proper distance. 
The equivalent proper distance, measured along any surface of constant-$t$, is:
	\begin{eqnarray}
\ell = \int_{2G_Nm}^{r_s} {dr\over \sqrt{1-2G_Nm/r}} &\approx&  \sqrt{2G_Nm} \int_{2G_Nm}^{2G_Nm+\epsilon} {dr\over \sqrt{r-2G_Nm}}
\nonumber\\ &&
=  \sqrt{2G_Nm} \left[ 2 \sqrt{r-2G_Nm}\right]_{2G_Nm}^{2G_Nm+\epsilon}.
\end{eqnarray}
That is 
\begin{equation}
\label{E:epsilon}
\ell \approx \sqrt{8G_Nm\epsilon} \gtrsim {L_P\over2\pi}.
\end{equation}
So, as long as the thin layer is more than a (proper distance) Planck length above the horizon, the trans-Planckian problem does not occur. 
Note that such results would not be changed  if working with a thick, instead of thin, shell --- as long as the Hawking radiation is emitted from some region more than a Planck length above the horizon, the trans-Planckian problem is avoided.

This, however, is not the only constraint that we might want to impose on $\epsilon$.  Another possible constraint comes from imposing the Unruh effect to \emph{quantitatively} explain the Hawking effect for observers at infinity, that is, $T_{U,\infty}\approx T_H$. Given that a thin shell held at fixed radial coordinate $r_s=2G_Nm+\epsilon$ undergoes a 4-acceleration of magnitude
\begin{equation}
A = {G_Nm/r_s^2\over\sqrt{1-2G_Nm/r_s}},
\end{equation}
this implies a locally measured Unruh temperature given by:
\begin{equation}
T_U = {A\over2\pi} = {G_Nm/r_s^2\over 2\pi\sqrt{1-2G_Nm/r_s}}.
\end{equation}
When redshifted to spatial infinity, this becomes
\begin{equation}
\label{E: unruh hawking}
T_{U,\infty} = {A\over2\pi Z} = {G_Nm/r_s^2\over 2\pi} = T_H \left(2G_Nm\over r_s\right)^2.
\end{equation}
From the definition of $r_s$, equation \eqref{E: unruh hawking} implies that $r_s\approx 2G_Nm$, or equivalently $\epsilon \ll 2G_Nm$.
Combining this result with \eqref{E:epsilon}, we have:
\begin{equation}
{1\over32\pi^2} \;{m_P\over m}\; L_P \lesssim \epsilon \ll 2 \,G_N m. 
\end{equation} 
In terms of proper distance above the horizon, this becomes:
\begin{equation}
\label{E: proper distance static}
{L_P\over2\pi}  \lesssim \ell \ll  4 \, G_Nm.
\end{equation} 
So, at least in the static approximation, and if you want the Unruh effect to \emph{quantitatively} explain the Hawking effect, the natural place to put the thin shell is only a few (proper) Planck lengths above the horizon.

There is an alternative that we shall point out but not further explore: To put the shell well above the horizon, say at the unstable photon orbit, $r_s=3G_Nm$, or at the ISCO (Innermost stable circular orbit),  $r_s=6G_Nm$. In this case the thermal flux reaching future null infinity is given by the modified temperature $T_{U,\infty} = T_H \left(2G_Nm/ r_s\right)^2 \leq T_H$ which is always (by construction) less than the Hawking temperature. This modified temperature is $4/9^{ths}$ of the usual Hawking temperature if the thin shell is placed at the unstable photon orbit, and $1/9^{th}$ of the usual Hawking temperature if it is placed at the ISCO.  This numerical difference is not surprising since, instead of null curves skimming along and peeling off from the horizon,  one is now interested in null curves emerging from the surface at $r_s$ --- and the key parameter is the 4-acceleration of that time-like surface. Taking $r_s$ to be macroscopically away from the horizon would then destroy the connection between the Hawking temperature and the ``peeling properties'' of near-horizon null geodesics. 
So, in this class of models, it is very difficult to see why the Hawking temperature should be universally related to the surface gravity. Also, with a shell only a few Planck lengths away from the horizon, quantum mechanics and uncertainty principles are enough to keep us genuinely not concerned with the presence of negative energy falling into the black hole. The same is not true for a large macroscopic region between the black hole and the shell, with any attempt to explain what happens with the negative energy flux in such places becoming more and more awkward. In this way, the result obtained from \eqref{E: proper distance static} is quite reassuring, since it shows that a few proper Planck lengths is already enough to evade trans-Planckian physics.

The task now is to partially and somewhat crudely include back-reaction effects by making the space-time geometry time-dependent. We shall do this by assuming $r_s \longrightarrow r_s(t)$ and $m \longrightarrow m(t)$. The thin shell will then connect two Vaidya space-times, as in Figure~\ref{F:Vaidya}. As we will see, much of the preceding analysis will survive the introduction of this partial back reaction.

\section{Piecewise Vaidya spacetime}
The Vaidya space-time metric consists of the simplest non-static generalization of the Schwarzschild black hole solution. It is obtained by allowing the mass parameter to evolve in time. 
It is sometimes known as the ``shining star'' space-time since, in its original version, it adds outgoing null radiation to Schwarzschild spacetime, and can be used as a good model for the exterior geometry of a non-rotating, spherically symmetric radiating star~\cite{Vaidya,Vaidya2}.
Note that this space-time is not a vacuum solution. The mass is not fixed, and its variation is assumed due to the absorption and emission of particles that travel throughout space-time along null rays.
We shall consider the concatenation of outgoing (radiating) and ingoing (absorbing) Vaidya space-times forming a kinematical model for Hawking radiation.

\vspace{-0.2cm}
\subsection{Vaidya spacetime in null coordinates}
Let us work in null coordinates $(w,r,\theta,\phi)$ and write the Schwarzschild space-time in the form
\begin{equation}
\d s^2 = -\left(1-{2G_Nm\over r}\right) \d w^2 \mp 2 \d w \d r + r^2(\d\theta^2+\sin^2\theta\;\d\phi^2).
\end{equation} 
To obtain, from this form of the metric, the usual Schwarzschild metric one can simply apply the coordinate transformation given by:
\vspace{-0.2cm}
\begin{equation}
\omega \longrightarrow t + f(r), 
\end{equation} 
with
\vspace{-0.2cm}
\begin{equation}
\frac{\d f(r)}{\d r} = \pm \frac{1}{1 - \frac{2m}{r}}.
\end{equation}
We can, then, extend the mass parameter $m$ to become time dependent $m \to m(w)$, obtaining the Vaidya space-time in the form
\begin{equation}
\d s^2 = -\left(1-{2G_Nm(w)\over r}\right) \d w^2 \mp 2 \d w \d r + r^2(\d\theta^2+\sin^2\theta\;\d\phi^2), 
\end{equation} 
(see for example~\cite{Vaidya, Vaidya2}).
The only non-zero component of the Einstein tensor for this metric is
\begin{equation}
G_{ww} =  \mp {2 \,G_N\dot m(w)\over r^2} \, , 
\end{equation}
where the overdot corresponds to a derivative with respect to $w$.
The upper ``$-$'' sign corresponds to outgoing Vaidya space-time while the lower $+$ sign corresponds to ingoing Vaidya space-time.
Let us now, for convenience, rewrite the metric in the following way:
\begin{equation}
\d s^2 = -f(w)^2 \left(1-{2G_Nm(w)\over r}\right) \d w^2 \mp 2 f(w) \d w \d r + r^2(\d\theta^2+\sin^2\theta\;\d\phi^2) \, , 
\label{GV}
\end{equation} 
which is equivalent to a coordinate transformation: 
\begin{equation}
w \to \int f(w)\; \d w ;\qquad  \d w \to f(w)\; \d w.
\end{equation}
Then the non-zero components of the Einstein tensor becomes
\begin{equation}
G_{ww} =  \mp {2 \,G_N f(w) \,\dot m(w)\over r^2}.
\end{equation}

So the set up is the following: We place a thin shell at a position $r_s(\omega)$. From the thin shell up to spatial infinity the metric is outgoing Vaidya, while from the horizon up to $r_s(\omega)$, the metric is ingoing Vaidya.
We then do the matching across the thin shell using the Israel--Lanczos--Sen formalism~\cite{Israel, Lanczos, Lanczos0, book}.

\subsection{Matching null coordinates outside/inside}

Using the metric in the form (\ref{GV}) there is no loss of generality in using a \emph{common}
coordinate  $w$ for \emph{both} inside and outside regions. To keep it continuous, though, we have to introduce
two matching functions $f_\pm(w)$. Then we join the two metrics
\begin{equation}
\small
\d s^2 = -f_\pm(w)^2 \left(1-{2G_Nm_\pm(w)\over r}\right) \d w^2 -\left(  \pm 2 f_\pm(w) \d w \d r \right)
+ r^2(\d\theta^2+\sin^2\theta\;\d\phi^2), \label{metf}
\end{equation} 
across the surface
\begin{equation}
(w, r_s(w), \theta,\phi) \, . 
\end{equation}
The subscript ``$+$'' functions correspond to the outside region and subscript ``$-$'' functions
to the inside. Recall again Figure \ref{F:Vaidya} for clarity.

Thus the (toy) model is completely specified
by the two mass functions $m_\pm(w)$, the two functions $f_\pm(w)$, and the location of the shell $r_s(w)$. 
More precisely it is the ratio $f_+(w)/f_-(w)$, rather than exact functions $f_\pm(w)$, that is physically relevant: Under a 
reparameterization $w\to h(w)$  we can modify both $f_\pm(w)$ but the ratio $f_0(w) = f_+(w)/f_-(w)$ remains fixed. The reader is invited to read about the junction conditions described in section \ref{S: General Relativity} in case the mathematics developed here is not clear enough.
\subsection{Thin-shell tangent and normal}

Now, let us understand better some characteristics of this thin shell. First, the (non-normalized) tangent and normal vectors are given by:
\begin{equation}
U^a= (1, \dot r_s(w), 0 ,0)^a ; \qquad\qquad  N_a = (-\dot r_s(w), 1, 0, 0)_a = \nabla_a(r-r_s(w)),
\end{equation}
with an overdot denoting $d/dw$. We now extend and normalize these vectors $U^a$ and $N_a$ to the entire manifold:
\begin{equation}
\label{E: u and n}
u^a = {U^a\over\sqrt{-g_{ab} U^a U^b}} = {U^a\over\|U\|};
\qquad\qquad
n_a  =  {N_a\over\sqrt{g^{ab}N_a N_b}} = {N_a\over\|N\|}.
\end{equation}
Note that by construction $U^a$ and $N_a$ depend only on $w$, not on $r$. The $r$-dependence in $u^a$ and $n_a$ rises only indirectly, via the normalizing functions. 
In order to explicitly rewrite \eqref{E: u and n}, note that:
\begin{equation}
g_{ab} =\left[\begin{array}{cc|cc}
-f_\pm(w)^2\; \left(1-{2G_Nm_\pm(w)\over r}\right)&\;\;\mp f_\pm(w)&0&0\\ \mp f_\pm(w)  &0&0&0\\ \hline 0&0&r^2&0\\0&0&0&r^2\sin^2\theta  
\end{array}\right],
\end{equation}
and
\begin{equation}
g^{ab} =\left[\begin{array}{cc|cc}
0&\mp \frac{1}{f_\pm(w)}&0&0\\ \mp \frac{1}{f_\pm(w)} &\;\;\left(1-{2G_Nm_\pm(w)\over r}\right)&0&0\\ \hline 0&0&1\over r^2&0\\0&0&0&1\over r^2\sin^2\theta  
\end{array}\right].
\end{equation}
In this way, we have:
\begin{equation}
\label{E: uuuu}
U_a = \left(-f_\pm(w)^2\;\left(1-{2G_Nm_\pm(w)\over r}\right)\mp f_\pm(w)\, \dot r_s(w) ,\; \mp f_{\pm} (w)\; ;\; 0, 0 \right),
\end{equation}
and
\begin{equation}
\label{E: nnnn}
\qquad
N^a = \left(\mp \frac{1}{f_\pm(w)} , \;\left(1-{2G_Nm_\pm(w)\over r}\right) \pm  \frac{\dot r_s}{f_\pm(w)}\;;\; 0, 0 \right),
\end{equation}
with the normalizing functions are then given by:
\begin{equation}
\label{E: u norm}
\|U\| = \sqrt{-g_{ab} U^a U^b} = \sqrt{f_\pm(w)^2\;(1-2G_Nm_\pm(w)/r)\pm 2 f_\pm \dot r_s(w)} \, , 
\end{equation}
and
\begin{equation}
\label{E: n norm}
\| N \| = \sqrt{g^{ab}N_a N_b} = \sqrt{(1-2G_Nm_\pm(w)/r)\pm 2 f_\pm(w)^{-1} \dot r_s(w) } \, ,
\end{equation}
from which we obtain:
\begin{equation}
\label{E: n and u}
\| N \| ={\| U \|\over f_\pm(w)}.
\end{equation}

\section{Exterior region --- outgoing Hawking radiation}

Let us now consider what happens in the outside region, between the thin shell at $r_s(w)$ and spatial infinity.
It is convenient (and implies no loss of generality) to choose the $w$ coordinate to set $f_+(w)\to1$, and set $m_+(w)\to m(w)$, so that in this exterior region the metric is simply:
\begin{equation}
\d s^2 = -\left(1-{2G_Nm(w)\over r}\right) \d w^2 - 2 \d w \d r + r^2(\d\theta^2+\sin^2\theta\;\d\phi^2).
\end{equation} 

\subsection{Blueshift/redshift}
As we have done before, in a dynamic space-time the general formula for the blueshift/redshift function is given by:
\begin{equation}
1+z ={(k_a V^a)_1\over (k_a V^a)_2}.
\end{equation}
Here we are looking along a null geodesic described by the affine null tangent $k_a$, while $(V^a)_1$ and $(V^a)_2$ are the 4-velocities of the emitter and observer. In the current context
\begin{equation}
1+z ={(k_a u^a)\over (k_a v^a)},
\end{equation}
where
\begin{equation}
k_a=(1,0,0,0), \quad v_\infty^a = (1,0,0,0), \; \; \mbox{and} \quad u^a = {(1, \dot r_s, 0 ,0)^a \over \| (1, \dot r_s, 0 ,0)^a \|}.
\end{equation}
In this way $v_\infty^a$ is the stationary observer at infinity and $u^a$ the shell velocity.

Thus, temporarily reinserting Newton's constant $G_N$ for clarity (and remembering that we are choosing $f(w)\to1$ in the exterior region) the blueshift/redshift from $r=r_s(w)$ to infinity is: 
\begin{equation}
 1+z_{\infty}(\omega) ={1 \over \| (1, \dot r_s(w), 0 ,0)^a \|} = {1\over\| U\|}.
\end{equation}
That is
\begin{equation}
\label{E: redshift infty}
 1+z_{\infty}(w)= {1\over\sqrt{1-2G_Nm(w)/r_s(w) + 2\dot r_s(w) }}.
\end{equation}
Note how naturally and cleanly this generalizes the static result
\begin{equation}
 1+z_{\infty}(w) ={1\over\sqrt{1-2G_Nm/r_s}}\; .
\end{equation}
Note that equation \eqref{E: redshift infty} presents  contributions both from the gravitational field itself as well as from the motion of the thin-shell. This computation of the redshift has significance beyond the thin-shell models considered here, and could be applied, for example, to a spherically-pulsating ``shining star'' space-time, as long as the star has a sharp surface at $r_s(w)$, and as long as the stellar exterior is pure outgoing null flux. As a consistency check we can set  $m(w)\to 0$, which means we are in flat space and, using the fact that now
\begin{equation}
\frac{\d w}{\d t} = \frac{\d (t-r)}{\d t} = 1 -\frac{\d r}{\d t},
\end{equation}
we then obtain: 
\begin{eqnarray}
1+z_{\infty}(\omega)
&\to& {1\over\sqrt{1+ 2\dot r_s(w) }} \equiv {1\over\sqrt{1+ 2(\d r_s(w)/\d w) }} = {1\over\sqrt{1+ 2{(\d r_s/\d t)\over(\d w/\d t)} }}
\nonumber\\
&&
= {1\over\sqrt{1+ 2\;{(\d r_s/\d t)\over1 - (\d r_s/\d t)} }} =\sqrt{1 - (\d r_s/\d t)\over1 + (\d r_s/\d t)}. 
\end{eqnarray}
This is the usual flat-space Doppler shift factor, as expected.

\subsection{Evading trans-Planckian physics}
Now, given that Hawking temperature seen by observers at infinity is given by:
\begin{equation}
T = \frac{m_P^2}{8\pi m}
\end{equation}
it seems fair to assume that on average, a Hawking photon will have energy given by $E = k_B T$. For the time evolving toy model being developed here, we can say that, as long as the black hole is ``slowly evolving'', we can use the adiabatic approximation to estimate the average energy of the Hawking photons reaching future null infinity as
\begin{equation}
E(w) = k_B T(w) = {m_P^2\over8\pi m(w)}.
\end{equation}
 This approximation is valid as long as the surface gravity satisfies $d\kappa/dw \ll \kappa^2$~\cite{Barcelo:2010a,Barcelo:2010b}, that is, as long as $d m(w)/dw \ll m_P/T_P$. There is a similar adiabaticity condition for the validity of Unruh radiation~\cite{barbado}. 

When back-tracked to the thin shell, the Hawking photons will have a blueshifted locally measured energy (in the rest frame $u^a$ of the thin shell) given by
\begin{equation}
E_s(w) = {m_P^2 \, Z(w)\over8\pi m(w)} = {m_P^2 \over8\pi m(w) \sqrt{1-2G_Nm(w)/r_s(w) + 2\dot r_s(w) }}
\end{equation}
If we impose that the energy $E_s(w)$ must be sub-Planckian, $E_s(w) \lesssim m_P$, then we have:
\begin{equation}
{m_P \over8\pi m(w) \sqrt{1-2G_Nm(w)/r_s(w) + 2\dot r_s(w) }} \lesssim 1.
\end{equation}
Expanding $r_s$ in terms of $\epsilon(w)$, that is $r_s(w) = 2G_Nm(w)+\epsilon(w)$, we have
\begin{equation}
{m_P\over8\pi m(w)} \sqrt{r_s(w)\over\epsilon(w)+ 2 r_s(w) \dot r_s(w)} \lesssim 1.
\end{equation}
Rearranging the terms we have
\begin{equation}
\epsilon(w)+ 2 r_s(w) \dot r_s(w) \gtrsim {m_P^2\over64\pi^2 m(w)^2} r_s(w).
\label{E:qq1}
\end{equation}
Since for an \emph{evaporating} black hole we must have $\dot r_s(w) < 0$,  this implies
\begin{equation}
\epsilon(w) \gtrsim {m_P^2\over64\pi^2 m(w)^2} r_s(w).
\label{E:qq2}
\end{equation}
Also, since we want the thin shell to lie outside the Schwarzschild radius, $r_s(w) > 2G_Nm(w) = 2 L_P m(w)/m_P$, we might rewrite \eqref{E:qq2} as
\begin{equation}
\epsilon(w) \gtrsim   {1\over32\pi^2} \;{m_P\over m(w)}\; L_P.
\label{E:qq3}
\end{equation}

\enlargethispage{0.3cm}

This is a $w$-dependent version of the result we previously obtained in the static approximation. Though similar, the two results actually have significant differences.
While certainly \eqref{E:qq3} is always true as long as evaporation overwhelms accretion, i.e. $\dot r_s(w)<0$, it is not the whole story since, looking at \eqref{E:qq1}, we see that a term is being neglected. Truly, we must write:
\begin{eqnarray}
\label{E: upsy}
\epsilon(w) &\gtrsim& {1\over32\pi^2} \;{m_P\over m(w)}\; L_P + 2\; r_s(w) |\dot r_s(w)|\\ &\approx& {1\over32\pi^2} \;{m_P\over m(w)}\; L_P + 4 \;G_Nm(w) |\dot r_s(w)|, \label{E: upsy2}
\end{eqnarray}
where, in \eqref{E: upsy2} we have considered only first order terms\footnote{For the full expression we have
\begin{eqnarray*}
\epsilon(w) &\gtrsim& {1\over32\pi^2} \;{m_P\over m(w)}\; L_P + 2\;(2 G_Nm(w) + \epsilon(w)) |\dot r_s(w)|
\end{eqnarray*}
which gives us
\begin{eqnarray*}
	\epsilon(w) &\gtrsim& \PC{{1\over32\pi^2} \;{m_P\over m(w)}\; L_P + 4G_Nm(w) |\dot r_s(w)|} \PC{1 + 2 |\dot r_s(w)|}.
\end{eqnarray*}
But, considering $|\dot r_s(w)|\ll 1$, \eqref{E: upsy} is a good approximation.}. 
It then becomes clear, looking at \eqref{E: upsy}, that how far away we must locate our shell depends not only on the mass of the black hole, but also  on the rate of evaporation. Hence, this already shows one of the limitations of the model proposed: it is valid only if both $|\dot r_s| \ll 1$ and $\epsilon(w) \ll 2G_N m(w)$. Now $|\dot r_s| \ll 1$ will certainly be true during most of the lifetime of the black hole, as long as it is slowly and adiabatically evaporating. Furthermore we shall soon see that $\epsilon(w) \ll 2G_N m(w)$ will hold if we want the Unruh effect to quantitatively explain the Hawking radiation.

Let us now estimate the proper distance between the location of the thin shell at $r_s(w)=2G_Nm(w)+\epsilon(w)$, and where the apparent horizon ``would have formed''. 
First, let us explain what we mean by ``would have formed''. The subtlety lies on the fact that we are matching two different metrics across a thin shell. We have already determined that $f_+(w)=1$, given that the outside region is where we and any possibility of measurements lies. We know nothing up to now about $f_-(w)$. It would, then, be problematic to measure the distance between the real horizon location and the shell with the inner metric. We, then, extrapolate our external metric to inside the shell to compare those distances. Note, then, that $r=2G_N m$  is a ``virtual'' location for the external metric. It is not actually part of the physical space-time.
Now, to measure the distance between such points, start by picking some arbitrary but fixed $w_*$ and considering the geometry
\begin{equation}
\d s^2 = -\left(1-{2G_Nm_+(w_*)\over r}\right) \d w^2 -\left(2  \d w \d r \right)
+ r^2(\d\theta^2+\sin^2\theta\;\d\phi^2). \label{E:frozen}
\end{equation} 
This ``freezes'' the external geometry at the moment $w_*$. We might, then, extrapolate such a metric to regions $r<r_s(w_*)$, so that we can say something about where the apparent horizon ``would have formed''. 
Indeed this ``frozen'' geometry is just Schwarzschild geometry in disguise, so all we need to do is to estimate the proper distance between $r_s(w_*)=2G_Nm(w_*)+\epsilon(w_*)$ and $2G_Nm(w_*)$:
\vspace{0.3cm}
\begin{equation}
\ell =  \int_{2G_Nm}^{2G_Nm+\epsilon} {dr\over\sqrt{1-2G_Nm(w_*)/r}}
\approx\int_{2G_Nm}^{2G_Nm+\epsilon} \sqrt{2G_Nm(w_*) \over r-2G_Nm(w_*)} dr,
\end{equation}
so that
\vspace{0.3cm}
\begin{equation}
\ell \approx \sqrt{8G_Nm(w_*) \epsilon(w_*)}. 
\end{equation}
Since this was calculated for any fixed but arbitrary $w_*$ we see
\begin{equation}
\ell \approx \sqrt{8G_Nm(w) \epsilon(w)}. 
\end{equation}
Now, using \eqref{E:qq3} for $\epsilon(w)$, we have:
\begin{equation}
\ell \gtrsim \sqrt{ {8G_Nm(w)\over32\pi^2} \;{m_P\over m(w)}\; L_P }\; \approx {L_P\over2\pi}.
\end{equation}

If, again, we wish to consider the more general relation between $\epsilon(w)$, $m(w)$ and $|\dot{r_s}|$, i.e., equation \eqref{E: upsy2}, we have:
\begin{eqnarray}
\ell \gtrsim \sqrt{8G_Nm(w) \PC{{1\over32\pi^2} \;{m_P\over m(w)}\; L_P  + 4G_Nm(w) |\dot r_s(w)|}}  .
\end{eqnarray}
This gives us:
\vspace{-0.3cm}
\begin{eqnarray}
\ell &\gtrsim& \sqrt{\frac{L_P^2}{4\pi^2}  + 32\; L_P^2 \PC{\frac{m^2(w)}{m_P^2}} |\dot r_s(w)|} \nonumber\\
&=& \frac{L_P}{2\pi}\sqrt{1 + 128 \pi^2 \PC{\frac{m^2(w)}{m_P^2}} |\dot r_s(w)|}. \label{E: big upsy}
\end{eqnarray}

So, in the presence of back-reaction and an evolving Vaidya space-time geometry, to avoid trans-Planckian physics we need the Hawking photons to be emitted from a region at least a (proper) Planck length above where the apparent horizon would be expected to form.
How far above the horizon, however, is very much dictated by the rate of evaporation of the black hole and, in this way, by  $|\dot{r_s}|$.
We see from equation \eqref{E: big upsy} that for $\ell$ to be located only a few Planck lengths away from the horizon, the rate of evaporation has to be such that
\begin{equation}
\label{E: upsy3}
 |\dot r_s(w)| \lesssim \frac{c}{128 \pi^2} \PC{\frac{m_P^2}{m^2(w)}},
\end{equation}
where we have recovered the factor of $c$. For a 10 solar masses black hole, this corresponds to:
\begin{equation}
|\dot r_s(w)| \lesssim \frac{c}{128 \pi^2} \PC{\frac{10^{-32}}{10^{62}}}  \approx \frac{10^{-94}\; 10^8}{10^4} = 10^{-90} m/s.
\end{equation}
Or, in a more convenient time frame:
\begin{equation}
|\dot r_s(w)| \lesssim 10^{-82} \; m/year = 10^{-37} \frac{L_p}{\text{(age of universe)}}
\end{equation}
This, without any doubt, is a small number. The question, though, is: how small, when compared to the average evaporation rate of a black hole? To find this out, let us first rephrase \eqref{E: upsy3} in terms of the black hole mass $m$:
\vspace{0.2cm}
\begin{align}
   |\dot r_s(w)| = \left\lvert\frac{\d r_s(w)}{\d w}\right\lvert = \left\lvert\frac{\d}{\d w}\PC{\frac{2 G_N m}{c^2}}\right\lvert = \frac{2 G_N }{c^2} \;\left\lvert\frac{\d m(w)}{\d w}\right\lvert.
\end{align}
Inserting this result into \eqref{E: upsy3}, we obtain:
\vspace{0.2cm}
\begin{equation}
\left\lvert\frac{\d m(w)}{\d w}\right\lvert	\lesssim \; \frac{c^2}{2 G_N } \;\frac{c}{128 \pi^2} \PC{\frac{m_P^2}{m^2(w)}} =\frac{1}{256 \pi^2} \frac{c^3}{G_N } \PC{\frac{m_P^2}{m^2(w)}},
\end{equation}
giving us:
\vspace{0.2cm}
\begin{equation}
\label{E: pred mass rate}
\left\lvert\frac{\d m(w)}{\d w}\right\lvert	\lesssim \; 3.96 \times 10^{-4} \PC{\frac{m_P}{t_P}} \PC{\frac{m_P^2}{m^2(w)}}.
\end{equation}

The approximated evaporation rate obtained by Frolov and Novikov (equation 10.1.19 of \cite{frolovnovikov}), on the other hand, is given by: 
\vspace{0.2cm}
\begin{equation}
\label{E: mass rate}
\left\lvert\frac{\d m(w)}{\d w}\right\lvert	\approx 2.59 \times 10^{-6} N \PC{\frac{m_P}{t_P}} \PC{\frac{m_P^2}{m^2(w)}},
\end{equation}
where $N$ is the number of states and species of particles that are radiated. By comparing \eqref{E: pred mass rate} and \eqref{E: mass rate} it then becomes easy to see that the requirement imposed by equation \eqref{E: upsy3} is not as restrictive as it seems. Basically all black holes which radiate approximately with a black body spectrum will satisfy it. Bearing this in mind we can now proceed in developing our model.

\subsection{From Unruh temperature to Hawking temperature}
\def\L{{\mathcal{L}}}

A shell holding a fixed position outside a black hole must have a 4-acceleration in order to keep it away from falling. 
In this way, if we have an observer sit on top of the shell, they would perceive a  thermal bath  due to the Unruh effect caused by such an acceleration. 
So, in order to obtain the Unruh temperature felt by the shell, we need to first calculate its 4-acceleration $A(w)$. Given the system parameters, we can expect $A(w)$ to be some function of $m(w)$, $r_s(w)$ and their derivatives. This calculation involves several steps and technical results. For the sake of fluidity, we have derived the result in Appendix \ref{C: appendix 6} and we will only present the acceleration formula in this chapter. It is given by:
\begin{equation}
A(w) = {1\over \|U\|} {G_Nm(w)\over r_s(w)^2}  +  {1\over \|U\|^2}  {d\|U\|\over dw}
= {1\over \|U\|} \left( {G_Nm(w)\over r_s(w)^2}  + {d\ln \|U\|\over dw}\right),
\end{equation}
where, we remember  \eqref{E: u norm}:
\begin{equation}
\|U\| = \sqrt{-g_{ab} U^a U^b} = \sqrt{1-2G_Nm_\pm(w)/r + 2 \dot r_s(w)}.
\end{equation}
This corresponds to a locally determined Unruh temperature of
\begin{equation}
T_U(w) = {A(w)\over2\pi} = {1\over2\pi\|U\|}   \left( {G_Nm(w)\over r_s(w)^2} +  {d\ln \|U\|\over dw}\right).
\end{equation}
When redshifted to spatial infinity, using the previously calculated redshift factor $Z(w) = 1 + z(\omega)$ given on \eqref{E: redshift infty}, this becomes
\begin{equation}
T_{U,\infty}(w) = {A(w)\over2\pi Z(w)} = {A(w)\;\|U\|\over2\pi} = {1\over2\pi} \left( {G_Nm(w)\over r_s(w)^2} + {d\ln \|U\|\over dw}\right).
\end{equation}
In terms of the adiabatically evolving Hawking temperature, $T_H(w) = 1/(8\pi G_N m(w))$, where we have set $\hbar=1$ and $c=1$, this is
\begin{equation}
\label{E: T infty}
T_{U,\infty}(w) = T_H(w) \;  \left\{ \left(2 G_N m(w)\over r_s(w)\right)^2 +  4 G_N m(w) \; {d\ln \|U\|\over dw}\right\}.
\end{equation}

Now, if we want the Unruh effect to \emph{quantitatively} explain the Hawking effect, we need $T_{U,\infty}(w) \approx T_H(w)$, or:
\begin{equation}
T_{U,\infty}(w) = {A(w)\over2\pi Z(w)} \approx {1\over 8\pi G_N m(w)}. 
\end{equation}
Looking at \eqref{E: T infty}, we see that this is equivalent to requiring the whole term inside the curly brackets to be approximately equal to one, or:
\begin{equation}
r_s(w) \approx 2G_Nm(w); \qquad  G_N m(w) \; {d \|U\|\over\ dw} \ll \|U\|.
\end{equation}
We might as well rewrite such conditions as:
\begin{equation}
r_s(w) \approx 2G_Nm(w); \qquad  G_N m(w) \; {d Z(w)\over\ dw} \ll Z(w).
\end{equation}

So as in the static case, also in this Vaidya context, if we want the Unruh effect of the accelerated thin shell to \emph{quantitatively} explain the Hawking effect, then we need the thin shell to hover just above the apparent horizon --- more precisely, just above where the apparent horizon would otherwise be expected to form --- at least one proper Planck length above the apparent horizon to avoid the trans-Planckian problem. Plus we need the ``slowly evolving'' adiabatic constraint on the evolution of the total redshift $Z(w)$. Note that, in order to obtain these results we only needed to consider the exterior region. Let us now see what results we will be able to derive from the interior geometry.

\section{Interior metric and the final fate of the Vaidya model black hole}

As previously mentioned, for the inside region, i.e. $r_s(\omega)<2 G_N m(\omega)$, the metric is given by the ingoing Vaidya space-time, described by some mass function $m_-(w)$. Can we then say anything reasonably explicit about the ingoing (negative energy) Hawking radiation and its impact on the central singularity? Can we say anything reasonably generic regarding the relevant Carter--Penrose diagrams?
Since now we are focused on analyzing the inner metric only, we might as well, for the time being, 
set $f_-(w)\to1$ and  $m_-(w)\to m(w)$. The inner metric then takes the form:
\begin{equation}
\d s^2 = -\left(1-{2G_Nm(w)\over r}\right) \d w^2 + 2 \d w \d r + r^2(\d\theta^2+\sin^2\theta\;\d\phi^2).
\end{equation} 
To obtain a better intuition about the inner geometry, we can calculate the Ricci tensor for this metric, which is given by:
\begin{equation}
R_{ww} = 2\;\frac{\dot{m}(w)}{r^2},
\end{equation}
the Kretschmann scalar, which is given by:
\begin{equation}
R_{abcd} R^{abcd} =  C_{abcd}C^{abcd} = 48 \PC{\frac{G_N\, m(\omega)}{r^3}}^2,
\end{equation}
and the orthonormal components of the Weyl tensor:
\begin{equation}
\small
\label{E: Weyl}
C_{\hat w\hat r\hat w\hat r} =-2C_{\hat w\hat \theta \hat w\hat \theta} = -2C_{\hat w\hat \phi \hat w\hat \phi} 
= 2C_{\hat r\hat \theta \hat r\hat \theta} = 2 C_{\hat r\hat \phi \hat r\hat \phi} 
= - C_{\hat \theta\hat \phi\hat \theta\hat \phi}  = -{2G_Nm(w)\over r^3}.
\end{equation}
So, the Weyl tensor is completely determined by the quantity $m(w)/r^3$, while the Ricci tensor is completely determined by $\dot m(w)/r^2$. 

Can we, with such information, say anything about the final evaporation state of our model? Let us start by recalling that the standard endpoints of the Hawking process are a naked singularity, a remnant, or complete evaporation~\cite{B&D}. Let us analyze case by case:

\paragraph{Naked singularity:}	Given that the only free parameter of a Schwarzschild spacetime is its mass, the only way to obtain a naked singularity is by imposing a negative mass. The same is basically valid for Vaidya spacetimes\footnote{Apart from instantaneous massless shell-focusing singularities at moments of black hole formation or final dispersal (see below), the only true naked singularities have negative mass. }. In this way, for us to obtain a naked singularity in the current setup, we need to have:
\begin{equation}
\lim_{w\to\infty} m(w) = m_\infty < 0.
\end{equation}
For this to happen, the black hole would technically have to ``continue to evaporate'' after all its mass is gone. Clearly this is a very unlikely physical situation. So, for the model adopted, naked singularities will be ruled out.

\paragraph{Remnant:}	A remnant corresponds to the situation in which, for whatever reason, the black hole stops (or asymptotically stops) its evaporation process, leaving some final mass eternally stuck inside the event horizon. Mathematically, this would mean
\begin{equation}
	\lim_{w\to\infty} m(w) = m_\infty > 0,
\end{equation}
or, at worst, a slow asymptotic approach to zero central mass. 
The black hole remnant, whatever its mass, would necessarily have a final temperature equal to zero for semi-classical theories, otherwise the evaporation process would continue.
For this to happen, of course, something would have to slow down and eventually stop the evaporation process. 
A way to do so would be to count on the action of some mysterious unknown charge, which would decrease the temperature until it drops to zero. Given the third law of thermodynamics, this scenario seems as, if not more, unlikely than the naked singularity case. 
Other possibilities, in which the remnant's behaviour is still far from clear, would be to include  higher curvature terms in the gravity action, or to consider the possibility of Planck size remnants as an effective approximation of some quantum gravitational principle \cite{remnants}.
So, as we know, physics is an experimental science and it is always possible that, with the new experiments being developed over the years, some set of new unexpected information might suggest to us some way out of violating the third law of thermodynamics and keeping black hole remnants. For now, however, the complete evaporation scenario easily stands out as the most plausible ending for a black hole.

\paragraph{Complete evaporation:}	The complete evaporation scenario happens when, in a finite time $w_*$, the black hole mass entirely ``evaporates'' via Hawking radiation. For this situation, we have:
\begin{equation}
	\lim_{w\to w_*} m(w) =  0.
\end{equation}
	The main question about this scenario is: what exactly happens at the instant $w=w_*$? We already know that for $w<w_*$ the geometry is certainly singular at $r=0$ and, for $w>w_*$ the geometry must be regular at $r=0$. We will address, in the following, some interesting points about this question.

Let us ask ourselves what a timelike observer will observe when $w\to w_*$. We can do that by expanding (under very mild conditions) the time dependent mass $m(w)$ in a so-called Puiseaux expansion~\cite{Puiseaux:1850,Puiseaux:1851}. The conditions for such an expansion are indeed much less  restrictive than those for a Taylor expansion. Doing so, we then have:
\begin{equation}
m(w) \sim  (w_*-w)^\gamma\; K_m \; H(w_*-w).
\end{equation}
Here $H(x)$ is the Heaviside step function, and the critical exponent $\gamma$ controls the behaviour of the final burst (of ingoing negative energy Hawking flux); $K_m$ is some fixed but arbitrary constant.
We impose $\gamma>0$ so that the mass goes to zero at $w=w_*$.

In the immediate vicinity of the final evaporation point, $(w_*,0,\theta,\phi)$, the 
null (causal) structure is determined by $0=-dw^2+2\,dw\,dr = dw(2dr-dw)$, so the outgoing null ray is $ r \sim {1\over2} (w_*-w)$, while the ingoing null ray is given by $dw=0$. Given that any future-directed timelike trajectory will have to lie inside the null cone, therefore, in between the outgoing and ingoing null rays, they can be expressed as
\begin{equation}
r_o(w) \sim  (w_*-w)\; K_r \; H(w_*-w); \qquad   K_r \in (1/2,\infty),
\end{equation}
where $K_r$ is some fixed but arbitrary constant.

Therefore, given equation \eqref{E: Weyl}, a timelike observer will see orthonormal Weyl components of the form
\begin{equation}
{m(w)\over r_o(w)^3} \sim {K_m\over K_r^3}\;  (w_*-w)^{\gamma-3},
\end{equation}
and orthonormal Ricci components of the form
\begin{equation}
{\dot m(w)\over r_o(w)^2} \sim {K_m\over K_r^2} \; \gamma \;(w_*-w)^{\gamma-3}. 
\end{equation}
We can, in this way, analyze what the observer will measure as a function of $\gamma$ case by case:
\begin{itemize}
	\item For $\gamma>3$ the orthonormal components smoothly approach zero, so Hawking radiation proceeds until the end with no final surprises.
	\item For $\gamma=3$ the orthonormal components at least remain bounded.
	\item For $0<\gamma<3$, the orthonormal components blow up.\footnote{Remember that by hypothesis $\gamma>0$.} This corresponds to so-called ``cosmic flashing'', an instantaneous glimpse of a naked singularity. This might not be too problematic since, also for general spherically symmetric spacetimes (instantaneous) naked massless shell-focusing singularities can also be visible at moments of black hole formation \cite{Lake:1992zz}.
\end{itemize}
Overall, in this framework, complete evaporation seems the most plausible outcome. Let us now study the relationship between $f_+(\omega)$ and $f_-(\omega)$ and how to model different possible evaporation processes.

\section{Models for evaporation scenarios}

The formalism we have developed up to this stage is quite generic. Given the incredible amount of information that can be extracted from a purely kinematical analysis, treating the exterior and interior regions independently, we have not yet made any specific choices about the internal physics of the thin shell. Let us now then link the exterior and interior regions by enforcing the most basic junction condition --- the continuity of the space-time metric (see section \ref{S: General Relativity} for further information). Adopting $G_N\to1$, this condition reads:
\begin{equation*}
\small
-\left(1-{2m_+(w)\over r_s}\right) \d w^2  - 2 \dot r_s\; \d w \d r  = -f_-(w)^2 \left(1-{2m_-(w)\over r_s}\right)\d w^2  + 2 f_-(w) \dot r_s \;\d w \d r,
\end{equation*}
here, without loss of generality, we have set $f_+(w)\to 1$. We can also rewrite this in a cleaner way:
\begin{equation}
\label{junction}
\left\{-\left(1-{2m_+(w)\over r_s}\right)  - 2 \dot r_s \right\} 
=
\left\{ -f_-(w)^2 \left(1-{2m_-(w)\over r_s}\right)  + 2 f_-(w) \dot r_s \right\}.
\end{equation}
Now, let us look at some interesting possibilities with different mass relations.

\subsection{Non equal masses case}
\begin{figure}
	\center
	\includegraphics[scale=1]{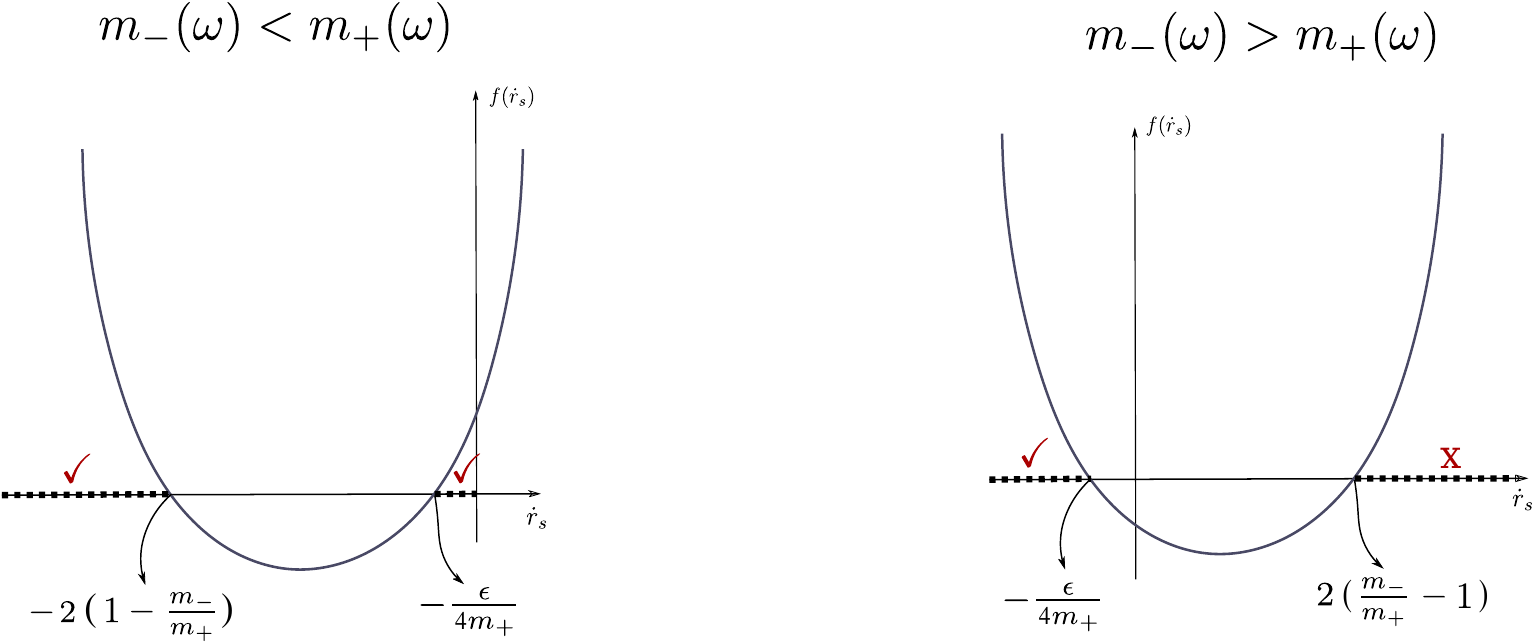}
	\caption[Possible scenarios for the radial velocity in the general case]{Possible scenarios for the radial velocity in the general case. The region marked with $X$ is eliminated since we are studying the $\dot{r}_s <0$ evaporation scenarios.}
	\label{F:parabolas}
\end{figure}

Rearranging equation \eqref{junction} we obtain a quadratic equation for $f_-(\omega)$:
\begin{eqnarray}
f_-^2(\omega) \left(1-{2m_-(\omega)\over r_s}\right)  - 2 f_-(\omega) \dot r_s
-\PR{\left(1-{2m_+(\omega)\over r_s}\right) + 2 \dot r_s} = 0.
\label{fmeq}
\end{eqnarray}
Solving this, we find:
\begin{eqnarray}
f_-(\omega) = \frac{\dot{r}_s \pm \sqrt{\dot{r}_s^2 + \PC{1- 2m_-/r_s}\PR{\PC{1 - 2m_+/r_s} +2\dot{r}_s}}}{(1 - 2m_-/r_s)}.
\end{eqnarray}
Given its physical meaning, we wish $f_-(\omega)$ to be real. This implies that the terms inside the square root must be positive:
\begin{equation}
\dot{r}_s^2 + \PC{1- \frac{2m_-}{r_s}}\PR{\PC{1 - \frac{2m_+}{r_s}} +2\dot{r}_s}>0,
\end{equation}
which, rearranging, gives us:
\begin{equation}
\dot{r}_s^2  + 2\dot{r}_s\PC{1- \frac{2m_-}{r_s}} + \PC{1 - \frac{2m_-}{r_s}}\PC{1 - \frac{2m_+}{r_s}} >0. 
\end{equation}
This places bounds on acceptable values of the model parameters $m_\pm(w)$ and $r_s(w)$. 
Finding the zeros of this quadratic, the edge of the physically acceptable region must satisfy
\begin{eqnarray}
\dot{r}_s^{\pm} =
-\PC{1- \frac{2m_-}{r_s}} \pm \sqrt{\frac{2}{r_s}\PC{1- \frac{2m_-}{r_s}}(m_+ - m_-)}.
\end{eqnarray}
\enlargethispage{20pt}
Substituting $r_s = 2m_+ + \epsilon$, this becomes
\begin{eqnarray}
\dot{r}^{\pm}_s &=& -1 +\frac{2m_-}{r_s} \pm \sqrt{\frac{2}{r_s^2}
	\PC{2m_+ +\epsilon -2m_-}(m_+ - m_-)}\,.
\end{eqnarray}
Making the strong assumption that $\epsilon \ll 2\; ||m_+-m_- ||, $, this can approximated by~\footnote{We had already argued $\epsilon \ll 2m_+$ in order for the Unruh effect to be qualitatively linked to the Hawking effect; this $\epsilon \ll 2 ||m_+-m_- ||$ assumption is considerably stronger.}:
\begin{align}
\dot{r}_s^{\pm} 
\approx  -1 +\frac{2m_-}{r_s} \pm \sqrt{\frac{4(m_+ - m_-)^2}{r_s^2}\PC{1 + \frac{\epsilon}{2(m_+ - m_-)}}},
\end{align}
which, expanding, gives us:
\begin{equation}
\dot{r}_s^{\pm}  \approx -1 +\frac{2m_-}{r_s} \pm \frac{2||m_+-m_- ||}{r_s}\PC{1 + \frac{\epsilon}{4(m_+ - m_-)}}
\end{equation}
Clearly, this approximation is not valid for the mass matching case, which will be evaluated next. But, for now,  the edges of the physically acceptable region are given by:
\begin{equation}
\dot{r}_s^- \approx -1 +\frac{4m_-}{r_s} - \frac{2m_+}{r_s} \approx  -2\left(1 - \frac{m_-}{m_+}\right)
\;\; \mbox{and}\;\;\;\;
\dot{r}_s^+ \approx  -\frac{\epsilon}{4m_+}
\end{equation}
for the $m_+ > m_-\;$ case. Symmetrically
\begin{equation}
\dot{r}_s^- \approx  -\frac{\epsilon}{4m_+}
\;\; \mbox{and}\;\;\;\;
\dot{r}_s^+ \approx  2\left(\frac{m_-}{m_+}-1\right)
\end{equation}
for the $m_+ < m_-\;$ case, which we will explore more closely in section \ref{ss:timematching}.
(For some scenarios, see figure \ref{F:parabolas}.) So, requiring only that $f_-(w)$ has to be real, we already obtain strong restrictions for regions where the model is valid. 

\subsection{Mass matching case}

The ``mass matching'' condition, $m_+(\omega) = m_-(\omega) = m(\omega)$,
corresponds to the interior and exterior Vaidya geometries having the same mass function.
If we choose to impose the ``mass matching'' condition, then by (\ref{junction}) either  
$f_-(\omega) = -1$ or
\begin{equation}
f_-(\omega) = 1 + \frac{2\dot{r}_s}{1 -2m(\omega)/r_s} = 1 - \frac{2|\dot{r}_s|}{1 -2m(\omega)/r_s} \label{712}
\end{equation}
where (assuming evaporation) we used $\dot{r}_s <0$.
The $f_-(w)=-1$ option can be safely discarded: By our metric set-up (\ref{metf}) and with $m_+=m_-=m$ this choice 
actually corresponds to attaching the outside metric to a copy of itself, and so represents a radiating white
hole spacetime, rather than an  evaporating black hole.

Now given that the matching surface is timelike and assuming that $\dot{r}_s < 0$ then it follows directly 
from the form of the induced metric that 
\begin{equation}
|\dot{r}_s| < \frac{1}{2}\PC{1 - \frac{2m(\omega)}{r_s}} \, . 
\end{equation}
Hence $f_-(w)$ in  (\ref{712}) is positive. Next defining $r_s = 2m(\omega) + \epsilon$, we have
\begin{eqnarray}
|\dot{r}_s| < \frac{1}{2}\PC{\frac{\epsilon}{2m + \epsilon}} 
\lesssim \frac{\epsilon}{4m(\omega)}
\end{eqnarray}
implying that a near-$2m$ transition surface is necessarily slowly evolving. 
\enlargethispage{10pt}
This already indicates that the mass matching condition can only be valid for extremely slow shell velocities and, and therefore, for a very slow evaporation. If we now substitute the value of $\epsilon$ that was previously found by analyzing the redshift condition,
\vspace{-0.2cm}
\begin{equation}
\epsilon \approx \frac{1}{32 \pi^2} \frac{m_P^2 }{m(\omega)},
\end{equation}
we obtain:
\vspace{-0.2cm}
\begin{eqnarray}
{|\dot{r}_s|}\lesssim \frac{1}{128 \pi^2}\;\; \PC{\frac{m_P}{m(\omega)}}^2.
\end{eqnarray}
This is the same result obtained before at equation \eqref{E: upsy3}. In this way, we see that the simple requirement of a timelike matching surface already imposes a condition of very small radial velocity, ensuring that we are dealing with an adiabatic evolution.

\vspace{-0.2cm}
\subsection{The ``time matching" case}
\label{ss:timematching}

If we enforce $f_+(\omega) = f_-(\omega) = 1$, so that coordinate time ``runs at the same rate'' on both sides of the shell, then the matching condition on the shell gives us:
\begin{equation}
m_+(\omega) = m_-(\omega) + 2r_s\dot{r}_s = m_-(w) -2r_s|\dot{r_s}|.
\end{equation}
From this we obtain $\; m_-(\omega) \geqslant m_+(\omega)$. We also want both $m_{\pm}(\omega) > 0$ individually. This gives us:
\vspace{-0.3cm}
\begin{eqnarray}
\label{E:eita}
m_- -2r_s|\dot{r_s}| > 0 \qquad \Rightarrow \qquad m_- > 2r_s|\dot{r_s}| . 
\end{eqnarray}
If we substitute $r_s = 2m_+ + \epsilon$ into \eqref{E:eita}, we obtain:
\vspace{-0.2cm}
\begin{equation}
m_- >\; 2(2m_+ + \epsilon)|\dot{r_s}| \;\approx \; 4m_+|\dot{r_s}|,
\end{equation}
giving us an upper limit for the speed of the shell:
\vspace{-0.2cm}
\begin{equation}
|\dot{r_s}| < \frac{m_-(\omega)}{4 m_+(\omega)}.
\end{equation} 
\begin{figure}
	\center
	\includegraphics[scale=0.5]{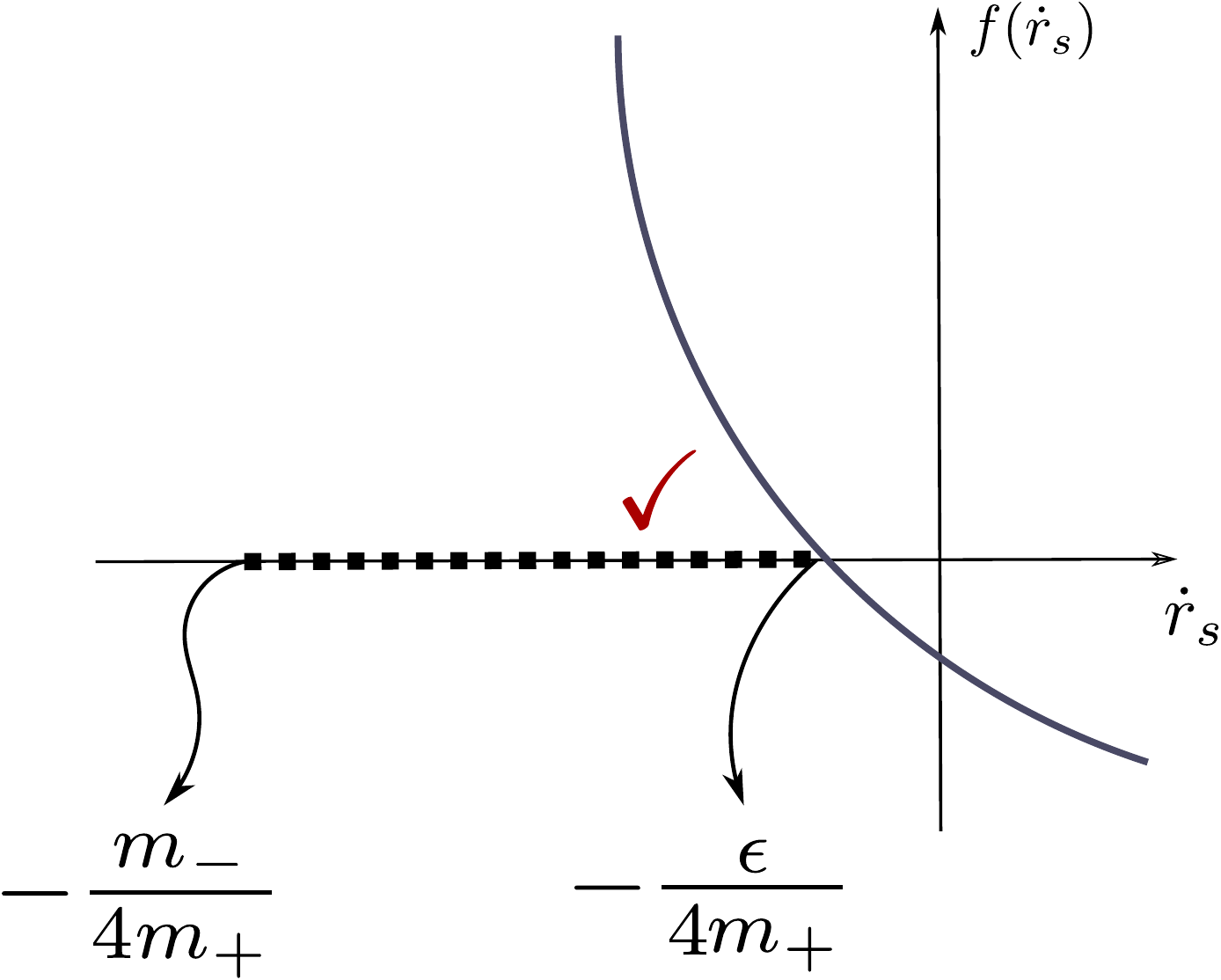}
	\caption{Allowed radial velocities for the time matching case.}
	\label{time matching}
\end{figure}
As $\;m_- > m_+$, we can use second scenario of figure~\ref{F:parabolas}:
\begin{equation}
\dot{r}_s >\; 2\PC{\frac{m_-}{m_+} -1} \qquad\;\hbox{or}\; \qquad \dot{r}_s < -\frac{\epsilon}{4m_+}\; .
\end{equation}
But the first condition gives us positive radial velocities (accretion dominating over the Hawking flux), so we just focus on the second condition. Requiring the interval on figure \ref{time matching} to have a non-zero length we obtain the relatively weak condition $m_-(\omega) > \epsilon(\omega)$.

The interesting feature here is actually the fact that there is a \emph{minimal} velocity for the evaporation rate of the black hole. Why does this happen? Having $m_- > m_+$ is something that is perhaps a bit unexpected and odd. For that to happen, the thin-shell will have to contain a negative-energy surface density. Thus, the black hole will have to keep evaporating, in order to equilibrate this otherwise unstable situation.

\subsection{The empty-interior massive shell (a consistency check)}
\label{ss:empty}
For the sake of completeness, we shall finally consider the extreme case of an “empty” interior; implying that the interior region will simply resume to a portion of Minkowski space. This model is somewhat different from the other models presented, and focuses attention on the exterior geometry. Its physical significance is dubious, since after the end of the evaporation of $m_-(w)$, the radiation process must stop, otherwise the interior mass would start to get more and more negative. It can be useful simply in the sense of checking that the radiation limits are still bounded and well behaved.   So, applying $m_-(\omega)=0$ in equation \eqref{junction}, we obtain:
\begin{equation}
-\left(1-{2m_+(\omega)\over r_s}\right)  - 2 \dot r_s = -f_-(\omega)^2  + 2 f_-(\omega) \dot r_s,
\end{equation}
implying
\vspace{0.2cm}
\begin{equation}
f_-(\omega)^2 - 2f_-(\omega) \dot{r}_s - \PC{1-{2m_+(\omega)\over r_s} + 2 \dot r_s} =0.
\end{equation}
Therefore:
\vspace{0.2cm}
\begin{eqnarray}
\small{
f_-(\omega) = \dot{r}_s \pm \sqrt{\dot{r}^2_s + \PC{1-{2m_+(\omega)\over r_s} + 2 \dot r_s}}
= \dot{r}_s \pm \sqrt{(1 + \dot{r}_s)^2 -{2m_+(\omega)\over r_s}}.}
\end{eqnarray}
We want $f_-(w)$ to be real, so substituting $r_s(\omega) = 2m_+(\omega) + \epsilon(\omega)$, we see
\vspace{0.2cm}
\begin{eqnarray}
(1 + \dot{r}_s)^2 > {2m_+(\omega)\over r_s} \approx 1 - \frac{\epsilon}{2m_+(\omega)}
;\qquad\qquad
|1 + \dot{r}_s| \;\gtrsim\; 1 - \frac{\epsilon}{4m_+(\omega)}.
\end{eqnarray}
As (per assumption) $\dot{r}_s<0$, and taking $|\dot r_s|<1$, so that the evaporation is not ultra-rapid, this implies
\vspace{0.2cm}
\begin{eqnarray}
\label{3}
\dot{r}_s \gtrsim -\frac{\epsilon}{4 m_+(\omega)}; \qquad\qquad
|\dot{r}_s| \lesssim \frac{\epsilon}{4 m_+(\omega)} \ll 1.
\end{eqnarray}
So in this case the velocity of the shell is extremely small, in accordance with the $|\dot{r}_s|$ limitations derived from adiabatic evaporation. It is important, however, to keep in mind that this is probably very different from any real ``final moments of evaporation'' scenario and this case was presented simply for the sake of completeness.

\section{Remarks}

So what have we learned from this exercise? 
First of all, it was interesting to know what is possible to learn simply from  a kinematical analysis of a space-time model. We were able to obtain a considerable amount of information simply by requiring the energies to remain bounded near the horizon. Furthermore, we have shown that, whereas (outgoing) Hawking radiation does not actually seem to need to \emph{cross} the horizon to be physically meaningful and correct, there are good quantitative reasons for believing that the Hawking radiation must arise from a region near the horizon --- since otherwise there is no good physical reason to connect the surface gravity to the Hawking temperature.

\enlargethispage{10pt}
We have sketched a number of scenarios for the evaporation process, and indicated how very general kinematic considerations can nevertheless lead to interesting constraints on the range of validity of these double-Vaidya thin-shell models. 

The toy model, being as simple as it is, does have limitations. The limit for evaporation rates were shown for different cases and, besides looking very small, they were shown to be in accordance with the evaporation rates for black holes with as masses as low as a solar mass. We do not believe and do not claim that this model will hold for black holes near their final evaporation moments, since what happens in those situations is still an unsolved question.

\newpage

\chapter{Famous Last Words}
A quote attributed Paul Valéry says: \emph{``A work is never finished; it is only abandoned''}. We believe the same to be valid for PhD theses~\footnote{The complete quote actually being: ``A work is never completed except by some accident such as weariness, satisfaction, the need to deliver, or death: for, in relation to who or what is making it, it can only be one stage in a series of inner transformations.''}. 

In this way, it is now the time to finally leave behind the work  developed  in this thesis. Before doing so, however, we would like to say some final words. This will not be a summary of everything that has been done, since this was presented in the abstract of the thesis. We will simply take a tour through some points that I believe deserve wrapping up, and we will mention possible future work.

During this thesis, we have revisited the gravitationally induced temperature gradients originally derived by Tolman, and extended this concept to fluids following generic four-velocities in general stationary space-times. Inspired by this generalization, we tackled the problem of the possibility of defining thermodynamic equilibrium for non-Killing flows. This was done by revising the current status of relativistic hydrodynamics for viscous fluids, and studying Born-rigid body motions. 

The Born-rigid, or rigid body, flow was shown to be one of the necessary conditions for a fluid to be in thermodynamic equilibrium --- and we have given several examples of congruences which move rigidly through space-time in Chapter 4. Some of the examples explicitly show non-Killing Born-rigid congruences. This is an interesting result on its own right, since it makes clear that the Herglotz--Noether theorem is not valid for general curved space-times. 

We have given examples of non-Killing Born-rigid congruences in Bianchi Type I  space-times with the specific forms:
\begin{equation}
\label{E:concl}
g_{\mu\nu} = \begin{pmatrix}
-1 & 0 & 0 & 0\\
0 & a(t) & 0 & 0\\
0 & 0 & b_1 & 0\\
0 & 0 & 0 & b_2
\end{pmatrix}; \quad 
g_{\mu\nu} = \begin{pmatrix}
-1 & 0 & 0 & 0\\
0 & a(t) & 0 & 0\\
0 & 0 & \frac{b_1{}^2}{a(t) - d} & 0\\
0 & 0 & 0 & b_2{}^2
\end{pmatrix};
\end{equation} 
and
\begin{equation}
\label{E:concl2}
g_{\mu\nu} = \begin{pmatrix}
-1 & 0 & 0 & 0\\
0 & e^{-t} & 0 & 0\\
0 & 0 & \frac{b_1{}^2\; (e^t -1)}{2d\; (d\;e^t -1)} & 0\\
0 & 0 & 0 & b_2{}^2
\end{pmatrix}.
\end{equation}
We could not find, however, any non-Killing solution for the general Bianchi Type I metric, given by:
 \begin{equation}
 \label{E:concl3}
\d s^2 = -\d t^2 + a(t)\; \d x^2 + b(t) \;\d y^2 + c(t)\; \d z^2,
\end{equation}
with $a(t)$, $b(t)$ and $c(t)$ non-zero and non-constant. This was somewhat surprising, given the existence of solutions for the specific metrics given by \eqref{E:concl} and \eqref{E:concl2}. On the other hand, given that these three metrics present expansion in one dimension and contraction in the other, it is not clear how a metric of the form \eqref{E:concl3} could satisfy such a constraint (in case that is indeed a necessary constraint). 
So, in order to further clarify this topic, one case that we wish to analyze in the future is the truly oblate universe, axisymmetric, expanding in the $z$ axis direction, for example, and contracting on the orthogonal $x$--$y$ plane. We would like to search for new non-Killing Born-rigid congruences and see whether they have any similarities.

Another point that we believe it is important to mention here is the fact that perfect equilibrium states can only be held for fluids following Killing flows. This has been shown in Chapter 4.  On the other hand, Killing flows are too restrictive, and a more complex and interesting analysis can easily be developed for fluids following non-Killing trajectories, by means of comparing the time-scales involved in the evolution of the system. Given, for example, the relaxation times obtained from Extended Irreversible Thermodynamics ($\tau_0, \;\tau_1$ and $\tau_2$ given by \eqref{E:tau}) and given the time-scales imposed by expansion, shear and vorticity:
\begin{equation}
\tag{\ref{E:timescales}}
\tau_{\theta} = \frac{1}{|\theta|}\, , \qquad \tau_{\sigma} = \frac{1}{\sqrt{\sigma_{\mu\nu}\,\sigma^{\mu\nu}}}\, , \qquad \tau_{\omega} = \frac{1}{\sqrt{\omega_{\mu \nu}\,\omega^{\mu\nu}}}\, ,
\end{equation}
it is clear that, as long as the time-scales given by the changes caused in the system are much smaller than the system's relaxation time, one can still analyze the evolution of such a system as an infinite sequence of quasi-equilibrium states. Furthermore, the temperature distribution in each of these equilibrium states must be given by 
\begin{equation}
\tag{\ref{e:euler3}}
a_b =  - \nabla_b \ln T,
\end{equation}
which is the generalized gravitationally induced temperature gradient presented in Chapter 3. We see, in this way, that there is no inconsistency in the assumption of equilibrium states along non-Killing flows made in some parts of this same chapter, given that the comparison between the time-scales of the changes and the relaxation times inherent to the system is what will dictate the possibility of temporary equilibrium states or not. The results presented in this thesis can be used for flows following general four-velocities, as long as its validity is initially checked in the time-scale sense here discussed.

We have also discussed in chapter 5 a few topics related to black hole thermodynamics, specifically addressing the trans-Planckian problem for Hawking radiation. We have adopted a simple toy model where we assumed the Hawking photons to be emitted not from the horizon, but from a shell located at a radial position $r_s(w) = 2G_N m(w) + \epsilon(w)$. We did this in order to investigate the values of $\epsilon(w)$ for which trans-Planckian energies do not occur. The result obtained, in terms of proper distance from the shell, was of the order of a Planck length. Some of the topics discussed in the chapters prior to chapter 5 were used in order to obtain this result. Chapter 5 then helps filling out the picture of how gravity and thermodynamics influence each other.

To conclude, we believe that the connections between thermodynamics, hydrodynamics and general relativity still have a fruitful future, and we hope that the results and discussions presented in the thesis will be able to clarify possible confusions and help us walk a tiny step in the right direction.

\newpage
\appendix
\chapter{Some technical results on the Vaidya model}
\label{C: appendix 6}

We will, in this appendix, derive the 4-acceleration of the thin shell treated in Chapter \ref{C: Black Holes Thermodynamics}. To do so, remember that the shell's 4-velocity and normal vector were given by:
\begin{equation}
u_a =\frac{1}{||U||} \left(-f_\pm(w)^2\;\left(1-{2G_Nm_\pm(w)\over r}\right)\mp f_\pm(w)\, \dot r_s(w) ,\; \mp f_{\pm} (w)\; ;\; 0, 0 \right),
\end{equation}
and
\begin{equation}
\qquad
n^a = \frac{1}{||N||}\left(\mp \frac{1}{f_\pm(w)} , \;\left(1-{2G_Nm_\pm(w)\over r}\right) \pm  \frac{\dot r_s}{f_\pm(w)}\;;\; 0, 0 \right),
\end{equation}
where $||U||$ and $||N||$ were given by:
\begin{equation}
\tag{\ref{E: u norm}}
\|U\| = \sqrt{-g_{ab} U^a U^b} = \sqrt{f_\pm(w)^2\;(1-2G_Nm_\pm(w)/r)\pm 2 f_\pm \dot r_s(w)} \, , 
\end{equation}
and
\begin{equation}
\tag{\ref{E: n norm}}
\| N \| = \sqrt{g^{ab}N_a N_b} = \sqrt{(1-2G_Nm_\pm(w)/r)\pm 2 f_\pm(w)^{-1} \dot r_s(w) } \, ,
\end{equation}

Remember now that, by definition $A^a = u^b \nabla_b u^a$ and that $A^a u_a =0$. So the four-acceleration $A^a$, whatever it is, is orthogonal to $u^a$. But we also have that $n^a u_a = 0$, which makes us conclude that $A^a = A\; n^a$, where $A= A(w)$ might have a time dependence. Now, using the fact that $n^a$ is normalized, we have:
\begin{equation}
A = n^a A_a = n^a  (u^b \nabla_b u_a) = n^a  (u^b \nabla_b u_a - u^b \nabla_a u_b) 
= n^a u^b (u_{a,b} - u_{b,a}).
\end{equation}
Note that on the third equality we have simply added zero to obtain a more convenient result. Now, notice that we might rewrite $A(w)$ as
\begin{eqnarray}
\label{E:4-acceleration}
A &=& {1\over2} (n^a u^b-n^b u^a)\, (u_{a,b} - u_{b,a})
\nonumber\\
&=& {1\over2\|U\|^2\|N\|} (N^a U^b-N^b U^a)\, \left(U_{a,b} - U_{b,a} - {\{U_a \|U\|_{,b} - U_b \|U\|_{,a}]\}\over \|U\|} \right) 
\nonumber\\
&=& {1\over2\|U\|^2\|N\|} (N^a U^b-N^b U^a)\, \left(U_{a,b} - U_{b,a} \right) 
- {{N^a \partial_a} \| U \|\over\|N\| \, \|U\| }.
\end{eqnarray}
To simplify this result any further, we will have to perform some side calculations. So, let us put this result aside for now, and come back to it later.

\vspace{-0.5cm}
\paragraph{The on-shell induced Levi--Civita tensor} \hfill

Our first goal will be to rewrite the term $(N^a U^b - N^b U^a)$ in the 4-acceleration. To do so, let us first note that
\begin{equation}
\sqrt{-g_{\pm}} = f_\pm(w)\, r^2\, \sin\theta  \, . 
\end{equation}
Also note that, since $\varepsilon^{ab}$ is the induced Levi--Civita tensor on the $w$-$r$ plane, we have: 
\begin{equation}
N^a U^b - N^b U^a = (N^w U^r - N^r U^w) \; f_{\pm} \; \varepsilon^{ab}.
\end{equation}
Specifically, $\varepsilon^{ab}$ is an antisymmetric 2-tensor, and in these particular $(w,r,\theta,\phi)$ coordinates we have   $\varepsilon^{wr} =  f_\pm(w)^{-1} = -\varepsilon^{rw}$. 
Then 
\begin{equation}
\label{E:N^U}
N^a U^b - N^b U^a = -\left([1-2G_Nm_\pm/r] \pm 2 f_\pm^{-1} \dot r_s\right) \; f_\pm\;  \varepsilon^{ab} = - \|N\|^2  \; f_\pm\;\varepsilon^{ab}.
\end{equation}

\enlargethispage{10pt}

\vspace{-0.8cm}
\paragraph{Exterior derivatives of tangent and normal vectors}\hfill

Similarly, let us now consider the exterior derivative
\begin{equation}
\partial_a U_b - \partial_b U_a =-(\partial_w U_r - \partial_r U_w)  \; f_\pm(w)^{-1} \;\varepsilon_{ab},
\end{equation}
where now $\varepsilon_{ab}$ is an antisymmetric 2-form, and in these particular $(w,r,\theta,\phi)$ coordinates we have   $\varepsilon_{wr} = - f_\pm(w) = -\epsilon_{rw}$, so that $\varepsilon^{ab}\varepsilon_{ab}=-2$.
We note that
\begin{equation}
\partial_w U_r - \partial_r U_w = \partial_w (\mp f_\pm) -    \partial_r (-f_\pm^2(1-2G_Nm_\pm/r)\mp f_\pm \dot r_s  ) = 
f_\pm^2 \; {2m_\pm\over r^2} \mp \dot f_\pm.
\end{equation}
That is
\begin{equation}
\label{E:dU}
\partial_a U_b - \partial_b U_a = \left(- f_\pm(w)\; {2G_Nm_\pm(w)\over r^2} \pm {\dot f_\pm(w)\over f_\pm(w)}\right)  \; \epsilon_{ab}.
\end{equation}
Meanwhile $N_a$ is surface-forming and so:
\begin{equation}
\partial_a N_b - \partial_b N_a = 0.
\end{equation}

\paragraph{Normal derivatives}\hfill

Now, on the last term of the acceleration formula \eqref{E:4-acceleration}, we need to calculate a normal derivative. Let us expand it:
\begin{equation}
N^a \partial_a = \mp \frac{1}{ f_\pm(w)} \partial_w + \left[\left(1-{2G_Nm_\pm(w)\over r}\right) \pm f_\pm(w)^{-1} \dot r_s \right] \partial_r,
\end{equation}
from which we obtain:
\begin{equation}
N^a \partial_a = \mp \frac{1}{f_\pm(w)}\left[\partial_w + \dot r_s \partial_r\right] + \left[\left(1-{2G_Nm_\pm(w)\over r}\right) \pm 2f_\pm(w)^{-1} \dot r_s \right] \partial_r,
\end{equation}
implying
\begin{equation}
\label{E:normal-derivative}
N^a \partial_a = \mp{1\over f_\pm(w)} \; U^a \partial_a + \| N \|^2 \partial_r 
= \mp{1\over f_\pm(w)} \; {d\over d w}  + \| N \|^2 \partial_r.
\end{equation}

\vspace{0.4cm}

\paragraph{Completing the acceleration calculation}\hfill 

We are now able to complete the 4-acceleration calculation by inputting the results obtained above, equations (\ref{E:N^U}) and (\ref{E:dU}) into \eqref{E:4-acceleration}:
\begin{equation}
{1\over2}(N^a U^b-N^b U^a)\, \left(U_{a,b} - U_{b,a} \right)  = \|U\|^2\;{2m(w)\over r^2 }.
\end{equation}
Here we also have used our choice of $f_+\longrightarrow1$ and the fact that $\|U\|=\|N\|$ for this case.

Similarly, in view of equation (\ref{E:normal-derivative}) we have
\begin{equation}
{N^a \partial_a} \| U \| = - {d\|U\|\over dw}+ \|U\|^2 \partial_r \|U\| =  - {d\|U\|\over dw}+ \|U\| {m(w)\over r^2}.
\end{equation}
Combining all these results, the 4-acceleration of the thin shell is given by the formula
\begin{equation}
A(w) = {1\over \|U\|} {G_Nm(w)\over r_s(w)^2}  +  {1\over \|U\|^2}  {d\|U\|\over dw}
= {1\over \|U\|} \left( {G_Nm(w)\over r_s(w)^2}  + {d\ln \|U\|\over dw}\right). 
\end{equation}

\vspace{0.3cm}

\paragraph{Constant-$w$ affine null vector}\hfill 

A particularly obvious and useful constant-$w$ null vector, to be used for defining affine parameters on the radial null geodesics,  is
\begin{equation}
k^a = (0,\pm f_\pm(w)^{-1},0,0);   \qquad k_a = (-1,0,0,0). 
\end{equation}
Here the $\pm$ is chosen to ensure that $k^a$ is future pointing in both regions. 
Now 
\begin{equation}
k^b \nabla_b k^a =  g^{ac} \; k^b\nabla_b k_c,
\end{equation}
and it is easy to see that
\begin{equation}
k^b \nabla_b k_c = k^b (\nabla_b k_c-\nabla_c k_b) =  k^b (\partial_b k_c-\partial_c k_b)  = 0.
\end{equation}
So $k^a = (0,\pm f_\pm^{-1},0,0)$ is the tangent to an \emph{affinely parameterized} null congruence. 

\vspace{0.4cm}
\paragraph{Constant-$r$ observer and constant-$r$ normal}\hfill 

A  ``constant-$r$ observer'' (to be used for defining some notion of ``distance'' to the evolving apparent horizon),  has 4-velocity
\begin{equation}
v^a = {\left(1,0,0,0\right)\over f_\pm \sqrt{1-2m_\pm/r}}; \qquad  v_a = {\left(-f_\pm^2(1-2G_Nm_\pm/r),\mp f_\pm,0,0\right)\over f_\pm \sqrt{1-2G_Nm_\pm/r}}.
\end{equation}
Near spatial infinity (where it makes sense to enforce $f\to1$), this reduces to
\begin{equation}
v^a = {\left(1,0,0,0\right)^a}; \qquad  v_a = {\left(-1,\mp1,0,0\right)_a}.
\end{equation}
In contrast, the non-normalized covariant vector normal to the surfaces of constant $r$ is $(\nabla r)_a=(0,1,0,0)_a$, and the unit normal to the constant $r$ surfaces is
\begin{equation}
\widehat{(\nabla r)}_a = {(0,1,0,0)_a\over\sqrt{1-2G_Nm_\pm/r}},
\label{E:normal-to-r}
\end{equation}
as one could possibly expect.


\newpage
\addchap{Publications related to the PhD}

\Large{\emph{Publications in Journals}}

\normalsize
\begin{itemize}
\item \textbf{Tolman temperature gradients in a gravitational field}\\
J. Santiago, M. Visser\\
European Journal of Physics \textbf{40} 2 (2019) 025604\\
	arXiv:1803.04106 [gr-qc]
	
\item \textbf{Gravity's universality: The physics underlying Tolman temperature gradients}\\
J. Santiago, M. Visser\\
International Journal of Modern Physics D \textbf{27} 14, 1846001 (2018)\\
Awarded first prize in the 2018 GRF essay contest.\\
		arXiv:1805.05583 [gr-qc]
		
\item \textbf{Tolman-like temperature gradients in stationary spacetimes}\\
J. Santiago, M. Visser\\
Phys. Rev. D \textbf{98} 064001 (2018)\\
arXiv:1807.02915 [gr-qc]
	\end{itemize}

\begin{itemize}		
\item \textbf{Evading the Trans-Planckian problem with Vaidya spacetimes}\\
I. Booth, B. Creelman, J. Santiago, M. Visser\\
J. Cosm. Astropart. Phys. \textbf{09} 067 (2019)\\
arXiv:1809.10412 [gr-qc]
	
\end{itemize}
\newpage

$\,$

\Large{\emph{Article not included in the thesis}}

\normalsize	
\begin{itemize}		
\item \textbf{``Twisted'' black holes are unphysical}\\
F. Gray, J. Santiago, S. Schuster, M. Visser\\
Mod. Phys. Lett. A \textbf{32} 18 (2017) 1771001\\
arXiv:1610.06135 [gr-qc]
	
\end{itemize}

\newpage
 	\addcontentsline{toc}{chapter}{Bibliography}
\bibliographystyle{plain}
\bibliography{tese_}

\begin{thebibliography}{100}

\bibitem{Adkins}
C.~J. Adkins.
\newblock {\em Equilibrium Thermodynamics}.
\newblock Cambridge University Press, 1983.

\bibitem{AnaeMatt}
A.~Alonso-Serrano and M.~Visser.
\newblock {Coarse Graining Shannon and von Neumann Entropies}.
\newblock {\em Entropy}, 19(5), 2017.

\bibitem{Anderson:1967}
J.~L. Anderson.
\newblock {\em Principles of relativity physics}.
\newblock Academic Press, 1967.

\bibitem{Andersson2007}
N.~Andersson and G.~L. Comer.
\newblock Relativistic fluid dynamics: Physics for many different scales.
\newblock {\em Living Reviews in Relativity}, 10(1):1, 2007.

\bibitem{barbado}
L.~C. Barbado and M.~Visser.
\newblock {Unruh-DeWitt} detector event rate for trajectories with
  time-dependent acceleration.
\newblock {\em Phys. Rev. D}, 86:084011, 2012.

\bibitem{Barcelo:2010a}
C.~Barcel{\'o}, S.~Liberati, S.~Sonego, and M.~Visser.
\newblock Hawking-like radiation from evolving black holes and compact
  horizonless objects.
\newblock {\em Journal of High Energy Physics}, 2011(2):1--30, 2011.

\bibitem{Barcelo:2010b}
C.~Barcel\'o, S.~Liberati, S.~Sonego, and M.~Visser.
\newblock Minimal conditions for the existence of a {H}awking-like flux.
\newblock {\em Phys. Rev. D}, 83:041501, Feb 2011.

\bibitem{LRR}
C.~Barcel{\'o}, S.~Liberati, and M.~Visser.
\newblock Analogue gravity.
\newblock {\em Living Reviews in Relativity}, 8(1):12, Dec 2005.

\bibitem{Bardeen:2017}
J.~M. Bardeen.
\newblock The semi-classical stress-energy tensor in a {S}chwarzschild
  background, the information paradox, and the fate of an evaporating black
  hole.
\newblock {\em arXiv:1706.09204 [gr-qc]}, 2017.

\bibitem{B&D}
N.~D. Birrell and P.~C.~W. Davies.
\newblock {\em Quantum Fields in Curved Space}.
\newblock Cambridge Monographs on Mathematical Physics. Cambridge University
  Press, 1982.

\bibitem{Jess-Vaidya}
I.~Booth, B.~Creelman, J.~Santiago, and M.~Visser.
\newblock Evading the trans-planckian problem with vaidya spacetimes.
\newblock {\em arXiv:1809.10412 [gr-qc]}, 2018.

\bibitem{Boyer}
R.~H. Boyer.
\newblock Rigid frames in general relativity.
\newblock {\em Series A. Mathematical and Physical Sciences}, 283:343--355,
  1965.

\bibitem{buchdahl:1949}
H.~A. Buchdahl.
\newblock Temperature equilibrium in a stationary gravitational field.
\newblock {\em Phys. Rev.}, 76:427--428, 1949.

\bibitem{Callen-book}
H.~B. Callen.
\newblock {\em Thermodynamics and an Introduction to Thermostatistics}.
\newblock New York : Wiley, 1985.

\bibitem{Carroll}
S.~M. Carroll.
\newblock {\em An introduction to general relativity: spacetime and geometry}.
\newblock Addison Wesley, 2004.

\bibitem{Carter}
B.~Carter.
\newblock Axisymmetric black hole has only two degrees of freedom.
\newblock {\em Phys. Rev. Lett.}, 26:331--333, 1971.

\bibitem{Kremer}
C.~Cercignani and G.~M. Kremer.
\newblock {\em The relativistic Boltzmann equation: theory and applications}.
\newblock Progress in mathematical physics; v.~22. Birkhäuser Verlag, 2002.

\bibitem{remnants}
P.~Chen, Y.C. Ong, and D.H. Yeom.
\newblock Black hole remnants and the information loss paradox.
\newblock {\em Physics Reports}, 603:1 -- 45, 2015.

\bibitem{Mariao2011}
M.~J. de~Oliveira.
\newblock Irreversible models with {Boltzmann{\textendash}Gibbs} probability
  distribution and entropy production.
\newblock {\em Journal of Statistical Mechanics: Theory and Experiment},
  2011(12):P12012, 2011.

\bibitem{doran}
C.~Doran.
\newblock New form of the {K}err solution.
\newblock {\em Phys. Rev. D}, 61:067503, Feb 2000.

\bibitem{Eckart}
C.~Eckart.
\newblock {The Thermodynamics of Irreversible Processes. III. Relativistic
  Theory of the Simple Fluid}.
\newblock {\em Phys. Rev.}, 58:919--924, 1940.

\bibitem{paradox1}
P.~Ehrenfest.
\newblock {Gleichförmige Rotation starrer Körper und Relativitätstheorie}
  ({U}niform rotation of rigid bodies and relativity theory).
\newblock {\em Physikalische Zeitschrift}, 10:918, 1909.

\bibitem{Einstein1}
A.~Einstein.
\newblock { Über das Relativitätsprinzip und die aus demselben gezogenen
  Folgerungen (On the relativity principle and the conclusions drawn from it)}.
\newblock {\em Jahrb. Radioakt. Elektron.}, pages 411--462, 1907.

\bibitem{Boo}
C.~Farias, V.A. Pinto, and P.~S. Moya.
\newblock What is the temperature of a moving body?
\newblock {\em Scientific Reports}, 7:17657, 2017.

\bibitem{frolovnovikov}
V.~P. Frolov and I.~D. Novikov.
\newblock {\em Black Hole Physics}.
\newblock Kluwer Academic Publishers, 1998.

\bibitem{frolov}
V.~P. Frolov and A.~Zelnikov.
\newblock {\em Introduction to black hole physics}.
\newblock Oxford University Press, 2011.

\bibitem{Gibbs}
J.~W. Gibbs.
\newblock {\em Elementary Principles in Statistical Mechanics: Developed with
  Especial Reference to the Rational Foundation of Thermodynamics}.
\newblock Cambridge Library Collection - Mathematics. Cambridge University
  Press, 2010.

\bibitem{Godreche}
C.~Godr{\`{e}}che.
\newblock Dynamics of the directed {Ising} chain.
\newblock {\em Journal of Statistical Mechanics: Theory and Experiment},
  2011(04):P04005, 2011.

\bibitem{Vaidya2}
J.~B. Griffiths and J.~Podolsky.
\newblock {\em {Exact Space-Times in Einstein's General Relativity}}.
\newblock Cambridge Monographs on Mathematical Physics. Cambridge University
  Press, Cambridge, 2009.

\bibitem{paradox2}
\O. Gr{\o}n.
\newblock Relativistic description of a rotating disk.
\newblock {\em American Journal of Physics}, 43(10):869--876, 1975.

\bibitem{gullstrand}
A.~Gullstrand.
\newblock Allgemeine {L}{\"o}sung des statischen {E}ink{\"o}rperproblems in der
  {E}insteinschen {G}ravitationstheorie ({G}eneral solution for the static
  one-body problem in {E}instein gravity).
\newblock {\em Ark.\ Mat.\ Astron.\ Fys}, 16(8):1--15, 1922.

\bibitem{river}
A.~J.~S. Hamilton and J.~P. Lisle.
\newblock The river model of black holes.
\newblock {\em American Journal of Physics}, 76(6):519--532, 2008.

\bibitem{Hartle}
J.~B. Hartle.
\newblock {\em Gravity: An introduction to Einstein’s general relativity}.
\newblock Pearson Education - Addison Wesley, 2003.

\bibitem{explosions}
S.~W. Hawking.
\newblock Black hole explosions.
\newblock {\em Nature}, 248:30--31, 1974.

\bibitem{Hawking1975}
S.~W. Hawking.
\newblock Particle creation by black holes.
\newblock {\em Communications in Mathematical Physics}, 43(3):199--220, 1975.

\bibitem{HawkingEllis}
S.~W. Hawking and G.~F.~R. Ellis.
\newblock {\em The Large Scale Structure of Space-Time}.
\newblock Cambridge Monographs on Mathematical Physics. Cambridge University
  Press, 1973.

\bibitem{Herglotz}
G.~Herglotz.
\newblock {Über den vom Standpunkt des Relativitätsprinzips aus als starren
  zu bezeichnenden Körper (On bodies that are to be designated as ``rigid"
  from the standpoint of the relativity principle)}.
\newblock {\em Annalen der Physik}, 336(2):393--415, 1910.

\bibitem{Hiscock}
W.~A. Hiscock and L.~Lindblom.
\newblock Stability and causality in dissipative relativistic fluids.
\newblock {\em Annals of Physics}, 151(2):466 -- 496, 1983.

\bibitem{InstabilityCIT}
W.~A. Hiscock and L.~Lindblom.
\newblock Generic instabilities in first-order dissipative relativistic fluid
  theories.
\newblock {\em Phys. Rev. D}, 31:725--733, 1985.

\bibitem{Hiscock1988}
W.~A. Hiscock and L.~Lindblom.
\newblock Nonlinear pathologies in relativistic heat-conducting fluid theories.
\newblock {\em Physics Letters A}, 131(9):509 -- 513, 1988.

\bibitem{Israel}
W.~Israel.
\newblock Singular hypersurfaces and thin shells in general relativity.
\newblock {\em Il Nuovo Cimento B (1965-1970)}, 44(1):1--14, 1966.

\bibitem{Israelvacuum}
W.~Israel.
\newblock Event horizons in static vacuum space-times.
\newblock {\em Phys. Rev.}, 164:1776--1779, 1967.

\bibitem{Israelelvacuum}
W.~Israel.
\newblock Event horizons in static electrovac space-times.
\newblock {\em Commun.Math. Phys.}, 8:245, 1968.

\bibitem{Israel:1981}
W.~Israel.
\newblock Thermodynamics of relativistic systems.
\newblock {\em Physica A: Statistical Mechanics and its Applications},
  106(1):204 -- 214, 1981.

\bibitem{Israel:1986}
W.~Israel.
\newblock Relativistic thermodynamics, thermofield statistics and super-fluids.
\newblock {\em J. Non-Equilibrium Thermodynamics}, 11:295 -- 316, 1986.

\bibitem{Israel:1980}
W.~Israel and J.~M. Stewart.
\newblock Progress in relativistic thermodynamics and electrodynamics of
  continuous media.
\newblock In {\em General Relativity and gravitation: 100 years after the birth
  of Albert Einstein, Vol.2}, pages 491--525. Plenum, New York, 1980.

\bibitem{Jacobson}
T.~Jacobson.
\newblock Thermodynamics of spacetime: {The Einstein} equation of state.
\newblock {\em Phys. Rev. Lett.}, 75:1260, 1995.

\bibitem{relax2}
J.~J. Kelly.
\newblock {\em Review of Thermodynamics}.
\newblock Notes - University of Maryland, 2002.
\newblock
  http://www.physics.umd.edu/courses/Phys603/kelly/Notes/\\ReviewThermodynamics.pdf.

\bibitem{Lake:1992zz}
K.~Lake.
\newblock Precursory singularities in spherical gravitational collapse.
\newblock {\em Phys. Rev. Lett.}, 68:3129--3132, 1992.

\bibitem{Lanczos0}
K.~{Lanczos}.
\newblock {{``Untersuching \"uber fl\"achenhafte verteiliung der materie in der
  Einsteinschen gravitationstheorie'' (Investigation of the matter distribution
  in Eintein's theory of gravity)}}.
\newblock {\em unpublished}, 1922.

\bibitem{Lanczos}
K.~{Lanczos}.
\newblock {{Fl{\"a}chenhafte Verteilung der Materie in der Einsteinschen
  Gravitationstheorie (Spatial distribution of matter in Einstein's theory of
  gravity) }}.
\newblock {\em Annalen der Physik}, 379:518--540, 1924.

\bibitem{Landsberg}
P.~T. Landsberg.
\newblock Does a moving body appear cool?
\newblock {\em Nature}, 212:571--572, 1966.

\bibitem{Lebon}
G.~Lebon, D.~Jou, and J.~Casas-Vázquez.
\newblock {\em Understanding Non-equilibrium Thermodynamics: Foundations,
  Applications, Frontiers}.
\newblock Springer Berlin Heidelberg, 2008.

\bibitem{Liu}
I-S. Liu, I.~Müller, and T.~Ruggeri.
\newblock Relativistic thermodynamics of gases.
\newblock {\em Annals of Physics}, 169(1):191 -- 219, 1986.

\bibitem{Lopez-Monsalvothesis}
C.~S. Lopez-Monsalvo.
\newblock Covariant thermodynamics \& relativity - {PhD} thesis, {University of
  Southampton}.
\newblock {\em arXiv:1107.1005 [gr-qc]}, 2011.

\bibitem{Maartens}
R.~Maartens.
\newblock Causal thermodynamics in relativity.
\newblock {\em arXiv:astro-ph/9609119}, 1996.

\bibitem{maxwell:1868}
J.~C. Maxwell.
\newblock On the dynamical theory of gases.
\newblock {\em The London, Edinburgh, and Dublin Philosophical Magazine and
  Journal of Science}, 35(235):129--145, 1868.

\bibitem{maxwell:1902}
J.~C. Maxwell.
\newblock {\em Theory of Heat}.
\newblock Republished by Cambridge University Press (2011), 1871.

\bibitem{dictionary}
Merriam-Webster.com.
\newblock ``{I}ntrinsic”.
\newblock Last accessed: 12-02-2019.
\newblock https://www.merriam-webster.com/dictionary/intrinsic.

\bibitem{MTW}
C.~W. Misner, K.~S. Thorne, and J.~A. Wheeler.
\newblock {\em Gravitation}.
\newblock Princeton University Press, 2017.

\bibitem{Noether}
F.~Noether.
\newblock {Zur Kinematik des starren Körpers in der Relativtheorie (On the
  kinematics of rigid bodies in the theory of relativity)}.
\newblock {\em Annalen der Physik}, 336(5):919--944, 1910.

\bibitem{Ott}
H.~Ott.
\newblock Lorentz-transformation der w{\"a}rme und der temperatur
  {(Lorentz-transformation of heat and temperature)}.
\newblock {\em Zeitschrift f{\"u}r Physik}, 175(1):70--104, 1963.

\bibitem{Pad1}
T.~Padmanabhan.
\newblock Statistical mechanics of gravitating systems.
\newblock {\em Physics Reports (Review Section of Physics Letters)},
  188(5):285--362, 1990.

\bibitem{Pad2}
T.~Padmanabhan.
\newblock {\em Statistical Mechanics of Gravitating Systems in Static and
  Cosmological backgrounds}, volume 602 of {\em Lecture Notes in Physics},
  pages 165--207.
\newblock Springer, 2002.

\bibitem{painleve}
P.~Painlev\'e.
\newblock La m\'ecanique classique et la th\'eorie de la relativit\'e
  ({Classical Mechanics and the theory of relativity}).
\newblock {\em C. R. Acad. Sci. (Paris)}, 173:677--680, 1921.

\bibitem{Pavani}
A.S. Parvan.
\newblock Lorentz transformations of the thermodynamic quantities.
\newblock {\em Annals of Physics}, 401:130 -- 138, 2019.

\bibitem{Pirani}
F.~A.~E. Pirani and G.~Williams.
\newblock Rigid motion in a gravitational field.
\newblock {\em Séminaire Janet}, 5(8):1--16, 1962.

\bibitem{Planck1}
M.~Planck.
\newblock { Zur Dynamik bewegter Systeme (On the dynamics of moving systems)}.
\newblock {\em Sitzungsberichte der Königlich-Preussischen Akademie der
  Wissenschaften}, pages 542--570, 1907.

\bibitem{Planck2}
M.~Planck.
\newblock {Zur Dynamik bewegter Systeme (On the dynamics of moving systems)}.
\newblock {\em Annalen der Physik}, 331(6):1--34, 1908.

\bibitem{Poisson}
Eric Poisson.
\newblock {\em A Relativist's Toolkit: The Mathematics of Black-Hole
  Mechanics}.
\newblock Cambridge University Press, 2004.

\bibitem{Puiseaux:1850}
V.~A. Puiseux.
\newblock Recherches sur les fonctions alg\'ebriques ({R}esearch on algebraic
  functions).
\newblock {\em J. Math. Pures Appl.}, 15:365--480, 1850.

\bibitem{Puiseaux:1851}
V.~A. Puiseux.
\newblock Nouvelles recherches sur les fonctions alg\'ebriques ({N}ew research
  on algebraic functions).
\newblock {\em J. Math. Pures Appl.}, 16:228--240, 1851.

\bibitem{Knudsen}
B.~E. Rapp.
\newblock Chapter 9 - fluids.
\newblock In {\em Microfluidics: Modelling, Mechanics and Mathematics}, Micro
  and Nano Technologies, pages 243 -- 263. Elsevier, Oxford, 2017.

\bibitem{Rezzolla}
L.~Rezzolla and O.~Zanotti.
\newblock {\em Relativistic Hydrodynamics}.
\newblock Oxford University Press, 2013.

\bibitem{Robinson}
D.~C. Robinson.
\newblock Uniqueness of the {Kerr Black Hole}.
\newblock {\em Phys. Rev. Lett.}, 34:905--906, 1975.

\bibitem{relax}
C.~M. Roland.
\newblock Characteristic relaxation times and their invariance to thermodynamic
  conditions.
\newblock {\em Soft Matter}, 4:2316--2322, 2008.

\bibitem{relax4}
T.~Roths, D.~Maier, C.~Friedrich, M.~Marth, and J.~Honerkamp.
\newblock Determination of the relaxation time spectrum from dynamic moduli
  using an edge preserving regularization method.
\newblock {\em Rheologica Acta}, 39(2):163--173, Mar 2000.

\bibitem{Sahai}
R.~Sahai, W.~H.~T. Vlemmings, P.~J. Huggins, L.-{\AA}. Nyman, and I.~Gonidakis.
\newblock {ALMA} observations of the coldest place in the universe: The
  {B}oomerang nebula.
\newblock {\em The Astrophysical Journal}, 777(2):92, 2013.

\bibitem{SalzmanTaub}
G.~Salzman and A.~H. Taub.
\newblock Born-type rigid motion in relativity.
\newblock {\em Phys. Rev.}, 95:1659--1669, 1954.

\bibitem{Santiago:2018-ijmpd}
J.~Santiago and M.~Visser.
\newblock {Gravity’s universality: The physics underlying Tolman temperature
  gradients}.
\newblock {\em International Journal of Modern Physics D}, 27(14):1846001,
  2018.

\bibitem{Santiago:2018-prd}
J.~Santiago and M.~Visser.
\newblock Tolman-like temperature gradients in stationary spacetimes.
\newblock {\em Phys. Rev. D}, 98:064001, 2018.

\bibitem{Santiago:2018-ejp}
J.~Santiago and M.~Visser.
\newblock Tolman temperature gradients in a gravitational field.
\newblock {\em European Journal of Physics}, 40(2):025604, 2019.

\bibitem{relax3}
P.~C. Sousa, E.~J. Vega, R.~G. Sousa, J.~M. Montanero, and M.~A. Alves.
\newblock Measurement of relaxation times in extensional flow of weakly
  viscoelastic polymer solutions.
\newblock {\em Rheologica Acta}, 56(1):11--20, Jan 2017.

\bibitem{tolman:1928}
R.~C. Tolman.
\newblock On the extension of thermodynamics to general relativity.
\newblock {\em Proceedings of the National Academy of Sciences of the United
  States of America}, 14(3):268--272, 1928.

\bibitem{weightofheat}
R.~C. Tolman.
\newblock On the weight of heat and thermal equilibrium in {General
  Relativity}.
\newblock {\em Phys. Rev.}, 35:904--924, 1930.

\bibitem{tolman:1933a}
R.~C. Tolman.
\newblock Thermodynamics and relativity.
\newblock {\em Science}, 77(1995):291--298, 1933.

\bibitem{tolman:1933b}
R.~C. Tolman.
\newblock Thermodynamics and relativity. {II}.
\newblock {\em Science}, 77(1996):313--317, 1933.

\bibitem{tolmanbook}
R.~C. Tolman.
\newblock {\em Relativity, Thermodynamics and Cosmology}.
\newblock Oxford: Clarendon Press, 1934.

\bibitem{tolman:1930}
R.~C. Tolman.
\newblock Static solutions of {E}instein's field equations for spheres of
  fluid.
\newblock {\em Phys. Rev.}, 55:364--373, 1939.

\bibitem{ehrenfest:1930}
R.~C. Tolman and P.~Ehrenfest.
\newblock Temperature equilibrium in a static gravitational field.
\newblock {\em Phys. Rev.}, 36:1791--1798, 1930.

\bibitem{TaniaMario}
T.~Tom\'e and M.~J. de~Oliveira.
\newblock Stochastic approach to equilibrium and nonequilibrium thermodynamics.
\newblock {\em Phys. Rev. E}, 91:042140, 2015.

\bibitem{unruh:1981}
W.~G. Unruh.
\newblock Experimental black-hole evaporation?
\newblock {\em Phys. Rev. Lett.}, 46:1351--1353, 1981.

\bibitem{Vaidya}
P.~C. Vaidya.
\newblock {The Gravitational Field of a Radiating Star}.
\newblock {\em Gen. Rel. Grav.}, 31:121--135, 1999.

\bibitem{Velasco}
R.~M. Velasco, L.~Scherer García-Colín, and F.~J. Uribe.
\newblock Entropy production: Its role in non-equilibrium thermodynamics.
\newblock {\em Entropy}, 13(1):82--116, 2011.

\bibitem{unexpected}
M.~Visser.
\newblock Acoustic propagation in fluids: An unexpected example of {L}orentzian
  geometry.
\newblock {\em arXiv:gr-qc/9311028}, 1993.

\bibitem{book}
M.~Visser.
\newblock {\em Lorentzian wormholes: from Einstein to Hawking}.
\newblock Springer--Verlag, New York, 1995.

\bibitem{gvp3}
M.~Visser.
\newblock Gravitational vacuum polarization. {III}. {E}nergy conditions in the
  (1+1)-dimensional schwarzschild spacetime.
\newblock {\em Phys. Rev. D}, 54:5123--5128, Oct 1996.

\bibitem{gvp4}
M.~Visser.
\newblock Gravitational vacuum polarization. {IV}. {E}nergy conditions in the
  unruh vacuum.
\newblock {\em Phys. Rev. D}, 56:936--952, Jul 1997.

\bibitem{acoustic}
M.~Visser.
\newblock Acoustic black holes: {H}orizons, ergospheres and {H}awking
  radiation.
\newblock {\em Classical and Quantum Gravity}, 15(6):1767, 1998.

\bibitem{Wahlquist}
H.~D. Wahlquist and F.~B. Estabrook.
\newblock Rigid motions in {E}instein spaces.
\newblock {\em Journal of Mathematical Physics}, 7(5):894--905, 1966.

\bibitem{conformaltheorem}
H.~D. Wahlquist and F.~B. Estabrook.
\newblock {Herglotz-Noether theorem in conformal space-time}.
\newblock {\em Journal of Mathematical Physics}, 8(4):919--919, 1967.

\bibitem{Wald}
R.~M. Wald.
\newblock {\em General Relativity}.
\newblock University of Chicago Press, 1984.

\bibitem{Wang:1998qx}
A.~Wang and Y.~Wu.
\newblock Letter: Generalized {V}aidya solutions.
\newblock {\em General Relativity and Gravitation}, 31(1):107--114, 1999.

\bibitem{Wheeler91}
J.~C. Wheeler.
\newblock Nonequivalence of the {Nernst-Simon} and unattainability statements
  of the third law of thermodynamics.
\newblock {\em Phys. Rev. A}, 43:5289--5295, 1991.

\bibitem{Wilks}
J.~Wilks.
\newblock {\em The Third Law of Thermodynamics}.
\newblock Oxford University Press, 1961.

\bibitem{Yano}
K.~Yano.
\newblock {\em {The theory of Lie derivatives and its applications}}.
\newblock North Holland -- Amsterdam, 1955.

\bibitem{Zemansky}
M.~W. Zemansky.
\newblock {\em Heat and Thermodynamics 5th ed.}
\newblock McGraw-Hill, New York, 1968.

\end{thebibliography}

\end{document}